\documentclass[10pt]{article}

\usepackage{a4wide}

\usepackage{pictex}
\usepackage{epsfig}

\usepackage{fancyheadings}
\usepackage{cite}
\usepackage{amsmath}
\usepackage{amssymb}

\begin{document}
%

%
%

\newcommand{\e}[1]{$\times10^{#1}$}

\def\check{{ \bf *{ \it check }* }}
\def\laplace{\nabla^2}
\def\grad{\nabla}

\def\figuremode{\small}
 
%
%

\title{\bf Relativistic Stars in Randall-Sundrum Gravity}

\author{{\bf Toby Wiseman}\thanks{e-mail: {\tt
      T.A.J.Wiseman@damtp.cam.ac.uk}} \\ \\
  Department of Applied Mathematics and Theoretical Physics,\\
  Center for Mathematical Sciences,\\
  Wilberforce Road,\\
  Cambridge CB3 0WA, UK}

\date{November 2001}

\maketitle

%
\begin{abstract}
%
  
  The non-linear behavior of Randall-Sundrum gravity with one brane
  is examined.  Due to the non-compact extra dimension, the
  perturbation spectrum has no mass gap, and the long wavelength
  effective theory is only understood perturbatively.  The full
  5-dimensional Einstein equations are solved numerically for static,
  spherically symmetric matter localized on the brane, yielding
  regular geometries in the bulk with axial symmetry.  An
  \emph{elliptic} relaxation method is used, allowing both the brane
  and asymptotic radiation boundary conditions to be simultaneously
  imposed. The same data that specifies stars in 4-dimensional
  gravity, \emph{uniquely} constructs a 5-dimensional solution.  The
  algorithm performs best for small stars (radius less than the AdS
  length) yielding highly non-linear solutions, core photons being
  redshifted by up to ${\cal Z} \simeq 12$. An upper mass limit is
  observed for these small stars, and the geometry shows no global
  pathologies.  The geometric perturbation is shown to remain
  localized near the brane at high densities, the confinement
  interestingly increasing for both small and large stars as the upper
  mass limit is approached. Furthermore, the static spatial sections
  are found to be approximately conformal to those of AdS. We show
  that the intrinsic geometry of large stars, with radius several
  times the AdS length, is described by 4-dimensional General
  Relativity far past the perturbative regime, the largest stars being
  tested up to a core redshift of ${\cal Z} \simeq 2.1$. This
  indicates that the non-linear long wavelength effective action
  remains local, even though the perturbation spectrum has no mass
  gap.  The implication is that Randall-Sundrum gravity, with
  localized brane matter, reproduces relativistic astrophysical
  solutions, such as neutron stars and massive black holes, consistent
  with observation.

\end{abstract}

\vspace{4.5cm}
\begin{flushright}
DAMTP-2001-99 \\
hep-th/0111057 
\end{flushright}

\newpage

\tableofcontents

\newpage

%
\section{Introduction}
%

Branes with matter confined upon them have now become an essential
component of string theory. Required by quantum theory on the
world-sheet, they have tremendous classical implications in the low
energy effective theory. New classes of compactifications are possible
where the matter is localized to the brane itself, unlike Kaluza-Klein
style compactifications where matter resides on the whole internal
space. The weakness of gravity then makes probing such dimensions
extremely difficult experimentally, leading to the idea that they may
be very large compared to Standard Model energy scales
\cite{Arkani-Hamed:1998rs,Antoniadis:1998ig}. A simple
`compactification' of this type is a model with just one non-compact
extra dimension, Randall-Sundrum gravity
\cite{Randall:1999vf,Lykken:1999nb}. With one asymptotically flat
brane, no moduli problem, and only a negative cosmological constant in
the bulk, it provides a very clean testing ground for gravitational
studies.  Intended as a toy model, work has shown it is possible to
embed this theory into higher dimensional super-gravities
\cite{Bergshoeff:2000zn,Duff:2000az}.

Linear and second order perturbation studies of one brane
Randall-Sundrum
\cite{Giddings:2000mu,Garriga:1999yh,Kudoh:2001wb,Giannakis:2000zx}
show that for localized objects much larger than the AdS length, a
brane observer views a local effective behavior which is simply
4-dimensional gravity. In Kaluza-Klein type compactifications, only
the homogeneous modes play a significant role for long wavelength
perturbations.  However, the non-linear behavior of models with
localized matter is a less tractable problem as the solutions can not
be homogeneous. Matter sources on the brane inevitably generate
inhomogeneity in the transverse space. Indeed, linear theory shows
that in one brane Randall-Sundrum, the inhomogeneous eigenmodes in the
transverse coordinate play a crucial role in ensuring a regular
horizon geometry.  With no mass gap in the perturbative spectrum,
non-linearity induced by brane matter is not yet understood
analytically. In particular, it is not clear how far into the
non-linear regime the effective 4-dimensional description, shown to
hold in the linear theory, remains valid.

The aim of this paper is to investigate the 5-dimensional non-linear
geometry of static stars in one brane Randall-Sundrum gravity. Stars
that are both small and large with respect to the AdS length are
studied. We study the geometry of small stars, where curvatures due to
the presence of matter are as large as the AdS curvature scale. We
expect that the qualitative behavior of these dense objects will be
similar in the one brane Randall-Sundrum model to other orbifold
models. For large stars, we examine whether a local effective
description remains in the non-linear regime. Our key results are;
\begin{itemize}
\item We find an \emph{elliptic} method to solve the full non-linear,
  5-dimensional, axi-symmetric Einstein equations. This elliptic
  approach enables us to simultaneously solve both the brane matching
  conditions and also the asymptotic AdS condition, ensuring regular,
  well defined horizon geometries. The same data for 4-dimensional
  star solutions \emph{uniquely} generates a regular bulk geometry.
\item The upper mass limit for stars of fixed radius is reproduced,
  both for small and large radii.  The brane is unable to stabilize
  ultra dense stars.
\item The effective description for long wavelength perturbations,
  corresponding to astrophysical objects, remains that of
  4-dimensional gravity far into the non-linear regime.
\item The perturbation of the geometry from AdS remains localized for
  non-linear stars of all radii, the decay of the perturbation
  steepening near the upper mass limit.
\item The spatial sections of both small and large stars are found to
  be approximately conformal to those of AdS.
\end{itemize}

This paper only studies the non-compact case of Randall-Sundrum
gravity. We later see that the ease of imposing boundary conditions in
one brane Randall-Sundrum makes this an attractive model to test. For
compact models the method could also be applied, with the boundary
conditions suitably altered, and is left for future investigation.

We organize the paper as follows. In the remainder of the first
section we briefly review the non-linear behavior of 4-dimensional
stars and then Kaluza-Klein and Randall-Sundrum gravities. We discuss why
non-linearity is not well understood in localized matter
compactifications. In the second section we highlight the main results
of the paper, in order to put the following methods and calculations
in context.

The third section discusses the method used to pose the bulk
5-dimensional Einstein equations in a framework suitable for elliptic
solution numerically. The regular metric is chosen to have a residual
conformal symmetry and results in elliptic second order derivatives of
the unknown metric functions in a sufficient subset of the Einstein
equations. Furthermore, it enables the brane to be fixed at constant
coordinate position whilst the asymptotic horizon metric takes a
simple form. This conformal symmetry allows the structure of the
constraint equations to be compatible with the elliptic relaxation,
implying that all the Einstein equations are satisfied even though
only a subset are relaxed.  The method is tested and found to work
well for configurations where the star radius is up to a few times the
AdS length. This is already a large enough scale to see the
4-dimensional effective behavior emerge.  As a consistency check, in
the fourth section, the 5-dimensional linear theory is also computed
numerically by an independent method and excellent agreement is found
in the low density regime.

The fifth section outlines the solutions obtained. The scheme is able
to generate fully non-linear regular solutions.  Strong evidence for
an upper mass limit is found by observing stars smaller than the AdS
length. For these the numerical method is most stable and allows
highly non-linear solutions to be calculated that trace the critical
behavior very close to this limit. Geodesics are constructed to probe
the global features of the static spatial geometry, which is found to
contain no pathologies, and interestingly is approximately conformal
to the spatial sections of AdS. For the larger stars, those that are
well described by 4-dimensional effective theory in the linear regime,
we again calculate non-linear configurations. We are able to calculate
solutions which are fully non-linear, where brane observables deviate
from the linear prediction by several times, and cannot therefore be
described in higher order perturbation theory. The confinement of the
geometric perturbation to the brane is found to \emph{increase} near
the upper mass limit for both small and large stars.  The large star
solutions are seen to very closely follow the 4-dimensional effective
gravity description for all levels of non-linearity tested, despite
this increase in confinement with stellar density.  This strongly
suggests that the correct description of long range behavior in
Randall-Sundrum gravity is simply 4-dimensional gravity, even in the
non-linear regime, provided the effective 4-dimensional curvature
scale remains below that set by the bulk cosmological constant, as for
neutron stars and large black holes away from the singularity. In
order to observe significant deviations from 4-dimensional physics one
would then need to observe short wavelength perturbations or possibly
dynamical effects over long time scales \cite{Sigurdsson:2001wz}.

%
\subsection{Stars in 4-dimensions}
\label{sec:stars_in_4d}
%

The behavior of static spherical matter in the non-linear regime of
general relativity is quite different from that in the Newtonian
limit. We briefly review the relevant features. Consider a star
composed of perfect fluid so that,
\begin{equation}
T_{\mu\nu} = \rho \, u_{\mu} u_{\nu} + P \left( g_{\mu\nu} + u_{\mu}
  u_{\nu} \right)
\label{eq:fluidbrane}
\end{equation}
with $u_{\mu} u^{\mu} = -1$. Assuming a static spherically symmetric
metric, one must specify the density profile, and together with metric
regularity and the requirement of asymptotically flat space, the
Einstein equations, $G_{\mu\nu} = 8 \pi G T_{\mu\nu}$, can be
integrated. An analytic solution exists for the simple top-hat density
profile,
\begin{equation}
\rho(r) = 
\begin{cases} 
\rho_0 & (r \leq R) \\ 
0 & (r > R) 
\end{cases}
\label{eq:tophat}
\end{equation}
where one finds a static
regular solution with ADM mass $M = m(\infty) = \frac{4}{3} \pi R^3
\rho_0$ provided the density is below a critical value. Specifically
the maximum attainable mass solution is
\begin{equation}
M_{\rm max} = \frac{4}{9} \, 8 \pi R 
\end{equation}
when units of $8 \pi G = 1$ are chosen, and therefore, for a given $R$
there is an upper bound to the mass of a static star, the core
pressure becoming infinite as $M \rightarrow M_{\rm max}$.  Note that
this is a purely non-linear effect. In the Newtonian theory there is
no upper mass limit for this equation of state, the core pressure
simply scaling quadratically with $\rho_0$.  The core pressure
characterizes this critical behavior, with $P_{\rm c} << \rho_0$ for
Newtonian stars, $P_{\rm c} \sim \rho_0$ being the regime where the
onset of non-linearity occurs and $P_{\rm c} \rightarrow \infty$ for
critical mass stars. Note that in these isotropic coordinates $R$ is
the proper angular radius of the star, giving the size of the $S^2$
associated with the spherical symmetry, at the edge of the star. When
we subsequently refer to a radius, we shall mean the proper angular
radius. It is interesting to note that for a general, physically
reasonable equation of state, $M_{\rm max} \leq \frac{4}{9} \, 8 \pi
R$ for fixed $R$ \cite{Wald}.  Thus uniform density,
\eqref{eq:tophat}, is the equation of state giving the largest
critical mass for a given radius.  It is for this reason that we use a
(slightly smoothed) top-hat as the density profile in the
5-dimensional solutions we construct later.

Observations indicate neutron stars are extremely relativistic objects
with $M \sim M_{\odot}$ and $R \sim 1 {\rm km}$ so that $GM/R \sim 1$
\cite{Wald}.  Taking a uniform density equation of state with $\rho_0$
to be a typical nuclear density and ignoring rotation, this provides
an upper mass limit of $\sim 5 M_{\odot}$ for neutron stars
\cite{Hartle:1978}.  If this result were considerably less then
stellar collapse would generically yield black holes and the
population of neutron stars of this radius would be zero.  Conversely
if the result were much larger then collapse to a black hole may be
difficult for the majority of stellar objects.

%
\subsection{Randall-Sundrum Gravity}
\label{sec:RS_gravity}
%

Now we review 5-dimensional Randall-Sundrum gravity with a single
$Z_2$ orbifold brane \cite{Randall:1999vf}. Whilst being a simple and
elegant theory that reproduces 4-dimensional linear gravity for long
wavelength perturbations, it has the crucial advantage that there are
no moduli fields to be stabilized as occurs generically for compact
extra dimensions. Unless otherwise stated, when referring to
Randall-Sundrum, it is implicit that we mean the one brane case, with
the brane being asymptotically flat.

The bulk matter is merely a negative cosmological constant and the
brane tension and matter localized on the brane are treated in a
distributional sense \cite{Israel:1966rt}. Varying the action,
\begin{equation}
S = \frac{1}{L^3} \int d^{5}x \sqrt{-g} \left( \frac{R}{2} - \Lambda
\right) + \frac{1}{L^3} \int_{\rm{brane}} d^{4}y
\sqrt{-h} \left( \sigma + {\cal L}_{\rm{brane}} \right)
\end{equation}
where the localized stress energy tensor $T_{\mu\nu} / L^3$ is derived as,
\begin{equation}
- \frac{1}{2} T_{\mu\nu} = \frac{\partial {\cal L}_{\rm{brane}}[h_{\mu\nu}]}{\partial h_{\mu\nu}} 
\end{equation}
giving the bulk Einstein equations,
\begin{equation}
G^{(5)}_{\mu\nu} = - \Lambda g_{\mu\nu} 
\end{equation}
and Israel thin shell matching conditions implementing the orbifold $Z_2$
symmetry,
\begin{equation}
2 \left[ K_{\mu\nu} - h_{\mu\nu} K \right]_{\rm{brane}} = \left( - \sigma h_{\mu\nu} + T_{\mu\nu} \right)
\label{eq:isreal}
\end{equation}
with $L$ the 5-dimensional Planck length, $\Lambda / L^3$ the bulk
cosmological constant, $K_{\mu\nu}$ the projection of the extrinsic
curvature of the brane hyper-surface with induced metric $h_{\mu\nu}$
and tension $\sigma / L^3$. Note the appropriate Gibbons-Hawking
boundary term is suppressed above.

The tension of the brane, $\sigma / L^3$, is chosen to admit static
solutions where the bulk metric is AdS and can be written as,
\begin{equation}
ds^2 = \frac{l^2}{z^2} \left( ds^2_{(\rm 4-Mink)} + dz^2 \right)
\label{eq:RSstaticmetric}
\end{equation}
where $ds^2_{(\rm 4-Mink)}$ is 4-dimensional Minkowski space, $l$ is
the AdS radius and is given by $\Lambda = - 6 / l^2$. More generally,
when the bulk is perturbed by matter, $l$ still gives the curvature
length associated with the cosmological constant, and is related to
the Ricci scalar as $l^2 = - 20 / {\cal R}$.  Localized matter on the
brane is taken to be of perfect fluid form, as in equation
\eqref{eq:fluidbrane}, so the physical density is $\rho / L^3$. The
hyper-surface $z=0$ is the conformal boundary of AdS, and $z
\rightarrow \infty$ the horizon.  A vacuum brane (vanishing
$T_{\mu\nu}$), intrinsically flat and with tension $\sigma / L^3$,
where, $\sigma = 6 / l$, can be located on the hyper-surface $z=z_0$.

Note that we can always choose $z = l$ by rescaling the coordinates as
$z \rightarrow \frac{l}{z_0} z$ together with $x_{\mu} \rightarrow
\frac{l}{z_0} x_{\mu}$ which is a natural scale invariance of this AdS
metric. Indeed we shall use a metric whose form is invariant under
conformal transformations in the $r, z$ plane, that allows the brane
to be placed at fixed $z$ independent of the matter on it, whilst
asymptotically towards the horizon the metric can still tend to the
simple form of \eqref{eq:RSstaticmetric}. Throughout the following
sections we choose units based on the AdS length, and therefore $l =
1$, implying that $- \Lambda = \sigma = 6$ and the brane location is
$z = 1$.

We may use a homogeneous ansatz, analogous to the Kaluza-Klein one,
namely,
\begin{equation}
ds^2 = \frac{l^2}{z^2} \left( g_{\mu\nu}(x^{\alpha}) dx^{\mu}
dx^{\nu} + dz^2 \right)
\label{eq:RSKKmetric}
\end{equation}
where the perturbation is homogeneous in $z$, only depending on
$x^{\mu}$, up to scaling by the warp factor. This does solve the bulk
Einstein equations provided $g_{\mu\nu}(x^{\alpha})$ is
4-dimensionally Ricci flat \cite{Brecher:1999xf}, as for example in
the black string solution
\cite{Chamblin:1999by,Gregory:1993vy,Gregory:2000gf}. Note that there
is no physical dilaton $\phi$ as now $z$ is not an angular coordinate
\cite{Charmousis:1999rg}.  Indeed, this ansatz is the non-linear
extension of the linearized zero mode as is the case for the
Kaluza-Klein ansatz \cite{Chamblin:1999cj}. However, the ansatz above
only applies to vacuum brane configurations.  Furthermore, consider
regularity for this geometry at large $z$. The 5-dimensional Weyl
components, $C_{\mu\nu}^{~~\alpha\beta}$, with indices as they would
appear in curvature invariants, are related to the 4-dimensional ones
of the metric $g_{\mu\nu}(x^{\alpha})$ by a factor of $( z / l )^2$
and thus diverge, along with curvature invariants, at large $z$.
Without a fundamental theory it remains unclear whether such
singularities are pathological \cite{Chamblin:1999by,Chamblin:1999cj}.

We will only be concerned with geometries that have a regular horizon
and thus are dynamically well defined.  Therefore, when matter sources
are present on the brane, boundary conditions must be imposed on the
geometry asymptotically far from the brane. More precisely we consider
the radiation boundary conditions of \cite{Giddings:2000mu}. Then for
static solutions, the scalar propagator in AdS decays towards the
horizon. The linear response of the brane geometry to matter localized
on it is found to be 4-dimensional standard General Relativity at long
distances compared to the AdS length $l$, \cite{Garriga:1999yh,
  Giddings:2000mu}. The relation between the 5-dimensional Planck
scale and the 4-dimensional one, $G = G_5/ l$ allows a range of $l$
from $\sim 1{\rm mm}$ (the limit of gravity measurement
\cite{Antoniadis:1998ig,Arkani-Hamed:1998nn,Hoyle:2000cv}),
corresponding to a 5-dimensional Planck length, $L \sim (10^8 {\rm
  GeV})^{-1}$, to the more conventional scheme where $l$ and $L$ are
both 4-dimensional Planck length valued. The scale $l$ allows us to
define the terms {\it large} and {\it small} for static objects in the
4-dimensional induced theory on the brane. Low density linear
astrophysical objects reside in the {\it large} regime, and are
indistinguishable in the 4-dimensional and 5-dimensional theory. {\it
  Small} objects are distinguished already at the linear level and may
have relevance at early times in the universe \cite{Argyres:1998qn}.

One way to characterize the induced geometry is to use the
Gauss-Codacci geometric decomposition
\cite{Shiromizu:1999wj,Sasaki:1999mi}, where the unknown bulk geometry
is parameterized in the projection of the bulk Weyl tensor onto the
brane. Some analytic progress has been made in the vacuum case. An
elegant solution for a black hole was obtained for a 3-dimensional
brane in a 4-dimensional bulk \cite{Emparan:1999wa}. No generalization
to 5-dimensions has so far been found. The solution is not of the
black string type \cite{Chamblin:1999by} having a regular horizon.
General restrictions on the horizon in 5-dimensions were considered in
\cite{Giannakis:2000ss}, and making some assumptions, the type of
asymptotic geometry was characterized using a `no hair' argument in
\cite{Shiromizu:2001jm}. Cosmological solutions
\cite{Kraus:1999it,Binetruy:1999hy,Binetruy:1999ut,Chamblin:1999ya}
show that a 4-dimensional effective description is recovered at late
times.  Integrable cosmological solutions \cite{Bowcock:2000cq} can be
analytically continued to give exact localized domain wall solutions
\cite{Gregory:2001xu,Gregory:2001dn}. These remarkably agree exactly
with a 4-dimensional effective gravity description. The AdS-CFT
conjecture allows one to write the form of the corrections to
effective 4-dimensional gravity as a contribution from a CFT on the
induced geometry with a mass cut-off
\cite{Giddings:2000mu,Verlinde:1999fy,Gubser:1999vj,deBoer:1999xf,Verlinde:1999xm,Shiromizu:2001ve,Shiromizu:2001jm}.

Two attempts have been made numerically to solve the bulk geometry
non-linearly. Time symmetric initial data was constructed for a black
hole spacetime indicating that the localization to the brane is
recovered non-linearly in this initial data \cite{Shiromizu:2000pg}.
Whether such localization survives dynamically remains an open and
difficult question. An attempt has also been made to solve the static
black hole geometry numerically by a hyperbolic evolution into the
bulk, making an ansatz for initial data on the brane
\cite{Chamblin:2000ra}. This again indicates localization non-linearly
with pancake like horizons, but generically the metric evolves into
singular configurations not far from the brane. This is a problem of a
hyperbolic approach, integrating the metric away from the brane, where
one must know exactly what initial data to take on the brane in order
to ensure that asymptotically the metric tends to AdS. A recent work
\cite{Germani:2001du} studied relativistic stars by making an ansatz
for the bulk Weyl tensor projection. Whilst some progress is possible,
allowing one to consider the intrinsic corrections to the effective
theory from terms quadratic in the stress tensor of the localized
matter, one has no reason to assume the form of the Weyl projection.
Furthermore, one has no control over the bulk and horizon geometry,
and the configurations studied are most likely pathological away from
the brane. As Randall-Sundrum gravity is only well defined when the
horizon boundary conditions are specified, we concentrate here on
solving for the full bulk geometry including both the asymptotic
properties and the brane boundary conditions. We only expect a well
defined, regular 5-dimensional geometry when the radiation boundary
conditions are imposed as in the linear theory.

\newpage

%
\section{Highlights of Results}
\label{sec:key_results}
%

We now highlight the main results obtained in the paper. The details
involved in the calculations can be found later, in particular the
numerical scheme that allows the full non-linear calculation of the
bulk geometry. However we wish to present the main results to give the
context in which the rather technical calculations were performed.

\begin{figure}
\centerline{\psfig{file=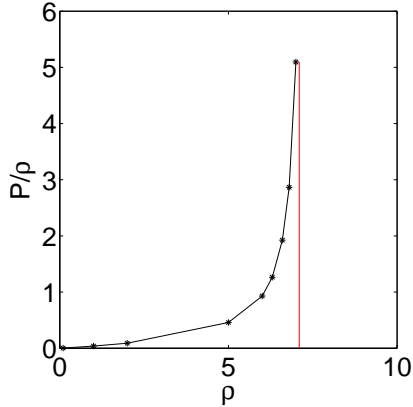,width=5.5cm}}
\caption[short]{ \figuremode  An illustration of core pressure, $P$, against core density,
  $\rho$, for configurations with $\xi = 0.3$ (see equation for
  density profile \eqref{eq:profile}). The proper angular radii vary
  monotonically from $R = 0.30 \, - \, 0.38$ from low to high density.
  The behavior strongly indicates a diverging core pressure for
  finite density, implying that for \emph{small} stars an upper mass
  limit for a given $R$ exists. The brane does not act to stabilize
  the large densities. The curve appears qualitatively similar to the
  usual 4-dimensional incompressible fluid star behavior. (all
  lattices: $dr = 0.02$, $r_{\rm max} = 2$, $dz = 0.005$, $z_{\rm max}
  = 4$.  systematic errors from comparison with linear theory in
  section \ref{sec:linear_check} are estimated at $\sim 2 \%$)
\label{fig:Adensitycurve} 
}
\end{figure}

\begin{figure}
\centerline{\psfig{file=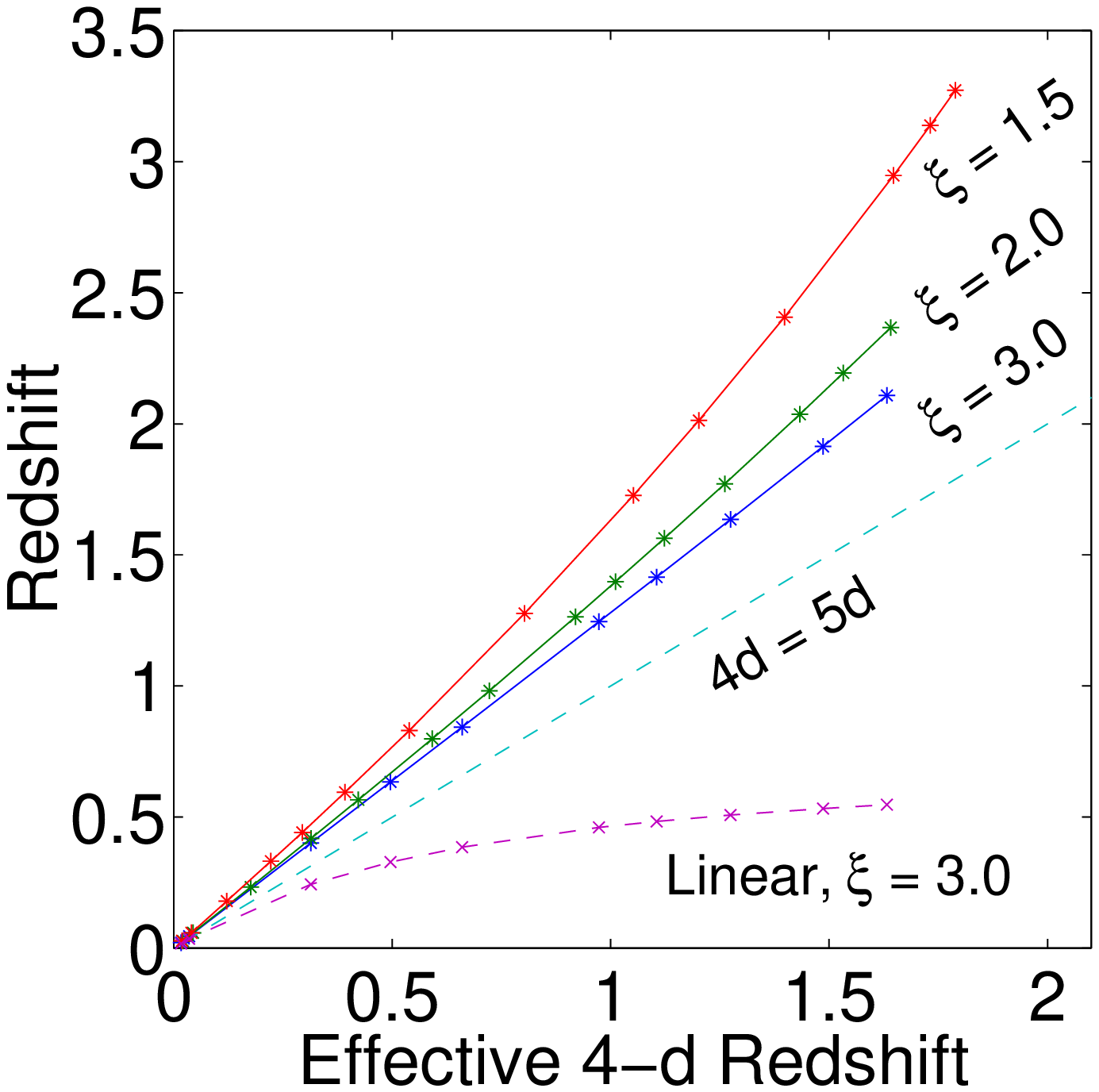,width=7.cm}
  \hspace{0.2cm} \psfig{file=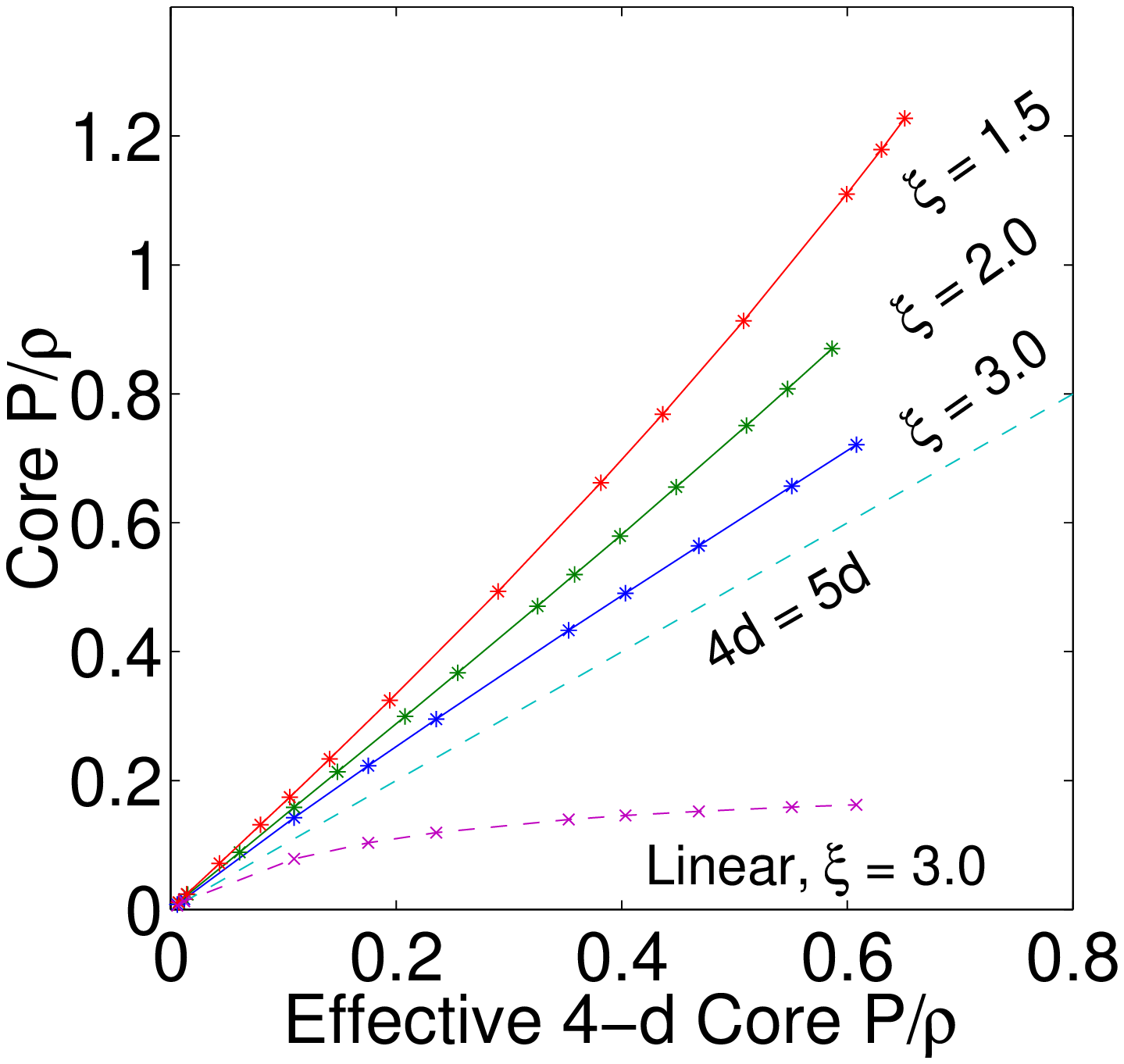,width=7.4cm} }
\caption[short]{ \figuremode  An illustration of 4-d against 5-d behavior; On the left the
  actual core redshift of a star, calculated from the 5-dimensional
  geometry, is plotted as a function of the same quantity calculated
  from the induced 4-dimensional effective theory. On the right, the
  core pressure divided by density is plotted in the same way.  Three
  values of, $\xi = 1.5, 2, 3$ are used to generate solutions for
  different core densities $\rho$. $\xi$ approximately corresponds to
  the proper radius, $R$, of the star.  One clearly sees that the
  larger the star, the closer the solutions lie to the `4d = 5d'
  $45^{o}$ line.  In the data presented moving vertically down towards
  this line the proper radius of the star increases.  Furthermore, for
  each $\xi$, the points fall approximately on straight lines. This
  indicates that the goodness of approximation of 5-dimensional theory
  by the effective 4-dimensional one is roughly independent of the
  core density, and hence non-linearity, over the range tested. The
  degree of approximation depends only on the star size. The most
  non-linear $\xi=3$ stars are at $\sim 75\%$ of their upper mass
  limit in the 4-dimensional effective theory. The last line plotted
  is the 4-dimensional linear theory prediction for the $\xi=3$ stars,
  again against the 4-dimensional non-linear theory. We see that the
  linear theory deviates strongly from this, showing that the
  solutions probed are fully non-linear, and beyond the reach of
  higher order perturbation theory. These graphs are strong evidence
  that the effective 4-dimensional description applies far into the
  non-linear regime, and probably right up to the upper mass limit.
  (lattices: $\xi = 1.5$: $dr = 0.10$, $r_{\rm max} = 10$, $dz = 0.02$
  and $0.04$, $z_{\rm max} = 21$, $\xi = 2.0$: $dr = 0.15$, $r_{\rm
    max} = 15$, $dz = 0.02$ and $0.04$, $z_{\rm max} = 31$, $\xi =
  3.0$: $dr = 0.20$, $r_{\rm max} = 20$, $dz = 0.03$ and $0.05$,
  $z_{\rm max} = 46$. two lattice $dz$ resolutions are used to
  extrapolate to $dz = 0$.  systematic errors from comparison with
  linear theory in section \ref{sec:linear_check} are estimated to be
  maximum for $\xi = 3.0$ at $\sim 10 \%$.)
\label{fig:compare_4d_5d} 
}
\end{figure}

\begin{itemize}
\item {\bf Upper Mass Limit for Small Stars} - Figure
  \ref{fig:Adensitycurve} \newline The numerical method we outline
  performs most stably for \emph{small} stars, with radius less than
  the AdS length. Extremely non-linear solutions can be found in this
  case. Figure \ref{fig:Adensitycurve} shows the ratio of core
  pressure to density for a (smoothed) top-hat density profile with
  fixed coordinate radius $\xi = 0.3$ (see equation
  \eqref{eq:profile}) yielding solutions with proper radius $R \sim
  0.3$. Newtonian theory predicts a linear dependence of $P/\rho$ on
  $\rho$. We clearly see a departure from this behavior and strong
  evidence that the core pressure diverges for finite core density,
  $\rho \simeq 7$. The numerical method does not give a convergent
  solution if a larger core density is used.  Note that for
  \emph{small} stars the behavior is not that of 4-dimensional GR.
  However the qualitative nature of the upper mass limit for fixed
  radius appears to persist to \emph{small} stars. For \emph{large}
  stars we cannot approach this limit so closely, but the indications
  are that again an upper mass limit would be found as the behavior
  follows 4-dimensional GR so closely (see below). A detailed
  description of this result is found in section
  \ref{sec:small_stars}.
\item {\bf Non-Linear Long Range Effective Theory} - Figure
  \ref{fig:compare_4d_5d} \newline In order to calculate the
  5-dimensional geometries for stars, we input a density profile and
  require isotropy on the brane. The matching conditions impose these
  requirements as boundary conditions. Once the solution is found we
  may read off the corresponding pressure profile on the brane. The
  redshift of photons propagating in the brane from the core of the
  star to some radius can be computed. The core pressure and the core
  redshift of a photon emitted to infinite distance on the brane are
  both coordinate scalar quantities for this static spherical
  symmetry. Given the density profile against proper distance for the
  5-dimensional solution one can compute exactly the same quantities
  in standard 4-dimensional gravity. A comparison then allows one to
  assess how good an effective description the 4-dimensional theory
  is.  Figure \ref{fig:compare_4d_5d} shows both the core redshift and
  pressure for Randall-Sundrum stars with various radii. The
  5-dimensional value is plotted against the same quantity calculated
  in the 4-dimensional theory for the same density profile. We see
  that for increasing $\xi$, which is approximately equal to the
  proper radius of the star, the difference between the 4-dimensional
  and 5-dimensional values decrease. Already for $\xi = 3$, only three
  times the AdS length, the predictions differ by only $\sim 20\%$.
  Note that a curve is also drawn to show how the linear approximation
  compares to full non-linear 4-dimensional GR for the $\xi = 3$ case.
  This shows that the solutions found clearly probe the fully
  non-linear regime, where one cannot meaningfully apply higher order
  perturbation theory.  The level of agreement depends on the proper
  size of the object, $R \simeq \xi$, as expected. The crucial result
  is that it does \emph{not} appear to depend on the core density.
  Perturbation theory predicts agreement for small density, but we see
  full non-linear agreement.  The implication in then that the full
  non-linear 4-dimensional effective theory is standard GR.
  Furthermore, by observing neutron star physics or massive black hole
  horizon geometries accessible through astrophysical measurements, we
  will be unable to differentiate between Randall-Sundrum and
  4-dimensional gravity. A detailed description of this result is
  found in section \ref{sec:non-linear}.
\end{itemize}

\newpage

%
\section{Solving By Elliptic Relaxation}
%

Our task is to construct solutions to Randall-Sundrum gravity sourced
by static spherically symmetric matter distributions on the
4-dimensional brane, such as those corresponding to stars. In order to
do this one must solve the full non-linear Einstein equations with
boundary conditions given by the matter localized on the brane and
that the asymptotic geometry is that of AdS as in the linear theory
with radiation boundary conditions.

A static spherical star in the induced brane geometry requires that
the metric in the bulk has an axial symmetry. This now becomes a
problem in two variables, with a radial, $r$, and an axial, $z$,
coordinate. In the linear theory \cite{Garriga:1999yh,
  Giddings:2000mu} one can choose a synchronous gauge with respect to
the background $z$ coordinate in \eqref{eq:RSstaticmetric} allowing
the metric perturbation components to decouple. There is no such
decoupling in the non-linear theory, and thus one expects to have to
solve a system of coupled non-linear partial differential equations.

\begin{figure}[htb]
\centerline{\psfig{file=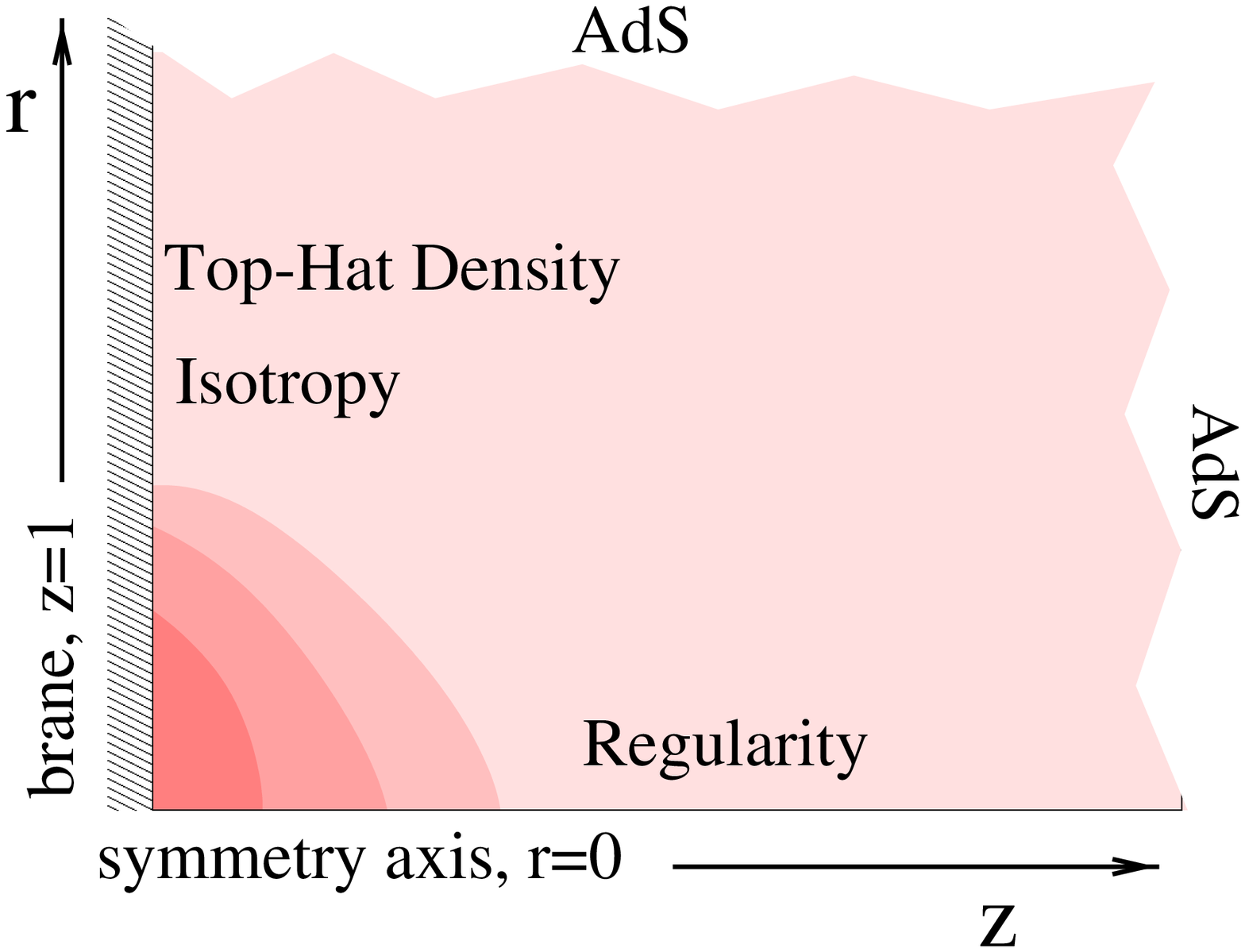,width=9cm}}
\caption[short]{ \figuremode  An illustration of the asymptotic and brane boundary data.
\label{fig:boundarydata} 
}
\end{figure}

Figure \ref{fig:boundarydata} schematically illustrates the boundary
data for the problem. The brane matching conditions are non-linear
equations relating normal derivatives of metric functions to the
functions themselves. Asymptotically we wish to recover AdS, again
placing constraints on the metric functions. We return to the nature
of these boundary conditions later but it is sufficient now to note
that conditions are specified on the brane and asymptotic boundaries
of the $r, z$ space. This is not data defined on a Cauchy surface as
in an ADM evolution, but rather is elliptic data.

One can always perform an ADM decomposition in the $z$ direction and
solve the bulk metric from initial data given on the brane, as in
\cite{Chamblin:2000ra}.  However, we see from figure
\ref{fig:boundarydata} that one has to supply initial data on the
brane that ensures asymptotically AdS behavior far from the brane.  In
the linear theory one can construct the Greens function only from
modes obeying this condition. In the full theory one has no such
luxury, and indeed it was found in \cite{Chamblin:2000ra} that data on
the brane generically evolves to give pathologies far from the brane.
If one then wishes to solve this 2 variable problem using a
`hyperbolic' evolution from one boundary, such as the brane, or from
large $z$ inwards, then one is confronted with a shooting problem as
the data is naturally defined on all the boundaries.  Furthermore this
is a shooting problem in 2 variables and therefore one is shooting
with functions rather than constants as in familiar 1 variable
shooting problems. Therefore the framework we will use to solve the
bulk equations is not that of a hyperbolic evolution, but rather by an
elliptic relaxation. A simple example where gravity can be solved
elliptically is in static 4-dimensional linear theory, which in a
Newtonian gauge results in a Poisson equation for the potential. One
more generally expects static GR to result in an elliptic problem. In
4-dimensions the vacuum axi-symmetric problem reduces to solving a
Laplace equation \cite{Weyl:1917}, although this does not generalize
to higher dimension \cite{Myers:1987rx}. In this paper we show that in
5-dimensions, the system can indeed be solved elliptically for this
symmetry.

%
\subsection{Relaxation for a brane star in Randall-Sundrum gravity}
%

The problem now is to identify a coordinate system where the equations
might admit a solution via a relaxation method with elliptic boundary
conditions.  We know of no general method to find such a choice for
static solutions in gravity. Instead we sketch a brief argument for
why the gauge we have chosen is suitable.

We wish to describe regular static geometries with axial symmetry
(induced spherical symmetry on the brane). We therefore consider a
manifold, topologically equivalent to that of the static vacuum Randall-Sundrum
solution and take the metric,
\begin{equation}
ds^2_5 = \frac{1}{z^2} \left[ - e^{2 T(r,z)} dt^2 + e^{2 R(r,z)}
  dr^2 + e^{2 S(r,z)} r^2 d\Omega_2^2 + e^{2 Z(r,z)} dz^2 + 2 e^{2
  v(r,z)} dr dz \right]
\label{eq:metric_general}
\end{equation}
which is the most general parameterization of such a geometry. We
still have 2 degrees of coordinate freedom in $r$ and $z$ which we
shall use to eliminate 2 metric functions and obtain a gauge suitable
for relaxation.

%

We wish to obtain bulk Einstein equations for the 3 remaining metric
functions that have elliptic differential operators in the second
derivative terms for $r, z$. Firstly we note that the off diagonal
term generically gives rise to hyperbolic second derivative terms in
Einstein equations of the form $v_{,rz}$ and therefore we eliminate
this with the residual coordinate freedom. For this diagonal metric,
$R$ is a lapse function with respect to the $r$ direction and
therefore no equation will contain $R_{,rr}$ derivatives.  Similarly,
$Z$ is a lapse function with respect to $z$ and there are no $Z_{,zz}$
derivatives entering the equations.  This can be seen if one
substitutes the diagonal metric into the Einstein bulk action, and
linearizes, giving, to lowest order in terms that give rise to second
derivatives in the Einstein equations,
\begin{align}
{\cal S}^{\rm{2^{{\rm nd}} \, Deriv}} \simeq & \int dr dz \frac{r^2}{z^3} \left( 
2\,\partial_{z} R\,\partial_{z} S + {({\partial_{z} S})^2} + \partial_{z} R\,\partial_{z} T + 
  2\,\partial_{z} S\,\partial_{z} T \right.
\notag \\ &
\qquad \qquad \left. + {({\partial_{r} S})^2} + 2\,\partial_{r} S\,\partial_{r} T + 2\,\partial_{r} S\,\partial_{r} Z + 
  \partial_{r} T\,\partial_{r} Z \right)
\end{align}
The full non-linear action can be varied and reproduces 4 of the 5
Einstein field equations, the $G^{r}_{~z}$ equation being missing, but
implied from the others by the Bianchi identities. The $R$ and $Z$
equations will, as mentioned, not contain second order elliptic
operators due to their lapse nature.

If the gauge choice $R = Z$ is also imposed using the remaining
coordinate degree of freedom then this linearized Lagrangian becomes
symmetric in these second derivative terms,
\begin{align}
{\cal S}^{\rm{2^{{\rm nd}} \, Deriv}}_{R=Z} \simeq & \int dr dz
\frac{r^2}{z^3} \left( 2\,\partial_{z} R\,\partial_{z} S + {({\partial_{z} S})^2} + \partial_{z} R\,\partial_{z} T + 
  2\,\partial_{z} S\,\partial_{z} T \right.
\notag \\ &
\qquad \qquad \left. + 2\,\partial_{r} R\,\partial_{r} S + {({\partial_{r} S})^2} + \partial_{r} R\,\partial_{r} T + 
  2\,\partial_{r} S\,\partial_{r} T \right)
\notag \\ 
\notag \\ 
= &
\int dr dz
\frac{r^2}{z^3} \left(  M_{i j} (\partial_{r} V^{(i)}) (\partial_{r}
  V^{(j)}) + \left[ \partial_{r} \leftrightarrow \partial_{z} \right] \right)
\end{align}
to lowest order, were the vector $V = \left\{ T, R, S \right\}$. Now
the $T, R, S$ equations contain second derivative terms which are
symmetric in $r$ and $z$ and thus are individually elliptic Laplace
operators.  The matrix $M_{i j}$ has one positive and two negative
eigenvalues indicating that the system is not positive definite.
Therefore we cannot guarantee $T, R, S$ will relax simultaneously, but
individually (ignoring singularities from $r, z = 0$ terms) the
equations for $T, R, S$ are elliptic to linear order and hence we can
attempt to relax them together.  Note that we are only considering a
subset of the Einstein equations as varying the action yields the
Einstein equations associated with the Einstein tensor components
$G^{t}_{~t}, G^{\theta}_{~\theta}$ and $G^{r}_{~r}+G^{z}_{~z}$, which
we term the `elliptic' Einstein equation components. We later consider
how the remaining equations, $G^{r}_{~z}$ and $G^{r}_{~r}-G^{z}_{~z}$,
are consistent with elliptic relaxation. We term these equations
`constraints', as they are implied from the other Einstein equation
components by the two non-trivial Bianchi identity components. Note
that they do contain second order derivatives, although they are not
elliptic. We use the term constraint as, in section
\ref{sec:brane_data}, we see the Bianchi identities imply that they
must be satisfied in the interior of the problem if they are satisfied
on the boundaries and the elliptic equations $G^{t}_{~t} = 6,
G^{\theta}_{~\theta} = 6$ and $G^{r}_{~r}+G^{z}_{~z} = 12$ are
satisfied in the interior. This is therefore analogous to the case of
hyperbolic evolution, where provided the constraints are satisfied on
a Cauchy surface, they will remain satisfied upon integration of the
evolution equations.

For future convenience we set,
\begin{align}
R & = A + B \notag \\
S & = A - B 
\label{eq:AB_defn}
\end{align}
and the metric is now,
\begin{equation}
ds^2_5 = \frac{1}{z^2} \left[ - e^{2 T(r,z)} dt^2 + e^{2 \left(A(r,z)+B(r,z)\right)} \left(
  dr^2 + dz^2 \right) + e^{2 \left(A(r,z)-B(r,z)\right)} r^2 d\Omega_2^2 \right]
\label{eq:metric_conf}
\end{equation}
and yields the full non-linear equations,
\vspace{5pt}
\begin{align}
\laplace T = & \left[
 - 2\,(\partial_{r} A)\,(\partial_{r} T) + 2\,(\partial_{r} B)\,(\partial_{r}
 T)  - {({\partial_{r} T})^2} \right] 
+ \left[ \, \partial_r \leftrightarrow \partial_z \, \right]
\notag \\ 
&  - \frac{2}{r} \, {\partial_{r} T} + \frac{1}{z} \left(  
  {2\,\partial_{z} A} - {2\,\partial_{z} B} +
  {4\,\partial_{z} T} \right) + 
  {\frac{4}{{z^2}}} \, \left( e^{2 \, (A + \,B)} - 1 \right)
\notag \\ 
\notag \\ 
\laplace A = &
\left[ - \frac{1}{2} (\partial_{r} A)^2  + 
  (\partial_{r} A)\,(\partial_{r} B) - \frac{1}{2} (\partial_{r} B)^2     + \frac{1}{2} (\partial_{r} A)\,(\partial_{r}
      T) - 
  \frac{1}{2} (\partial_{r} B)\,(\partial_{r} T) \right] + \left[ \, \partial_r \leftrightarrow \partial_z \, \right]
\notag \\ &  
+ \frac{1}{r} \left( - \partial_{r} A +
 \partial_{r} B + \frac{1}{2} \partial_{r} T \right) 
+ \frac{1}{z} \left( \frac{1}{2} \partial_{z} A - \frac{1}{2} \partial_{z} B - 
  \frac{1}{2} \partial_{z} T \right) 
+ \frac{1}{z^2} \, \left( e^{2 \, (A +
 \,B)} - 1 \right)
\notag
\end{align}
\begin{align}
\laplace B = &
\left[ \frac{3}{2} \, (\partial_{r} A)^2 - 3\,(\partial_{r} A)\,(\partial_{r} B)
 + \frac{3}{2} \, (\partial_{r} B)^2 + \frac{3}{2} \,(\partial_{r}
 A)\,(\partial_{r} T) - \frac{3}{2} \,(\partial_{r} B)\,(\partial_{r} T)
 \right] + \left[ \, \partial_r \leftrightarrow \partial_z \, \right]
\notag \\ &  
+ \frac{1}{r} \left( 3 \,\partial_{r} A  - 3\,\partial_{r} B   + 
  \frac{3}{2} \,\partial_{r} T \right) + \frac{1}{z} \left( -
 \frac{9}{2} \,\partial_{z} A + \frac{9}{2} \,\partial_{z} B - 
  \frac{3}{2} \,\partial_{z} T
 \right)
\notag \\ &  
 - \frac{1}{r^2} \left( e^{4\,B} -
 1 \right) -
 \frac{3}{z^2} \left(
   e^{ 2 \,\left( A + B \right) } - 1 \right)
\label{eq:TAB_poisson} 
\end{align}
with $\laplace = \partial_r^2 + \partial_z^2$. These equations for $T,
A, B$ are linear combinations of the elliptic Einstein equations,
denoted $\{tt\}, \{rr+zz\}, \{\theta\theta\}$ after removing a
homogeneous $z^2$ blue-shifting factor for convenience, as described
in appendix \ref{app:einstein_eqns}. These must be supplemented with
the remaining Einstein equations, $G^{r}_{~z} = 0$ and
$G^{r}_{~r}-G^{z}_{~z} = 0$, the constraints, similarly rescaled and
denoted $\{rz\}$ and $\{rr-zz\}$,
\begin{align}
\{rz\} = \partial_{r} \partial_{z} & \left( 2 B - 2 A - T \right) 
+ \frac{4}{r} \,\partial_{z} B
+ \frac{1}{z} \left( - 3\,\partial_{r} A - 3\,\partial_{r} B \right)
\notag \\ 
& + 2\,(\partial_{r} A)\,(\partial_{z} A) + 
  2\,(\partial_{z} A)\,(\partial_{r} B) + 
  2\,(\partial_{r} A)\,(\partial_{z} B) - 
  6\,(\partial_{r} B)\,(\partial_{z} B) 
\notag \\ 
& + 
  (\partial_{z} A)\,(\partial_{r} T) + 
  (\partial_{z} B)\,(\partial_{r} T) + 
  (\partial_{r} A)\,(\partial_{z} T) + 
  (\partial_{r} B)\,(\partial_{z} T) - 
  (\partial_{r} T)\,(\partial_{z} T)  = 0
\label{eq:constrz} 
\end{align}
and $\{rr-zz\}$,
\begin{align}
  \{rr-zz\} = & \left[ - 2 \partial^2_r A + 2 \partial^2_r A - 2
    \partial^2_r T \right] - \left[ \partial_r \leftrightarrow
    \partial_z \right]
  \notag \\
  & + \frac{8}{r} \partial_r B + 2 ( \partial_r A )^2 + 4 ( \partial_r A
  ) ( \partial_r B ) - 6 ( \partial_r B )^2 + 2 ( \partial_r A ) (
  \partial_r T ) + 2 ( \partial_r B ) ( \partial_r T ) - ( \partial_r
  T )^2
  \notag \\
  & + \frac{6}{z} \left( \partial_z A + \partial_z B \right) - 2 (
  \partial_z A )^2 - 4 ( \partial_z A ) ( \partial_z B ) + 6 (
  \partial_z B )^2 
- 2 ( \partial_z A ) ( \partial_z T ) 
\notag \\
& - 2 (
  \partial_z B ) ( \partial_z T ) + ( \partial_zT )^2 
= 0
\label{eq:constrr-zz}
\end{align}
The metric functions $T, A, B$ enter these two constraint equations
$\{rz\}$ and $\{rr-zz\}$ with hyperbolic second derivatives
$\partial_r \partial_z$ and $\partial_r^2 - \partial_z^2$
respectively.  For reference the Einstein tensor components are given
in appendix \ref{app:einstein_eqns}.

It is important at this point to raise the issue that one might be
able to solve for remaining metric functions algebraically or by
integration of a constraint, and therefore have to relax fewer metric
functions in the $r, z$ plane. We have no reason to suggest that such
a scheme could not be used but were unable to find such a scheme that
used the constraints directly.  Integrating a function over the
lattice using the hyperbolic nature of the constraints is extremely
non-local compared to one iteration of a local Poisson equation solver
such as Gauss-Seidel.  This non-locality was generically found not to
yield convergent schemes.  In fact, $R = A + B$ can be thought of as a
lapse function, and can actually be algebraically determined directly
from the $\{rr+zz\}$ Einstein equation, the corresponding Hamiltonian
constraint.  This could be used directly to eliminate this metric
function, but again the remaining variables could not be relaxed. The
only scheme we found to work was one where the 3 metric functions were
all elliptically relaxed together.

Each bulk equation, \eqref{eq:TAB_poisson}, appears to contain second
order elliptic operators, but there are singular terms as $r
\rightarrow 0$.  It is certainly true is that away from $r = 0$ the
second order operators are non-singular and therefore can individually
be solved by elliptic relaxation. However, whilst each equation
individually appears elliptic, when avoiding singular points, the
three taken together are not necessarily so.  Experimentally we do fortunately
find that for a straightforward numerical scheme the three can indeed
be consistently relaxed together. The scheme to deal with the singular
terms is discussed in detail in the later section \ref{sec:origin}.

The following sections consider the boundary data for the relaxation.
We examine,
\begin{itemize} 
\item the boundary data that must be specified on the brane and
  asymptotically.
\item how the constraint equations are satisfied through the boundary
  data when only the elliptic bulk equations are relaxed.
\item how to specify data at the origin where singular terms are
  present.
\end{itemize}

%
\subsection{Local Conformal Symmetry and Brane Coordinate Position}
\label{sec:brane_pos}
%

With the gauge choice discussed in the previous section, the metric
\eqref{eq:metric_conf} still has residual coordinate freedom, namely
2-dimensional conformal transformations, $\bar{r} = f(r,z), \bar{z} =
g(r,z)$,
\begin{equation}
d\bar{r}^2 + d\bar{z}^2 = \Omega(r,z) \left(
  dr^2 + dz^2 \right)
\label{eq:static_rs}
\end{equation}
so that $f, g$ satisfy the usual Cauchy-Riemann relations. The data
for such a transformation can be taken as specifying $g$ on all
boundaries and $f$ at one point on any boundary. This always allows
the brane to be moved by such a transformation from $\bar{z}_{\rm
  brane} = 1$ to $z_{\rm brane} = 1 + h(r)$ where $h$ is an arbitrary
function. For example, we could take $g(r, 1) = 1 + h(r)$, $g(r =
\infty, z) = z + h(r = \infty)$, and $g(r, z \rightarrow \infty) = z +
h(r = \infty)$, and in addition, $\partial_r g = 0$ at $r = 0$
implying $f(r = 0, z) = 0$, taking $f(r = 0, z = \infty) = 0$. 

Note that a crucial feature of this transformation is that provided
$h(r)$ decays to zero as $r$ increases, the metric is asymptotically
unaffected by such a transformation. The solution to the Laplace
equation for $g$ with the data above is then $g = z$ plus a
perturbation from $h(r)$ that dies away as $\frac{1}{z}$ far away from
the brane, and so $g \rightarrow z, f \rightarrow r$ as $z \rightarrow
\infty$. For example, consider the static vacuum Randall-Sundrum
metric in variables $\bar{r}, \bar{z}$ after such a conformal
transformation to $r, z$, which yields,
\begin{equation}
ds^2_5 = \frac{1}{z^2} \left[ \left(\frac{z}{g}\right)^2 \left( - dt^2 + \left( g_{,r}^2 + g_{,z}^2 \right) \left(
  dr^2 + dz^2 \right) + \left(\frac{f}{r}\right)^2 r^2 d\Omega_2^2
  \right) \right] 
\label{eq:transformed_conf}
\end{equation}
As $f \rightarrow r, g \rightarrow z$ and the derivatives $g_r
\rightarrow 0$, $g_z \rightarrow 1$ for large $r$ or large $z$,
asymptotically the metric will tend to the static Randall-Sundrum
solution, \eqref{eq:RSstaticmetric}. Thus, \emph{local}
transformations of the coordinate position of the brane only effect
the metric \emph{locally}. We are able to consistently position the
brane at $z = 1$ and again the asymptotic behavior is that $T, A, B
\rightarrow 0$ as $r$ or $z \rightarrow \infty$. This is shown
explicitly in appendix \ref{app:asym_lin} for the linear theory.

The significance of this is considerable. Contrast this for instance
with the synchronous gauge used in the linearized analyses
\cite{Giddings:2000mu,Garriga:1999yh} where the residual gauge
transformations $\xi_5 = \xi_5(x)$, and $\xi_{\mu} = \xi_{\mu}(x)$,
can be used to move the brane coordinate position.  A relaxation
scheme in this gauge would either have to remain in the
`Randall-Sundrum' transverse gauge, where the horizon metric remains
simple, and include a new degree of freedom in the relaxation which
would represent the position of the brane, or alternatively place the
brane at a fixed coordinate location, using a Gaussian normal gauge,
which perturbs the metric asymptotically and therefore requires a
complicated and non-local boundary condition asymptotically that would
encode this degree of freedom. Then the metric would no longer decay
to the simple form of \eqref{eq:RSstaticmetric}. Either case is
complicated and with no guarantee of convergent relaxation, may be
unlikely to work.

In summary, the conformal gauge was chosen to yield equations
\eqref{eq:TAB_poisson} that have elliptic second order operators
allowing relaxation methods to be applied. Conveniently we see it also
allows the brane to be consistently placed at fixed coordinate
location, say $z = 1$, and the asymptotic behavior is simply $T, A, B
\rightarrow 0$ as $r$ or $z \rightarrow \infty$ for radiation boundary
conditions.

%
\subsection{Brane Boundary Data}
\label{sec:brane_data}
%

In order to solve the system we must specify the matter on the brane
by satisfying the brane matching conditions \eqref{eq:matching_eqns}
of appendix \ref{app:einstein_eqns}. In addition, we will also show
that only one of the two constraint equations must be enforced on the
brane itself. It will be shown in the subsequent section
\ref{sec:asym_data} that the condition $T,A,B \rightarrow 0$ as $r, z
\rightarrow \infty$ is sufficient to ensure the constraints are then
satisfied everywhere.

In 4-dimensional gravity, static spherical symmetry requires two
metric functions to parameterize the geometry. Two conditions are
required to fix these degrees of freedom in the solution. One can take
these to be specifying a density profile, and requiring
isotropy, so that $G^{r}_{~r} = G^{\theta}_{~\theta} = P(r)$ thus
fixing the radial and angular pressure component to be equal.
Together with asymptotic boundary conditions in $r$, the metric can be
solved for.

On the brane in the 5-dimensional case we have the same two
conditions, a density profile and isotropy. However we now have 3
metric functions $T, A, B$. Using the matching conditions
\eqref{eq:matching_eqns} these become
\begin{equation}
\rho = -\sigma  - 2\,\left( -3 + 3\,z \,\partial_{z} A - 
     z \, \partial_{z} B \right) \, e^{- ( A + B )}
\label{eq:density}
\end{equation}
and for isotropy the simpler linear condition,
\begin{equation}
B_{,z} = 0
\label{eq:isotropy}
\end{equation}
This fixes two metric components, say $A$ and $B$, leaving the
remaining component $T$. There are also additional constraint
equations. Since all the matter dependent data will be specified on
the brane, the asymptotic boundary data being simply that the metric
tend to AdS at the horizon, these constraints must fix $T$ in order to
have agreement between the physical data of 4-dimensional and
Randall-Sundrum gravity.

Calculating the non-trivial components of the Bianchi identities
$\grad_{\mu} G^{\mu}_{~\nu} = 0$ for the metric
\eqref{eq:metric_conf}, and assuming that the elliptic Einstein
equations, $\{tt\}, \{rr+zz\}, \{\theta\theta\}$ are satisfied,
yields,
\begin{align}
\frac{1}{g} \, \left[ \, \partial_r \left( g \, G^{r}_{~z} \right) +
    \partial_z \left( \frac{g}{2} \, \left(G^{r}_{~r} -
    G^{z}_{~z}\right) \right) \, \right] & = 0 \notag \\
\frac{1}{g} \, \left[ \, \partial_z \left( g \, G^{r}_{~z} \right) -
    \partial_r \left( \frac{g}{2} \, \left(G^{r}_{~r} -
    G^{z}_{~z}\right) \right) \, \right] & = 0
\label{eq:constpois}
\end{align}
where $g = \sqrt{-\det{g_{\mu\nu}}}$. Thus the quantities $g \, G^{r}_{~z}$
and $g \left(G^{r}_{~r} - G^{z}_{~z}\right)$ satisfy Cauchy-Riemann
relations and therefore separately satisfy Laplace equations. An
example of data for the system is to specify $g \, G^{r}_{~z}$ on all
boundaries, and $g \left(G^{r}_{~r} - G^{z}_{~z}\right)$ at only one
point. 

Thus if $G^{r}_{~z} = 0$ is used to determine $T$ on the brane, and $g
\, G^{r}_{~z}$ is also zero asymptotically away from the brane then
provided $g \left(G^{r}_{~r} - G^{z}_{~z}\right)$ vanishes at one
point, say asymptotically, the pair of constraints will also be solved
everywhere, provided the elliptic Einstein equations are satisfied in
the bulk. The vanishing of these quantities asymptotically is
discussed in the later section \ref{sec:asym_data}. Note also that as
$g \, G^{r}_{~z}$ satisfies a Laplace equation, this zero data on all
boundaries has the \emph{unique} solution that $\{rz\}$ and
$\{rr-zz\}$ are true everywhere within.

On the brane we choose to use $\{rz\}$ to determine $T$. One can see
from equation \eqref{eq:constrz} that this constraint has a linear
second order differential operator acting on $T$ which is hyperbolic
with characteristics in the $r$ and $z$ directions, so that it can be
integrated in from $r=\infty$ to $r=0$ along the brane.  Now all 3 of
the metric functions are determined on the brane by the constraints,
density and isotropy conditions and the 5-dimensional brane data is
consistent with that of the 4-dimensional system. Thus we expect, and
indeed find that the same stellar data as for standard 4-dimensional
GR \emph{uniquely} specifies the 5-dimensional bulk geometry.

%
\subsection{Linearized Equations and Their Solution Numerically}
\label{sec:linear}
%

We now construct the solution to the linear theory in the conformal
gauge described above. As there is no matter in the bulk except for
the cosmological constant, using the synchronous transverse traceless
gauge for the linear perturbations one finds that the perturbing
metric components decouple and can be solved using a Greens function.
Such solutions are given in \cite{Giddings:2000mu,Garriga:1999yh}.  In
this section we explicitly coordinate transform back to a metric of
the form \eqref{eq:metric_conf} for a spherical static brane source.
One must now only solve decoupled equations with simple boundary
conditions, and this is used to provide an independent check of the
full non-linear method in section \ref{sec:linear_check}.

Firstly perturb the static Randall-Sundrum metric,
\eqref{eq:RSstaticmetric} as follows,
\begin{align}
  ds^2 & = \frac{1}{\bar{z}^2} \left( ds^2_{\rm (pert)} + d\bar{z}^2
  \right)
  \notag \\
  \notag \\
  ds^2_{\rm (pert)} & = - \left( 1 - 6 a - 2 \bar{r}
    \partial_{\bar{r}} a \right) dt^2 + \left( 1 + 2 a \right)
  d\bar{r}^2 + \bar{r}^2 \left( 1 + 2 a + \bar{r} \partial_{\bar{r}} a
  \right) d\Omega_{\rm 2}^2
\label{eq:lin_metric}
\end{align}
where $a = a(\bar{r},\bar{z})$, which is the synchronous transverse
traceless gauge. The linearized constraint equations $\{rz\},
\{rr-zz\}$ are satisfied and the 3 bulk equations reduce to,
\begin{equation}
\nabla^2_{\rm{(AdS)}} a = \left( \partial_{\bar{r}} \partial_{\bar{r}} + \partial_{\bar{z}} \partial_{\bar{z}} + 
  \frac{4}{\bar{r}} \,\partial_{\bar{r}} - \frac{3}{\bar{z}} \,\partial_{\bar{z}} \right) a = 0
\label{eq:lin_laplace}
\end{equation}
an elliptic operator acting on $a$. As discussed in
\cite{Giddings:2000mu,Garriga:1999yh} the transverse traceless
condition does not allow coordinate freedom to place the brane at
$\bar{z} = 1$.

The coordinate transformation to bring the metric into the form
\eqref{eq:metric_conf} is then $\bar{r} = r + f(r,z)$, $\bar{z} = z +
g(r,z)$ with,
\begin{align}
\partial_r g & = - \partial_z f
\notag \\ 
\partial_z g & = \partial_r f + a
\label{eq:lin_metric_transform}
\end{align}
which yield Poisson equations for $f, g$,
\begin{align}
\nabla^2 f & = - \partial_r a
\notag \\ 
\nabla^2 g & = \partial_z a
\label{eq:lin_metric_transform2}
\end{align}
where $\nabla^2 = \partial^2_r + \partial^2_z$. Equation
\eqref{eq:lin_laplace} for $a$ is unchanged to linear order as $\bar{r},
\bar{z} \rightarrow r, z$. The coordinate transformed metric
components $T,A,B$ are then,
\begin{align}
T & = -3\,a - r \, \partial_{r} a - {\frac{g}{z}}
\notag \\ 
A & = a + \frac{1}{2}\,\partial_{r} f + {\frac{f}{2 r}} + 
  \frac{r}{4}\,\partial_{r} a - {\frac{g}{z}}
\notag \\ 
B & = \frac{1}{2}\,\partial_{r} f - {\frac{\,f}{2 r}} - 
  \frac{r}{4}\,\partial_{r} a
\label{eq:lin_metric_cmpts}
\end{align}
in terms of $a, f, g$, and now the boundary conditions on the brane
can be calculated from the brane matching conditions, equations
\eqref{eq:matching_eqns} in appendix \ref{app:einstein_eqns}, if one
places the brane at $z=1$ in the `conformal metric' coordinates,
giving,
\begin{align}
  \nabla^2_{\rm{(3-d)}} g & = \partial^2_{r} g + \frac{2}{r}
  \partial_{r} g = \frac{1}{6 \, z} \rho
  \notag \\
  \partial_z a & = - \frac{2}{r} \partial_r g
  \notag \\
  P & = O( a^2 )
\label{eq:lin_matching}
\end{align}
which apply on the brane at $z=1$. The last equation results from a
Bianchi identity and shows that in the Newtonian approximation the
leading contribution to the pressure is second order. The first and
second relations above give Neumann boundary data on the brane for the
elliptic equation for $a$ \eqref{eq:lin_laplace}. Asymptotically $a$
is chosen to be zero as the AdS scalar propagator decays as
$\frac{1}{r}$ and $\frac{1}{z}$ at large $r$ and $z$ respectively when
the radiation boundary conditions are imposed. On the $r=0$ axis the
function is taken to be even.

Now consider the boundary conditions for $f$ and $g$ which must be
compatible with \eqref{eq:lin_metric_transform}. We must impose $f =
0$ at $r = 0$ for regularity implying $\partial_r g = 0$ at $r = 0$.
Then take $g$ as determined by \eqref{eq:lin_matching} on the brane
and choose $g \rightarrow 0$ as $r \rightarrow \infty$, which we are
allowed to choose providing $\rho \rightarrow 0$ asymptotically.

For large $z$ we must specify that $f, g$ behave as in equation
\eqref{eq:lin_afg} in the appendix \ref{app:asym_lin}, where these
functions are calculated in the asymptotic regime. Numerically we take
the large $r, z$ boundaries at finite coordinate position. Thus on a
finite lattice there is data to specify on the asymptotic boundaries.
We choose that $f = 0$ on the large $z$ boundary, and $g = 0$ at large
$r$, which we expect to be a reasonable approximation to the gauge
chosen by the non-linear method. The normal derivative for the other
function is then determined from \eqref{eq:lin_metric_transform}. The
complete linear boundary conditions are shown in figure
\ref{fig:lin_bc}.

In appendix \ref{app:asym_lin} we solve the linear theory
asymptotically for a point density source on the brane. The result is
that $T, A, B \rightarrow 0$ in the linear regime for large $z$ and
the Weyl components decay asymptotically. Thus in the full non-linear
theory, provided the perturbation is asymptotically small, which we
indeed do later see in the numerical solutions, it can be treated
perturbatively in the large $z$ region, and $T, A, B = 0$ is the
correct boundary condition.

\begin{figure}[htb]
\begin{center}
\input{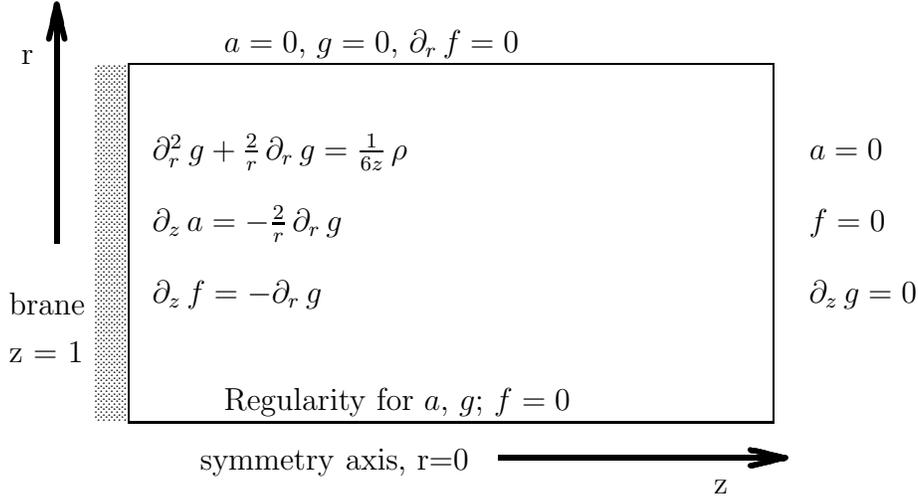}
\end{center}
\caption[short]{ \figuremode  An illustration of the linear boundary
  conditions. The matching conditions and constraints specify $a, f,
  g$ on the brane itself, with the asymptotic AdS condition specifying
  the functions asymptotically.
\label{fig:lin_bc} 
}
\end{figure}

%
\subsection{Asymptotic Data}
\label{sec:asym_data}
%

We have shown above that in our gauge $T,A,B \rightarrow 0$
asymptotically for large $r$ and $z$. It is important to note that
this alone does not guarantee that the constraint equations are
satisfied at large $z$. In the linear theory the constraints were
satisfied by construction. In the relaxation method we wish to impose
the boundary condition $T,A,B \rightarrow 0$ and have no further
freedom to explicitly enforce the constraints at large $r$ and $z$.
Only the elliptic bulk equations are solved by the relaxation, with
one constraint $\{rz\}$ being enforced on the brane itself. We now
show that requiring $T, A, B \rightarrow 0$ asymptotically does indeed
imply the constraints are satisfied.

Consider the constraint structure, \eqref{eq:constpois} which applies
when the elliptic Einstein equations are satisfied. The constraint
$\{rz\}$ obeys a Laplace equation,
\begin{equation}
\laplace \phi = (\partial_r^2 + \partial_z^2) \phi = 0
\end{equation}
where $\phi \equiv g \, G^{r}_{~z}$. The measure $g =
\sqrt{-\det{g_{\mu\nu}}} \sim \frac{1}{z^5}$ at large $z$ and $\sim
r^2$ at small $r$. In the scheme we outline this constraint is exactly
enforced on the brane itself, and thus $\phi = 0$ at $z = 1$ for all
$r$. Provided $T, A, B$ are finite at $r = 0$ then the form of
$G^{r}_{~z}$ guarantees that at small $r$ it can diverge no faster
than $\sim \frac{1}{r}$. Then $\phi$ which includes the measure $g
\sim r^2$ is forced to zero. Thus with finite $T, A, B$ we must also
find $\phi = 0$ at $r = 0$.

In a finite box it is possible to have relevant boundary data on the
large $r$ and $z$ boundaries compatible with $\phi = 0$ on the
boundaries $z = 1$ and $r = 0$. However, as the boundary is moved to
infinity, the general solution to the Laplace equation must simply be
linear in both $r$ and $z$. Imposing $T, A, B = 0$ at large $z$, and
assuming $T, A, B$ behave smoothly asymptotically, then $\phi \sim
\frac{1}{z^5} G^{r}_{~z} \sim \frac{1}{z^3} \{rz\}$, and $\{rz\}$ will
decay to zero at large $z$ and finite $r$.  Thus $\phi$ cannot scale
linearly in $z$ and so must be identically zero.

The second constraint $\{rr-zz\}$ satisfies Cauchy-Riemann relations
as in \eqref{eq:constpois} with the first constraint $\{rz\}$ which
implies that if $\phi = 0$, as shown above, then $\xi \equiv g (G^{r}_{~r}
- G^{z}_{~z}) = c$, a constant. Consider that $\xi \sim \frac{1}{z^5}
(G^{r}_{~r} - G^{z}_{~z} ) \sim \frac{1}{z^3} \{rr-zz\}$ at large z.
Again, if $T, A, B \rightarrow 0$ smoothly as $z \rightarrow \infty$,
then from the form of $\{rr-zz\}$, $\xi \rightarrow 0$ at least as
fast as $\frac{1}{z^3}$. This determines that the constant $c = 0$.

Thus we have shown that provided $T, A, B \rightarrow 0$ smoothly as
$z \rightarrow \infty$, and that they are \emph{finite} at $r = 0$,
and that the elliptic Einstein equations are satisfied, and in
addition the constraint $\{rz\}$ is satisfied exactly on the brane,
this guarantees that the two constraints will we satisfied on both the
$r = \infty$ and $z = \infty$ boundaries. We have made the above
arguments assuming an infinite lattice. In the numerical scheme the
boundaries will actually be enforced at finite $r, z$. We assess the
accuracy of this necessary approximation by varying the physical
lattice size and showing that the solutions are insensitive to this
(appendix \ref{app:testing}). 

%
\subsection{The Origin and Relaxation}
\label{sec:origin}
%

Observe that equation \eqref{eq:metric_conf} contains singular terms
as one approaches $r=0$ which are more severe than those of the usual
cylindrical coordinate system. These singular terms, going as
$\frac{1}{r^2}$ when $r \rightarrow 0$, occur only in the $B$ equation.
The requirement of no boundary at $r=0$ implies $T_{,r} = A_{,r} = 0$.
The function $B$ must also be even about $r=0$, but in addition $B=0$
must be true at $r=0$ in order to have a regular solution.  Now we
must consider whether this regularity condition is consistent with
elliptic data, as we specify $B$ and its normal derivative at $r = 0$,
yet in addition specify $B$ on all the other boundaries.

We Taylor expand the functions $T, A, B$ about $r=0$ as,
\begin{align}
T & = t_0(z) + r^2 t_2(z) + O(r^4) \notag \\
A & = a_0(z) + r^2 a_2(z) + O(r^4) \notag \\
B & = r^2 b_2(z) + O(r^4) 
\end{align}
and substituting into the `Poisson' equations \eqref{eq:TAB_poisson}.
Taking the leading order behavior of these equations in $r$, we have 3
ordinary differential equations in $z$ involving the functions
$a_0(z), t_0(z)$ and $t_2(z), a_2(z), b_2(z)$. The functions $t_2(z),
a_2(z), b_2(z)$ are determined by the next to leading order equations
in an elliptic relaxation. Thus we have three equations and only two
functions, $a_0(z), t_0(z)$, to satisfy them with. However, one finds
that the three ordinary differential equations are not independent.
Indeed, the $z$ derivative of the one resulting from the leading
behavior in the $B$ `Poisson' equation is a linear combination of the
others and the constraint $\{rz\}$. This is a direct result of the
Bianchi identities. Thus if the elliptic equations are solved at
$r=0$, with the condition that $T, A$ are even and $B=0$ there, then
provided that the constraint $\{rz\}$ is satisfied then $\partial_r B
= 0$ is also implied, as required for regular geometric behavior.

In the previous section we showed that provided the elliptic equations
were relaxed, $\{rz\}$ is satisfied on the brane, and importantly $T,
A, B$ are finite at $r = 0$, then indeed the constraint equation
$\{rz\}$ is satisfied \emph{everywhere}, obviously including $r = 0$.
We therefore conclude that if a finite solution to the relaxation
problem is found, complete with finite and even $T, A$, with $B = 0$
at $r = 0$, then it must not only satisfy the constraint $\{rz\}$, but
following from this, also satisfy geometric regularity $\partial_r B =
0$.

%
\subsection{Numerical Scheme}
\label{sec:method}
%

We use an iterative convergence scheme to relax the bulk equations
\eqref{eq:TAB_poisson}. The finite differencing, boundary conditions
and scheme details are stated in appendix \ref{app:numerical}. The
boundary conditions for $A, B$ on the brane are given by the density
and isotropy matching conditions, and $T$ is determined from $\{rz\}$.
$T,A,B$ are required to vanish on the large $r, z$ boundaries.

We now discuss the main technical difficulty, namely that the equation
for $B$ contains singular terms at $r = 0$. Setting $B=0$ as a
boundary condition, relaxing the elliptic Einstein equations and
satisfying the constraints will imply $B_{,r}=0$ in the final solution,
but will not guarantee $B_{,r}=0$, during the early stages of the
relaxation. Therefore the solutions fail to converge almost
immediately having highly singular behavior. It is important to note
that this is simply a problem of using an iterative relaxation scheme
with the coordinate system chosen, and does not reflect any physical
divergence along the symmetry axis.  $B$ must go quadratically in $r$
to ensure all terms are finite during relaxation.

Note that the usual spherical coordinate system singular terms
involving $X_{,r} / r$, where $X$ is an even function, pose no
problem for our relaxation scheme.  Specifically it is only the term
$\left(e^{4 B} - 1\right) / r^2$ in the $B$ equation that
requires the correct $B \sim r^2$ behavior.

A solution is provided by the constraints. Note that we could
determine $B$ from the constraint equation $\{rz\}$ by integration.
The reason that we do not do this and relax for $T, A$, is that the
scheme is extremely non-local and we could not implement it in a
convergent manner. However, determining $B$ from $\{rz\}$ is
attractive as there are no singular terms even if $B$ goes linearly
and not quadratically near $r = 0$.

We have a situation where full determination of $B$ from the
constraint $\{rz\}$ is incompatible with relaxation. On the other
hand, determining $B$ by relaxation is impossible as $B_{,r} \ne 0$
gives highly singular terms during the early stages of this
relaxation. Thus we use a combination, finding that calculating the
singular $B$ terms in the equations \eqref{eq:TAB_poisson} from a
solution for $B$ integrated using the constraint is a good compromise.
We term this solution $B2$, integrating along the $r$ and $z$
directions, out from $r = 0$ and in from large $z$, where the boundary
condition $B2 = 0$ is employed, consistent with the boundary
conditions for $B$. The appendix \ref{app:numerical} describes exactly
which terms are determined from $B2$.  The integration of $\{rz\}$
means that $B2$ has the correct quadratic behavior near $r = 0$ as
$T, A$ behave as even functions there due to the boundary conditions
imposed on them. The singular source terms, calculated using $B2$, are
suppressed at large $r$ and thus the scheme is not too non-local for
the relaxation procedure, and is found to work extremely well.  In
appendix \ref{app:testing} we compare the solution $B2$, as integrated
from $\{rz\}$, with that relaxed using the bulk $B$ equation for a
global consistency check on the error in the solution.  The two are
found to be in close agreement. Furthermore, the comparison with the
numerical linear solution in the low density regime (section
\ref{sec:linear_check}) again confirms that the metric solution is
correct on the $r = 0$ symmetry axis.

We cut off the lattice at finite $r, z$ and then, in the later section
\ref{app:testing}, ensure that solutions are insensitive to the cut
off.  The brane is chosen to be at $z=1$ as discussed in section
\ref{sec:brane_pos}.  Finally the constraint $\{rz\}$ is implemented
by integrating in from the large $r$ boundary at $z=1$ to solve for
$T$ on the brane. The relaxation and constraint integration are
iterated together in a loop.

\begin{figure}[htb]
\centerline{\psfig{file=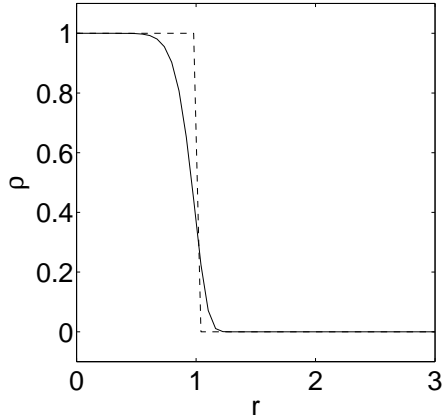,width=6cm}}
\caption[short]{ \figuremode  An illustration of the deformed top-hat defined in equation
  \eqref{eq:profile} for $\rho_0 = 1, \xi = 1$. A top-hat density
  corresponds to an incompressible fluid and in standard GR gives the
  largest mass possible for a given proper angular radius. We slightly
  deform the top-hat to avoid numerical artifacts associated with
  discretization.
\label{fig:deformedtophat} 
}
\end{figure}

There are two physical scales in the problem. Firstly the AdS length
which we have chosen to be of unit magnitude in our units. Then there
is the radial size of the density profile, which we take to be a
deformed top hat function.  We smooth the top hat function to avoid
numerical artifacts at the edge of the star and take the density
profile
\begin{equation}
\rho(r) = \rho_0 e^{-\left(\frac{r}{\xi}\right)^{10}}
\label{eq:profile}
\end{equation}
which is illustrated in figure \ref{fig:deformedtophat} and closely
approximates a top-hat function with core density $\rho_0$. A
characteristic width is defined by $\xi$, although $\xi$ is
the coordinate distance rather than a coordinate independent measure
of radius. We define the proper radius of the star, $R$, to be the
proper angular radius at coordinate distance $r = \xi$,
\begin{equation}
R = r e^{A - B} \Big|_{r = \xi, \, z = 1}
\end{equation}

When we choose to solve for stars with fixed $\xi$ for several
different core densities $\rho_0$, the proper radii of these stars
will vary slightly. For linear stars, $R \simeq \xi$, and for
non-linear ones we find that generically the angular radius is still
similar to $\xi$, but a little larger.

We find numerically that for $R << 1$ we have very good convergence
properties, but for $R >> 1$ one requires extremely large numbers of
grid points.  However $\xi \sim 3$ is large enough for our purposes to
see the 4-dimensional limit emerge. For $\xi \lesssim 3$ we find
solutions in the non-linear regime approaching the limit of stability
for a static star. We find that for the \emph{smallest} stars tested,
$\xi = 0.3$, the code converges for configurations thought to be
extremely close to the critical point, with photons emitted from the
stellar core having redshifts of ${\cal Z} \simeq 15$. For the
\emph{largest} stars tested, $\xi = 3$, the code converges for
solutions with ${\cal Z} \simeq 2$, reaching at least $75 \%$ of the
estimated critical density.

%
\section{Numerical Comparison with Linear Theory in the Low
  Density Regime}
\label{sec:linear_check}
%

Detailed numerical tests and consistency checks of the non-linear
method described above are presented in appendix \ref{app:testing}.
However a powerful independent test can be performed by simply
comparing the solutions of the full non-linear scheme with those of
the linear scheme, outlined in section \ref{sec:linear}. In the regime
where the density perturbation on the brane is sufficiently small that
the metric perturbation is much less than unity everywhere the two
methods must agree.

The test is extremely valuable as the linear theory automatically
satisfies the constraints and also has well understood asymptotic
boundary conditions. The close agreement found, and described below,
between the non-linear and linear methods in the low density regime
indicates that both points are indeed satisfied well in the non-linear
case. In addition, the non-linear method uses regularization for
singular terms in $B$ at $r = 0$, and the agreement implies the
quality of this approximation is very good.  Finally it means that
finite boundary and resolution effects, which are inevitable in a
numerical method, are likely to be small at the resolutions and
lattice sizes used, and an estimate of absolute error in the metric
functions and physical brane observables can be made from the
comparison.

\begin{figure}
\centerline{\psfig{file=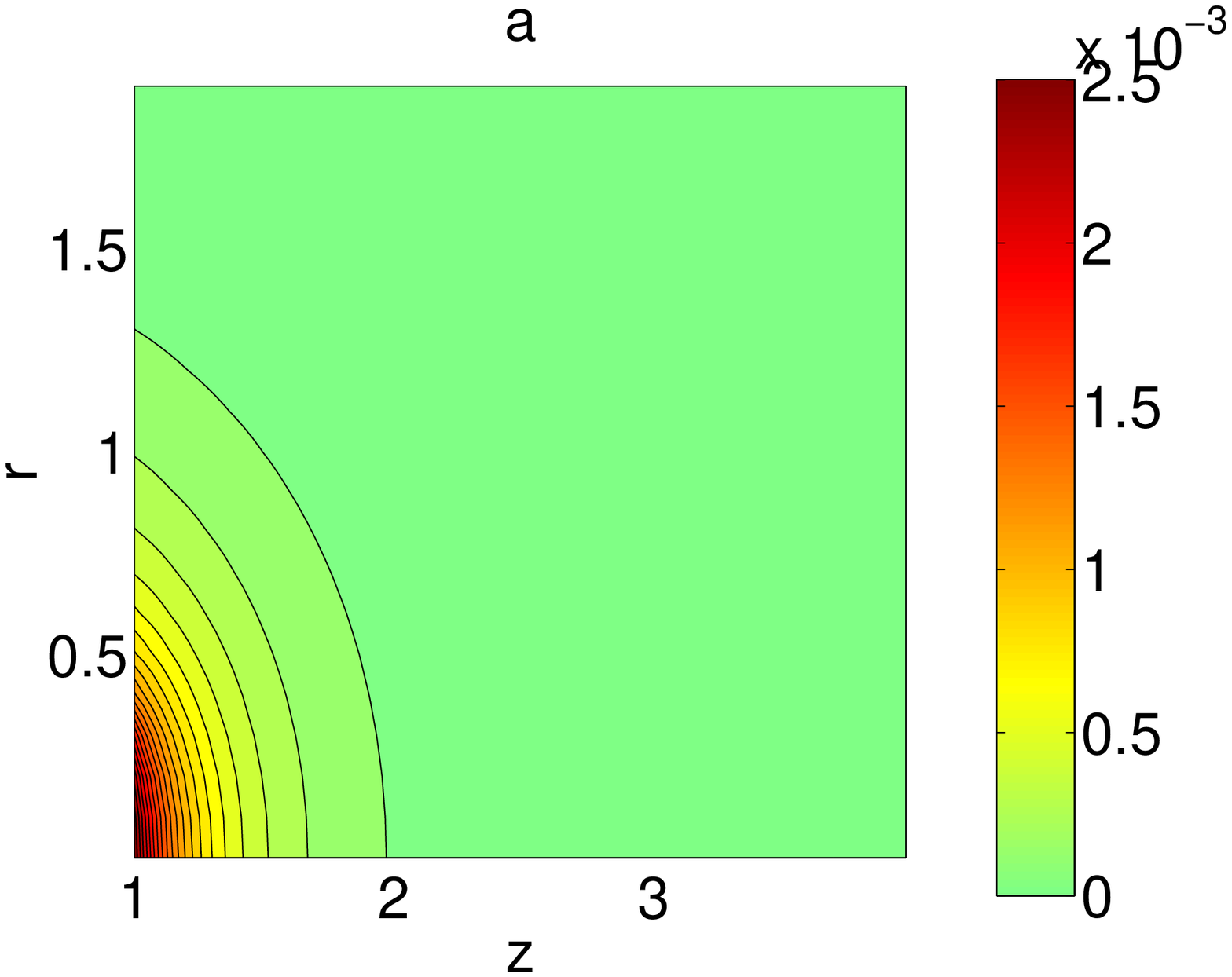,width=6cm}
  \psfig{file=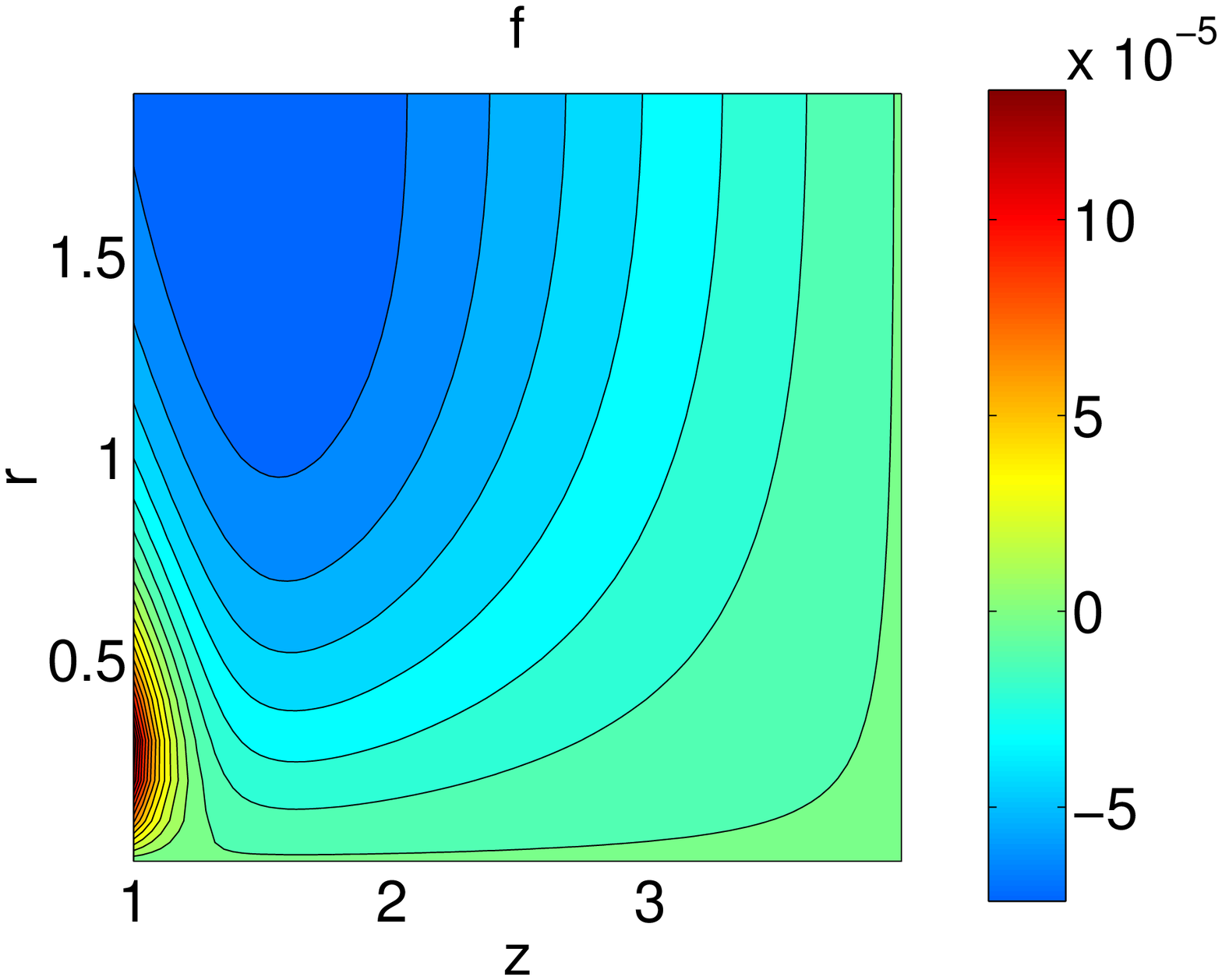,width=6cm}  \psfig{file=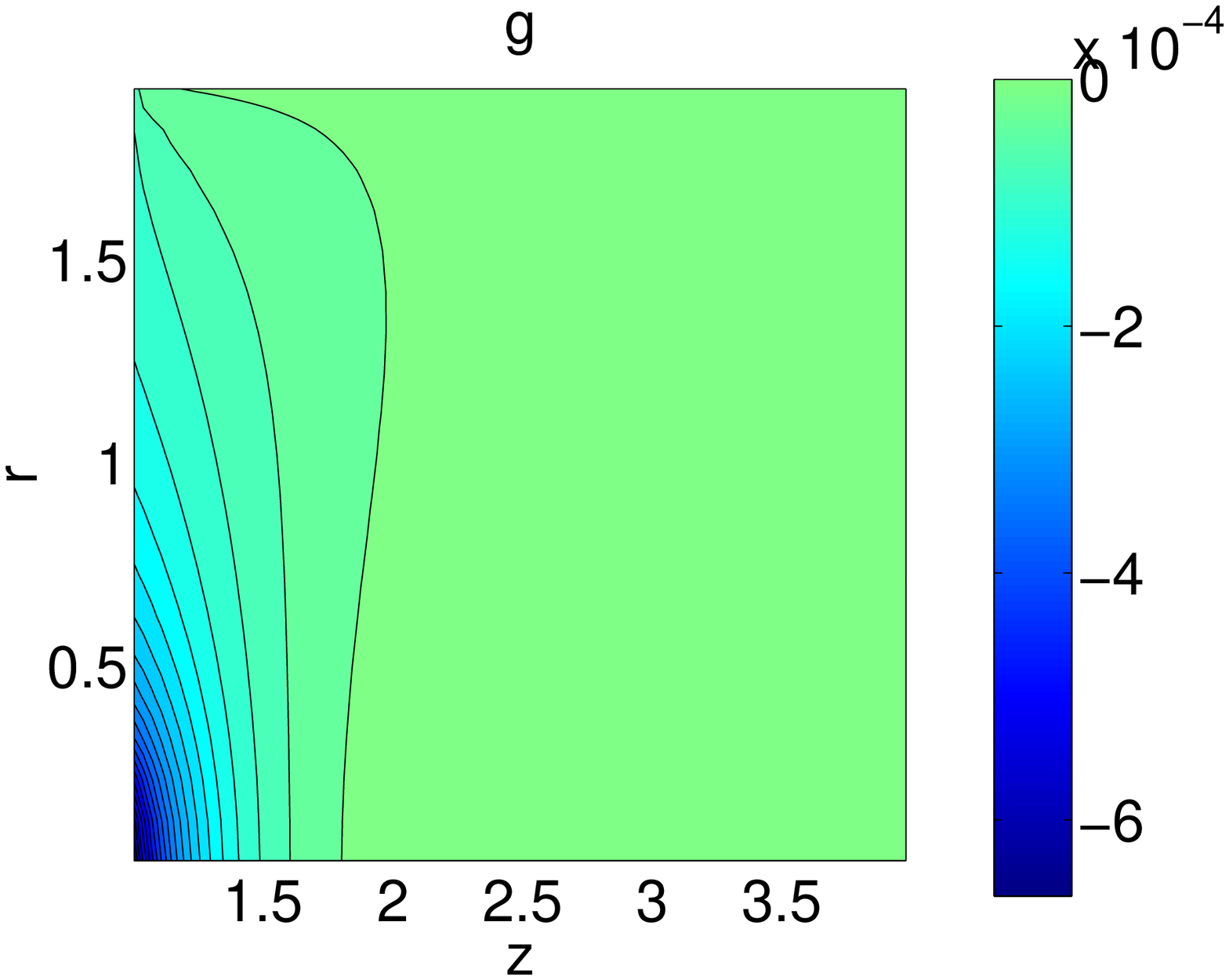,width=6cm}}
\caption[short]{ \figuremode  An illustration of the variable $a$ in the transverse
  traceless metric of the linear theory, and coordinate transform
  functions $f, g$ for a solution with $\xi = 0.3$ generated using
  the linear method. The core density is chosen to be $\rho_0 = 0.1$.
  The decay of $a$ to zero away from the brane is clearly seen. The
  boundary conditions for $f, g$ are also clear; $f = 0$ at r = 0 and
  on the large $z$ boundary, $g = 0$ at large $r$, and $f, g$ are
  fixed on the brane by the matter. Figure \ref{fig:linear_sol3} then
  displays the result of this coordinate transform to the metric of
  form \eqref{eq:metric_conf}, allowing comparison with the non-linear
  method. (lattice: $dr = 0.02$, $r_{\rm max} = 2$, $dz = 0.005$,
  $z_{\rm max} = 4$)
\label{fig:linear_sol2} 
}
\end{figure}

\begin{figure}
\centerline{\psfig{file=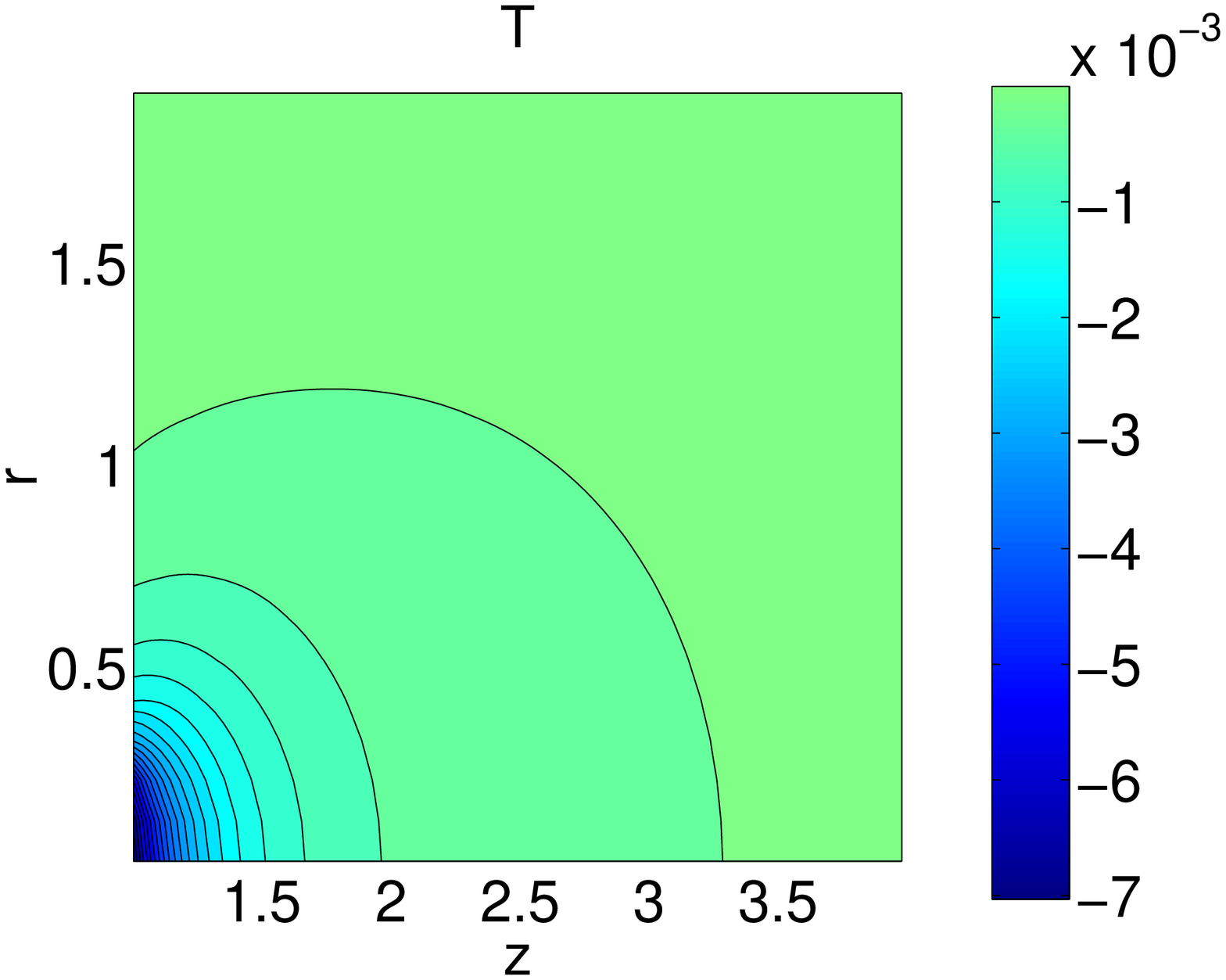,width=6cm} 
  \psfig{file=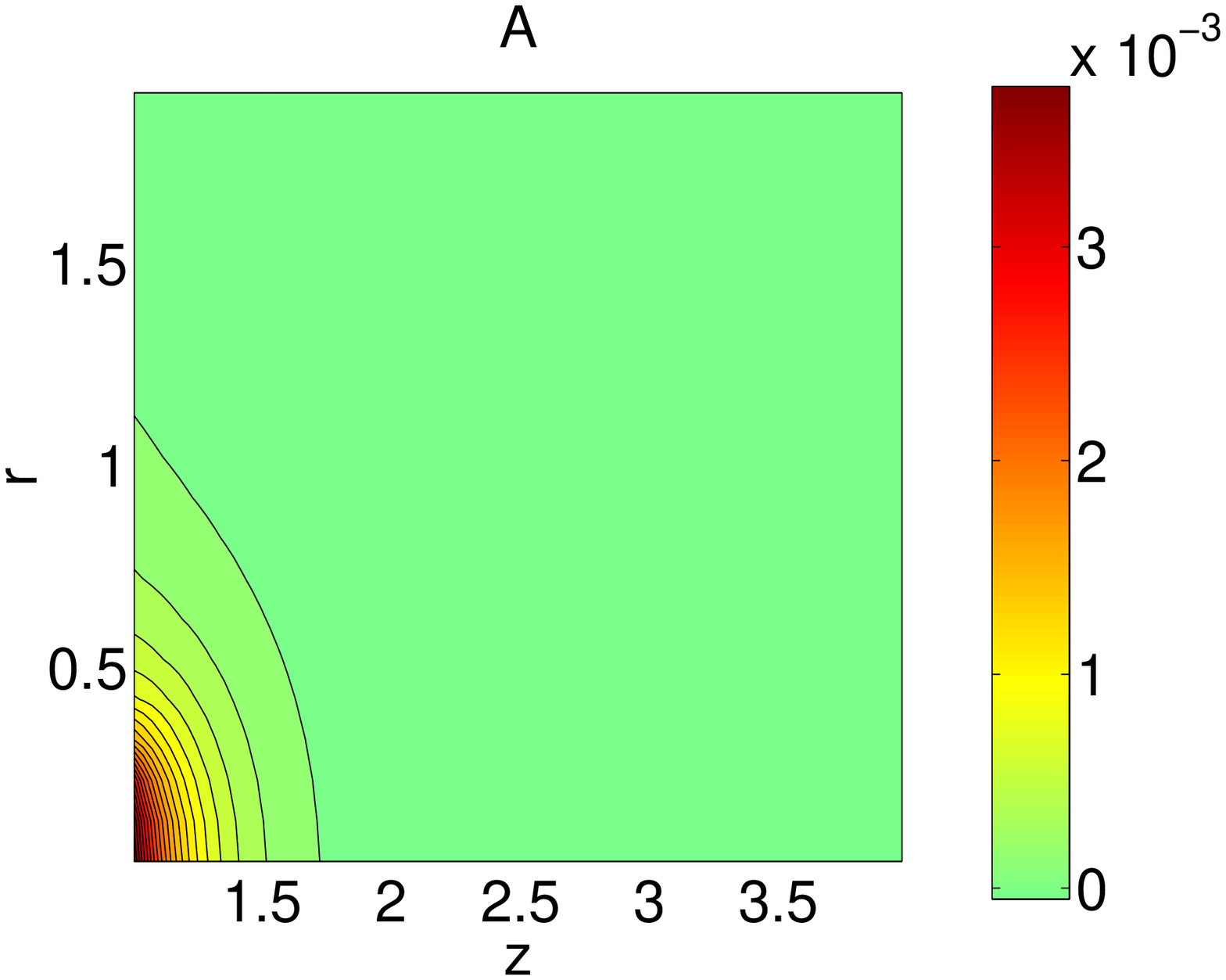,width=6cm} \psfig{file=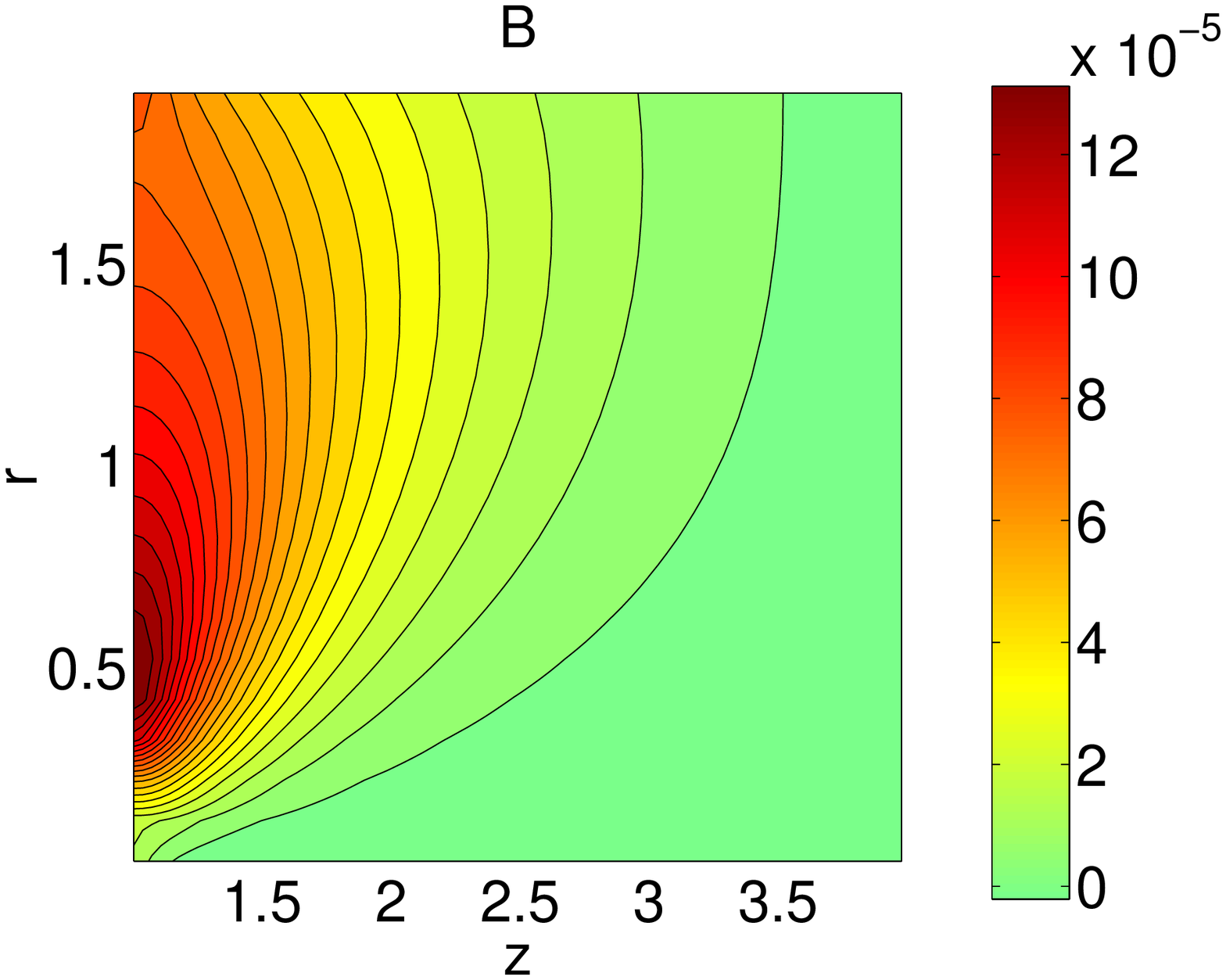,width=6cm}}
\centerline{\psfig{file=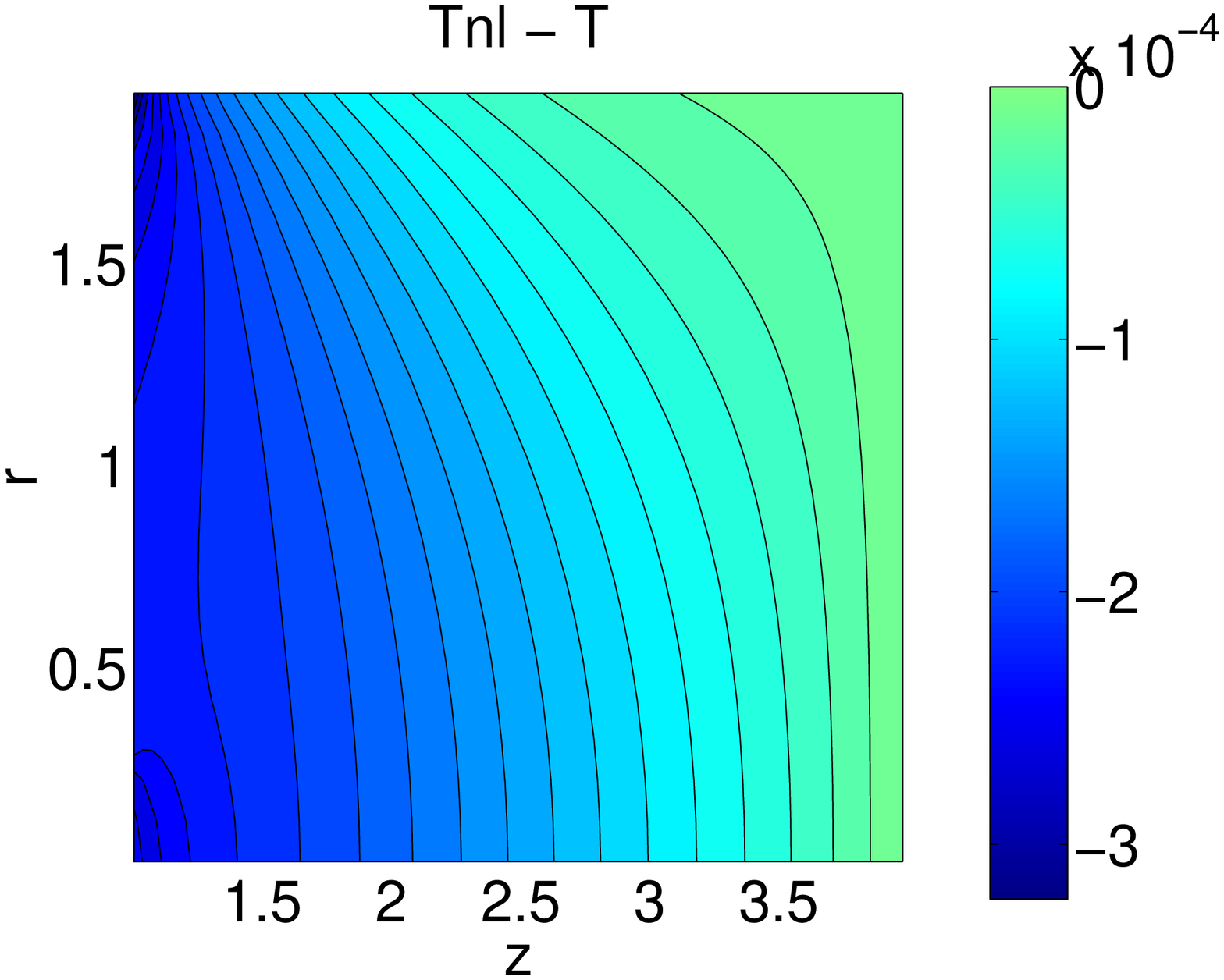,width=6cm} \psfig{file=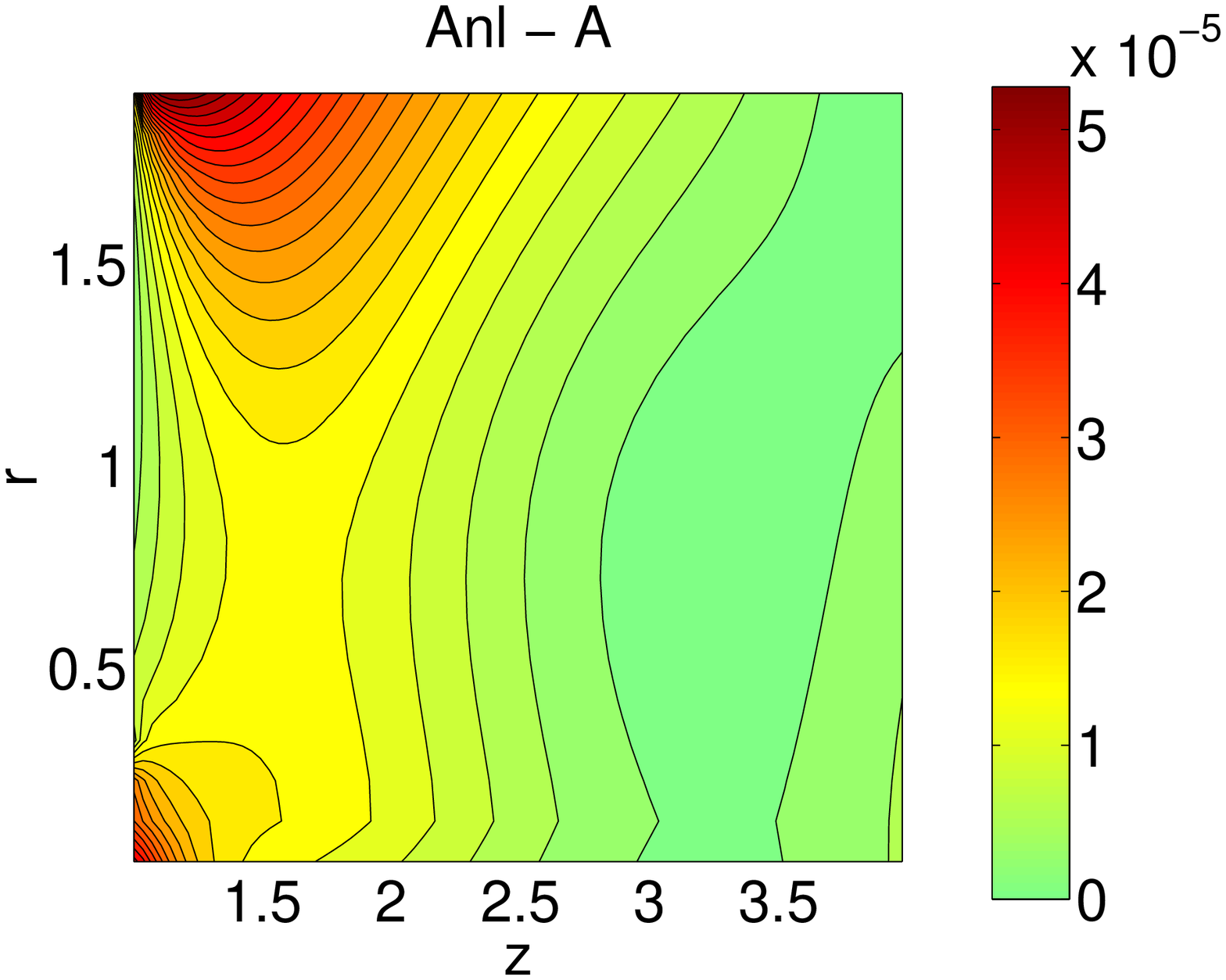,width=6cm} \psfig{file=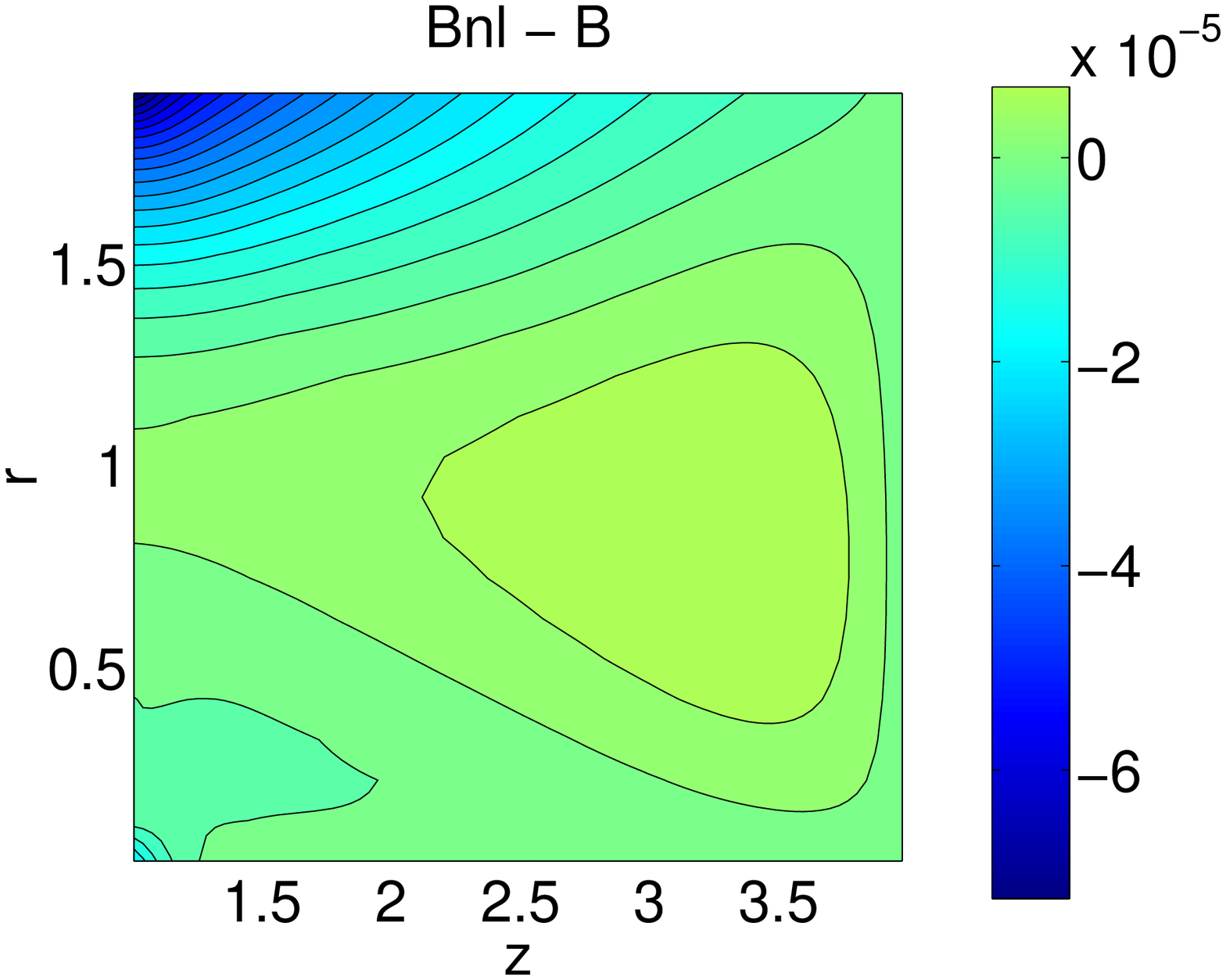,width=6cm}}
\caption[short]{ \figuremode  Upper plots; An illustration of $T, A, B$ generated from the linear
  method, for $\xi = 0.3$, $\rho_0 = 0.1$ with $a, f, g$ shown in
  figure \ref{fig:linear_sol2}. The form of the functions agrees
  extremely well with the non-linear method (see for example the later
  figure \ref{fig:A8_metric}). In particular we see that the metric
  functions do indeed decay towards the asymptotic boundaries, as we
  require in our non-linear boundary conditions. Note that the
  function $B$ is much smaller in magnitude than the other functions
  $T, A$. Although $B$ appears not to decay at large $r$, in fact it
  is so small everywhere, that at the large $r$ boundary it is of the
  same magnitude as the decayed $T, A$. The smallness of $B$ implies
  that the spatial geometry of the metric is effectively described as
  a conformal deformation of hyperbolic space by a factor $e^{2 A}$.
  Lower plots; The full non-linear method was used to calculate
  solutions for the same brane matter, and the difference in the
  metric functions is shown.  The density is sufficiently small that
  the non-linear method should reproduce the linear solution as $T, A,
  B << 1$. The non-linear method gives the quantities $Tnl, Anl, Bnl$
  as a solution. The agreement is extremely good. For $T$ the peak
  deviation is only $\simeq 4\%$ of the peak value of the function.
  For $A$ this difference is even less at $\simeq 1\%$.  For $B$ the
  difference is again very small for the lower half range of $r$.
  Near the maximum $r$ boundary the error becomes larger but as $B$ is
  so small, the absolute differences are tiny.  The remarkable
  agreement between linear and non-linear shows that the boundary
  conditions imposed for the non-linear elliptic relaxation are indeed
  reproducing the asymptotic AdS geometry well.  The following plot
  shows some brane observables calculated from the matching conditions
  and these have similar, if not even better agreement.
\label{fig:linear_sol3} 
}
\end{figure}

We calculate a solution using the full non-linear method for the
\emph{smallest} sized star considered elsewhere in this paper, with
$\xi = 0.3$. A sufficiently small energy density $\rho_0 = 0.1$ is
chosen so that the metric perturbations are everywhere small. The
lattice size and resolution are the same as used in later sections of
the paper which examine the physical behavior of \emph{small} stars.
Thus estimates of error calculated here are directly applicable to later
results. Figure \ref{fig:linear_sol2} shows the functions $a, f$ and
$g$ of section \ref{sec:linear}, calculated for the same size and
density of star.  The functions $T, A, B$ are given in figure
\ref{fig:linear_sol3}, the bottom row showing the numerical difference
between the two methods for the metric functions. The agreement is
strikingly good. The functions $T, A$ agree extremely well, the
maximum difference being for $T$, occurring at the core of the star,
the fractional difference between the linear and non-linear methods
being only $4 \%$. For $A$ the difference is less at only $1 \%$.
Although the relaxed function $B$ is set to zero at the large $r$
boundary in the non-linear method, and from the linear method we see
this is not quite true, we find that the magnitude of $B$ is much less
than that of both $T, A$ and this tiny absolute difference
appears to have no effect on the solutions in the interior of the
lattice.  This is very strong evidence that the non-linear method is
indeed finding the correct physical solution and the boundary
condition are effective. Furthermore the error for $B$ at small
$r$ is tiny and indicates that the singular term regularization scheme
(section \ref{sec:method}) performs extremely well.

\begin{figure}
  \centerline{\psfig{file=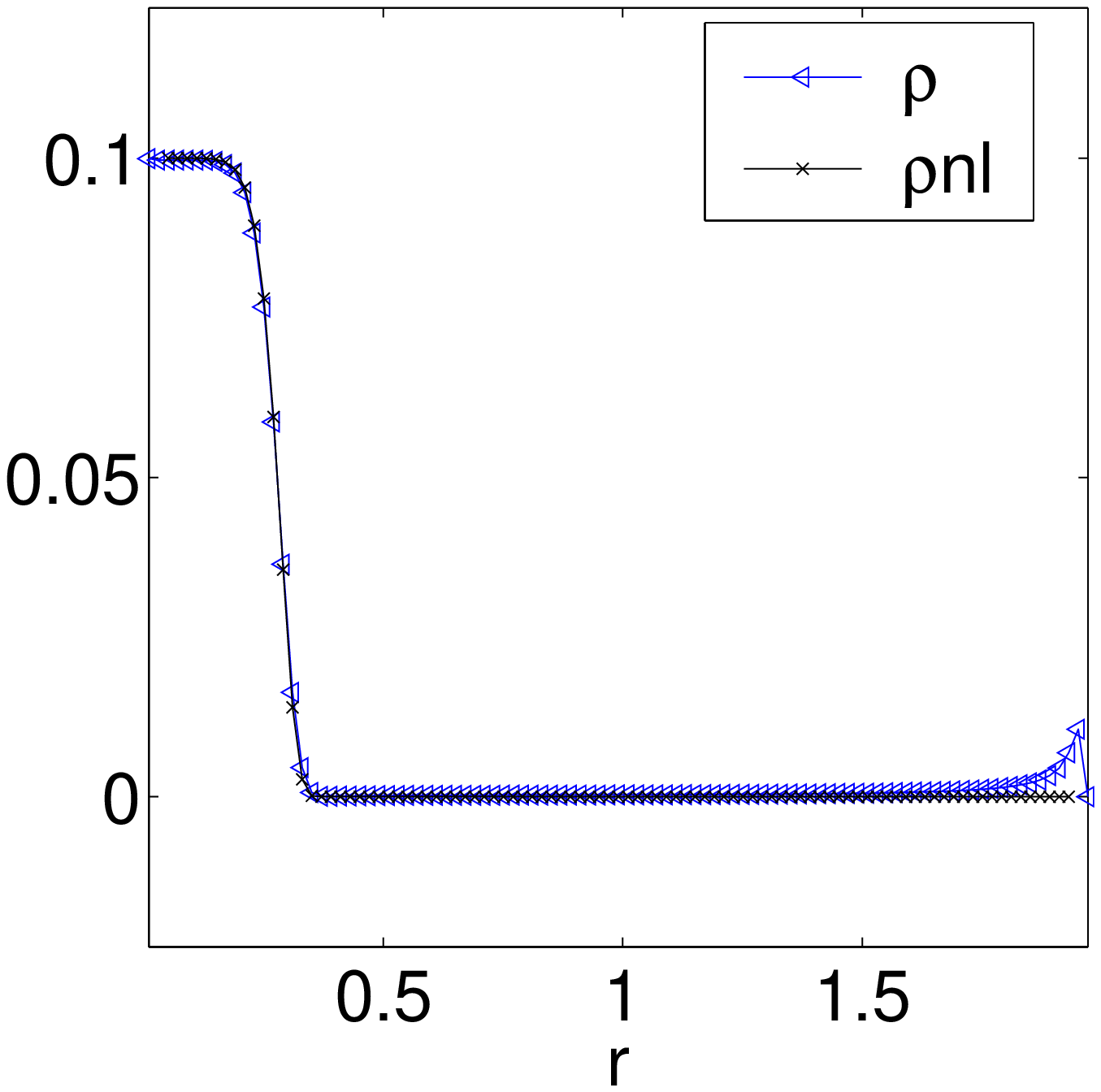,width=5cm} \hspace{0.1cm}
    \psfig{file=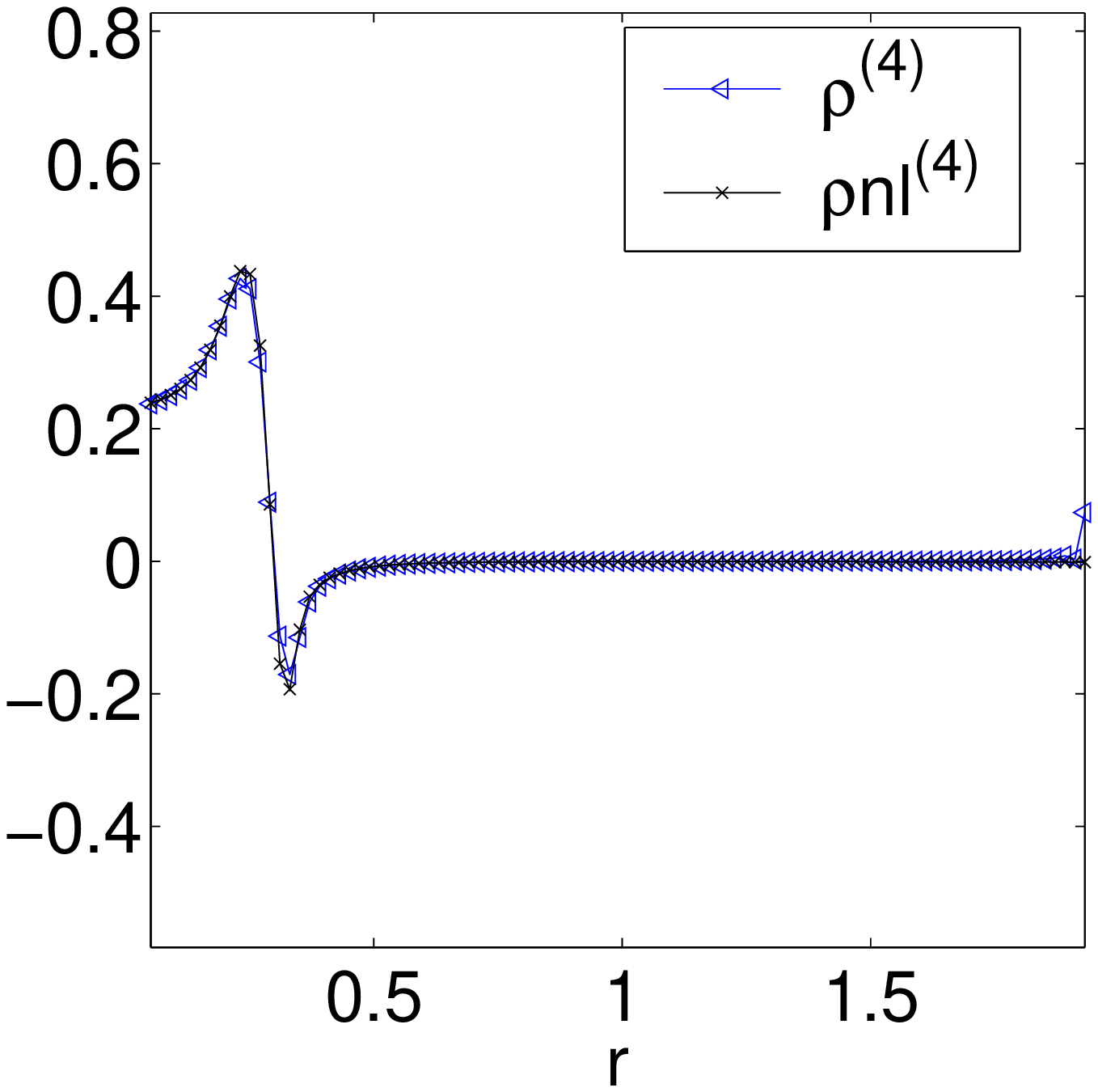,width=5cm} \hspace{0.1cm}
    \psfig{file=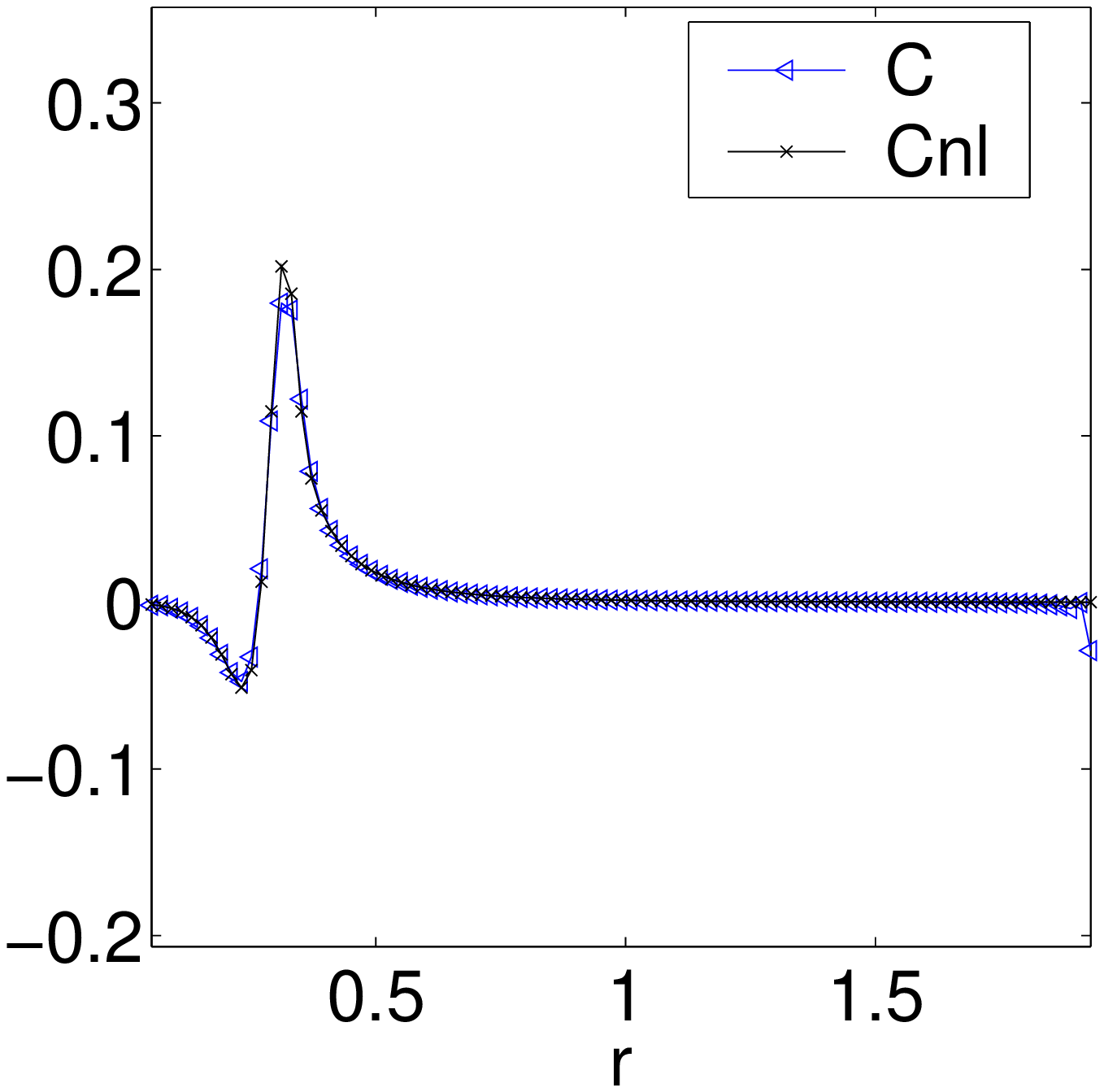,width=5cm}} 
\caption[short]{ \figuremode  An illustration of the density, induced density and induced
  Weyl component on the brane generated from the linear method, for
  $\xi = 0.3$, $\rho_0 = 0.1$, as in figures \ref{fig:linear_sol2}
  and \ref{fig:linear_sol3}. On the left, the density is calculated
  from the brane matching conditions. In the non-linear method (here
  ${\rho}nl$) the density profile is an input boundary condition. For
  the linear method ($\rho$ in the figure) the density is less direct
  as an input. The middle plot shows the effective 4-dimensional
  density on the brane, ie. the density required to produce the
  induced 4-geometry actually found in the 5-dimensional solution. The
  rightmost plot shows the 4-dimensional Weyl component of the induced
  metric, as described in appendix \ref{app:einstein_eqns}. This gives
  a measure of the curvature on the brane. In all the linear
  quantities shown, we see that there are errors at the large $r$
  boundary, although such errors are not so noticeable in the
  non-linear solution. We expect such errors from the finite size
  lattice and note that the errors appear to have no physical effect
  on the inner points where extremely good agreement is found between
  the linear and non-linear methods, the deviations being hardly
  visible on the plots and the maximum differences being of order $1
  \%$ of the typical core values.
\label{fig:linear_observables} 
}
\end{figure}

A further check is to compute the actual density and induced density
and Weyl curvature component for the two solutions. The actual density
is enforced as a boundary condition in the non-linear method.  For the
linear method it is implemented as a boundary condition for $a$, but
the density plotted is computed from the brane matching conditions
after the coordinate transform to the gauge \eqref{eq:metric_conf}.
Hence we can see small finite boundary effects. The induced density is
that required to support the induced 4-dimensional geometry in
standard GR, and is therefore the $tt$ component of the Einstein
tensor of the induced metric, given explicitly in appendix
\ref{app:einstein_eqns}. The induced Weyl component is similarly a
measure of curvature of the induced metric, detailed in the same
appendix. Such induced quantities are extensively used in this paper
to compare a 4-dimensional effective theory with actual 5-dimensional
solutions.  These quantities are plotted in figure
\ref{fig:linear_observables} and excellent agreement is found between
the two methods. An interesting point is that the density in the
linear method is a little distorted at large $r$ due to the presence
of the boundary. This does not appear to effect the interior solution
which indicates that small boundary effects do not degrade the
solution significantly. For all these quantities, differences in the
core values of $\sim 1 \%$ are found between the methods. 

\begin{figure}
\centerline{\psfig{file=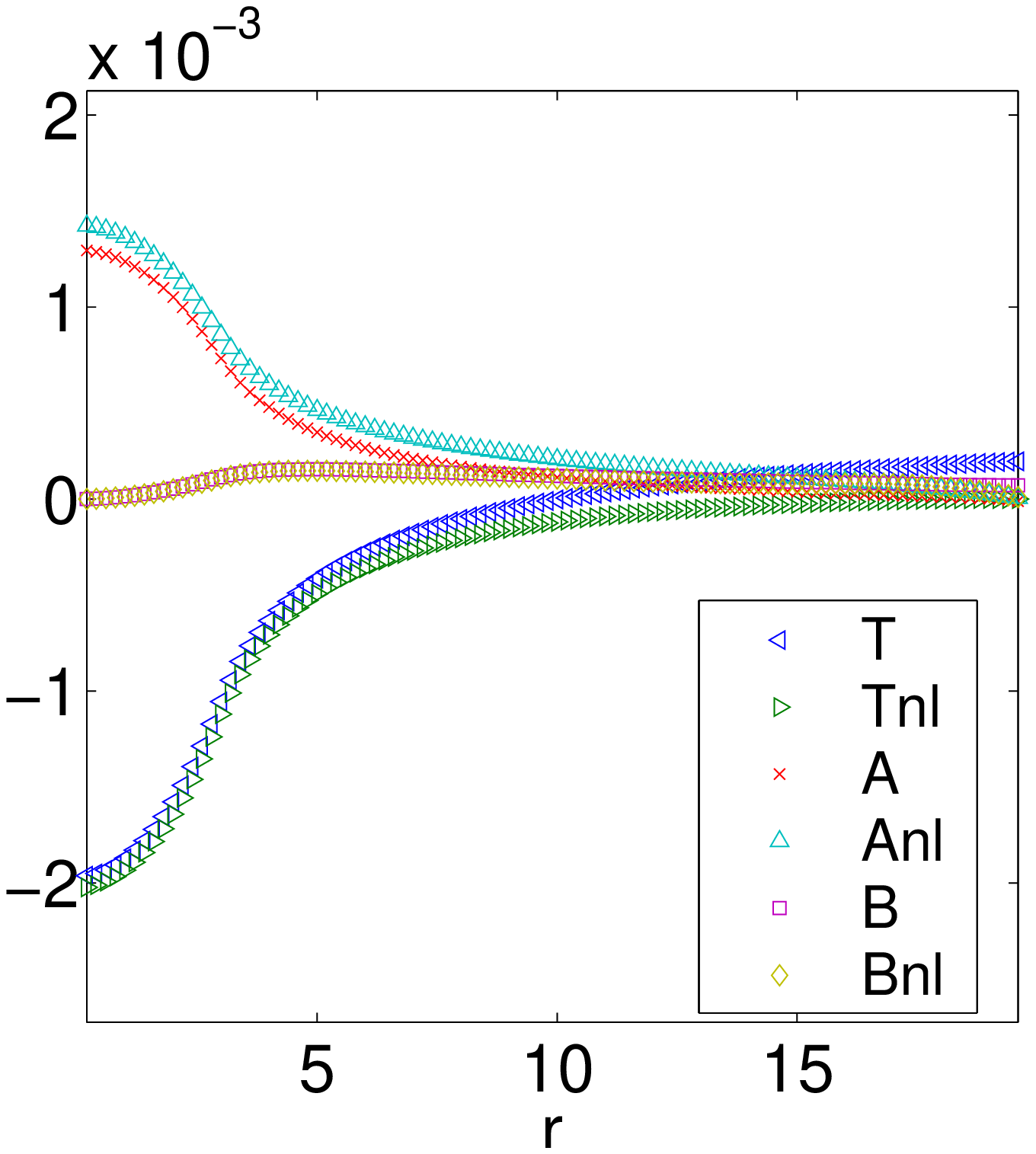,width=6cm} \hspace{0.1cm}
  \psfig{file=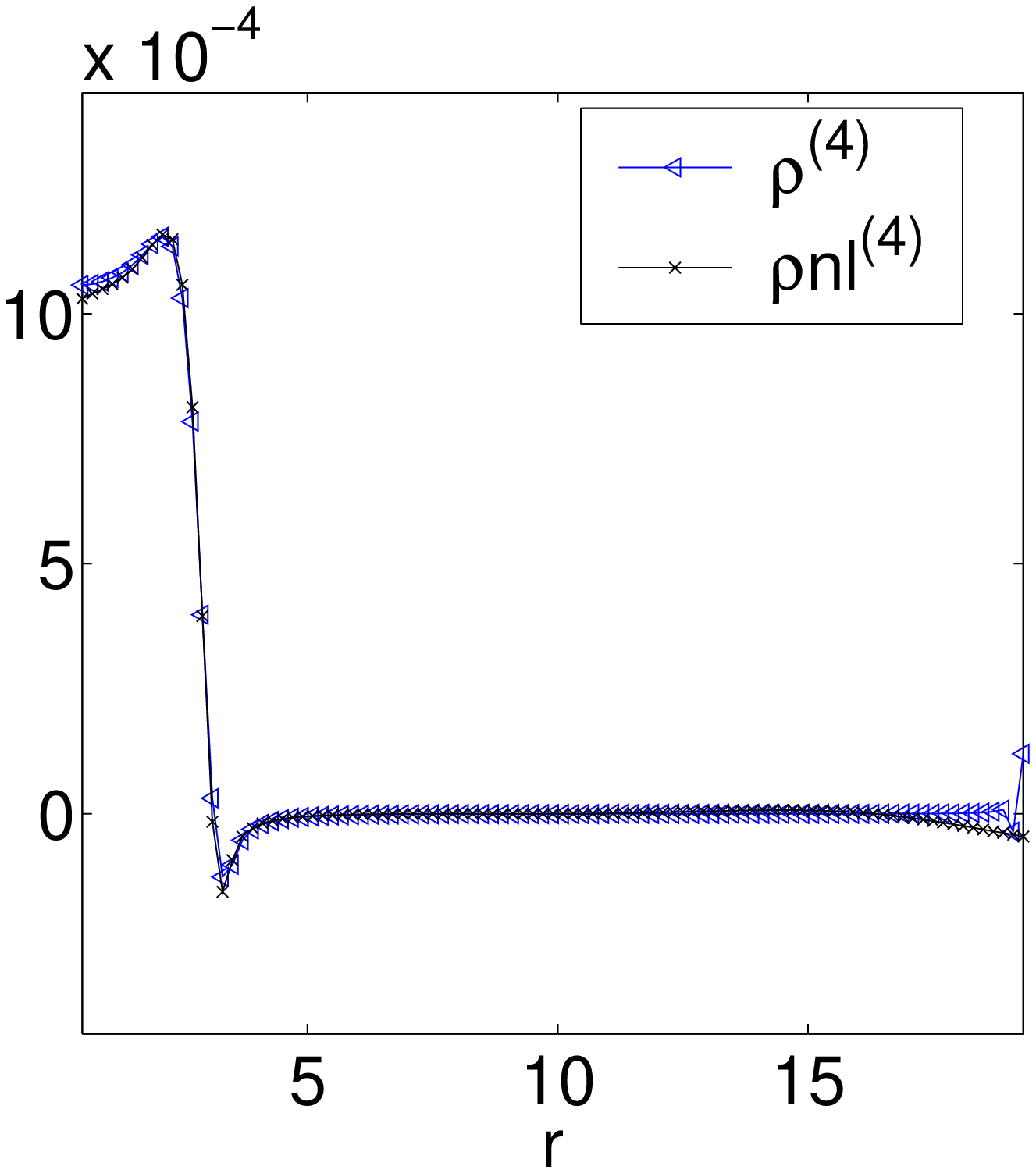,width=6cm}}
\caption[short]{ \figuremode  On the left, the metric functions $T, A, B$ are shown, now
  just on the brane, for the linear and non-linear methods for a
  \emph{large} star, $\xi = 3$, $\rho_0 = 0.001$. The density is
  sufficiently small that the linear method should reproduce the
  non-linear method solutions. $Tnl, Anl, Bnl$ are quantities from the
  non-linear method, $T, A, B$ are from the linear. The maximum
  differences for the linear and non-linear are on the brane itself,
  as for the \emph{small} star case shown in the earlier figure
  \ref{fig:linear_sol2}. We see that the peak differences at $r = 0$
  are $10\%$ for both $T$ and $A$ compared to the peak function values
  themselves with very small absolute difference for $B$. Thus the
  agreement between the non-linear and linear method in this low
  density regime is again extremely good, for \emph{large} stars as
  well as \emph{small}. Whilst the function $T$ is a coordinate scalar
  for static spherical geometries, $A, B$ are not. An example of
  another coordinate scalar is the 4-dimensional effective density.
  This is the density profile that would reproduce the induced
  geometry in a purely 4-dimensional theory. The non-linear
  ${\rho}nl^{(4)}$ and linear $\rho^{(4)}$ are plotted for the same
  solution and again very good agreement is found, $\sim 3 \%$ peak
  difference compared to peak value. (all lattices: $dr = 0.2$,
  $r_{\rm max} = 20$, $dz = 0.03$, $z_{\rm max} = 46$)
\label{fig:linear_solF} 
}
\end{figure}

Finally, in figure \ref{fig:linear_solF} we plot some of the same
quantities for a star with $\xi = 3$, the \emph{largest} size of star
considered in this paper, again using the same resolution and lattice
size as are used later in the paper. The $T, A, B$ metric functions
are compared on the brane, the location on the lattice where the two
methods give the greatest difference. Also the induced density is
plotted.  Excellent agreement is again found between the methods,
particularly in the induced density, where differences of only $\sim 3
\%$ are found. For the metric functions themselves, maximum
differences of $\sim 10 \%$ are seen relative to the peak values of
the functions.

To conclude, in the low density regime the non-linear method performs
extremely well at the resolutions and lattice sizes used in this
paper. Comparison with the linear theory shows that the constraints
are correctly imposed and the asymptotic geometry is indeed that of
AdS at the horizon, consistent with the linear theory analysis of
section \ref{sec:asym_data}. This also implies that our method to
implement the singular terms at the $r = 0$ axis in the elliptic
relaxation works extremely well, as we see no obvious artifacts
associated with $r = 0$ in the comparisons. The \emph{maximum}
differences in the metric functions for \emph{small} and \emph{large}
stars are $\sim 4 \%$ and $\sim 10 \%$ respectively at the star cores.
Much smaller errors are found in actual brane observables, such as
induced density and redshift, which we will in fact be using later,
with typically $\sim 1-3 \%$ differences for both \emph{small} and
\emph{large} stars.

%
\section{Physical Solutions and Results}
\label{sec:results}
%

Appendix \ref{app:testing} shows in detail that the method outlined
does indeed solve the Einstein equations to a good accuracy using the
resolutions and lattice sizes considered in this paper. The previous
section \ref{sec:linear_check} shows that in the low density regime,
the non-linear method very closely reproduces the linear theory
results which can be numerically computed by a simpler, independent
method. Confident that the method gives solutions of good quality,
allowing physical tests and comparisons to 4-dimensional effective
theory, we now proceed to investigate the non-linear behavior of
Randall-Sundrum stars.

Firstly we study \emph{small} stars, showing typical solutions for
$\xi = 0.3$ (so $R$ is several times less than the AdS length, $l =
1$, in our units), describing their geometry and showing the upper
mass limit is reproduced. These are the first calculations of high
energy density non-linearity on branes, from localized matter. We
expect that the qualitative phenomenon found here, are not specific to
the Randall-Sundrum model. Then \emph{large} stars are considered,
results being shown for $\xi = 3$, so $R \simeq 3$, the largest sizes
that could be relaxed in a reasonable time. The induced geometry on
the brane is shown to be well described by a 4-dimensional effective
theory.  The confinement of the geometric deformation to the brane is
then confirmed for both \emph{large} and \emph{small} stars, in both
the linear and non-linear regime, consistent with a pancake like
scaling predicted in
\cite{Chamblin:1999by,Giddings:2000mu,Garriga:1999yh}. In fact the
degree of confinement is interestingly found to increase for highly
non-linear stars near their upper mass limit, indicating non-linearity
does effect the bulk geometry, even in the case of \emph{large} stars.
Finally we consider in detail how the transition to the 4-dimensional
effective theory proceeds for increasing $R$.

%
\subsection{The Geometry of \emph{Small} Stars}
\label{sec:small_stars}
%

{\bf (Some results
  presented in section \ref{sec:key_results}, `Highlights of Results') \\ }

In this section we consider stars generated with a density profile of
coordinate radius $\xi = 0.3$, a few times smaller than the AdS
length.  We present a series of configurations, all of the same
$\xi$, ranging from the linear to the non-linear regime, the most
non-linear example being the densest star we could numerically compute
for $\xi = 0.3$ with proper radius $R = 0.38$. As we shall see, it is
a highly relativistic object, the numerical method performing most
stably for \emph{small} stars.

\begin{figure}
\centerline{\psfig{file=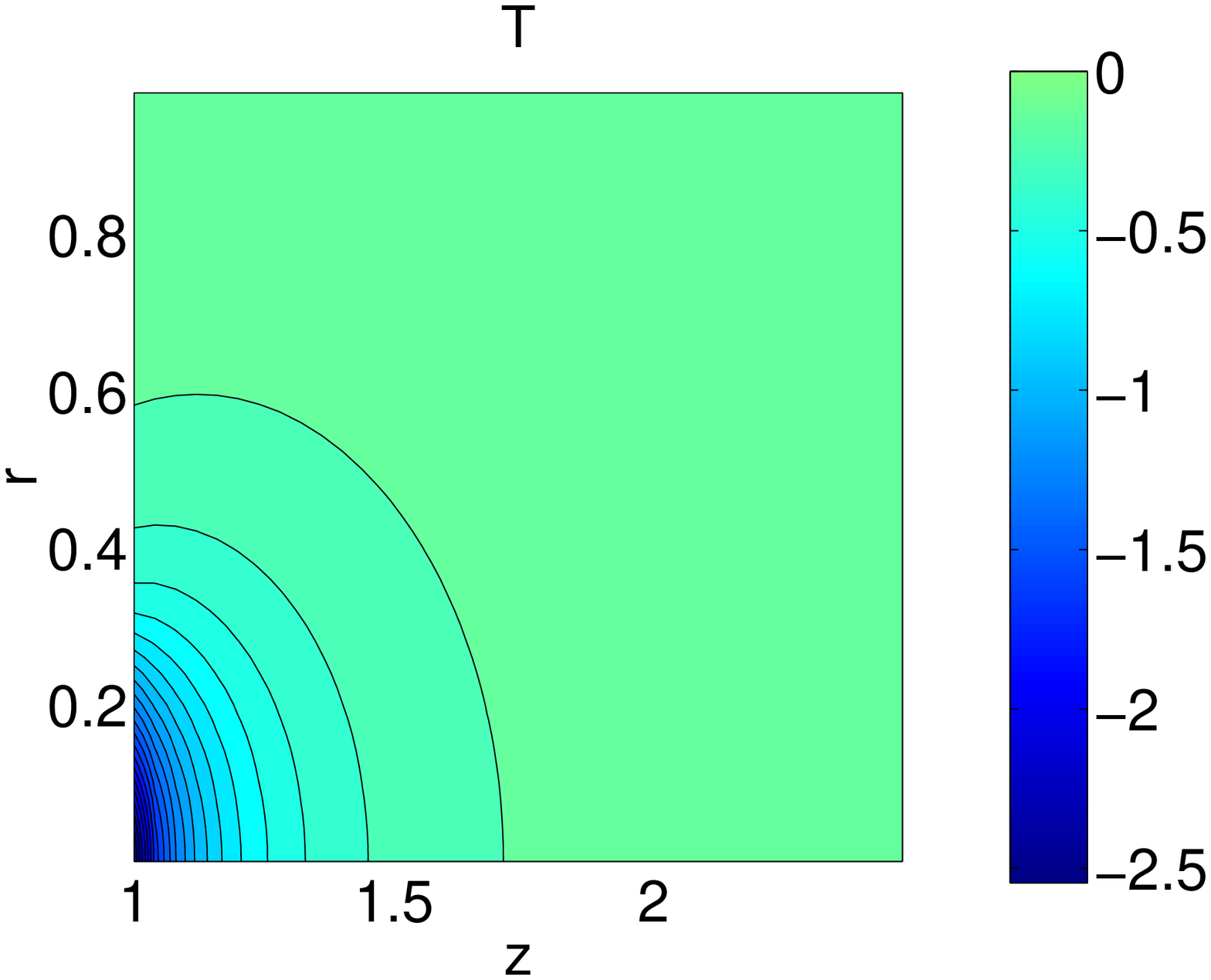,width=6cm} \hspace{0.2cm} \psfig{file=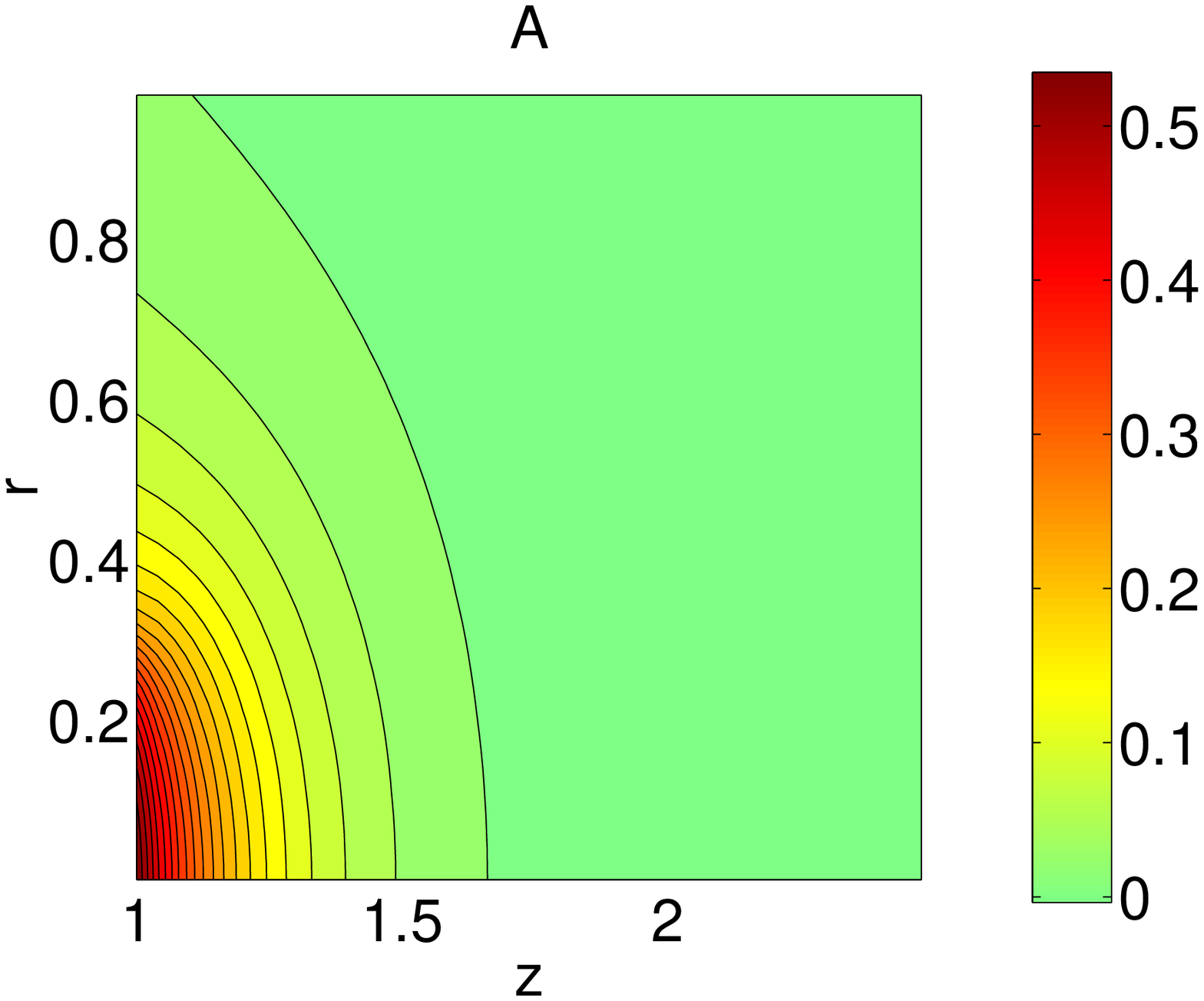,width=6cm} \hspace{0.2cm}
  \psfig{file=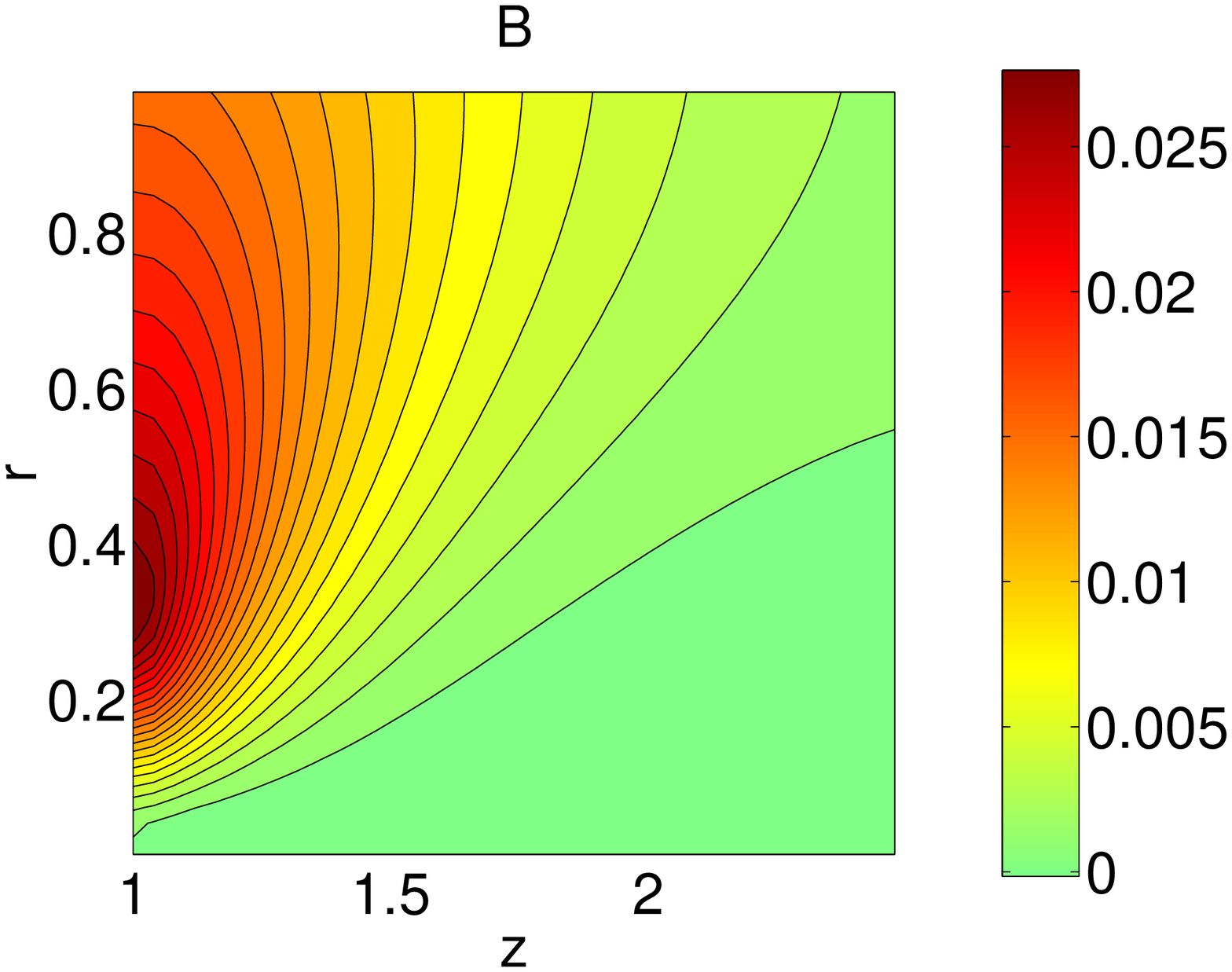,width=6cm}}
\vspace{0.2cm}
\caption[short]{ \figuremode  An illustration of the metric functions $T, A, B$ for a
  configuration with $\xi = 0.3$ and core $\rho_0 = 7.0$. This is an
  extremely relativistic `micro-star', whose proper angular radius is
  $R = 0.38$, approximately a third of the AdS length.  The core value
  of the metric function $T = -2.5$ corresponds to a redshift of
  ${\cal Z} = 12$ for photons created in the star core traveling to $r
  = \infty$ in the brane. The metric function $B$ is much smaller than
  $T$ and $A$ indicating the spatial sections are approximately
  conformal to those of AdS. In addition, the maximum of $|A|$ is
  considerably less than that of $|T|$ indicating the red-shifting
  effects are more pronounced than spatial deformations from the
  unperturbed AdS. The solution is thought to be close to the upper
  mass limit which appears to be present for both \emph{small} and
  \emph{large} stars.  The value of $\rho_0$ is actually slightly
  larger than the brane tension $\sigma = 6$.  (lattice: $dr = 0.02$,
  $r_{\rm max} = 2$, $dz = 0.005$, $z_{\rm max} = 4$)
\label{fig:A8_metric} 
}
\end{figure}

\begin{figure}
\centerline{\psfig{file=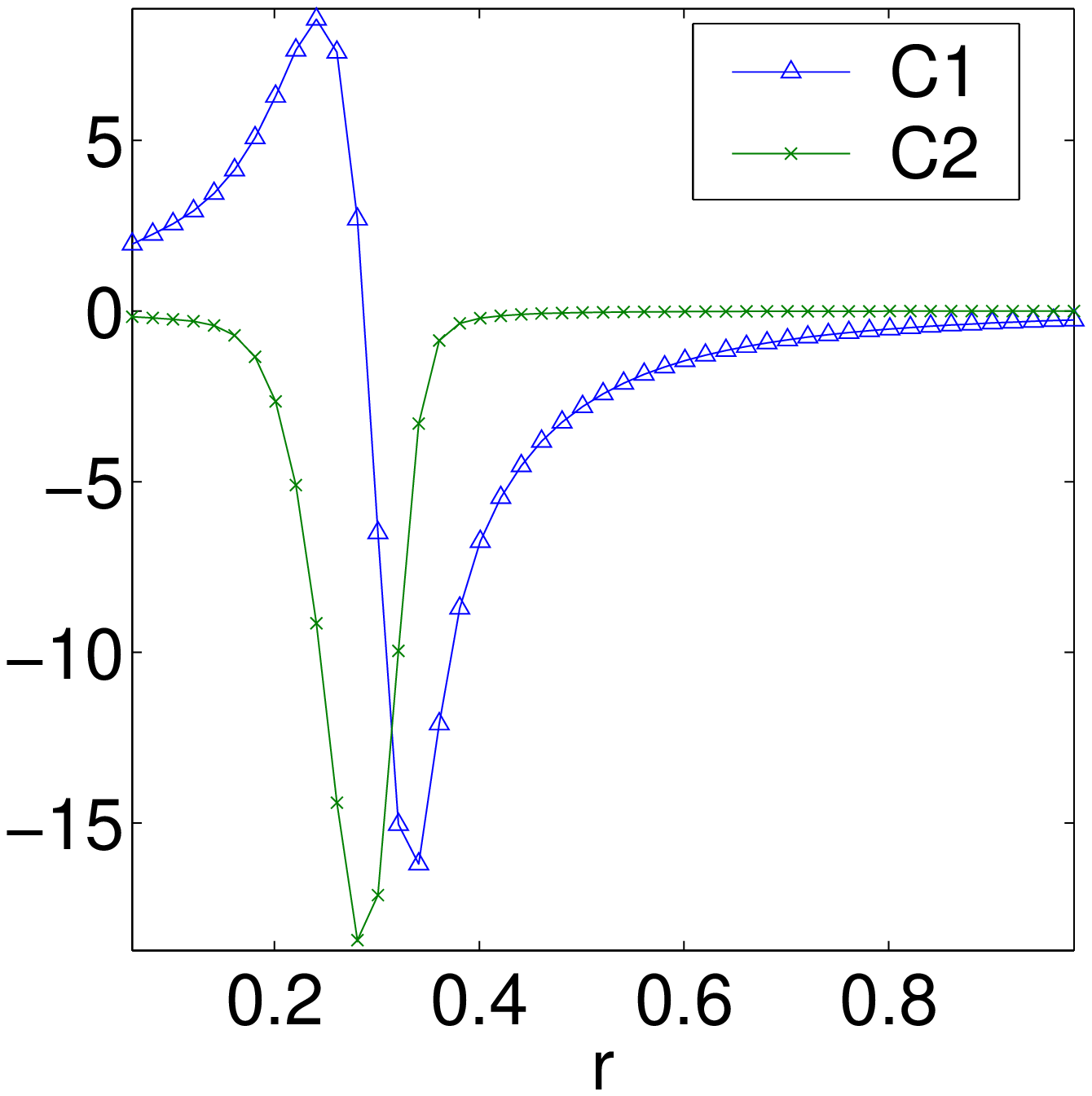,width=6cm}
  \hspace{0.2cm} \psfig{file=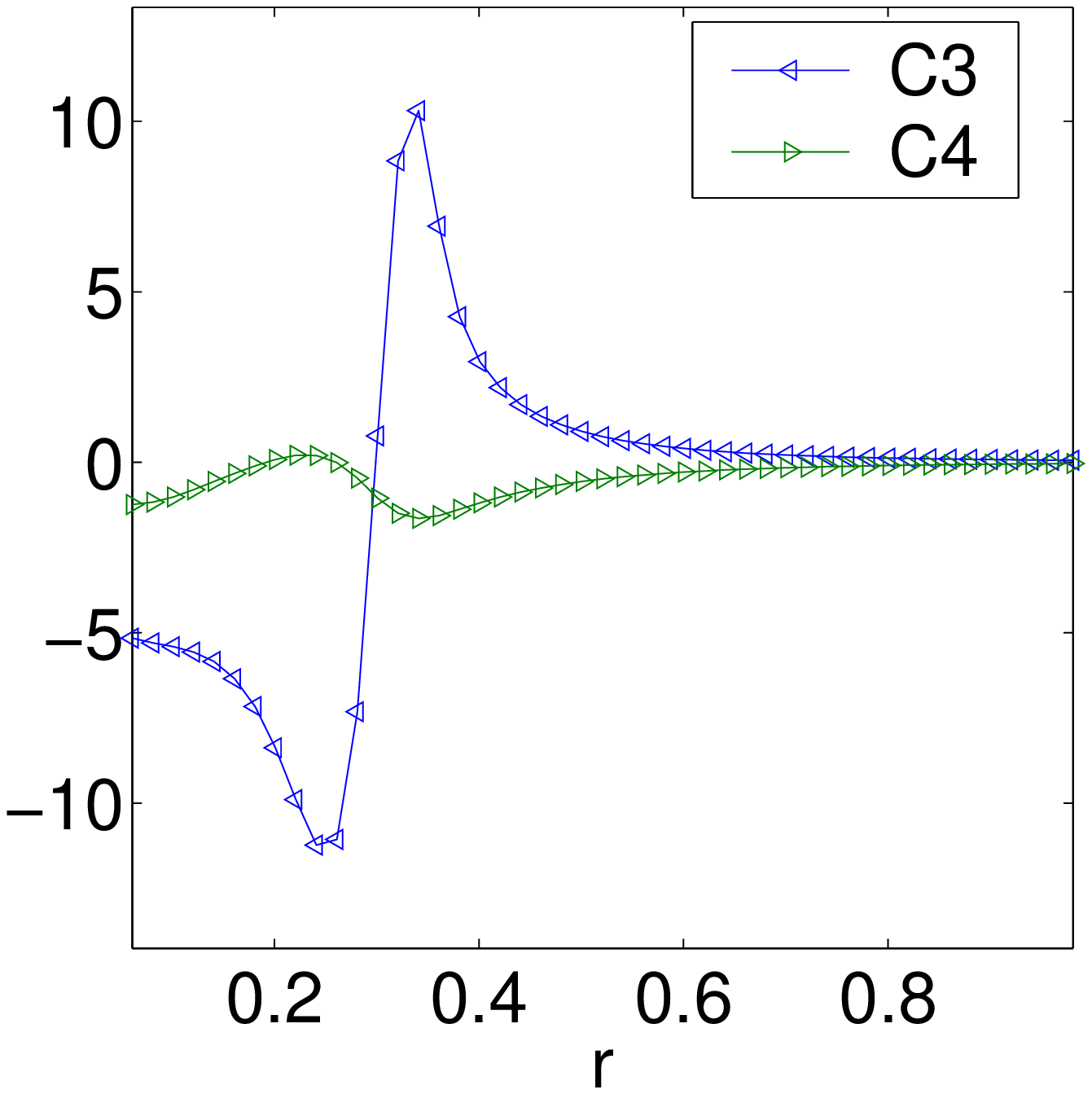,width=6cm}} 
\caption[short]{ \figuremode  An illustration of the Weyl components (defined in appendix
  \ref{app:einstein_eqns}) calculated on the brane for the same
  configuration as in figure \ref{fig:A8_metric}. The curvatures
  measuring up to $\sim 10-15$ are comparable to the characteristic
  curvature from the negative bulk cosmological constant which gives a
  Ricci scalar $|{\cal R}| = 20$.
\label{fig:A8_weyl} 
}
\end{figure}

We start by showing the form of the metric for the most non-linear
solution corresponding to a core density $\rho_0 = 7.0$, which is
larger than the brane tension, $\sigma = 6$. The metric functions are
shown in figure \ref{fig:A8_metric}. The most striking feature is that
the configuration shows confinement of the perturbation, just as for
the linear solutions (the metric functions of less dense stars also
with $\xi = 0.3$ are found in figure \ref{fig:linear_sol3} of section
\ref{sec:linear_check} and figure \ref{fig:origin_check1} in appendix
\ref{app:testing}).  The metric function $B$ is much smaller than the
other two functions and thus, from the metric \eqref{eq:metric_conf},
the spatial sections are well approximated by a conformal deformation
of those of AdS. The metric function $T$ has a peak magnitude of
$2.5$, greater than one, giving rise to large redshift effects. The
peak value of $A$ is still less than one, indicating the non-linearity
is less pronounced in the spatial perturbations. In figure
\ref{fig:A8_weyl} we obtain a measure of the curvatures on the brane
by plotting the scaled Weyl components given in appendix
\ref{app:einstein_eqns} for the most non-linear configuration. It is
clear that the curvatures generated in the solution are large, even
compared to the bulk Ricci scalar, $|{\cal R}| = 20$.

\begin{figure}
\centerline{\psfig{file=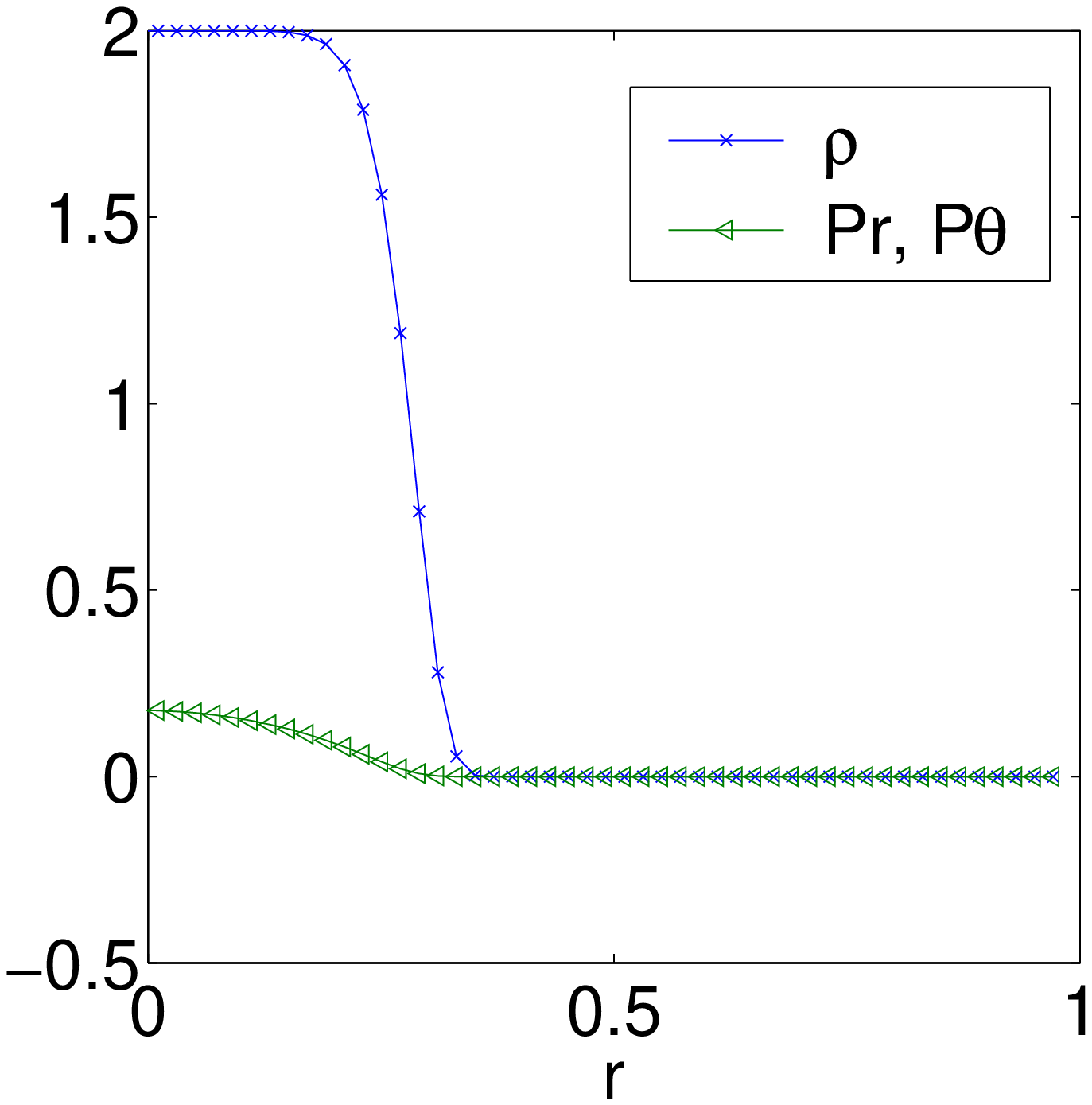,width=6cm} \hspace{0.2cm}
  \psfig{file=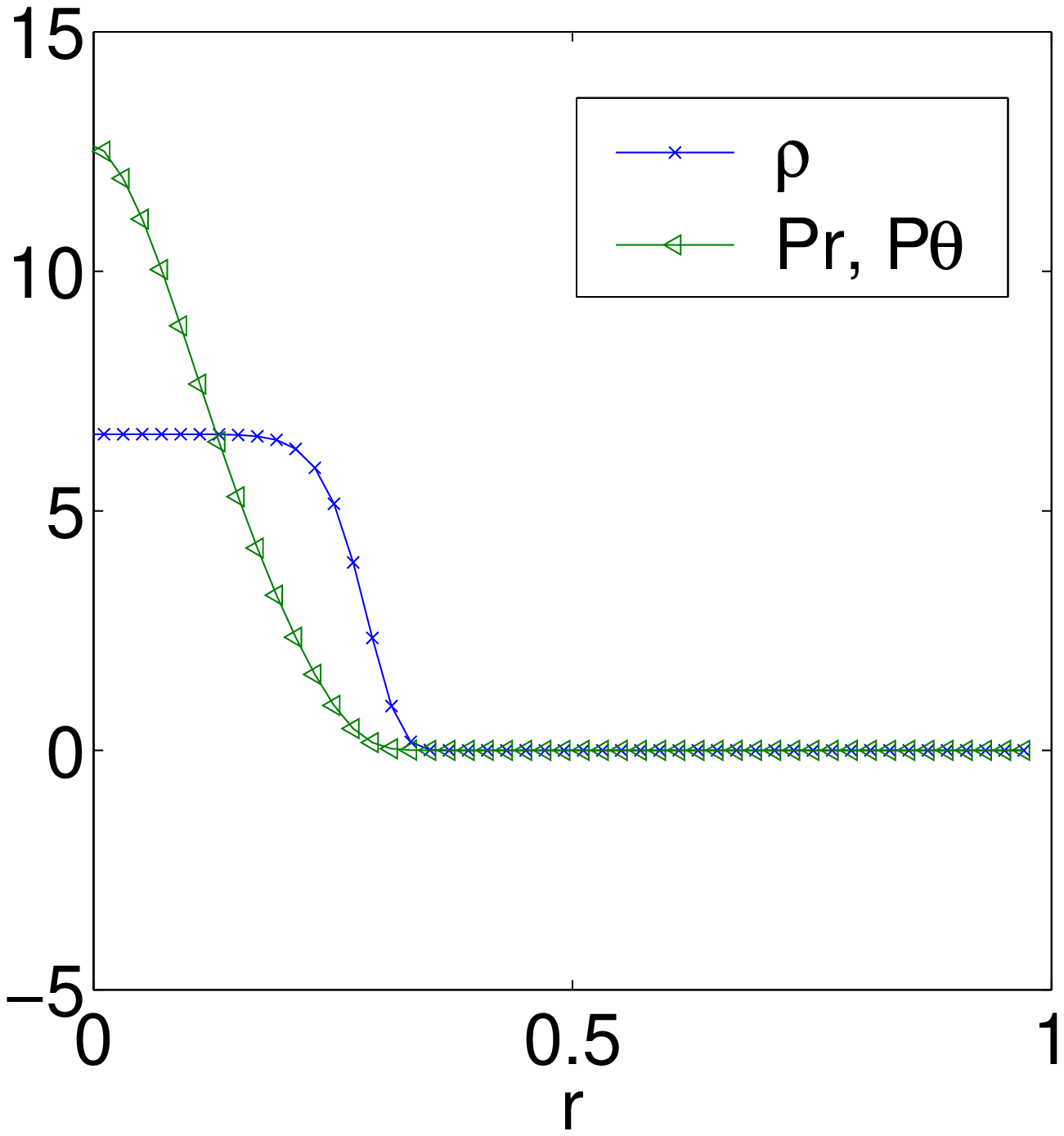,width=5.8cm} \hspace{0.2cm} \psfig{file=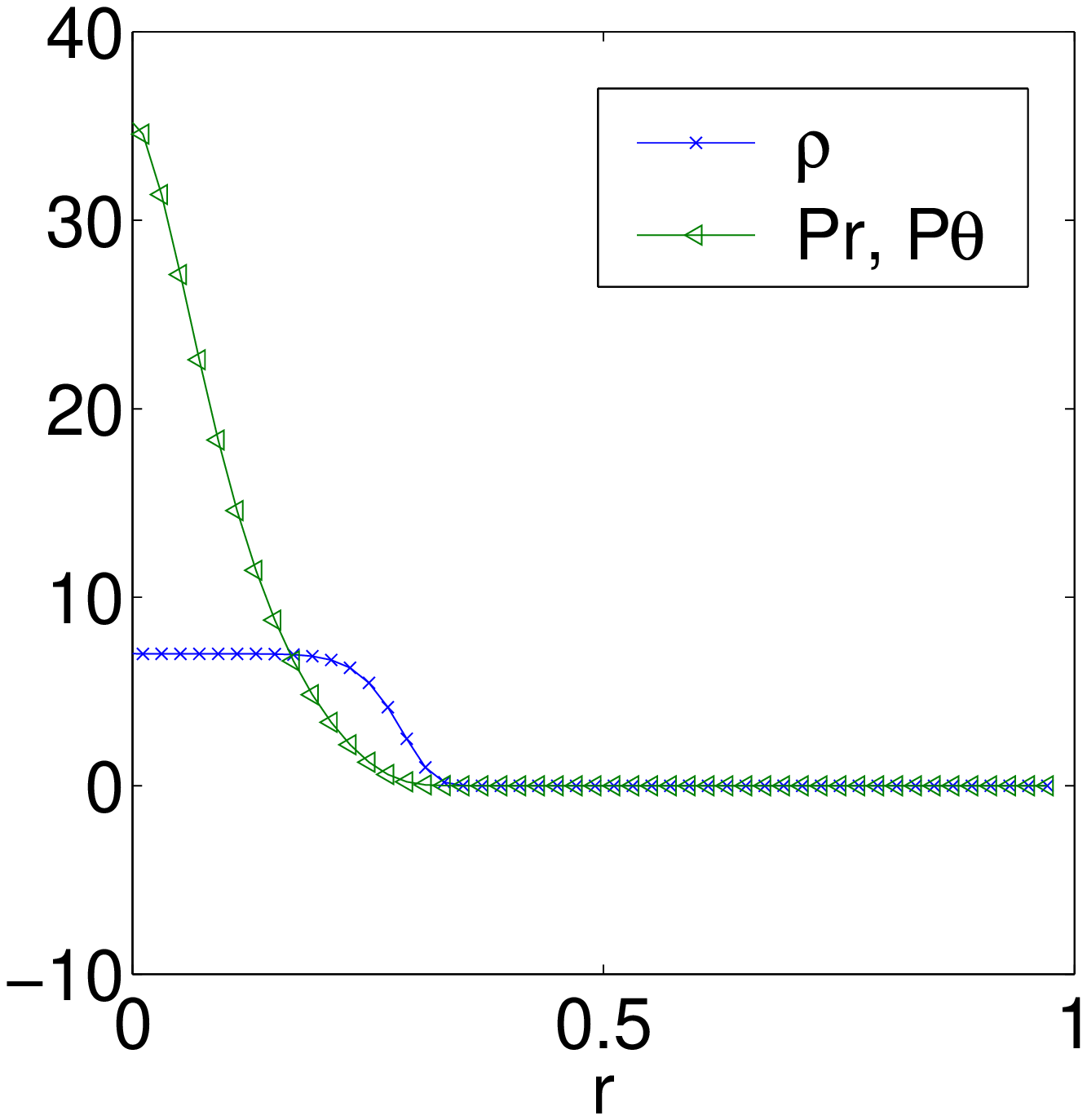,width=6cm}}
\caption[short]{ \figuremode  An illustration of $\rho, P_r,
  P_{\theta}$ for configurations with $\xi = 0.3$. The core densities
  $\rho_0 = 2.0, \, 6.6, \, 7.0$ from left to right. The lowest
  density, still relatively large compared to the brane tension
  $\sigma = 6$, yields a pressure visibly much smaller than the
  density. The middle and right solutions have core pressures larger
  than the core density indicating the extreme non-linearity of the
  solutions.  Note that for just a $6 \%$ change in density from $\rho
  = 6.6$ to $\rho = 7.0$ the core pressure over density increases by a
  factor of $\sim 2.5$.  The density and pressures are measured from
  the brane matching conditions.  Whilst the density is input as a
  boundary condition, the pressures are derived from the solution. The
  isotropy boundary condition functions well, the radial and angular
  pressure components being indistinguishable in such a plot.  (all
  lattices: $dr = 0.02$, $r_{\rm max} = 2$, $dz = 0.005$, $z_{\rm max}
  = 4$)
\label{fig:A8_act_density} 
}
\end{figure}

The form of the density profile is an input to the solution, and from
the remaining brane matching conditions we can extract the pressure on
the brane. There are two components of pressure, the radial and
angular components, but the boundary condition of isotropy ensures
that these are equal. Figure \ref{fig:A8_act_density} shows the
pressure calculated for various density profiles, each with $\xi =
0.3$ but with different $\rho_0$. The solutions range from the near
linear where $P << \rho$ to the highly non-linear where $P >> \rho$.
The most nonlinear solution that was relaxed (corresponding to the
solution in figure \ref{fig:A8_metric}) is plotted and gives an
extremely large central core pressure, with $P / \rho \simeq 5$.

The limiting behavior can be seen by plotting the ratio of core
pressure to density against the core density. In the Newtonian theory
there is a quadratic dependence of the pressure on the density. In
figure \ref{fig:Adensitycurve} (found in the `Highlights of Results'
section \ref{sec:key_results}), we explicitly see a departure from the
linear dependence of $P / \rho$ on $\rho$, which holds for very low
densities. Instead we find diverging behavior as $\rho_0 \rightarrow
7$. For stars with higher core density, no convergent numerical
solution was found. The apparent divergence in $P / \rho$ strongly
suggests that the reason we cannot relax denser stars is that the
static solutions do not exist, in analogy with the 4-dimensional case.
Thus even for \emph{small} stars we have the same qualitative behavior
of an upper mass limit as in standard GR. The task of investigating
the dependence of the limiting mass on radius for $R << 1$ is left for
future work, and may have interesting implications for micro black
hole formation.

\begin{figure}
\centerline{
  \psfig{file=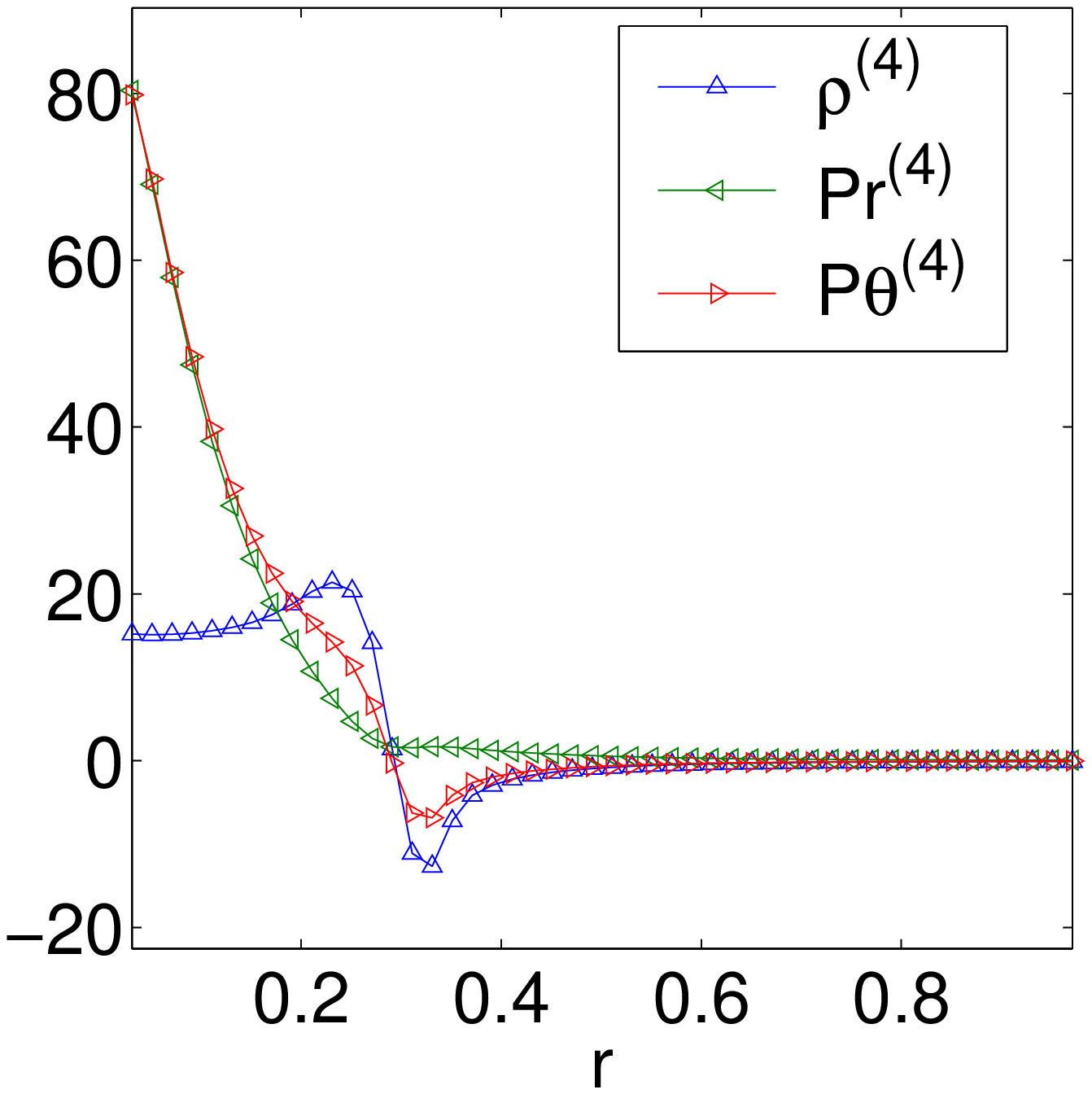,width=6cm} \hspace{0.1cm}
  \psfig{file=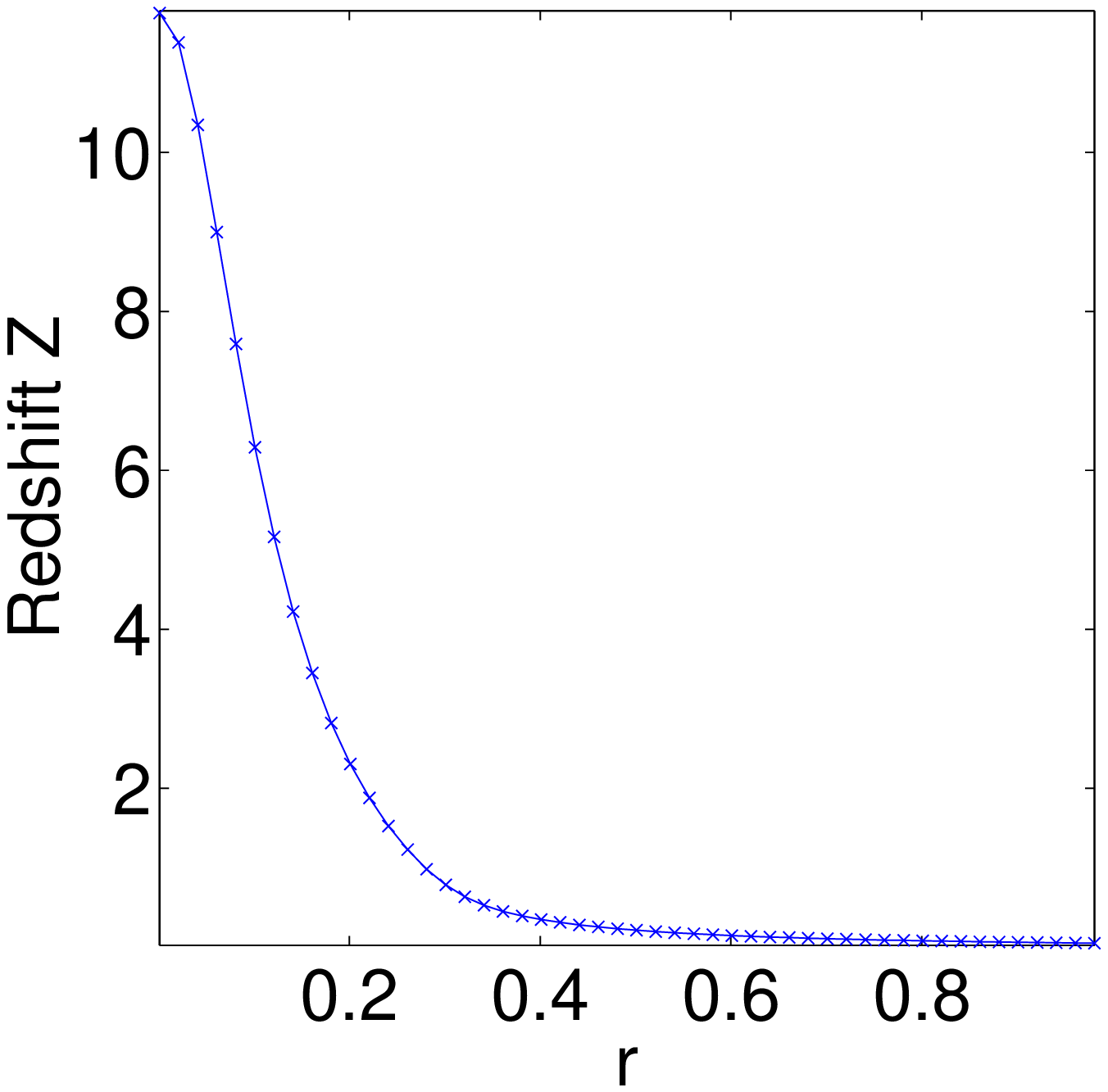,width=6.5cm}}
\caption[short]{ \figuremode  On the left, an illustration of induced $\rho^{(4)}, P_r^{(4)},
  P_{\theta}^{(4)}$ for the $\xi = 0.3$, $\rho_0 = 7$ solution shown in
  figure \ref{fig:A8_metric}. The curves have a similar qualitative
  form to the right hand plot of figure \ref{fig:A8_act_density}, the
  actual 5-dimensional density and pressures for the solution, but
  quantitatively deviate by a factor $\sim 2$. As we expect,
  4-dimensional gravity does not describe the 5-dimensional induced
  geometry well for \emph{small} stars.  On the right, for the same
  solution an illustration of the redshift of photons emitted from
  coordinate position $r$ on the brane and observed asymptotically at
  large $r$ on the brane. The redshift of photons emitted from the
  stellar core is ${\cal Z} \sim 12$ indicating the extremely
  relativistic nature of this `micro-star' solution.
\label{fig:A8_observables} 
}
\end{figure}

For the purposes of this paper we are primarily interested in how
closely the induced brane geometry is described by a purely
4-dimensional local description, and therefore we wish to calculate
intrinsic properties of this brane geometry. The
intrinsic metric on the brane at $z=1$ is,
\begin{equation}
ds^2_{\rm induced} = \left. - e^{2 T(r,z)} dt^2 + e^{2 \left(A(r,z)+B(r,z)\right)}
  dr^2 + e^{2 \left(A(r,z)-B(r,z)\right)} r^2 d\Omega_2^2 \, \right|_{z=1}
\label{eq:metric_induced}
\end{equation}
and then we may calculate a complete set of induced 4-geometric
quantities such as the Einstein tensor and the Weyl tensor which
together specify the geometry. In 4-dimensions with static spherical
symmetry there is one independent Weyl component and 3 Einstein tensor
components.  We characterize the Einstein tensor components in terms
of the effective density, $\rho^{(4)}$, radial pressure,
$P_{r}^{(4)}$, and angular pressure component, $P_{\theta}^{(4)}$,
that would result in such a geometry for 4-dimensional gravity. Their
explicit form is given in appendix \ref{app:einstein_eqns}. This is a
useful characterization of the curvature as, later, in the
\emph{large} star case, we see that the induced density and pressure
agree with the actual quantities calculated from the brane matching
conditions. Another quantity we compute is the metric component $T$,
which due to the static symmetry is a scalar function under $r, z$
coordinate transformations. Thus for static configurations the
redshift, ${\cal Z}$, of photons emitted from the core of a star to
infinity on the brane,
\begin{equation}
e^{- T(r=0,z=1)} = 1 + {\cal Z} 
\end{equation}
is a well defined physical quantity. Such intrinsic quantities are
plotted for the most non-linear solution with $\xi = 0.3$, and are
found in figure \ref{fig:A8_observables}. We see that the induced
density and pressure have similar forms to the 5-dimensional
quantities but have approximately twice the value.  

\begin{figure}[htb]
\centerline{\psfig{file=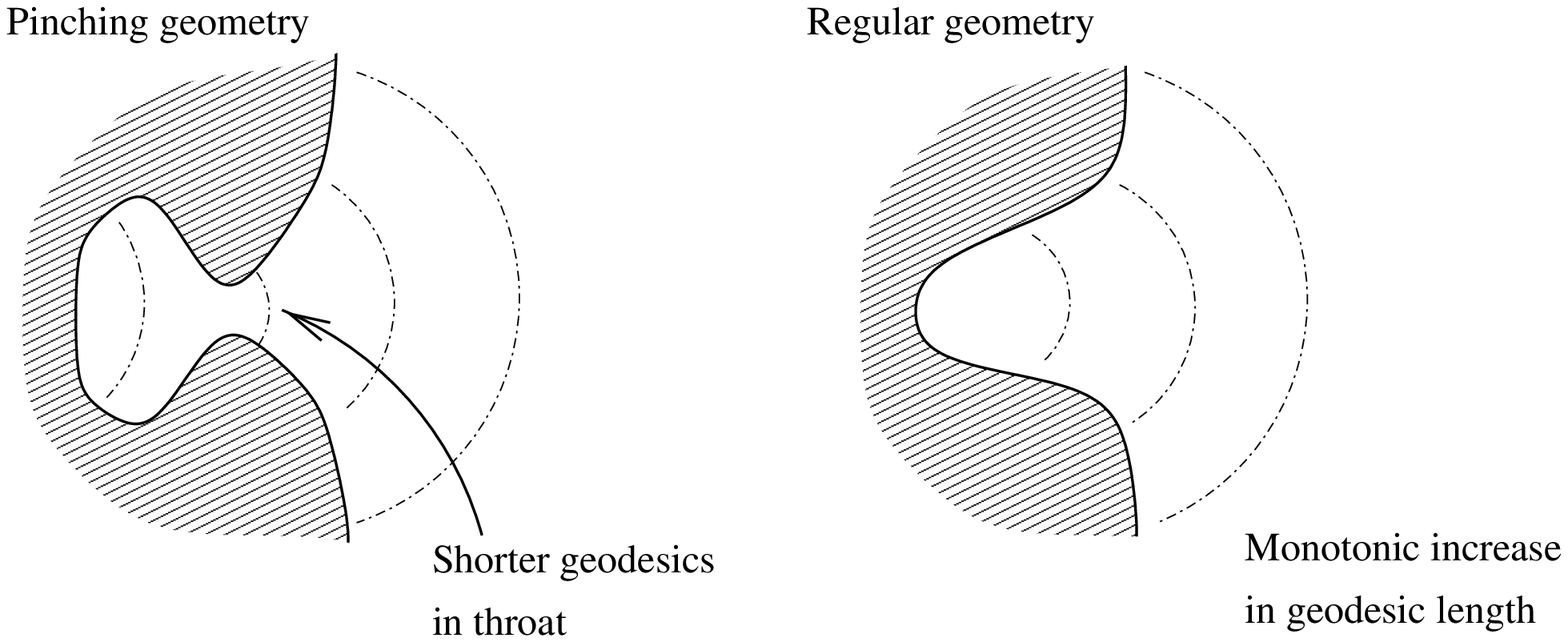,width=14cm}}
\caption[short]{ \figuremode  A schematic illustration of a brane
  `pinching off' a stellar region. Plotting the length of symmetric
  spatial geodesics against radial coordinate distinguishes the
  left-hand from the right-hand cases. A pinched geometry could lead to
  pathologies at finite stellar density, the matter region becoming
  entirely trapped. Instead we find our solutions have monotonically
  increasing spatial geodesic length corresponding to the right-hand
  case.
\label{fig:pinched} 
}
\end{figure}

\begin{figure}
\centerline{\psfig{file=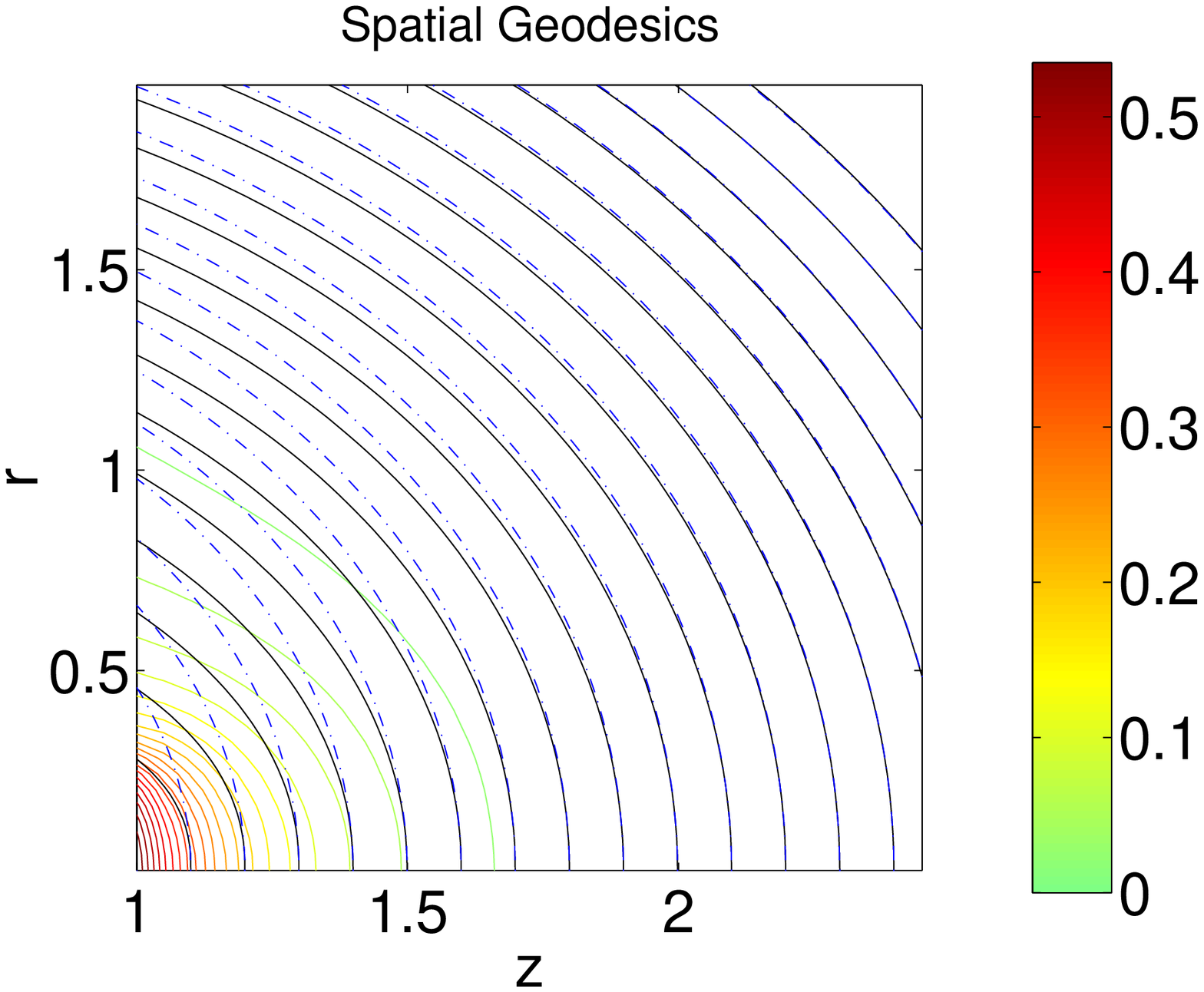,width=8.5cm} \hspace{1cm}
  \psfig{file=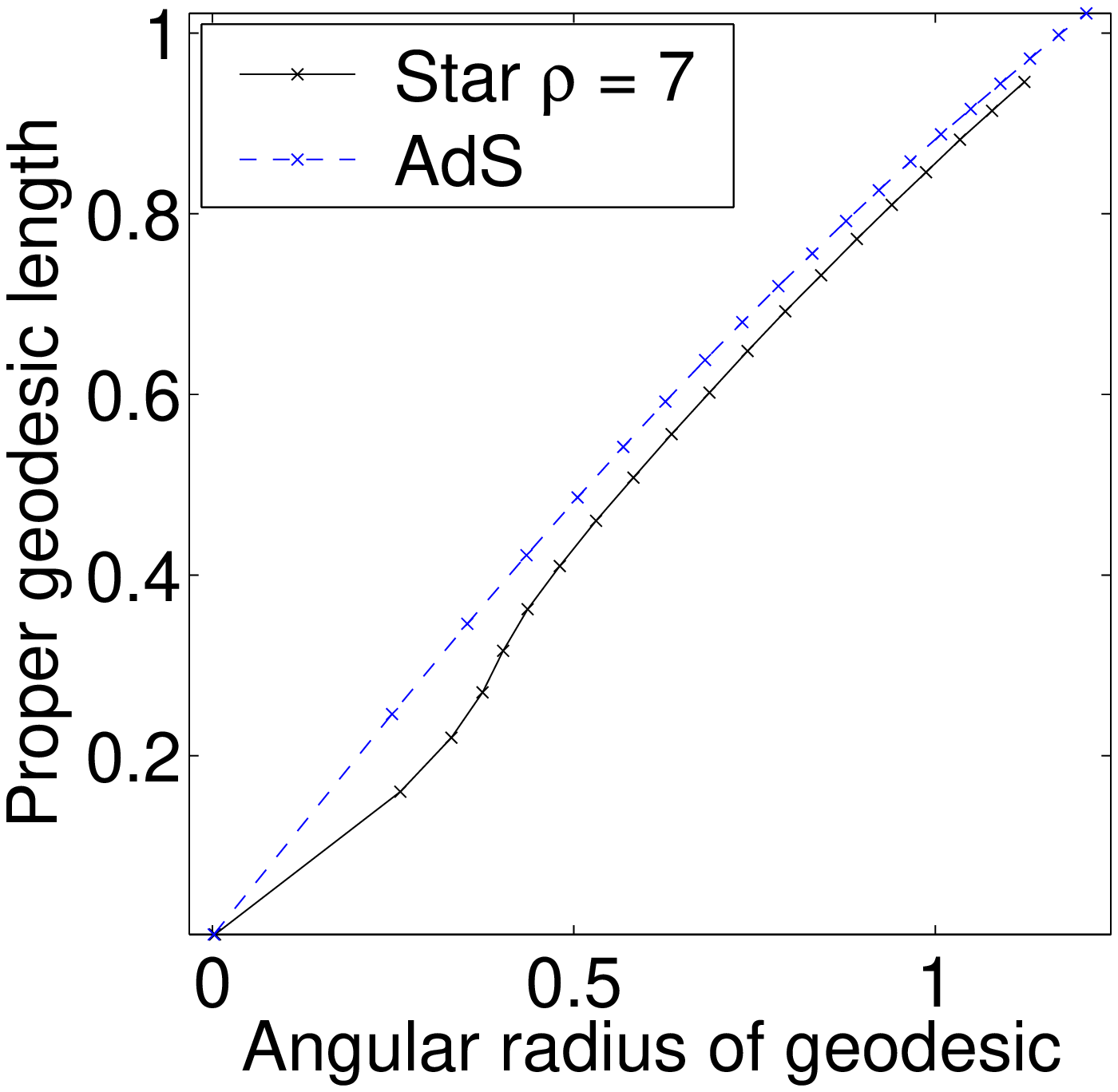,width=7cm}}
\vspace{0.1cm}
\caption[short]{ \figuremode  An illustration of spatial geodesics for the solution $\xi =
  0.3$, $\rho_0 = 7$ shown in figure \ref{fig:A8_metric}. Note that
  the geodesics (blue and black) are superimposed on a contour plot of
  the metric function $A$ (red - green). As the metric function $B
  \sim 0$ the spatial geometry is simply conformal to hyperbolic
  space, the conformal factor being $e^{2 A}$. The dotted blue lines
  are the AdS geodesics, which are simply circles centered on the
  boundary $z=0$ in these coordinates. The black solid lines are the
  geodesics calculated for the star geometry. We see clear deviations
  in the star geometry from the unperturbed AdS. The right plot shows
  the path length against proper angular radius of interception with
  the brane. The function is smooth and monotonically increasing
  indicating that there is no `pinching' off of the section of the
  brane containing the star.
\label{fig:A8_geo} 
}
\end{figure}

The largest metric deviation is in $T$. This is best characterized in
terms of the core redshift of photons which gives a value of ${\cal Z}
\simeq 12$, indicating the non-linearity of the solution.  The spatial
curvature of the metric is large too.  A striking feature of the
solutions is that the metric function $B$ is very small compared to
the other two functions.  If we now approximate $B \sim 0$ then the
spatial metric becomes,
\begin{equation}
ds_{\rm spatial}^2 = e^{2 A} \frac{1}{z^2} \left( dr^2 + r^2 d\Omega^2
+ dz^2 \right) 
\end{equation}
which just corresponds to a conformal transformation of flat sliced
hyperbolic space. This allows us to understand the spatial geometry by
considering the metric function $A$ in figure \ref{fig:A8_metric} as
this conformal factor. The positive $A$ implies that the volume in the
bulk near the star on the brane is `more' than in the unperturbed
case.  This indicates that the solution does not pinch off the portion
of the brane containing the matter, an idea illustrated in figure
\ref{fig:pinched}. We more rigorously show this by plotting, in figure
\ref{fig:A8_geo}, the geodesics of this spatial geometry which are
symmetric about $r = 0$. The relevant AdS spatial geodesics are
circles centered on $z=0$ in these coordinates. We see the actual
distortion of the geodesics from the non-linear curvature. The proper
distance of the curved geodesics does indeed increase monotonically
for increasing $r$ intersection with the brane indicating that no
pinching or geometric pathologies are occurring.

%
\subsection{The Geometry of \emph{Large} Stars}
\label{sec:large_stars}
%

We now consider the geometry of the \emph{largest} stars relaxed,
having coordinate size $\xi = 3$, the characteristic proper radius
being a little larger, and therefore several times the AdS length. For
these larger stars we were able to relax configurations with core
redshifts of ${\cal Z} \simeq 2.1$. The key result of this section is that
for \emph{large} stars, the effective theory on the brane is indeed
4-dimensional General Relativity even when the configuration becomes
non-linear.

Note that ${\cal Z} \simeq 2.1$ is not as close to the upper mass
limit as for the densest \emph{small} star of the previous section.
For higher densities the numerical scheme gave no convergent solution,
although this is almost certainly an artifact of the scheme and not an
indication that higher density solutions do not exist. We later
estimate that the ${\cal Z} \simeq 2.1$ solution has a core density
that is $\sim 75 \%$ of the top-hat upper limit for its proper radius.

\begin{figure}
\centerline{\psfig{file=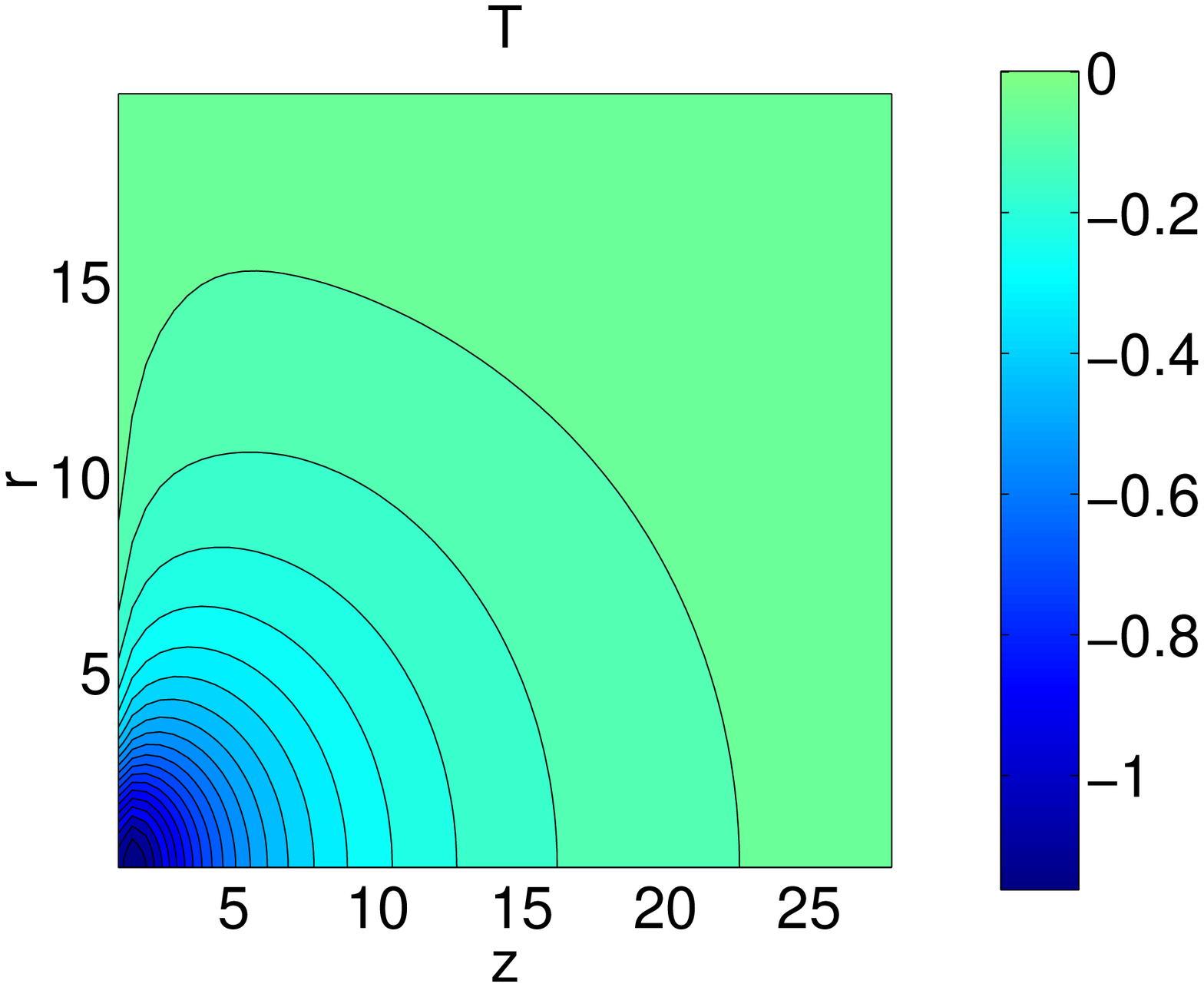,width=6cm} \hspace{0.1cm} \psfig{file=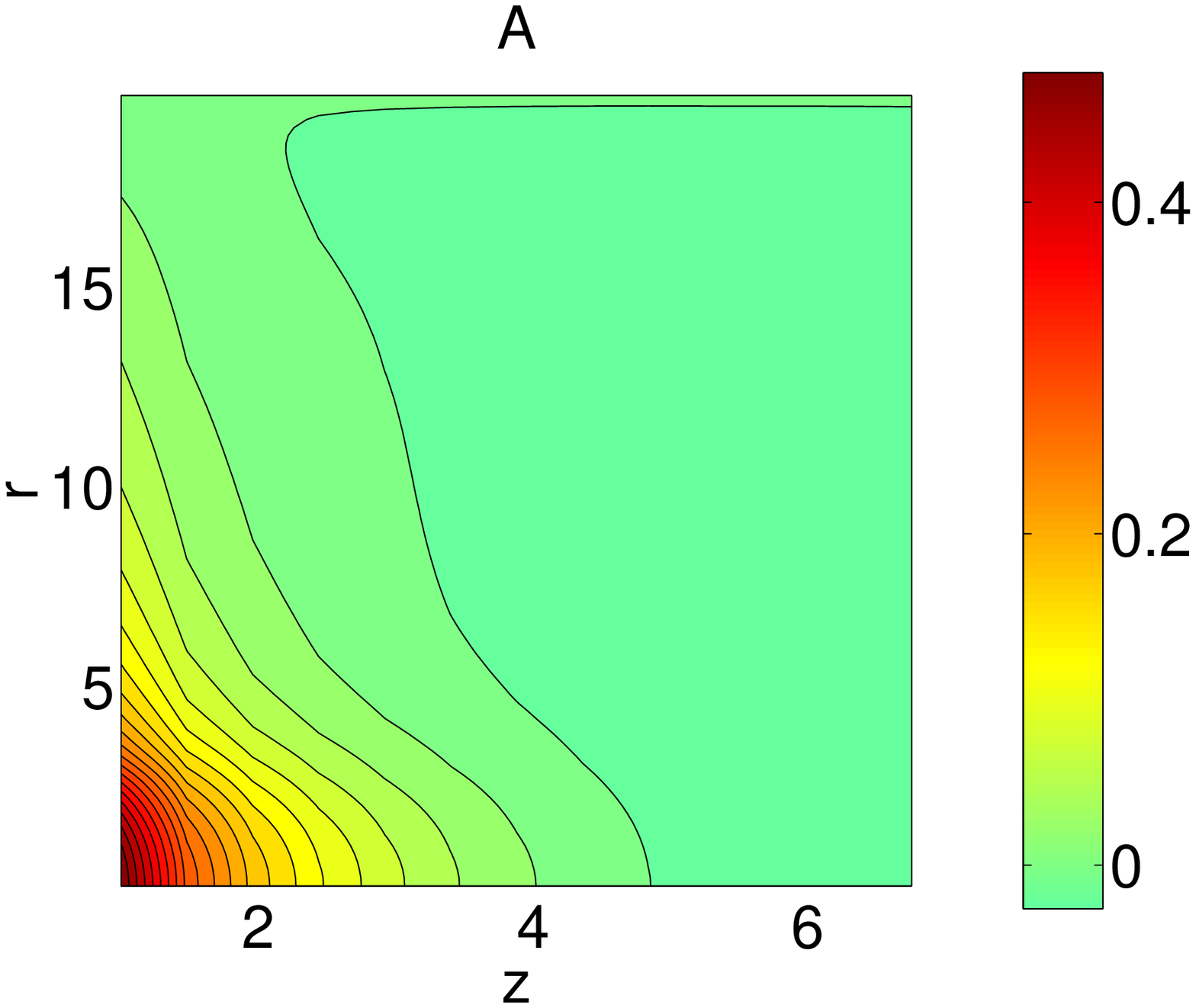,width=6cm} \hspace{0.1cm}
  \psfig{file=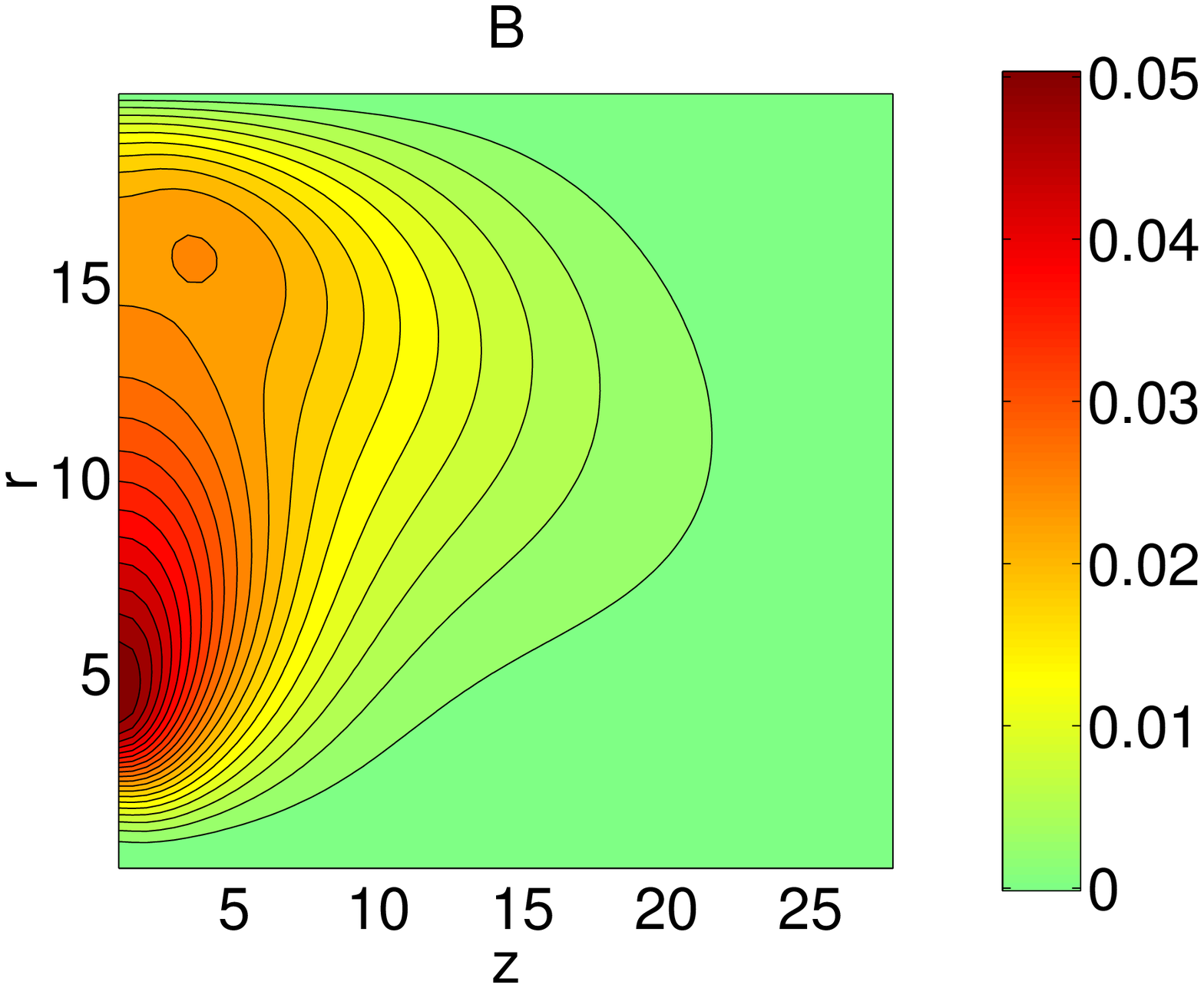,width=6cm}}
\vspace{0.2cm}
\caption[short]{ \figuremode  An illustration of $T, A, B$ for a \emph{large} star with $\xi
  = 3$ and $R = 3.7$. The core density $\rho_0 = 0.15$ corresponds to
  the densest star our method relaxed stably for this $\xi$. We later
  compute that this core density is at least $75 \%$ of the top-hat
  upper limit for a star of such a radius.  Again the characteristic
  $z$ confinement of $T$ and $A$ is seen.  A reduced $z$ range is
  plotted for $A$ to graphically resolve the detail. As with the
  \emph{small} stars, $B$ is much smaller than $T, A$ and indicates
  that the spatial geometry is simply conformal to flat sliced
  hyperbolic space. The maximum perturbation from AdS appears in $T$
  where a peak value of $|T| \simeq 1.1$, which gives rise to a
  redshift of ${\cal Z} \simeq 2.1$ for photons emerging from the core
  to infinity on the brane. This is not a small perturbation and is
  later (section \ref{sec:non-linear}) shown to be beyond the reach of
  higher order perturbation theory.  Note that the $z$ depth of the
  lattice is actually $z_{\rm max} = 46$ in order to ensure the
  asymptotic behavior is good (only $z < 25$ is actually plotted).
  Thus many points are required in the z-direction to maintain AdS
  length resolution. (lattice: $dr = 0.2$, $r_{\rm max} = 20$, $dz =
  0.03$, $z_{\rm max} = 46$)
\label{fig:F8_metric} 
}
\end{figure}

\begin{figure}
\centerline{\psfig{file=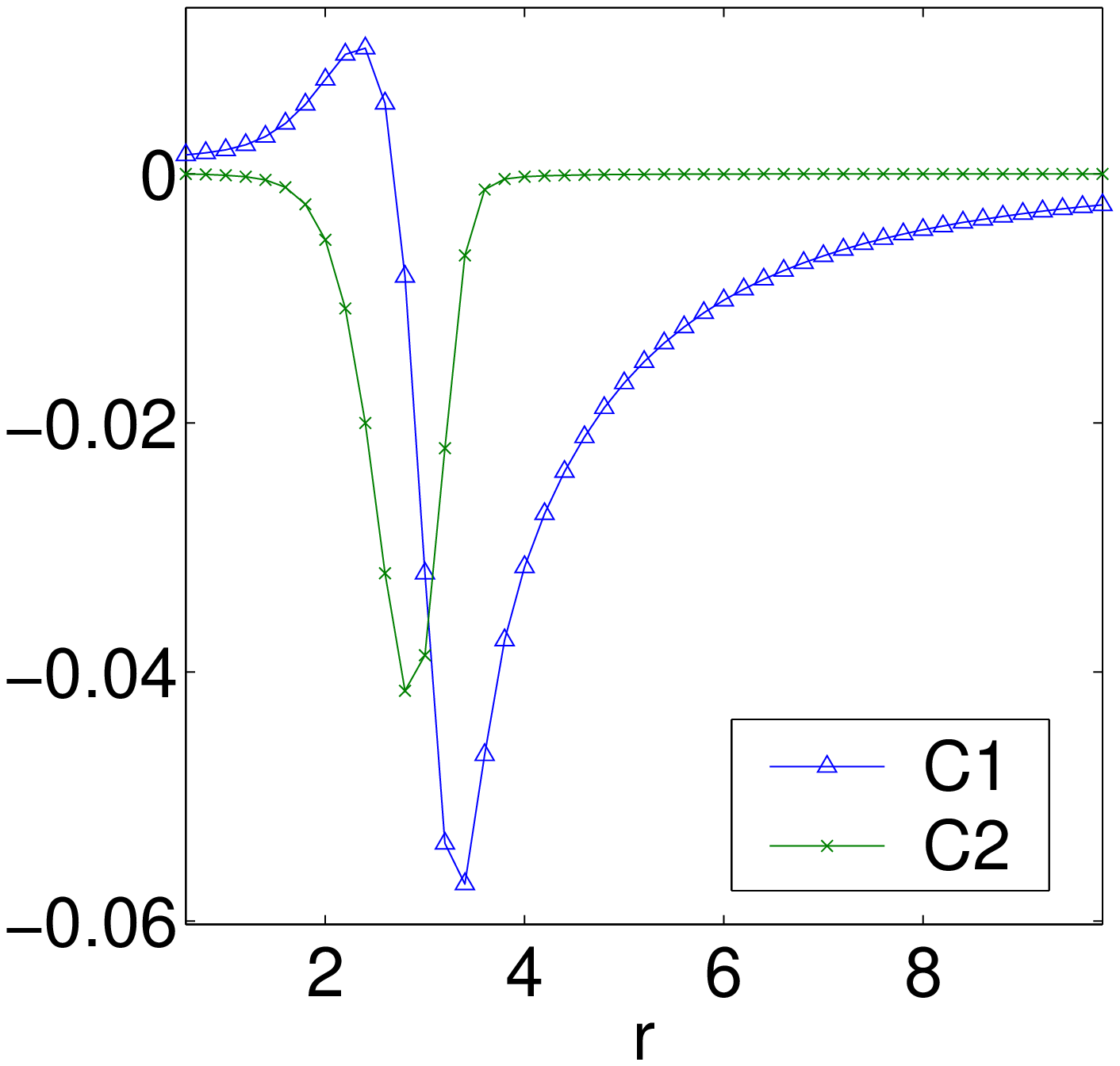,width=6cm}
  \hspace{0.2cm} \psfig{file=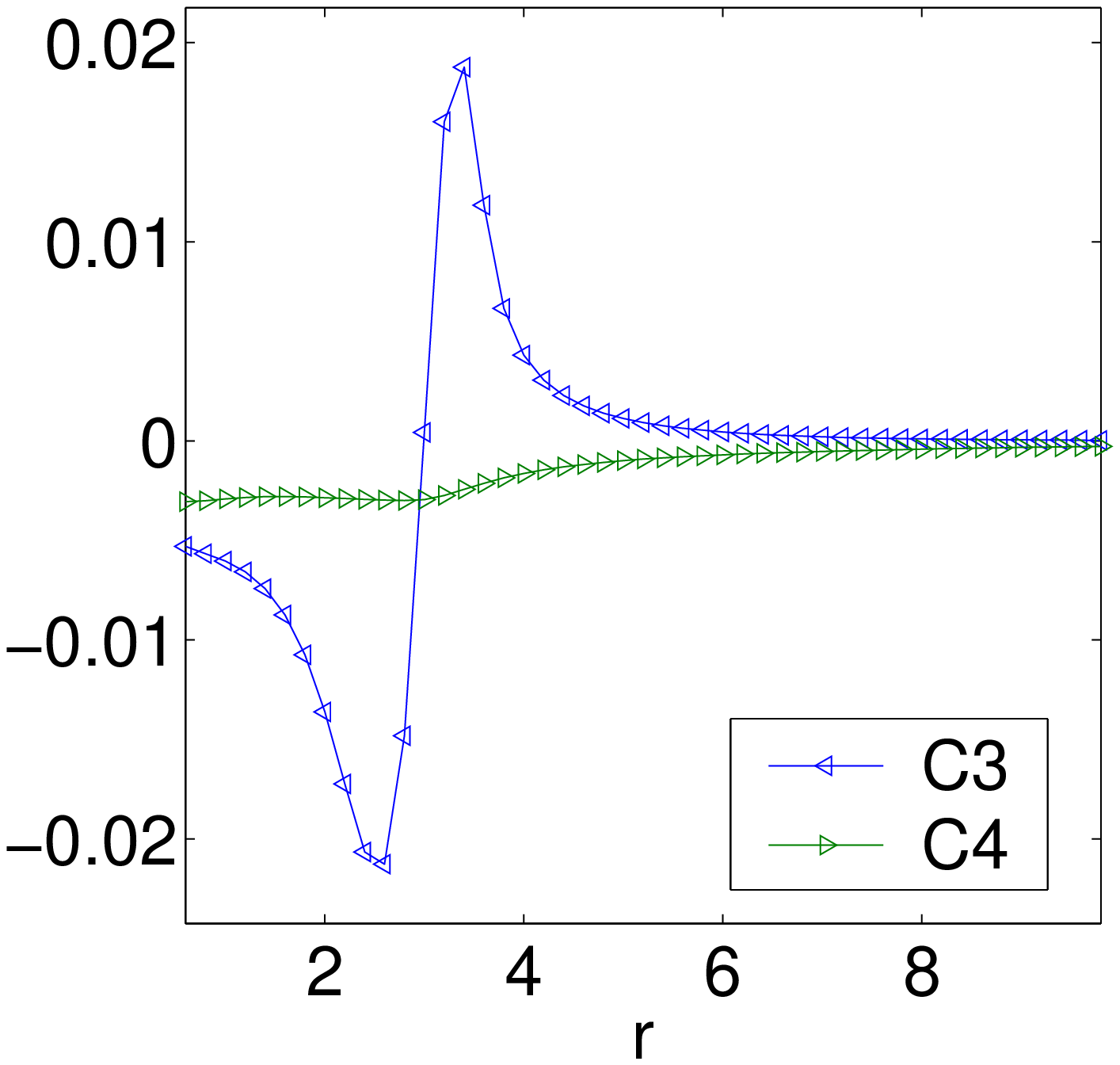,width=6cm}} 
\caption[short]{ \figuremode  An illustration of the Weyl components $C1, C2, C3, C4$
  (described in appendix \ref{app:einstein_eqns}) on the brane. These
  indicate the magnitude of the curvature perturbation. For the
  \emph{small} stars the densities and curvatures are large compared
  to the brane tension and unperturbed AdS curvature. Now for
  \emph{large} stars both the density ($\rho_0 = 0.15$) and typical
  curvature is small compared to the fundamental bulk and brane
  scales. This implies that the solution, whilst non-linear, does not
  probe the scales associated with the higher dimensions, the metric
  functions remaining bounded and well behaved. If a 4-dimensional
  effective theory is to be reproduced, this separation of curvature
  scales between the fundamental ($\Lambda, \sigma$), and star matter
  scales ($\rho_0$, $R$), should be observed for \emph{large} objects.
  (lattice: $dr = 0.2$, $r_{\rm max} = 20$, $dz = 0.03$, $z_{\rm max}
  = 46$)
\label{fig:F10_weyl} 
}
\end{figure}

Figure \ref{fig:F8_metric} shows the metric functions for the most
dense star relaxed. We see that as for the \emph{small} stars, again
the function $B$ is very small compared to $T, A$. $T$ is large in
value, reaching a peak on the brane at $r=0$ of $|T| \simeq 1.1$. The
form of the metric functions appears qualitatively similar to those of
the \emph{small} stars, although less localized in the $r$ direction.
This is to be expected from the linear theory \cite{Giddings:2000mu},
which shows that the asymptotic behavior of the propagator on the
brane is $\sim \frac{1}{r^2}$ at \emph{small} scales and $\sim
\frac{1}{r}$ for \emph{large}. Again the metric functions are
localized in $z$. In fact $A$, whilst positive near the brane, falls
off quickly and becomes slightly negative (just visible in the plot),
before asymptotically decaying to zero, consistent with the asymptotic
behavior predicted in the linear theory \eqref{eq:lin_TAB}. The
relation between the typical proper distance that the metric functions
protrude, and the stellar radius $R$, is discussed later in section
\ref{sec:confinement}. For the linear stars we expect such
localization, but the configuration shown is not a small perturbation
and again localization is exhibited. The 5-dimensional Weyl components
are plotted in figure \ref{fig:F10_weyl} to show the magnitude of the
characteristic curvature which is seen to be much less than the AdS
curvature scale. To recover 4-dimensional effective behavior
non-linearly it is crucial that the characteristic scales of the
solution are insensitive to the 5-dimensional scale, and this is
exactly what is seen here.

\begin{figure}
\centerline{\psfig{file=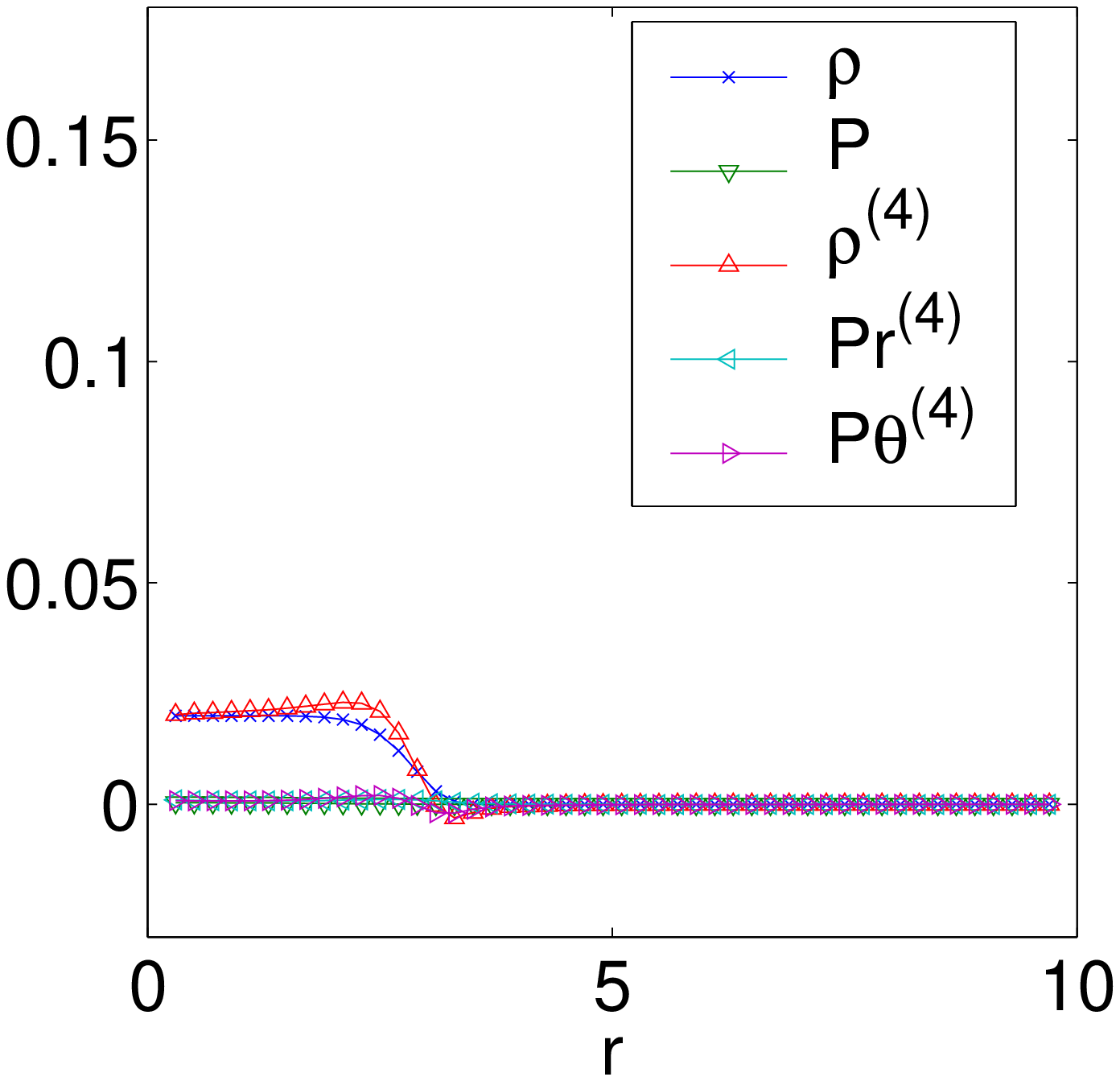,width=6cm} \hspace{0.2cm}
  \psfig{file=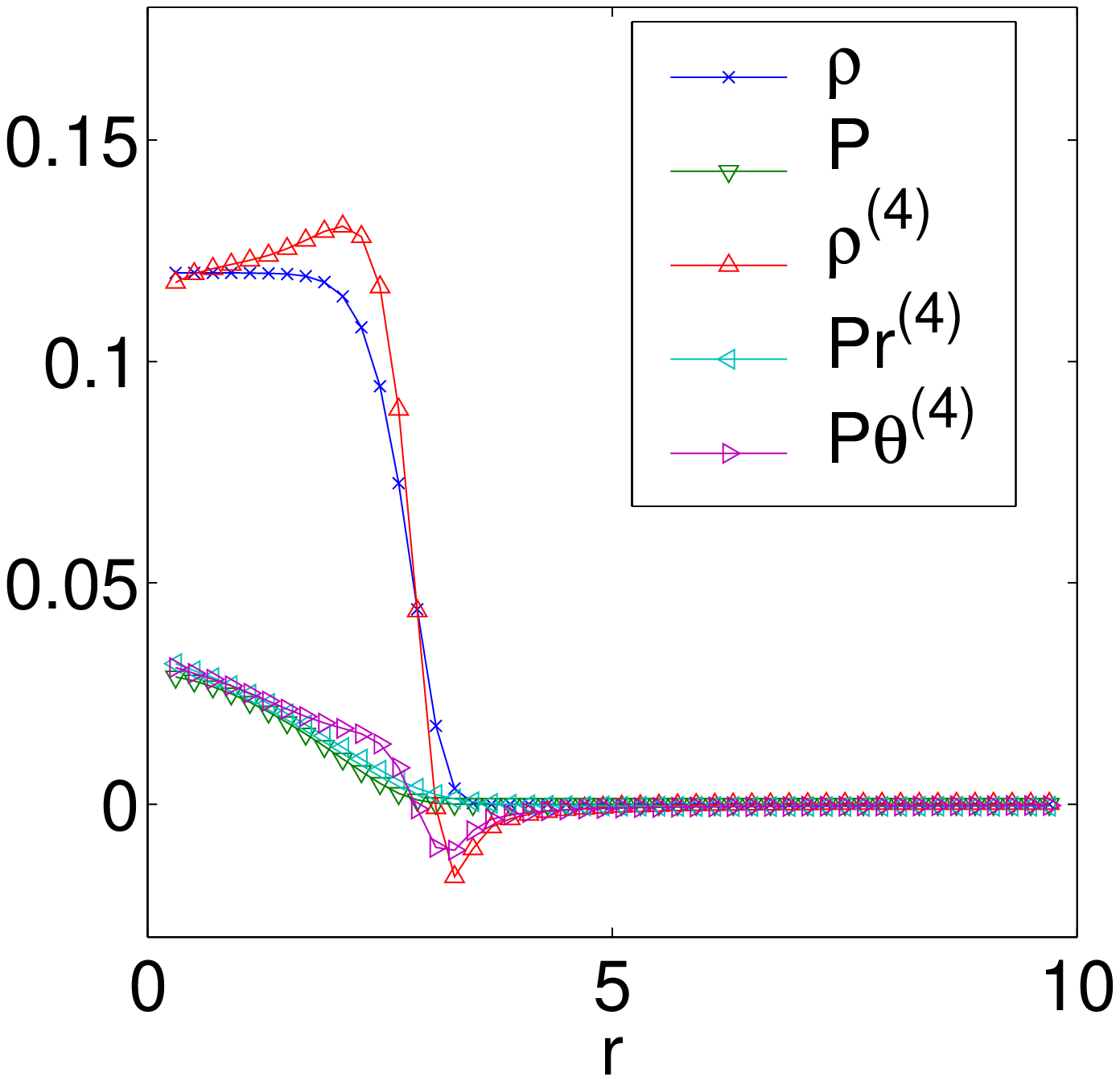,width=6cm} \hspace{0.2cm} \psfig{file=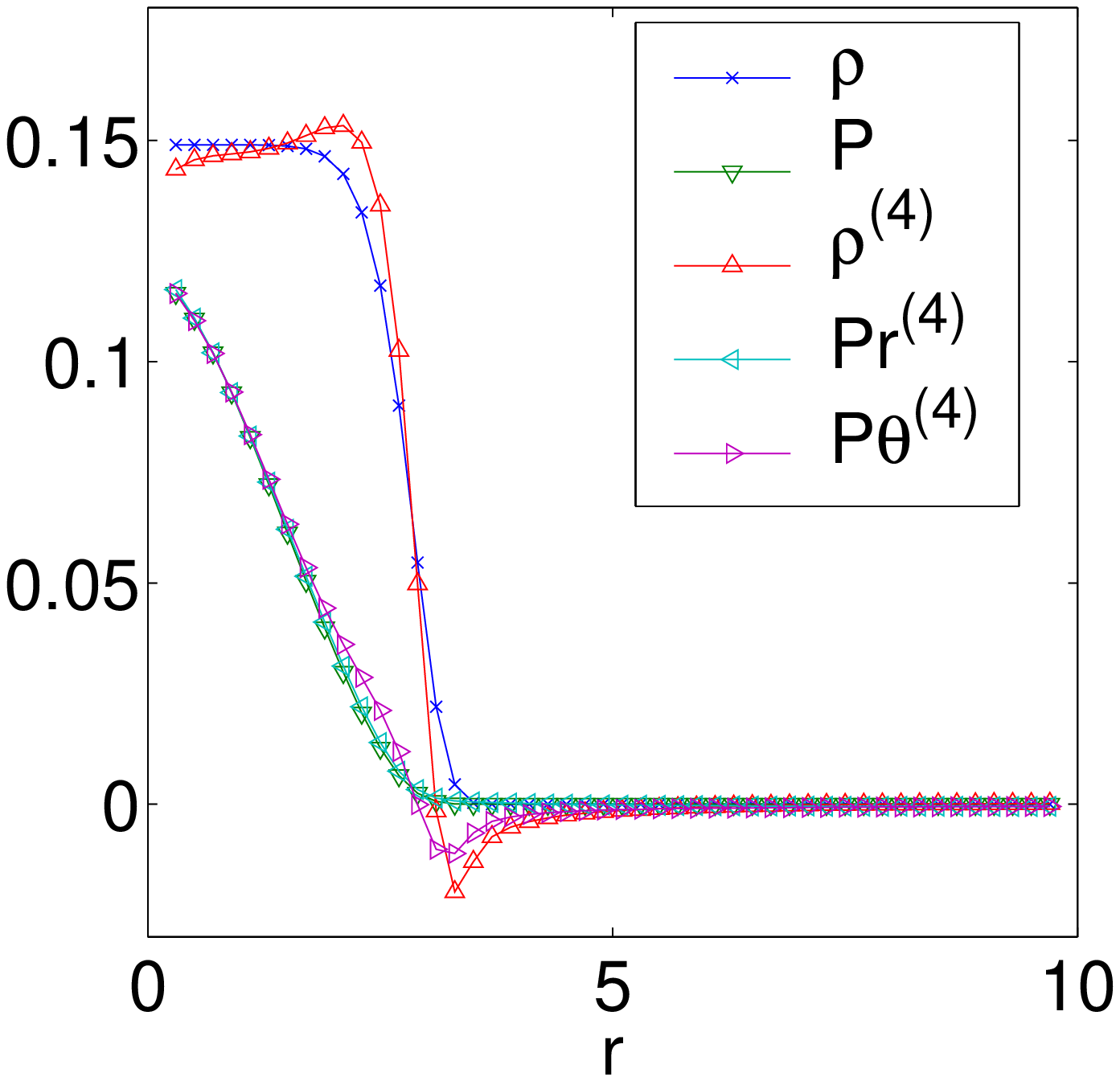,width=6cm}}
\caption[short]{ \figuremode  An illustration of actual density and pressure $\rho, P$ together with 4-dimensional effective density and
  pressure $\rho^{(4)}, Pr^{(4)}, P\theta^{(4)}$, for $\xi = 3$
  \emph{large} stars with various core densities, $\rho_0 = 0.020,
  0.120, 0.149$ from left to right. The actual density and pressure
  are calculated from the brane matching conditions. Note that
  isotropy is maintained as a boundary condition and the two pressure
  components $Pr, P\theta$ are equal to high precision, plotted here
  as simply $P$. Moving from the linear (left) to non-linear (right)
  we see the pressure becomes significant compared to the density. The
  key result of the plot is the small differences found between the
  actual and effective quantities. The effective quantities are the
  density and pressure components required to derive the induced brane
  geometry in 4-dimensional GR. We see extremely close agreement
  between both the density and pressure components of the actual
  5-dimensional solution and the effective theory. Note that isotropy
  is not input into the induced pressure components which are simply
  calculated here from the induced metric, yet $Pr^{(4)} \simeq
  P\theta^{(4)}$ is indeed found. For linear configurations it is
  expected from theory that the 5-dimensional solution on the brane
  looks 4-dimensional. However the key result of this paper is that
  even for highly non-linear solutions, (the rightmost plot), the
  agreement with a 4-dimensional effective theory is just as good.
  This is further characterized in the section \ref{sec:non-linear}
  and in the `key results' figure \ref{fig:compare_4d_5d}. (all
  lattices: $dr = 0.2$, $r_{\rm max} = 20$, $dz = 0.03$, $z_{\rm max}
  = 46$. no extrapolation to $dz = 0$ is performed leading to a
  maximum $\sim 4 \%$ systematic in $\rho^{(4)}$, estimated by
  comparison with $dz = 0.05$)
\label{fig:F_density_compare} 
}
\end{figure}

Now we examine the intrinsic geometry on the brane itself. Figure
\ref{fig:F_density_compare} shows the 4-dimensional induced density,
and radial and angular pressures that would give rise to such an
induced metric configuration in 4-dimensions. Plotted with them are
the actual 5-dimensional density, an input for the system, together
with the 5-dimensional pressure measured from the brane matching
conditions.  We see extremely close agreement for both linear and
non-linear stars.  The degree of non-linearity can be seen in the
rightmost plot as the pressure is becoming large compared to the
density. For this \emph{large} star the agreement is striking. Of
course linear theory states that a 4-dimensional intrinsic behavior
should be observed. A key result of this paper is that this applies
far beyond the linear regime. In the graphs we see that the agreement
between the actual 5-dimensional density and the 4-dimensional
effective density appears approximately independent of the density
over the range tested. This is seen more clearly in figure
\ref{fig:compare_4d_5d} of the `Highlights of Results' section
\ref{sec:key_results} and in section \ref{sec:non-linear}.

%
\subsection{Non-Linear Confinement}
\label{sec:confinement}
%

We now examine the transverse extent of the solutions. The function
$T$ is a scalar under residual coordinate transformations which
preserve the static spherical symmetry of the metric. The value of $T$
decays asymptotically to zero along the axis $r = 0$ away from the
brane, having its largest magnitude at (for \emph{small} stars), or
near (for \emph{large} stars) the brane.  Photons emitted at the core
of the star may propagate along this axis. A static observer at some
point on this line will then observe a redshift in the received
photons. We could characterize the geometry by considering the proper
distance along the line $r=0$ for the redshift to take a certain
value. For convenience, we equivalently choose to consider the point
where the value of $T$ is a half that on the brane.  This is a unique
point for all the configurations tested and characterizes the
confinement of the perturbation to the brane.

\begin{figure}
  \centerline{\psfig{file=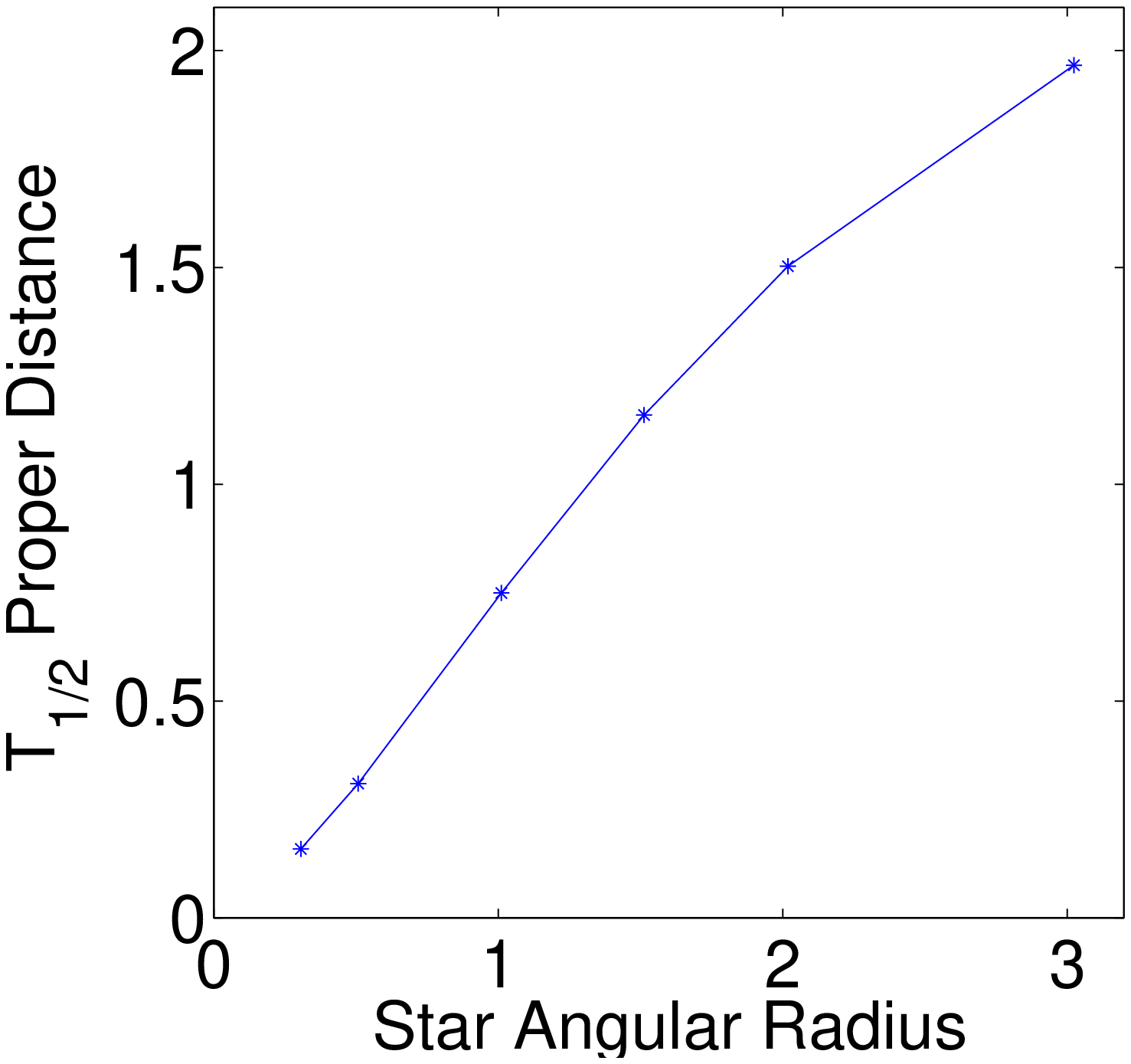,width=6cm}
    \hspace{0.1cm} \psfig{file=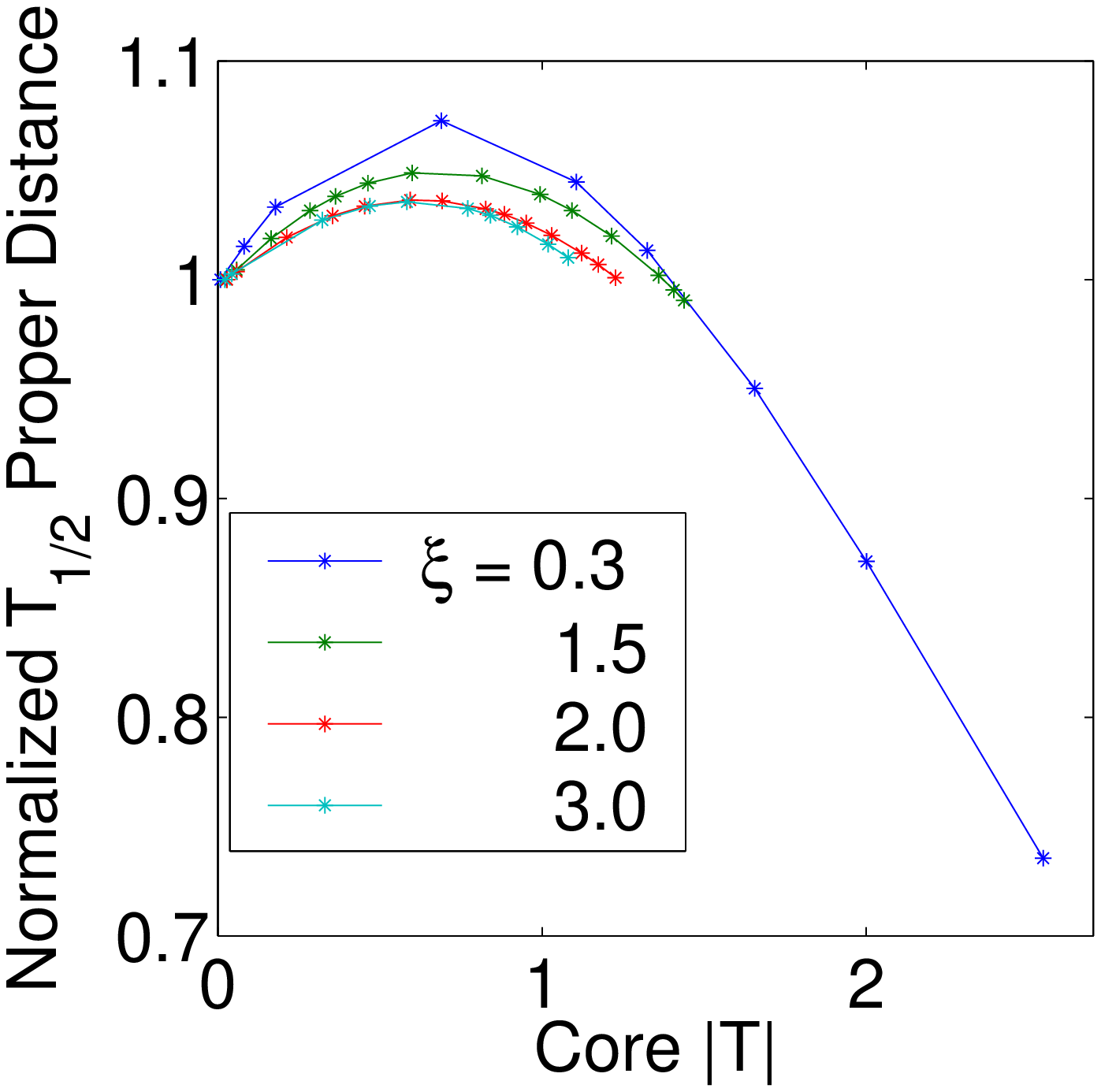,width=6cm}
    \hspace{0.1cm} \psfig{file=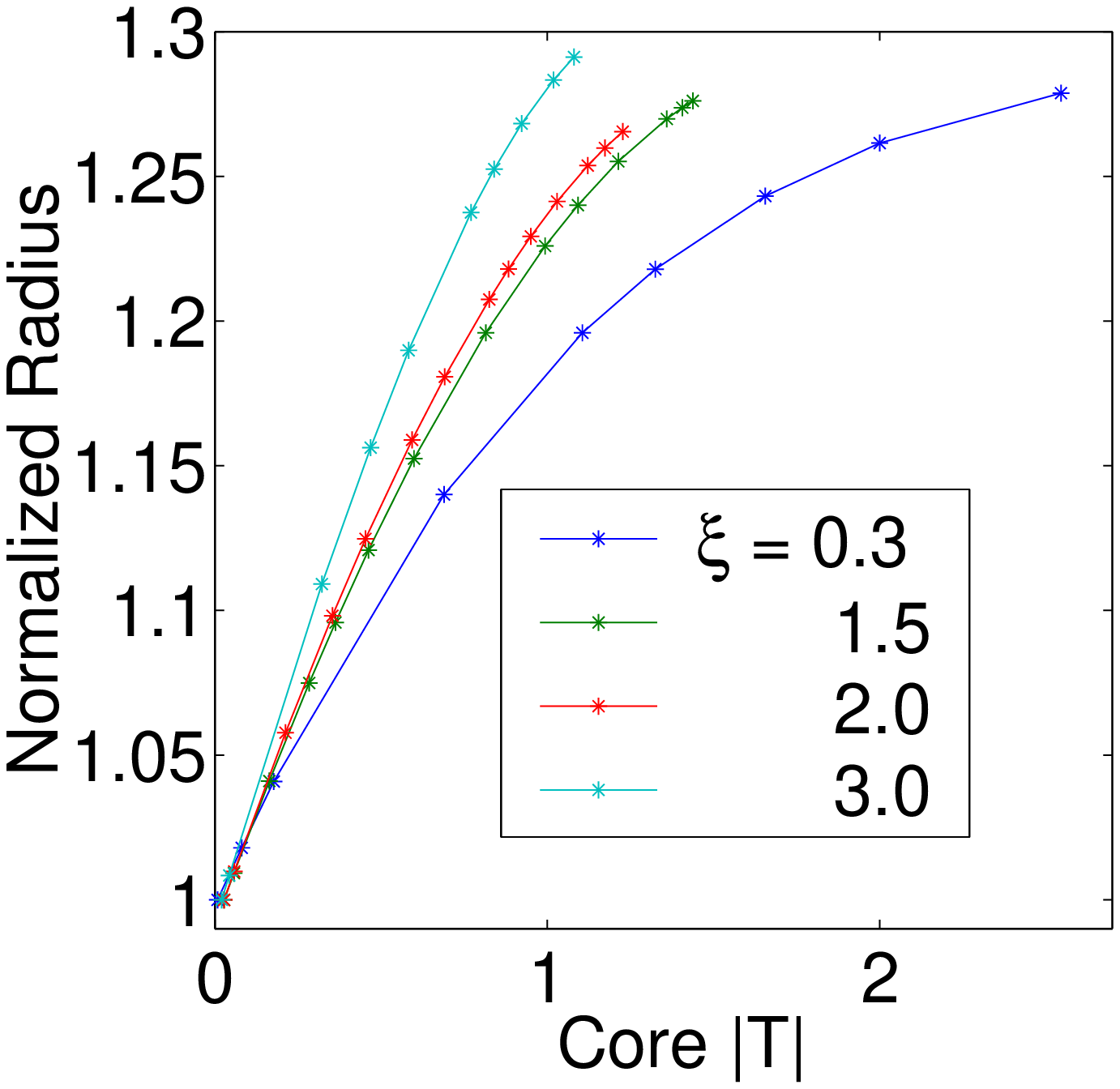,width=6cm}}
  \vspace{0.1cm}
\caption[short]{ \figuremode  An illustration of confinement for
  various star radii. The left plot is of proper distance along the
  axis $r = 0$ to $T_{1/2}$, the point where $T$ takes half the value
  it has on the brane, against the proper star radius. The results
  plotted are extrapolated to stars of zero density.  The middle plot
  then shows how the same quantity changes with increasing density,
  the core value of the metric function $T$ being the measure of the
  geometric response to this density.  Curves are plotted for
  different $\xi$, and the $T_{1/2}$ distance is normalized using the
  zero density values of the leftmost plot.  The \emph{smallest} stars
  relaxed have the highest non-linearity and we observe first an
  increase in $T_{1/2}$ distance, with increasing core $|T|$, and thus
  density, and then a decrease for larger densities.  The increase is
  expected as the proper radius, plotted in the right-hand graph
  (normalized by $\xi$), is increasing monotonically for increasing
  density - note, the coordinate width of the density profile, $\xi$,
  is fixed in these solutions but the actual proper angular size then
  must be found from the solution itself. The leftmost plot shows
  exactly this increase of $T_{1/2}$ distance for increasing radius.
  However the decrease in $T_{1/2}$ distance for high core $|T|$ is
  not expected and indicates that the physical perturbation is
  becoming more peaked near the brane for higher density stars. This
  appears to be a purely non-linear effect. With the solutions
  available for $\xi = 0.3$, it is unclear whether the $T_{1/2}$
  distance reaches zero at finite $T$, asymptotically at the upper
  mass limit, or not at all.  This remains an interesting topic for
  future work.  The \emph{larger} stars, whilst further from their
  upper mass limit, still show similar behavior, the peak value of
  the $T_{1/2}$ distance being at approximately the same core value of
  $|T|$, or alternatively core redshift. We conclude that the
  confinement of the physical solutions in fact increases near the
  upper mass limit, even for \emph{large} stars.  It implies that
  non-linearity does effect the geometry of the bulk perturbation for
  \emph{large} stars, although the modification from linear behavior
  only occurs on long wavelengths. (lattices: $\xi = 0.3$: $dr =
  0.02$, $r_{\rm max} = 2$, $dz = 0.005$, $z_{\rm max} = 4$, $\xi =
  1.5$: $dr = 0.10$, $r_{\rm max} = 10$, $dz = 0.02$, $z_{\rm max} =
  21$, $\xi = 2.0$: $dr = 0.15$, $r_{\rm max} = 15$, $dz = 0.02$,
  $z_{\rm max} = 31$, $\xi = 3.0$: $dr = 0.20$, $r_{\rm max} = 20$,
  $dz = 0.03$, $z_{\rm max} = 46$.)
\label{fig:confine} 
}
\end{figure}

The first plot in figure \ref{fig:confine} shows how the extent of the
star depends on its angular radius on the brane. The stars shown are
all in the linear regime. Several low density solutions are used to
extrapolate results to zero core density. We see an approximately
linear relation over the range of $\xi$ tested. At larger $\xi$, the
relation deviates from linear, appearing to become flatter, consistent
with the `pancake' scalings predicted in
\cite{Giddings:2000mu,Chamblin:1999by}. Thus the characteristic fall
off distance of the redshift along the axis increases for increasing
star radius. The \emph{larger} the star, the shallower this function
is as it decays away from the brane. This is to be expected as the
asymptotic behavior of the Greens function clearly depends on the
radial scale probed.  Numerically this slow fall off is a reason that
very \emph{large} stars are difficult to simulate. One requires very
large physical lattice size in the $z$ direction whilst maintaining
resolution better than the AdS length.

The remainder of the graphs show the variation of this confinement
distance with increasing density for a fixed $\xi$. The
\emph{smallest} stars simulated, with $\xi = 0.3$ relax with the
greatest degree of non-linearity. For these we see the very
interesting feature that the confinement distance first increases in
the linear regime as the density, and thus core $|T|$, increases, but
then begins to decrease as the configuration becomes highly
non-linear. As discussed earlier, fixed coordinate $\xi$ does not fix
the angular radius of the star, which is also plotted in the figure
and is seen to monotonically increase with core $|T|$. Thus for the
linear configurations, where $|T| << 1$, the increase in confinement
distance follows the radial increase in star size on the brane as one
would expect. However the decrease in transverse extent for the very
dense stars appears to be a significant and purely non-linear effect.
For the most non-linear star tested, $\xi = 0.3$ and this extent is
$\simeq 75\%$ of the zero density one with the same $\xi$, whilst the
angular radius is $\simeq 25\%$ larger.  It is important to note that
the absolute extent of the star clearly does not decrease, but rather
the decay distance decreases implying the function $T$ falls more
steeply near the brane for very non-linear configurations than for
linear ones.

The maximum density star for $\xi = 0.3$ is very close to what
appears to be a critical mass limit, as discussed in section
\ref{sec:small_stars}.  However with the solutions available, we
cannot determine whether the transverse size actually tends to zero at
the critical point, or whether it remains finite. For the larger stars
we are unable to probe so far into this non-linear regime for reasons
of numerical stability, but the same curves for $\xi = 1.5, 2, 3$
stars are plotted, where qualitatively the behavior appears to be
similar over the range of core redshifts available.  Again the turn
around in transverse extent is observed for the more non-linear stars
where the angular radius remains increasing, the turn around points
being at approximately the same value of core $|T|$ as for the $\xi =
0.3$ case. The densest $\xi = 3$ star has approximately the same
confinement distance as a zero density star of the same $\xi$, but
has a proper radius $\sim 30 \%$ larger. Thus this confinement
distance dependence on density does not appear to become less
pronounced with increasing $R$. It would be very interesting future
work to see the scaling of this effect for very \emph{large} stars, if
such solutions could be computed.

The fact that this behavior is still seen for non-linear $\xi = 3$
stars is important.  As we have seen in section \ref{sec:large_stars},
non-linearity does not introduce AdS scale curvature perturbations
into the \emph{large} star bulk geometry. The source length scale and
AdS length scale remain separated. However, we see here that the
non-linearity does change the nature of the bulk geometry, though the
modification of behavior appear to be only on \emph{large}
wavelengths. Despite the non-linearity modifying the bulk response,
the 4-dimensional effective GR description appears to hold as well for
the \emph{large} non-linear stars as for the linear ones.

It is worth noting that if the confinement distance does go to zero at
the upper mass limit, as is possibly indicated in the $\xi = 0.3$
behavior, then the same may also occur for \emph{large} stars
extremely close to the upper mass limit.  This zero confinement
distance would indicate that the AdS length scale was entering the
geometry of the perturbation to the bulk, and would give rise to a
deviation from 4-dimensional behavior in the induced geometry.
Presumably for very \emph{large} stars one would have to be extremely
close to the critical point in order to see such effects.

The increase in confinement is reminiscent of the change in the sense
of deflection of the brane, relative to the `Randall-Sundrum'
transverse-traceless gauge coordinates in the linear theory. This
would occur if the sign of the trace of the stress energy changes, as
happens for incompressible fluid matter in the strong gravity regime.
Thus, although there is no `radion' in the one brane case at low
energies \cite{Charmousis:1999rg}, one can heuristically think of this
confinement of the perturbation as an analogous, although
non-dynamical, quantity.  Furthermore, this would confirm the
suspicion that for very \emph{large} stars, the normalized confinement
of figure \ref{fig:confine}, would indeed change by an order one
amount, for strong gravity configurations, where all curvatures remain
small.

%
\subsection{Upper Mass Limits and 4-dimensional Effective Theory}
\label{sec:non-linear}
%

{\bf (Some results
  presented in `Highlights of Results' section \ref{sec:key_results}) \\ }

In the previous section \ref{sec:large_stars} we observed that
4-dimensional behavior was recovered on the brane for relativistic
\emph{large} stars with $\xi = 3$. In this section we characterize the
transition from 5-dimensional to effective intrinsic 4-dimensional
behavior.

In order to compare like with like, we use the 5-dimensional solutions
to generate 4-dimensional density profiles as a function of proper
distance in the induced geometry. Assuming isotropy we numerically
integrate the 4-dimensional Einstein equations given in appendix
\ref{app:einstein_eqns} using this generated density profile with the
relevant boundary conditions for asymptotic flatness at large $r$. The
parameters we compare to the actual 5-dimensional solutions are the
core value of $T$, related to the redshift of photons from the star's
core, and also the core pressure. With the static, spherical symmetry
both these quantities are coordinate scalars under the $r, z$
coordinate freedom.  Note that obviously the core density agrees by
construction.

In figure \ref{fig:compare_4d_5d} (found in the `Highlights of
Results' section \ref{sec:key_results}) we plot the deviation of these
quantities in the 4-dimensional effective theory from the actual
measurements made on the 5-dimensional solutions. Three sets of stars
are shown with different $\xi$. For each set a range of core densities
are presented. The lattices used are generated at two values of $dz$
and quadratic extrapolation is used to calculate the $dz = 0$
continuum value, as described in appendix \ref{app:numerical}.

The results are striking. We have already seen that the $\xi = 3$
stars give good agreement with the 4-dimensional Einstein equations
acting on the induced metric. This is again observed in these plots,
where the actual 5-dimensional quantity is plotted against the induced
effective 4-dimensional one, and the points for $\xi = 3$ lie close to
the $4d = 5d$ straight line for both redshift and core $P / \rho$.
$\xi$ is not the actual proper radius of the star, but rather a
coordinate radius. However $\xi$ is approximately the proper radius,
and this proper radius increases as one moves vertically down (to
larger $\xi$) towards the $4d = 5d$ line.  Thus, the \emph{larger}
stars do lie nearer the $4d = 5d$ line, indicating a better
approximation by the 4-dimensional induced theory.  This is true over
the range of solutions, for both linear and non-linear densities. For
finite sized stars, the effective theory consistently underestimates
both 5-dimensional quantities plotted.

Furthermore the points for $\xi = 3$ approximately lie on a straight
line.  For $\xi = 3$ perfect agreement with the 4-dimensional
effective theory is not expected as the proper radii are only between
$3 - 4$ times larger than the AdS length. However the fact that the
points lie on a straight line indicates that the degree of deviation
from the intrinsic description appears to be independent of the
non-linearity over the range of densities tested. We discuss this
further shortly. 

We also plot a line indicating the core redshift and $P / \rho$ for
$\xi = 3$, from 4-dimensional linear theory, integrated in a similar
fashion to the 4-dimensional non-linear theory, using the same density
profile. This is graphed against the non-linear 4-dimensional
quantities to indicate the degree of non-linearity.  Already for
redshifts above $0.2$ we see that the linear and non-linear
4-dimensional theory have very poor agreement. In fact for the most
dense $\xi = 3$ configurations, the linear theory underestimates the
core redshift and pressure by a factor of about three. Thus for the
large redshift $\xi = 3$ stars, the non-linear corrections to these
linear quantities are much larger than the quantities themselves.
Therefore the regime tested is fully non-linear, and as such, is far
beyond the reach of second or higher order perturbation theory.

\begin{figure}
  \centerline{\psfig{file=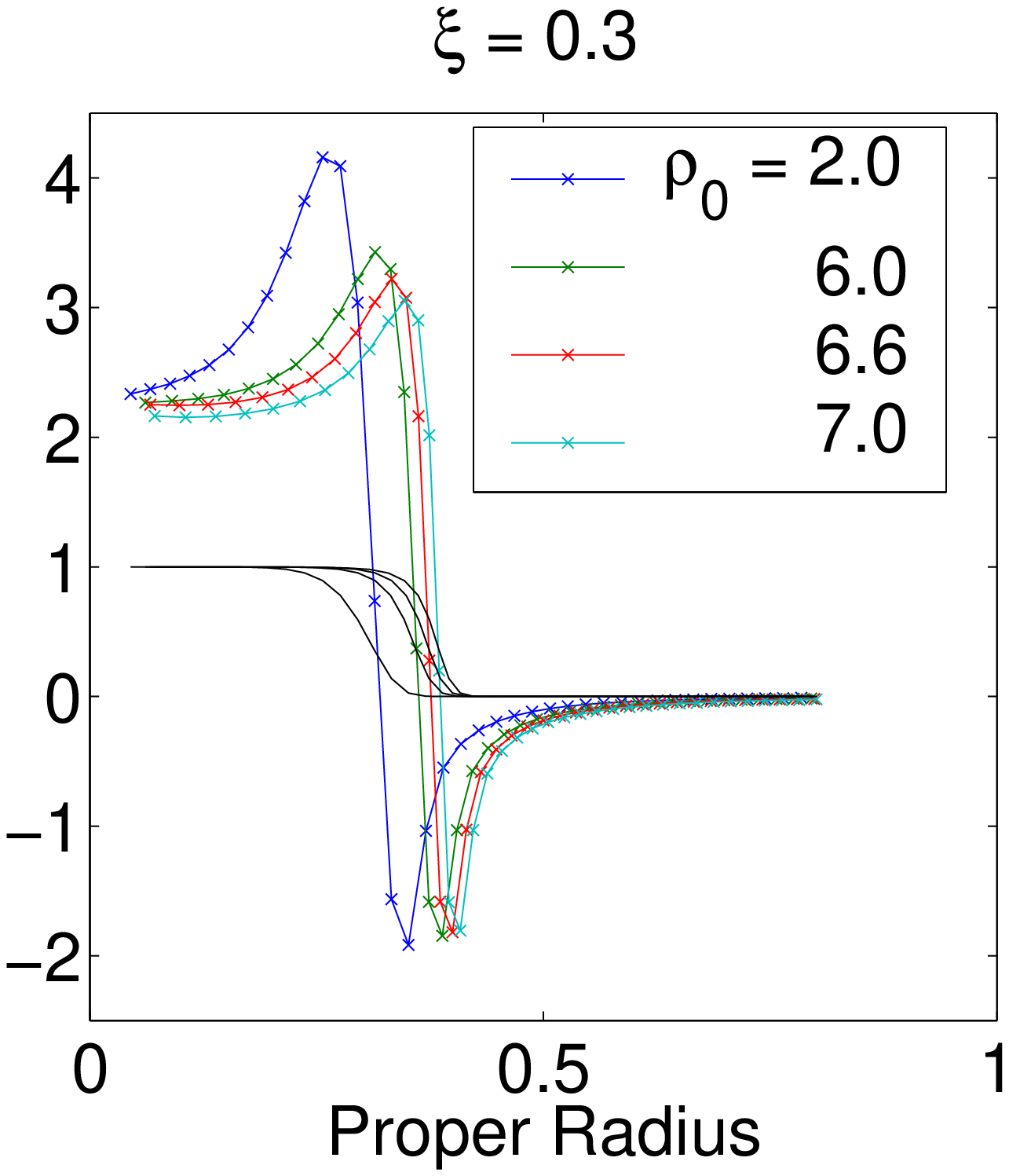,width=5.8cm}
    \hspace{0.1cm} \psfig{file=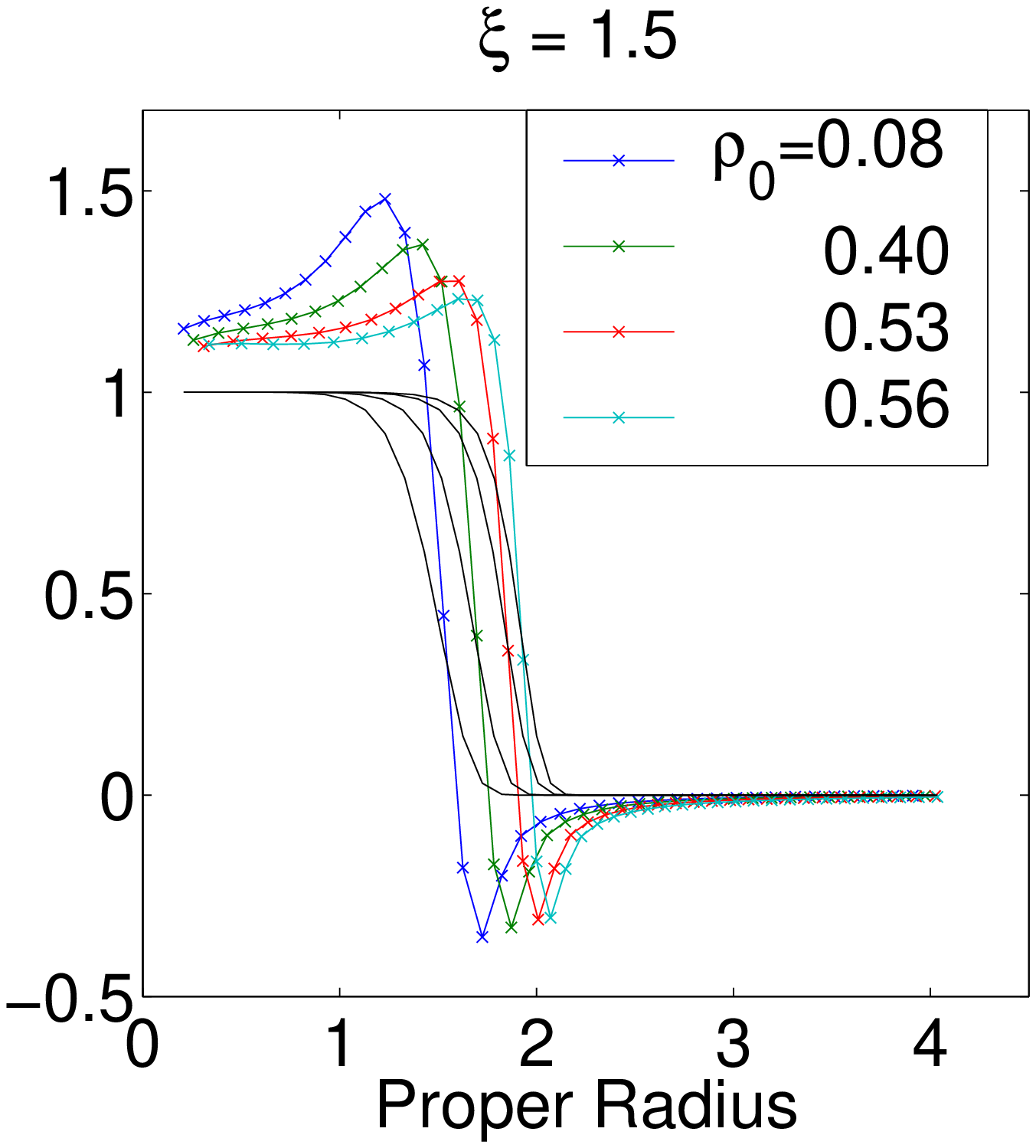,width=6cm}
    \hspace{0.1cm} \psfig{file=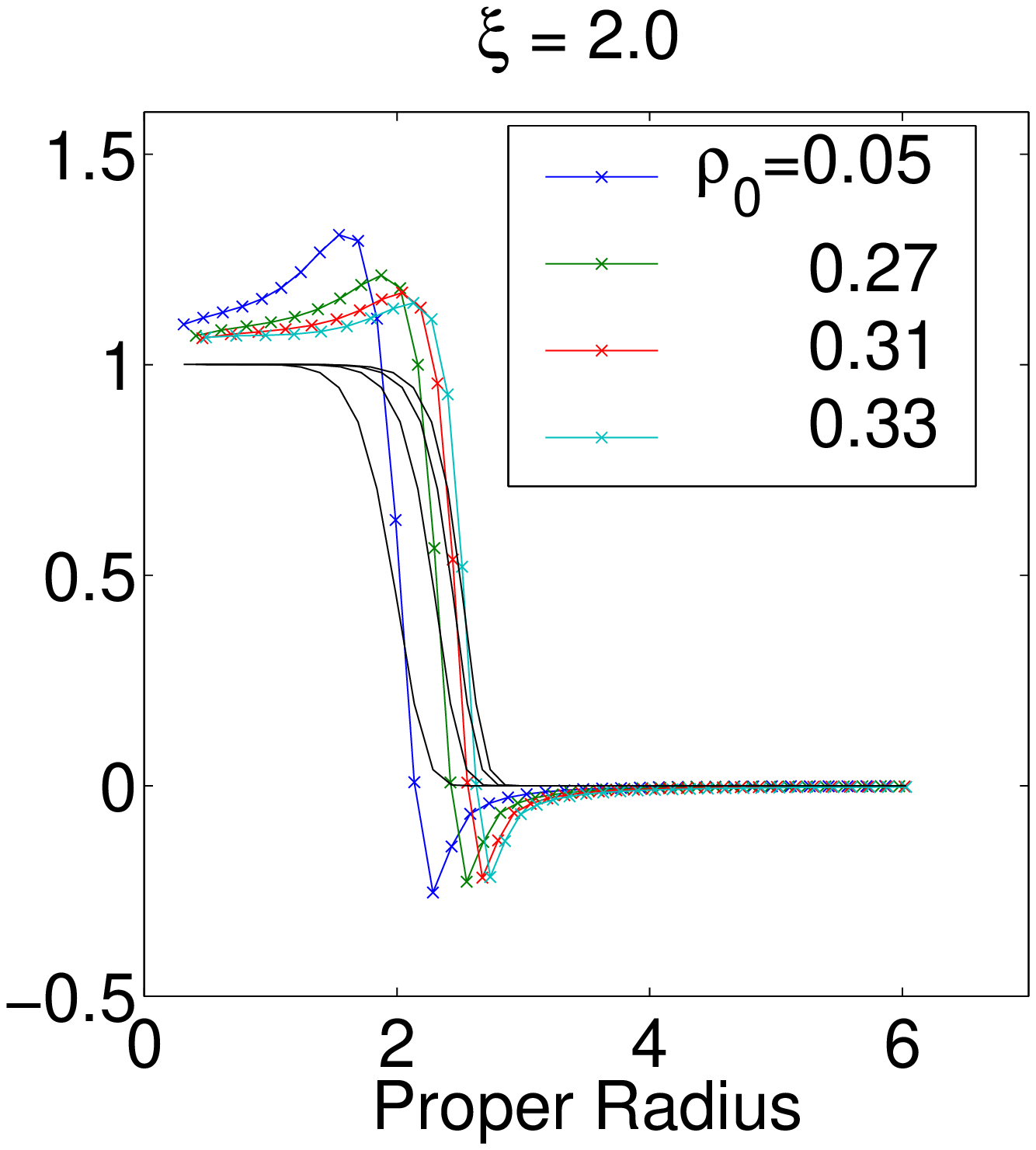,width=6cm}}
  \vspace{0.1cm}
\caption[short]{ \figuremode  An illustration of induced density
  normalized by the actual density at the core as a function of proper
  radius on the brane for several values of $\xi$. Also plotted are
  the corresponding actual density profiles (in black), again
  normalized to unit value at the core. We see increasing agreement,
  for both linear and non-linear densities, between the actual and
  induced curves as $\xi$, and thus the proper radius of the star,
  increases. For each value of $\xi$ the lowest density star has a
  very small core redshift. The densest stars have the following core
  redshifts; $\xi = 0.3$: ${\cal Z} = 11.7$, $\xi = 1.5$: ${\cal Z} =
  3.2$, $\xi = 2.0$: ${\cal Z} = 2.4$, all far into the non-linear
  regime. These profiles confirm the results of figure
  \ref{fig:compare_4d_5d}, showing the \emph{local} agreement of
  quantities.
\label{fig:effective_rho}
}
\end{figure}

It is important to note that these quantities merely indicate an
agreement of a global nature, and thus for completeness we also
include figure \ref{fig:effective_rho}. This confirms the intrinsic
description gives an increasingly good approximation to the induced
geometry \emph{locally}, as the star radius increases, for both linear
and non-linear core densities. Note also the previous figure
\ref{fig:F_density_compare} in section \ref{sec:large_stars}.

We see from the \emph{small} stars, in section \ref{sec:small_stars},
that there appears to be an upper mass limit for a given radius, as in
the 4-dimensional theory. For the larger stars our method currently
does not allow us to stably relax configurations very near the
critical mass. However, we infer this mass to be close to its value in
the 4-dimensional theory as we see such close agreement leading up to
the limit. For the most non-linear solution with $\xi = 3$, the
induced angular radius is $R = 3.7$. The upper mass limit in
4-dimensional theory, derived for a top hat density profile, is given
as $M_{\rm max} = 8 \pi \frac{4}{9} R$ in our units, which corresponds
to $M_{\rm max} = 41$, whereas the effective 4-dimensional mass
computed for our most non-linear solution is $M = 30$ which is
approximately $\simeq 75 \%$ of this 4-dimensional critical mass for
this radius. Thus the density $\rho_0$ is similarly $\simeq 75 \%$ of
the critical density. Note that this 4-dimensional density limit will
overestimate the 5-dimensional density limit for a star of this
radius, as we have seen the effective theory underestimates the
5-dimensional core $P / \rho$. Thus the core density of this solution
is \emph{at least} $\simeq 75 \%$ of the 5-dimensional density
corresponding to the upper mass limit for a star of the same radius.

\begin{figure}
\centerline{\psfig{file=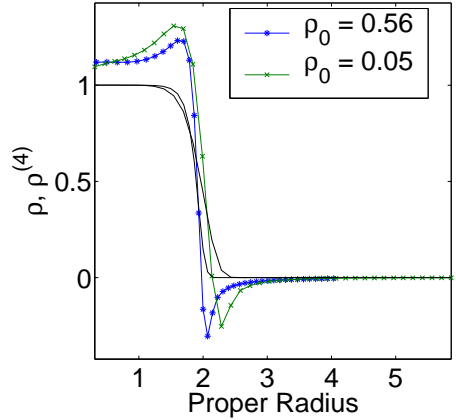,width=6cm}
}
\vspace{0.1cm}
\caption[short]{ \figuremode  
  In this figure we compare a highly non-linear ($\rho_0 = 0.56$, the
  densest $\xi = 1.5$ star available) and a low density star ($\rho_0
  = 0.05$, $\xi = 2.0$) with the same proper radius $\sim 2.0$. The
  induced (coloured) and actual (black) densities are plotted for both
  stars, normalized by the actual core density, as in the previous
  figure \ref{fig:effective_rho}. The black curves, the actual
  densities, agree by construction, as a result of choosing stars with
  the same proper radius. The $\xi = 1.5$ star has a core redshift of
  ${\cal Z} = 3.2$, whilst the $\xi = 2.0$ has ${\cal Z} = 0.06$.
  Note also that the confinement proper distances are $1.15$ and
  $1.51$ for the $\xi = 1.5$ and $\xi = 2.0$ stars respectively.
  Despite one star being linear and the other highly non-linear near
  its upper mass limit, the denser one being considerably more
  confined to the brane, the similarity between the induced curves
  shows that this has little effect on the degree of approximation of
  the effective theory.
\label{fig:compare_stars}
}
\end{figure}

Agreement for \emph{large} stars was expected for low densities. We
also find it for high densities. We should not compare linear and
non-linear stars of the same $\xi$ directly as these have different
proper radii. Thus in figure \ref{fig:compare_4d_5d}, because the
stars of the same $\xi$ have slightly varying proper radius, it
merely allows us to say larger stars, both linear and non-linear, are
better approximated by a 4-dimensional effective theory, since
vertical lines in the figure approaching the $4d = 5d$ line from above
are increasing in proper radius. We must use different $\xi$ values
to find stars with the same proper radius but different density. In
figure \ref{fig:compare_stars} we do exactly this, and quantify how
the core density effects the degree of agreement between the induced
and actual brane quantities. In this figure, two stars of the same
proper radius $\simeq 2$, are shown overlaid using a suitable
normalization.  One star is highly non-linear, near its upper mass
limit with a core redshift of ${\cal Z} \sim 3$. The other is a low
density solution. We see very similar induced profiles for both, after
normalization by the actual core density, indicating that the
approximation of the induced theory to the actual is roughly
independent of the degree of core density of the star, over the
density range tested. We find similar behavior for the induced
pressures. Note that the confinement proper distance to $T_{1/2}$,
discussed in section \ref{sec:confinement}, is quite different for the
two stars, being $1.15$ and $1.51$ for the high and low density stars
respectively. Thus even though the 5-dimensional geometry is different
near the upper mass limit than at low density, for a fixed proper
radius, the induced behavior appears to be approximated by the
4-dimensional theory equally well.

%
\section{Conclusion}
%

We have outlined a scheme which allows the full non-linear Einstein
equations to be numerically solved \emph{elliptically} for
Randall-Sundrum gravity, with static, spherically symmetric matter on
the brane, giving rise to regular 5-dimensional solutions with axial
symmetry.  Radiation boundary conditions imply that the horizon
geometry is that of AdS. Due to the elliptic nature of our scheme, the
method allows both an asymptotically AdS horizon geometry, and the
brane matching relations to be simultaneously imposed as boundary
conditions. We explicitly show that data specifying a star geometry in
4-dimensions \emph{uniquely} determines the full 5-dimensional bulk
solution.

Using this numerical scheme we study \emph{small} and \emph{large}
stars in Randall-Sundrum gravity. Highly non-linear configurations are
calculated for \emph{small} stars, whose radius is less than the AdS
length, having core red-shifts up to ${\cal Z} \simeq 12$. An upper
mass limit is found that is qualitatively similar to that in
4-dimensions, implying that the brane does not stabilize highly
non-linear static configurations.  No pathologies are found in the
spatial geometry or the red-shifting behavior for configurations near
this mass limit.  The spatial geometry is found to be approximately
conformal to hyperbolic space for stars of all radii. Despite the
continuum of modes in the perturbation spectrum, \emph{large} stars,
with radius greater than the AdS length are found to be described well
by a local effective theory which is simply 4-dimensional gravity for
the intrinsic geometry and matter.  Whilst this was known in linear
and second order perturbation theory, we have demonstrated this for
highly non-linear configurations, inaccessible in perturbative
approaches, the largest dense star studied having proper radius
$\simeq 3.7$ times the AdS length and a core red-shift of ${\cal Z}
\simeq 2.1$.

In order to recover a \emph{local} long wavelength 4-dimensional
effective description, the confinement of the perturbation to the
brane must persist non-linearly, allowing the curvature scale of the
perturbation to remain separated from the compactification curvature
scale.  Interestingly, the confinement of the geometric perturbation
is actually found to \emph{increase} for densities near the upper mass
limit.  This is seen for all stellar sizes tested, and appears to be a
purely non-linear phenomenon. Thus the bulk geometry for the
\emph{large} stars is quantitatively different for a linear and
non-linear source, and yet for both, the induced geometry is simply
described by 4-dimensional gravity.

This elliptic method of solving the Einstein equations could be
extended to the compact extra dimension case, both with localized
matter and without. Due to the mass gap in the perturbative spectra,
the long wavelength theories, and thus \emph{large} stars, are
analytically tractable.  It is exactly in the opposite, short
wavelength regime, that this elliptic method works most effectively,
and could provide a powerful tool to study the behavior of
\emph{small} stars, where no general analytic solutions exist.
However, we expect that the qualitative features of \emph{small} dense
star behavior seen in this one brane model, will be common to other
types of model.

We have focussed on regular static geometries. Rotating configurations
and singular black hole solutions are also of utmost interest
astrophysically. We believe the implication of our static result is
that the long range behavior in one brane Randall-Sundrum gravity is
simply 4-dimensional gravity. We expect it to also hold for non-static
and non-regular cases which provides much scope for future work, both
analytic and numerical.

%
\section*{Acknowledgements}
%

The author would like to thank Neil Turok for much support and advice
on this work, and also Andrew Tolley and Adam Ritz for useful
discussions. Computations presented were performed on COSMOS at the
National Cosmology Supercomputing Center in Cambridge. The author was
supported by a PPARC studentship, and now by Pembroke College,
Cambridge.

\newpage

%
\section{Appendix A: Testing the Method}
\label{app:testing}
%

The numerical scheme outlined in this paper does indeed converge to a
stable solution after iteration. The most important test of the method
is performed in section \ref{sec:linear_check}, comparing low density
solutions with those generated by an independent method based on the
linear theory outlined in section \ref{sec:linear}. The two methods
are shown to be consistent, the maximum discrepancy between the metric
functions being worst for \emph{large} stars at $\sim 10 \%$ for $\xi
= 3$, with much lower differences ($\sim 2 \%$) in actual and induced
density and pressure. This indicates that the asymptotic AdS behavior
is reproduced by the boundary condition that $T, A, B \rightarrow 0$
asymptotically, that the method of regularizing the singular terms in
$B$ works effectively, and that the constraint structure that requires
only $\{rz\}$ to be imposed on the brane does indeed guarantee that
the constraints are well satisfied in the bulk and asymptotically,
even for finite lattice size.

Although this linear check strongly indicates that the method behaves
as expected, it does not directly test each component issue.  It is
also important to test how the scheme performs for the large
perturbations that form the basis of this paper. In this appendix we
address these points, performing consistency checks of the scheme for
such non-linear solutions. We study,
\begin{itemize}
\item the behavior of the singular term approximation at the origin
\item the degree of constraint and elliptic Einstein equation
  violation
\item the convergence behavior for varying resolution, and the effect
  on the constraints and solution of varying the physical lattice size
\item the return of the geometry asymptotically to AdS at the horizon
\end{itemize}

%
\subsection{Origin Regularization}
%

\begin{figure}
\centerline{\psfig{file=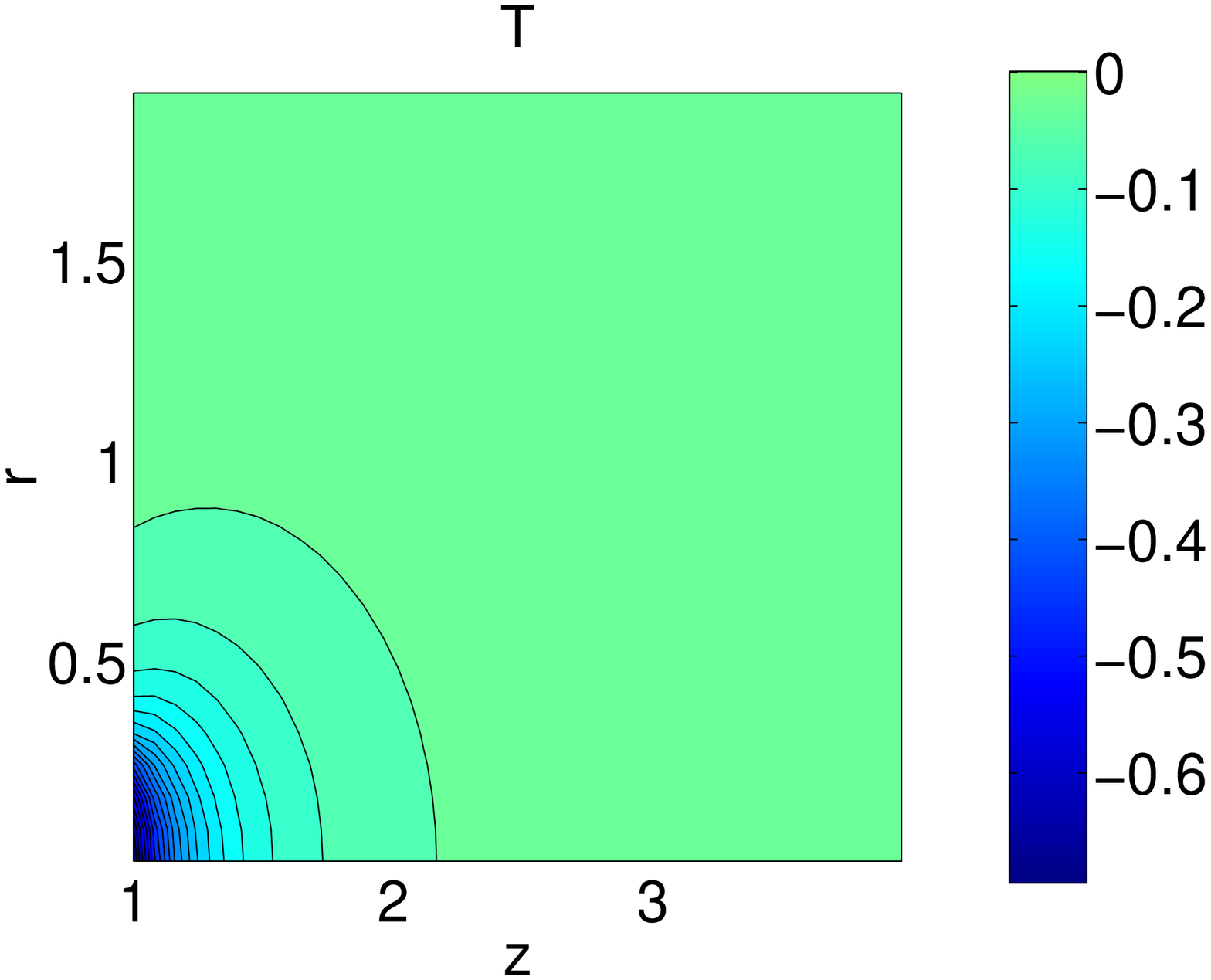,width=6cm} \hspace{0.1cm} \psfig{file=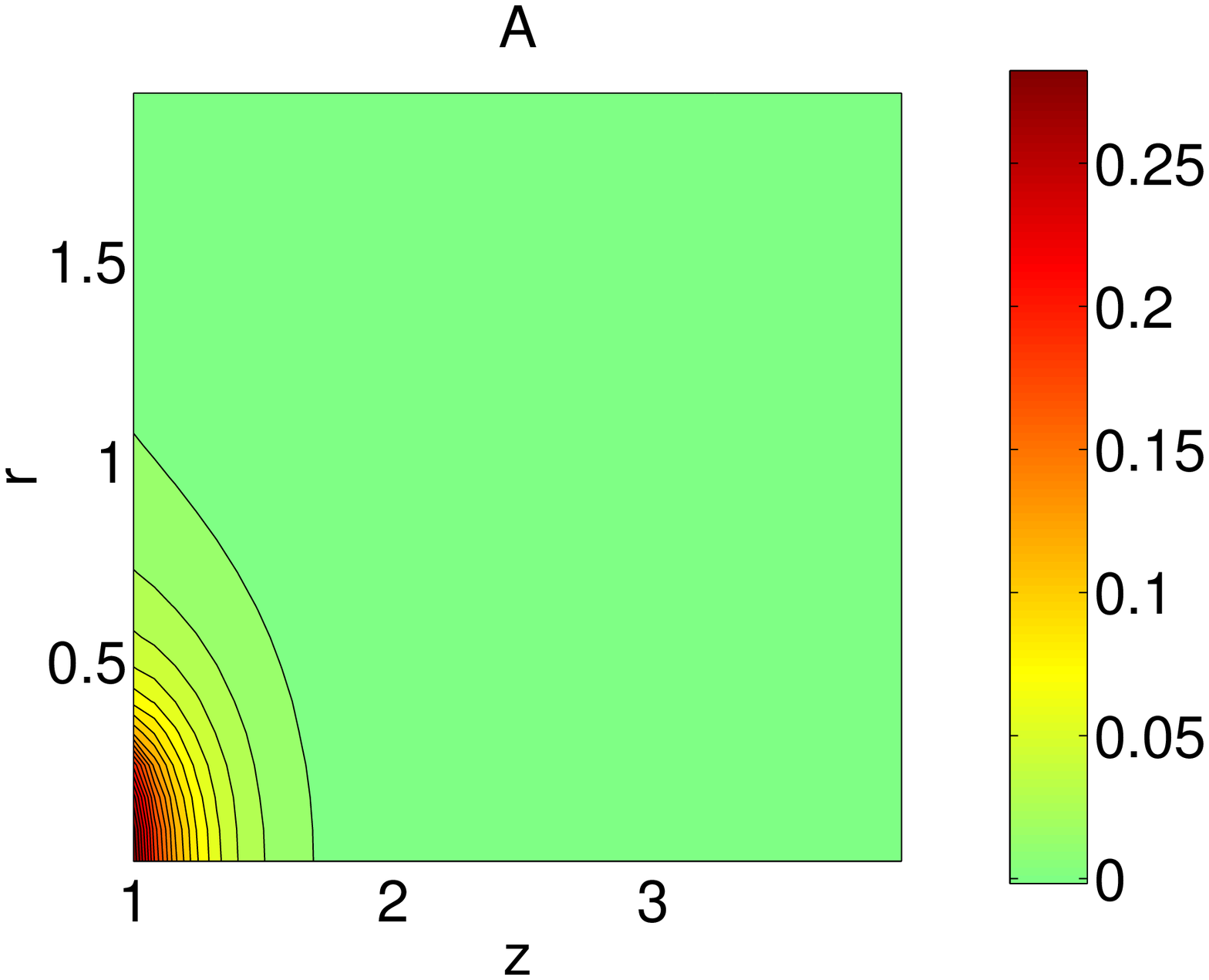,width=6cm}}
\centerline{\psfig{file=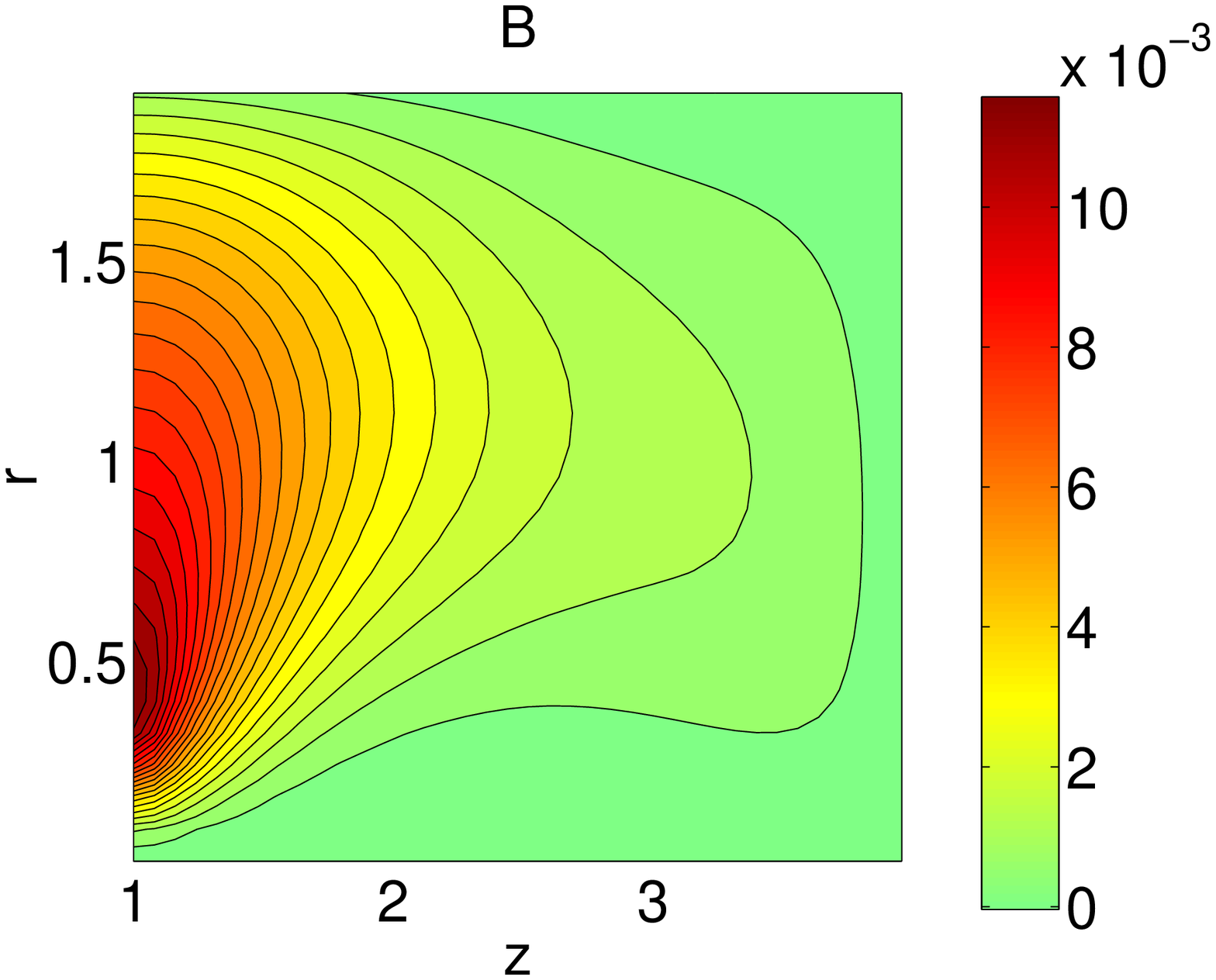,width=6cm} \hspace{0.1cm} \psfig{file=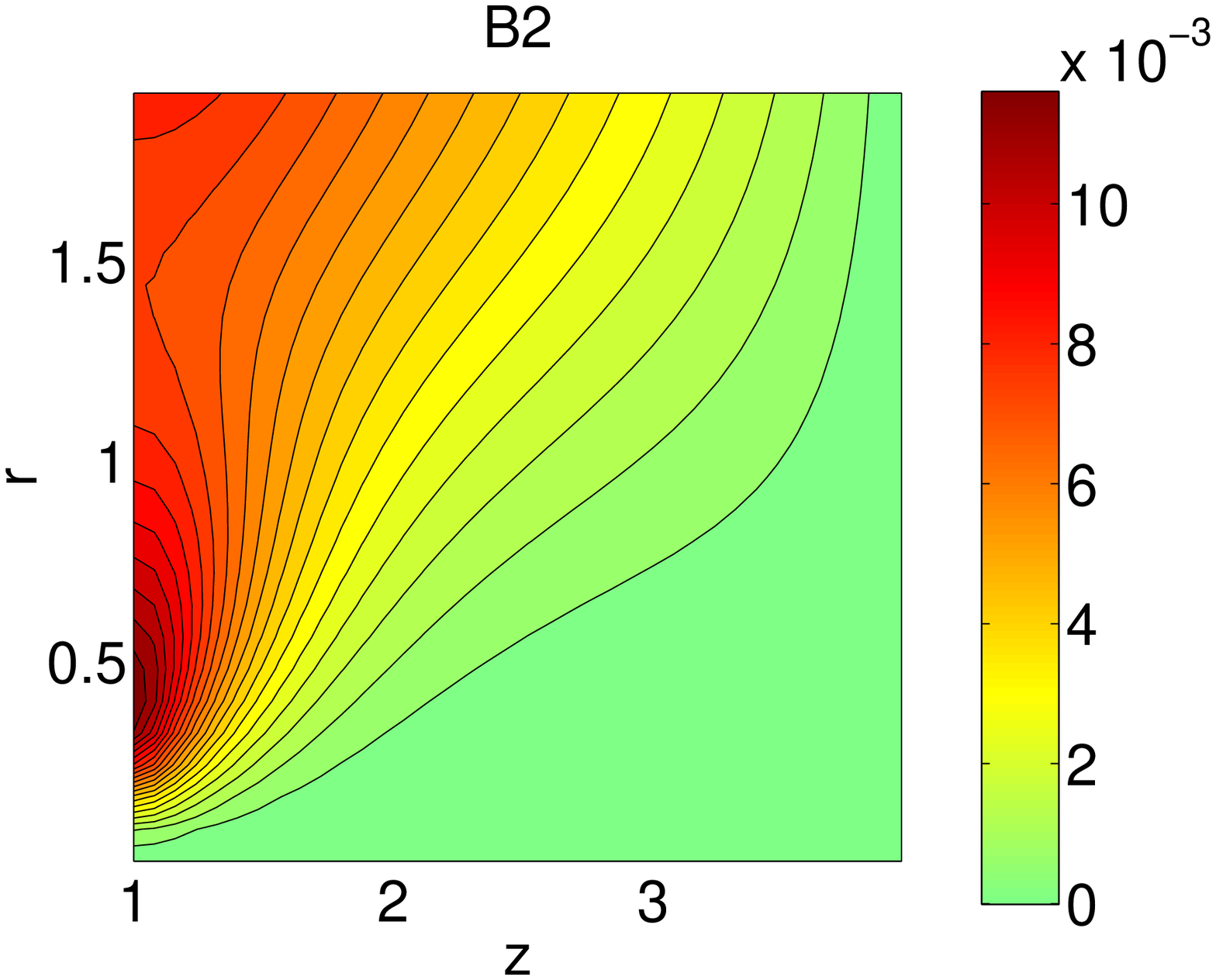,width=6cm}}
\vspace{1cm}
\caption[short]{ \figuremode  An illustration of $T, A, B$ and $B2$ for a star with $\xi =
  0.3$ and $\rho_0 = 5.0$. The star is dense enough that the
  configuration is non-linear, the peak value of $|T|$ being $\simeq
  0.7$. As discussed earlier the metric contains singular terms
  involving $B$. Not only can we relax $B$ using the bulk equations,
  we can also integrate $B$ out from the origin $r=0$ using the
  constraint $\{rz\}$ to find another function, $B2$, which would
  equal $B$ if the Einstein equations were satisfied exactly. $B2$ has
  cleaner quadratic behavior at $r=0$ than the relaxed $B$, and is
  therefore used to calculate some of the singular source terms in the
  $B$ bulk `Poisson' equation, rendering the source term more stable.
  The key point we see in the plot here is that the two values of $B$
  and $B2$ agree very well over the lower half of the lattice, $r <
  0.5 \, r_{\rm max}$. In the upper half of the lattice $B$ is forced
  to zero by the boundary conditions imposed whereas the integrated
  $B2$ is not.  Even there, the difference between the functions is
  small compared to the values of the other metric functions $T, A$.
  It is only over the lower half of the lattice that the singular
  source terms have significant contribution (see figure
  \ref{fig:origin_check2}) due to their suppression by $\frac{1}{r}$
  factors, and therefore the plots suggest that the singular term
  approximation is very good. They also show that whilst $B$ and $B2$
  differ at large $r$, the difference is very small in absolute terms
  showing the constraints are well satisfied. (lattice: $dr = 0.01$,
  $r_{\rm max} = 2$, $dz = 0.01$, $z_{\rm max} = 4$)
\label{fig:origin_check1} 
}
\end{figure}

Section \ref{sec:method} discussed how the function $B$ is integrated
from $\{rz\}$ giving $B2$ which is used to calculate singular terms in
the bulk equations. We compare the $B$ relaxed from the bulk equations
and $B2$ integrated from the constraint. The two functions are
extremely close in the lower half of the $r$ range, as shown in figure
\ref{fig:origin_check1} for a typical, dense \emph{small} star
solution. The integration of $B2$ starts from $r=0$ and thus we expect
the largest difference at the asymptotic $r$ boundary.  For both
\emph{small} and \emph{large} stars, we find the value of both $B$ and
$B2$, and hence their difference, are extremely small compared to the
metric functions $T, A$ indicating that the constraints are very well
satisfied for the resolutions and boundary locations used.

\begin{figure}
\centerline{\psfig{file=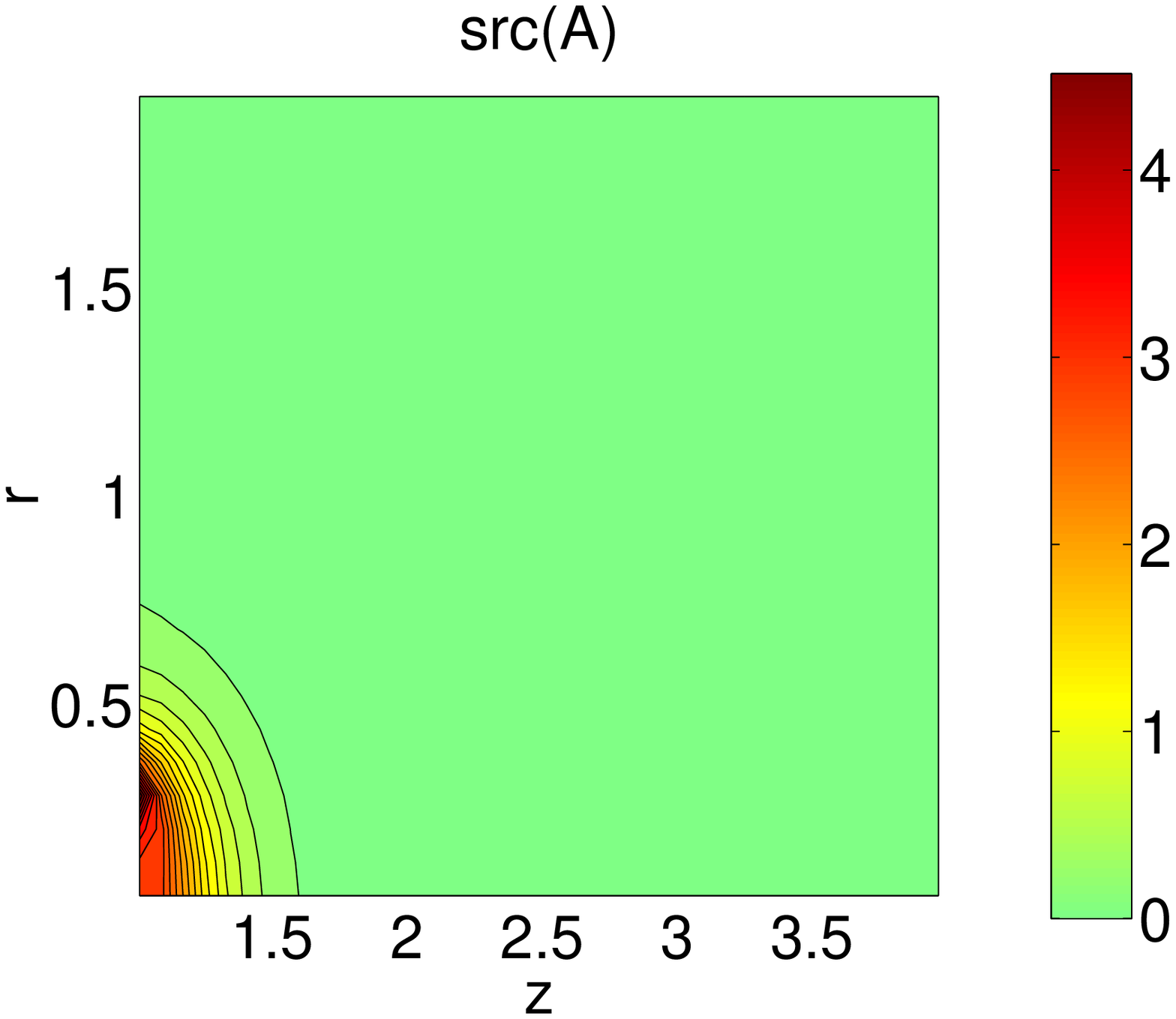,width=6cm}
  \hspace{0.1cm} \psfig{file=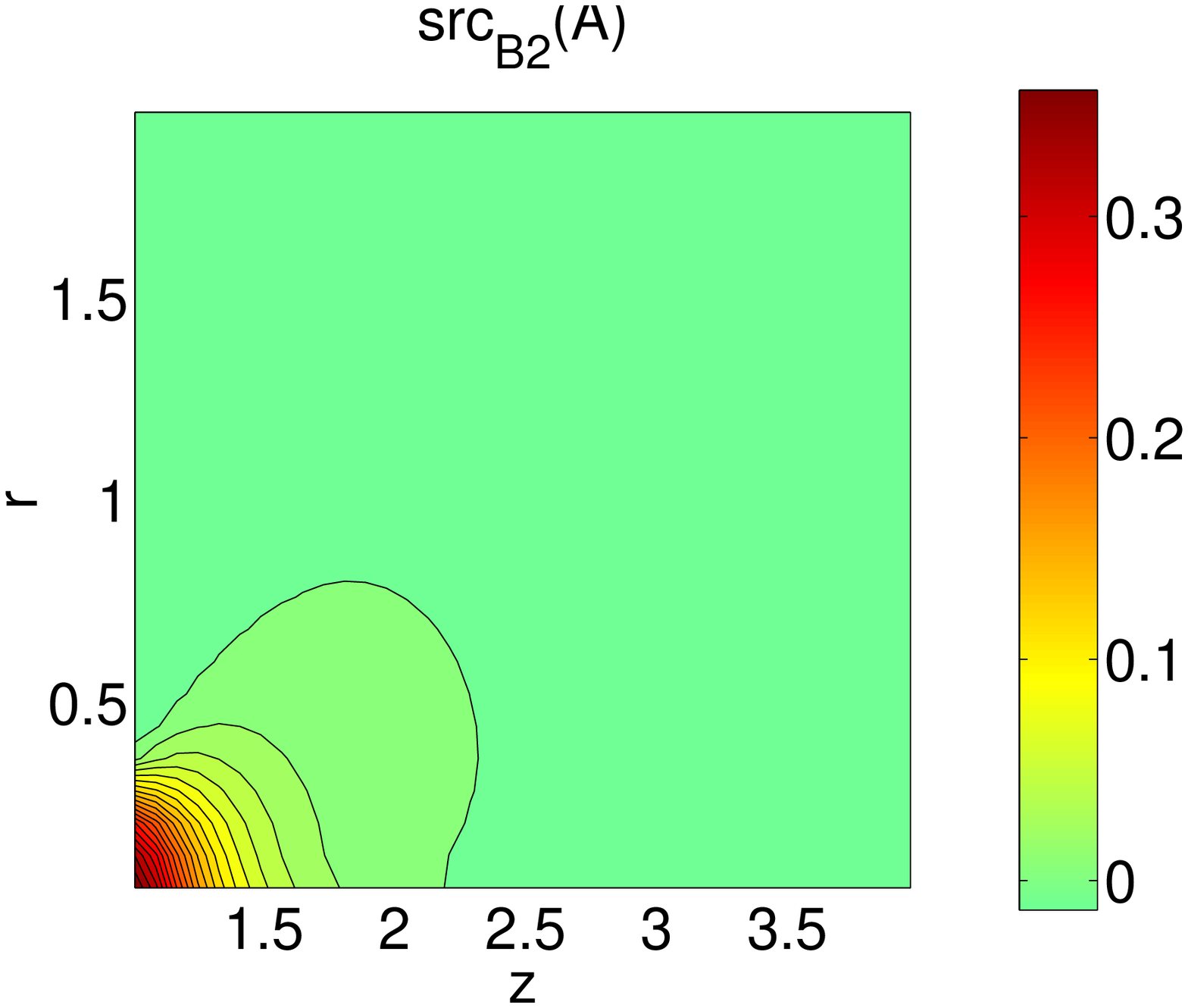,width=6cm} \hspace{0.1cm} \psfig{file=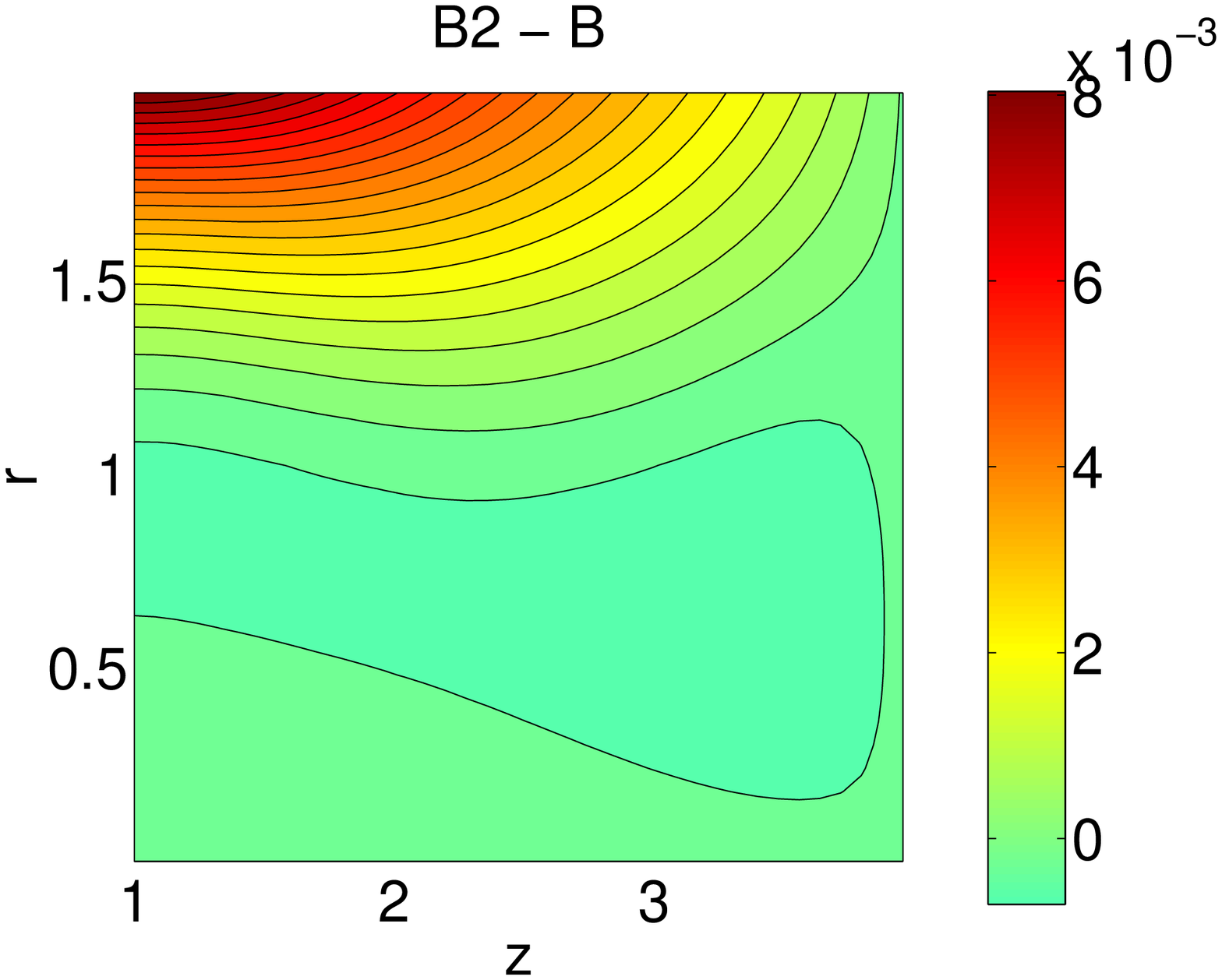,width=6cm}}
\centerline{\psfig{file=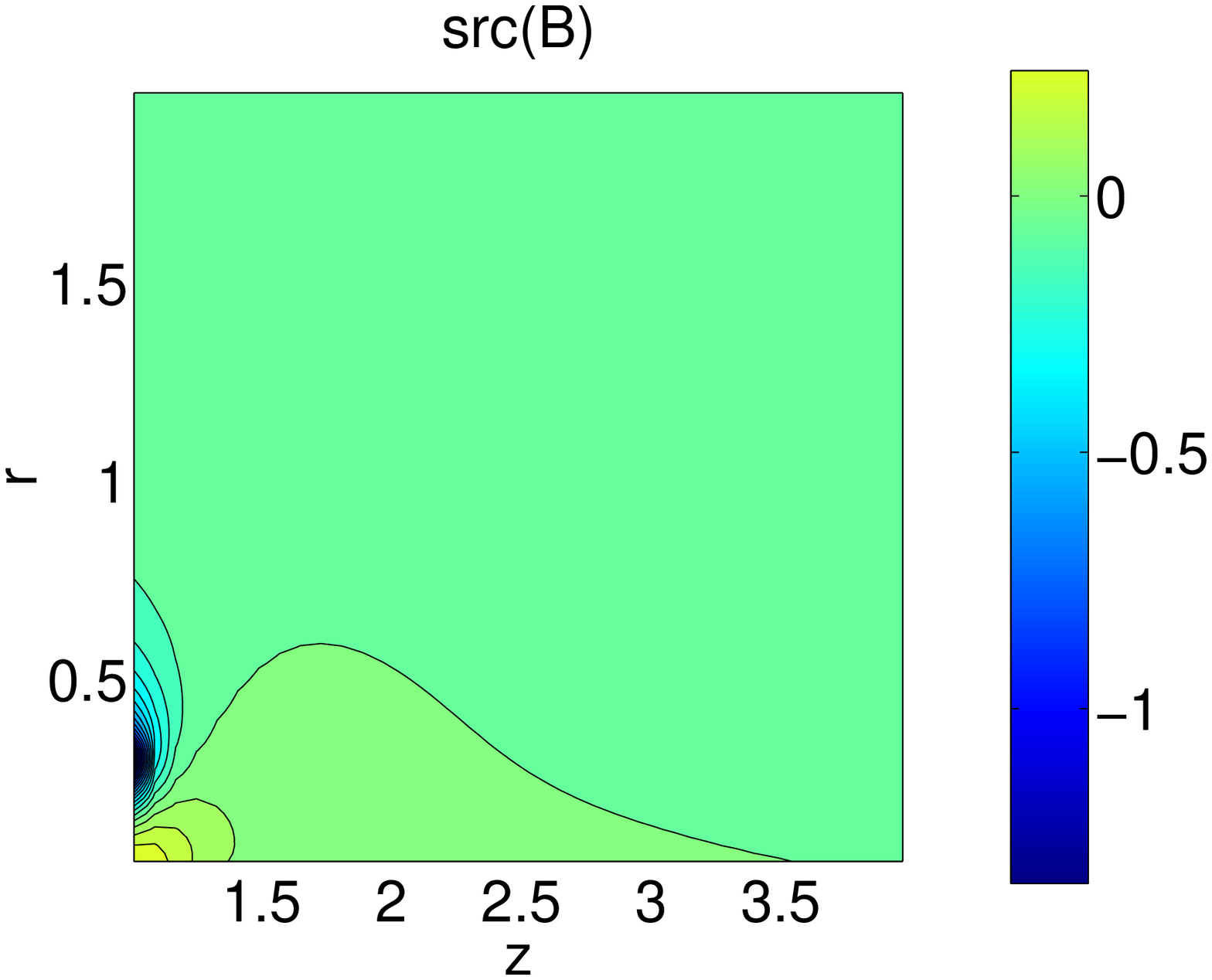,width=6cm}
  \hspace{0.1cm} \psfig{file=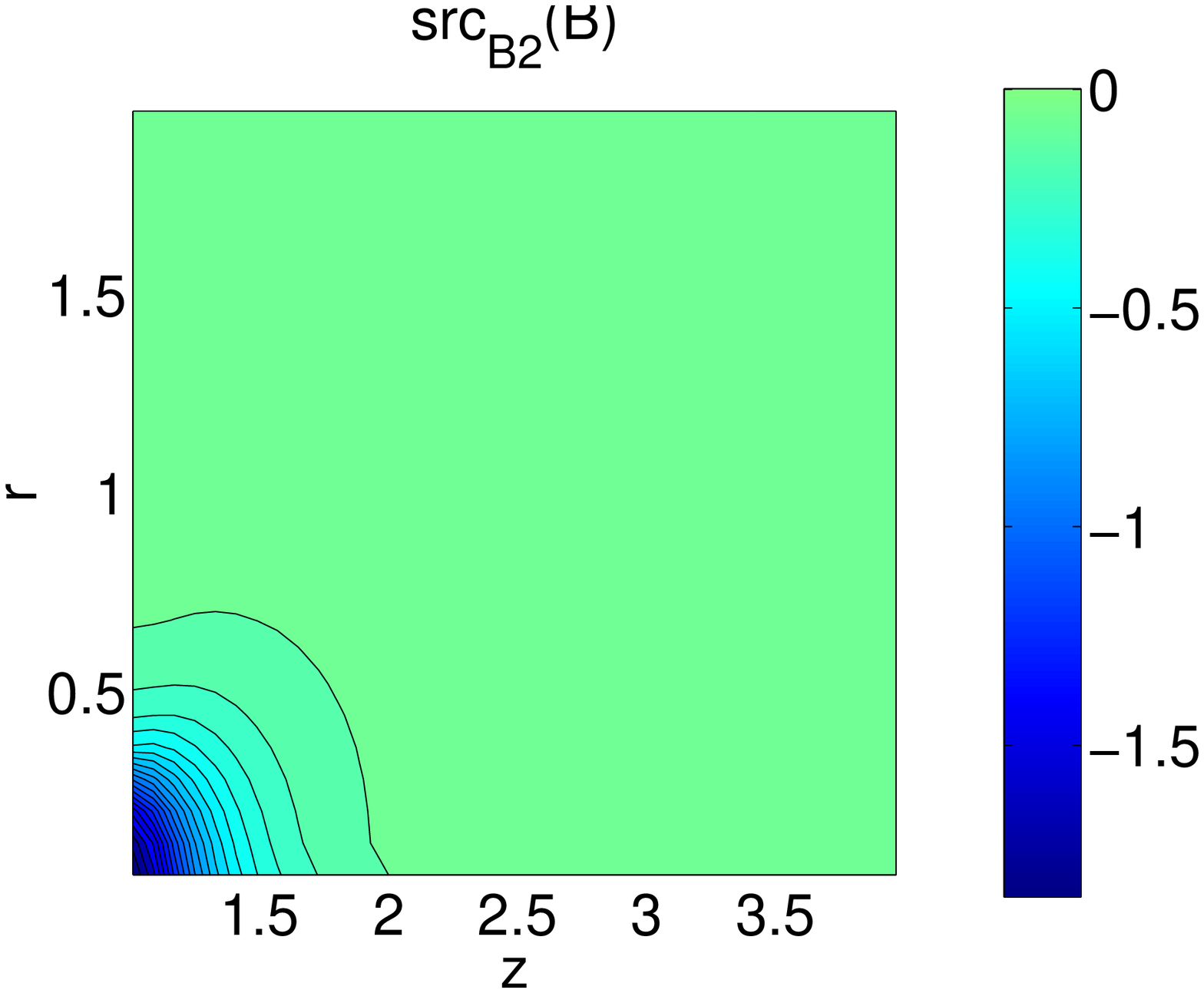,width=6cm}}
\vspace{1cm}
\caption[short]{ \figuremode  An illustration of $src(A), src_{B2}(A), src(B),
  src_{B2}(B)$ and $B2-B$ for the configuration $\xi = 0.3$ and $\rho_0
  = 5.0$ shown in figure \ref{fig:origin_check1}. The top right frame
  shows the difference between $B$ and the integrated $B2$, used for
  the singular source terms in the $B$ equation, as integration out
  from the origin gives cleaner quadratic behavior at $r=0$. There is
  very good agreement between $B$ and $B2$, with $B2-B$ being very
  small, showing the constraints are well satisfied. The divergence at
  large $r$ is still very small in absolute terms (see figure
  \ref{fig:origin_check1}).  On the top left and bottom left are the
  plots $src(A), src(B)$ which show the full source terms for the $A$
  and $B$ bulk `Poisson' equations \eqref{eq:TAB_poisson}. On their
  right are $src_{B2}(A), src_{B2}(B)$ which are the contributions
  from the singular terms where $B$ is replaced by $B2$ to get
  improved regular behavior at $r=0$. We see that for $A$ these
  contributions are small compared to the other non-singular terms,
  but for $B$ they are significant. We also see that all the source
  terms are localized about the origin, exactly where we see $B2$ is
  an excellent approximation to $B$. This justifies the use of the
  method and also indicates why it works so effectively.  Although
  integrating $B2$ is highly non-local, as the source terms are
  confined near the origin, the non-locality is not transfered into
  the relaxation procedure and we obtain a stable method.
\label{fig:origin_check2} 
}
\end{figure}

How good an approximation is it to use $B2$ to replace singular source
terms in the $A, B$ bulk equations? $B2-B$ is extremely small in the
lower half of the $r$ range and figure \ref{fig:origin_check2} shows
the contributions of the singular terms calculated using $B2$ in these
equations.  The key feature is that the contributions are localized
near $r=0$ due to the inverse $r$ factors in these singular terms.
Both the $A$ and $B$ sources are localized near $r=0$, as are the
contributions from the singular terms. Thus the sources only give
significant contributions in the region where $B2$ extremely well
reproduces $B$, the lower half of the $r$ range. Therefore we can
expect this to be a very good approximation to make in the equations.

%
\subsection{Degree of Einstein Equation Violation}
%

\begin{figure}
\centerline{\psfig{file=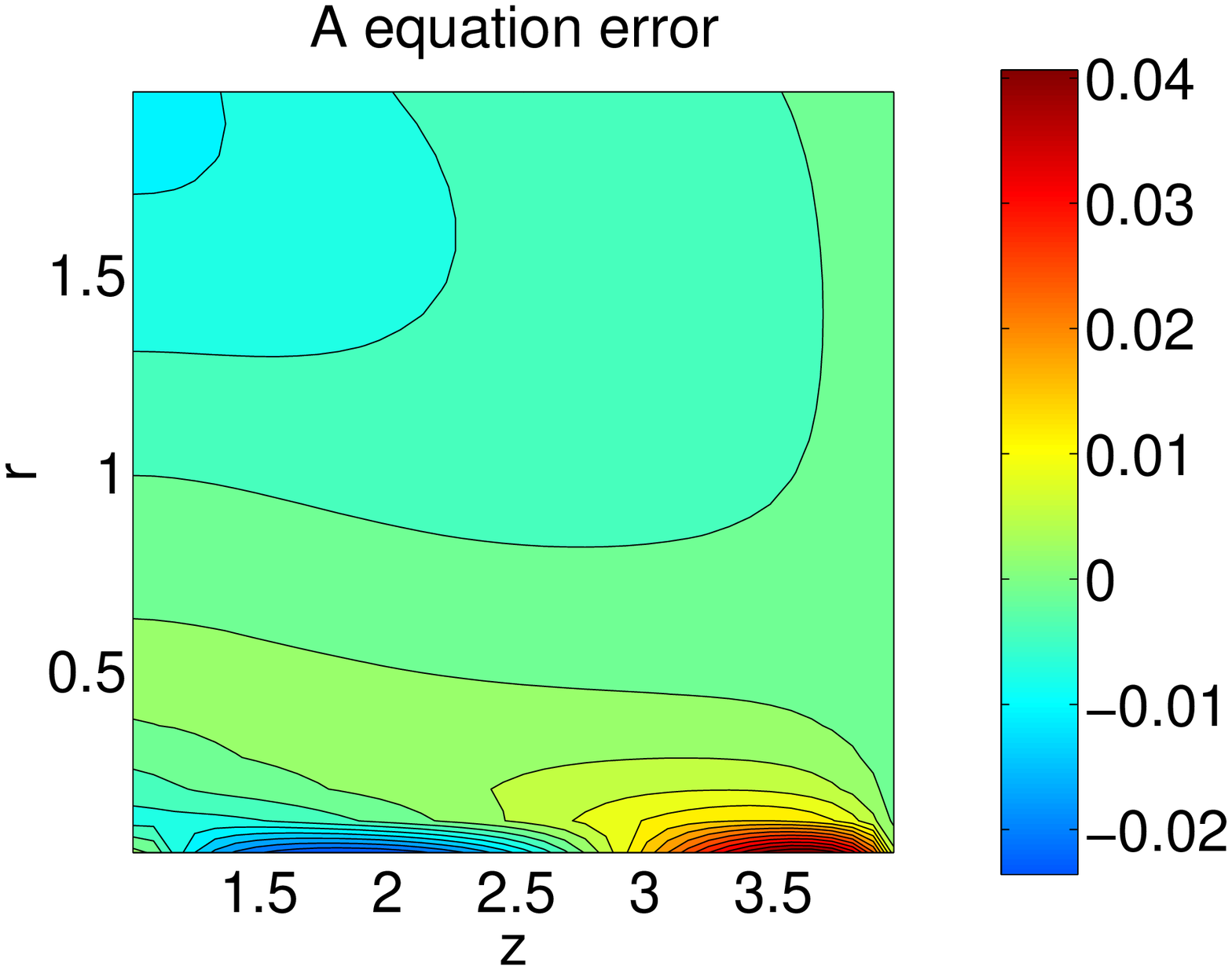,width=6cm}
  \hspace{0.1cm} \psfig{file=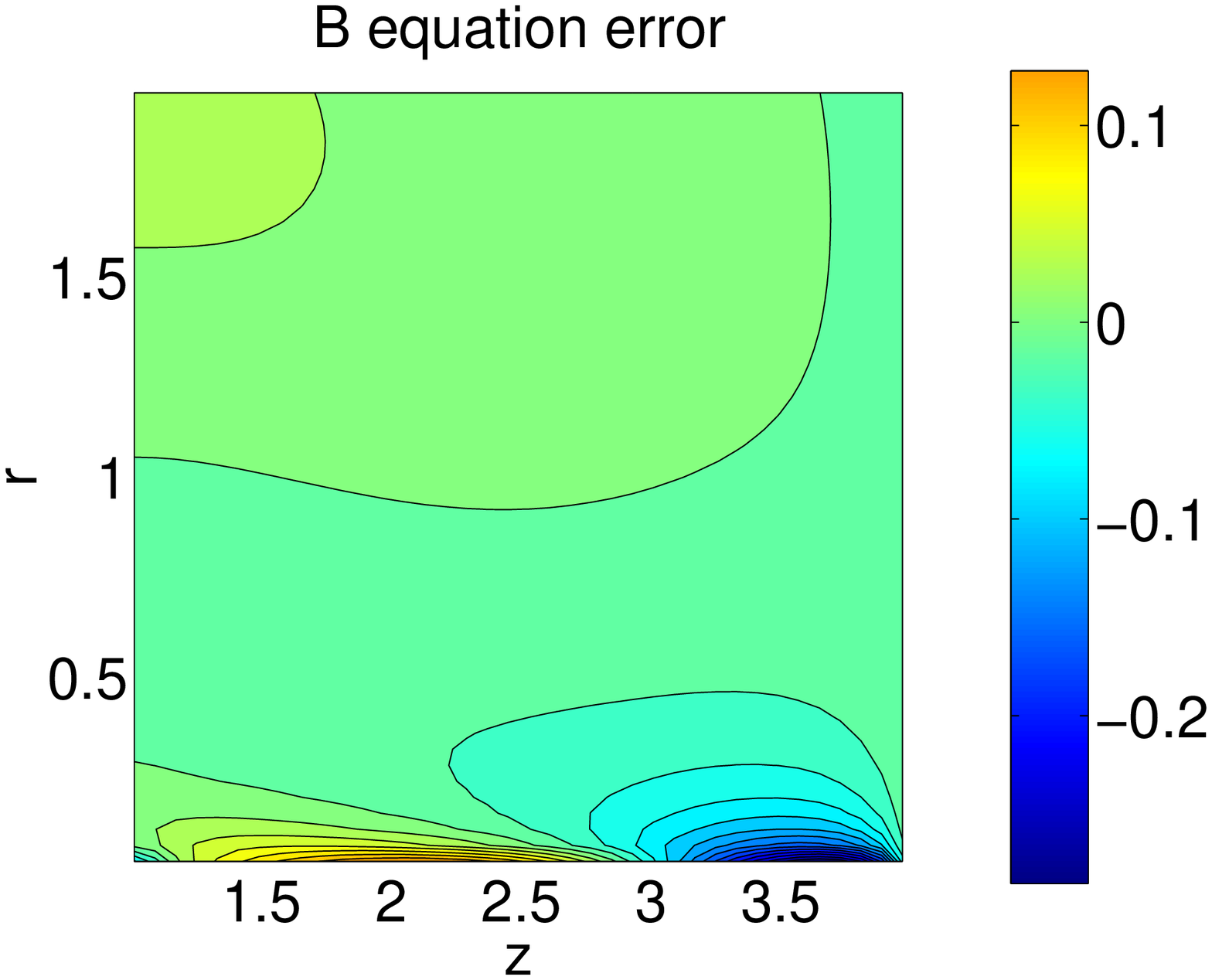,width=6cm}}
\centerline{\psfig{file=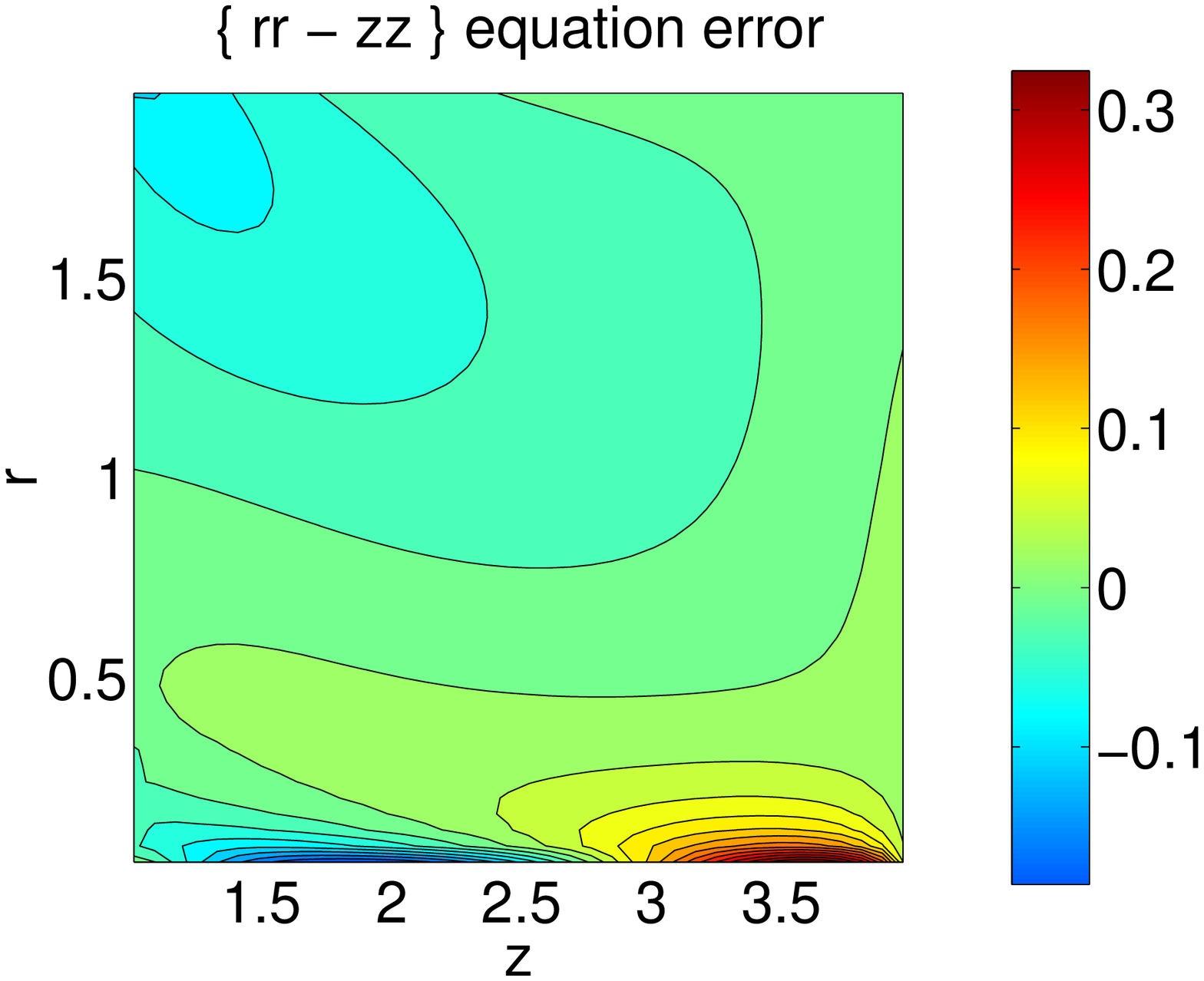,width=6cm}
  \hspace{0.1cm} \psfig{file=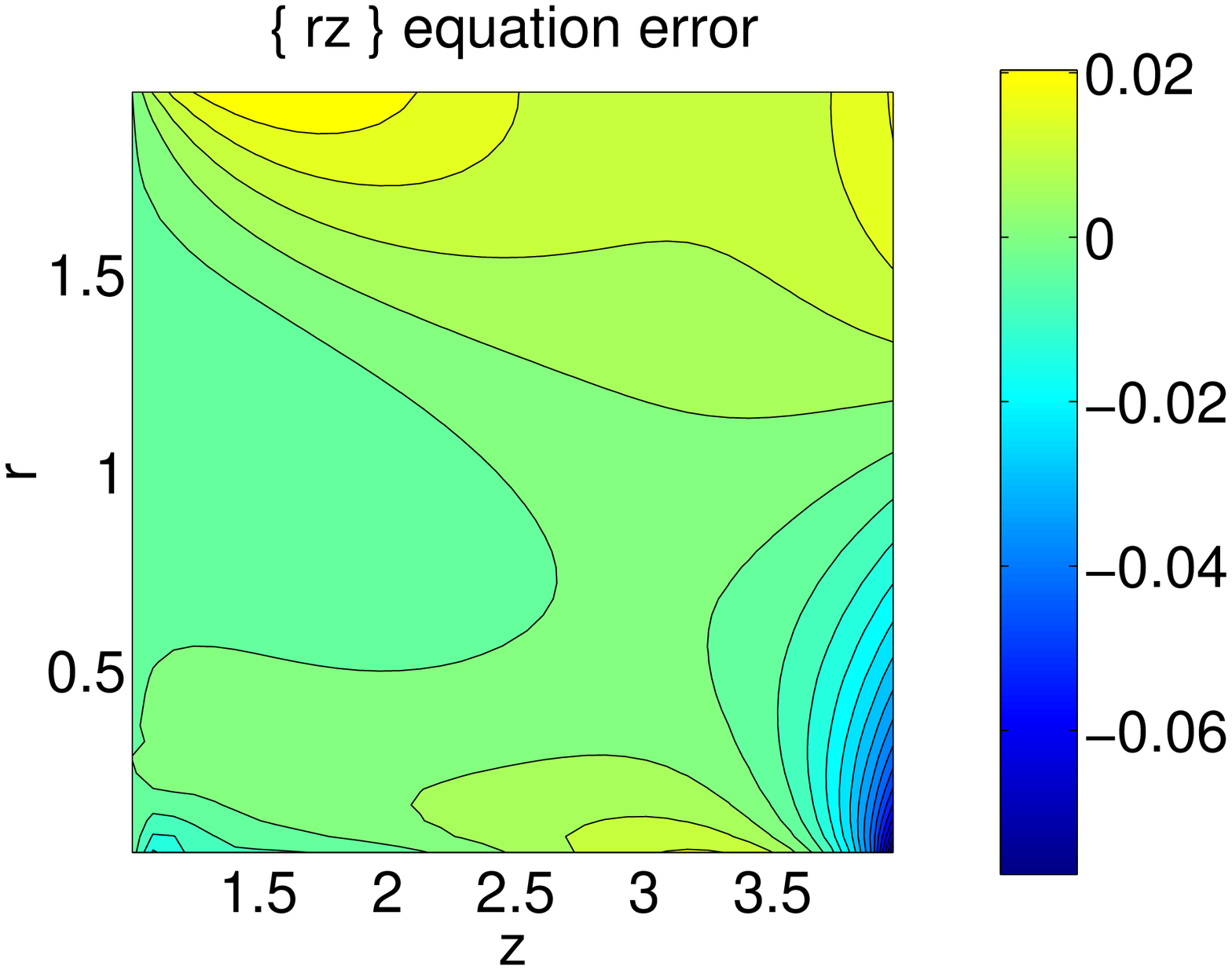,width=6cm}}
\caption[short]{ \figuremode  
  Plots of all the Einstein equations for $\xi = 0.3$, with non-linear
  source $\rho_0 = 5.0$. Top; an illustration of error in the $A, B$
  equations when using the true $B$ for the singular source terms
  rather than $B2$. Bottom; the constraints $\{ rz \}, \{ rr-zz \}$.
  Note that the $T$ equation error is effectively zero as there are no
  singular source terms where $B$ is substituted for $B2$. As one
  would expect, the largest errors are found near $r=0$. It must be
  noted that the peak absolute value of the Weyl components
  $C1,C2,C3,C4$ is $\simeq 14$ and these peak values occur away from
  the symmetry axis and thus are not effected by the singular term
  regularization, (eg. see figure \ref{fig:weyl} for the components on
  the brane and symmetry axis for the slightly denser $\rho_0 = 7.0$
  star). Thus the peak Einstein equation error is $\simeq 2 \%$ of
  this peak physical curvature value, implying the Einstein equations,
  both elliptic and constraints are more than adequately satisfied at
  the resolution and lattice size used. (lattice: $dr = 0.01$,
  $r_{\rm max} = 2$, $dz = 0.01$, $z_{\rm max} = 4$)
\label{fig:origin_check3} 
}
\end{figure}

We graphically illustrate the constraints $\{rr-zz\}$ and $\{rz\}$ in
figure \ref{fig:origin_check3}, together with the errors in the
equations \eqref{eq:TAB_poisson} for $A, B$. The bulk equations for
$T, A, B$ are relaxed to machine precision, but in the $A, B$
equations some singular terms in the source are calculated from $B2$
rather than $B$. Here we plot the error in these $A, B$ equations by
differencing the left- and right-hand sides of the `Poisson' equations
\eqref{eq:TAB_poisson}, but using only $B$ to calculate the source.
This difference then gives the error due to the singular term
regularization scheme. 

We firstly note that the errors in the $A$ source are much less than
in $B$. This is evident from the previous section where we saw the
contribution to the $A$ source from singular terms is correspondingly
less. As expected, the errors are localized mainly about the origin.
Now consider the constraint $\{rz\}$. This has no singular terms and
has no large violations at the origin. The maximum values of violation
occur at large $z$ and the error is small compared to that of
$\{rr-zz\}$, which is localized at the origin and is $\sim 5$ times
greater. We conclude from this that the constraints are indeed well
satisfied, as indicated by $\{rz\}$, the major contribution to the
error in the $\{rr-zz\}$ equation appearing not to come from
asymptotic violations due to finite lattice size but rather from the
singular term regularization.

We must compare these values to the physical curvatures of the
solutions. The Weyl components provide just such a measure. In fact we
compare with the quantities $C1, C2, C3, C4$ which have the
blue-shifting factor of $\frac{1}{z^2}$ removed, discussed in appendix
\ref{app:einstein_eqns}, as the $A, B$ equations and constraints
similarly have this factor removed relative to the Einstein tensor
components. The peak value of these quantities is $|C2|_{\rm peak}
\simeq 14$. In addition, the peak curvatures are not located on the $r
= 0$ axis (eg. figure \ref{fig:weyl}) and thus are not sensitive to
the singular term regularization procedure there. It is not physically
sensible to make a direct numerical comparison, but these physical
curvatures are clearly much larger than the errors induced from the
origin regularization scheme by a factor of $\sim 50$. Indeed,
comparing to the constraint $\{rz\}$ which we saw above appears not to
suffer directly from the origin singular terms, the factor is even
larger $\sim 250$, indicating that the asymptotic boundary conditions
and Bianchi identities are well satisfied.  This is consistent with
the conclusions of the comparison with the independent linear method
of section \ref{sec:linear_check}, the linear method satisfying the
asymptotic behavior, and constraints, and agreeing extremely well
with the non-linear method in the low density regime. Furthermore, the
close agreement of $B$ in the solutions of the linear and non-linear
method confirms the origin regularization scheme performs well. This
section shows the quality of solution remains high for non-linear
configurations.

%
\subsection{Convergence and Finite Size Properties}
\label{sec:converge}
%

\begin{figure}
\centerline{\psfig{file=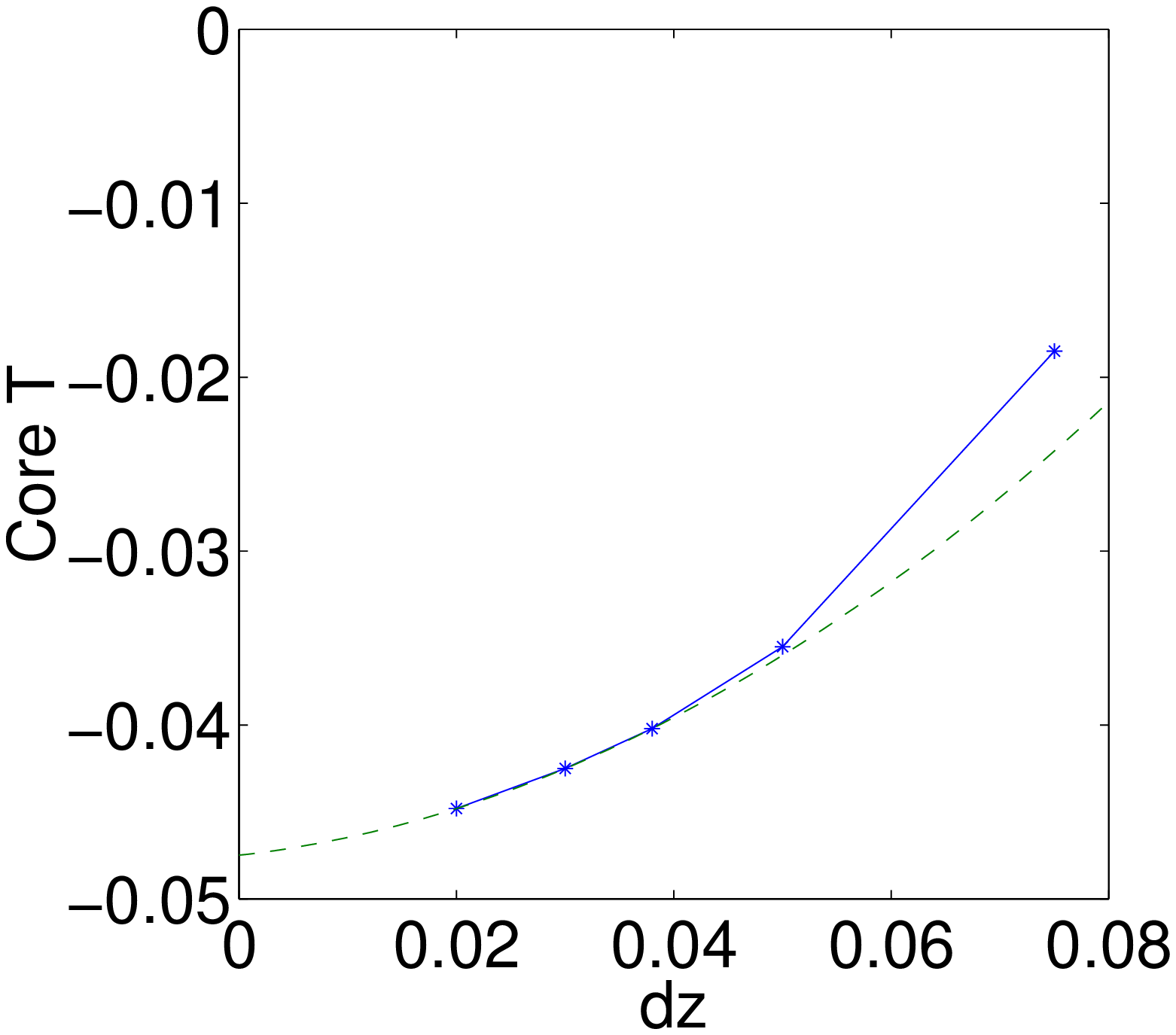,width=6cm}
  \hspace{0.1cm} \psfig{file=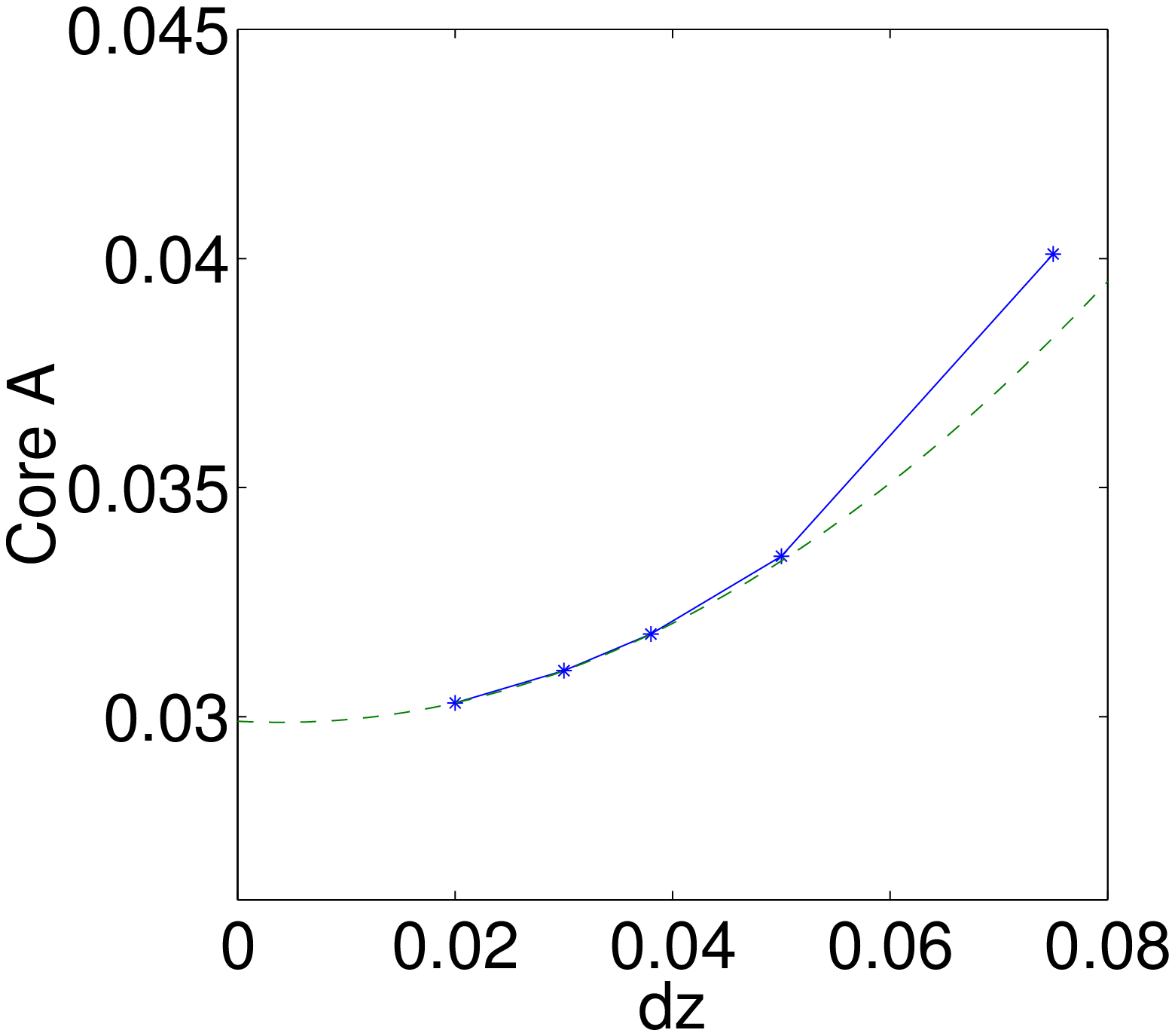,width=6cm}}
\vspace{0.2cm}
\centerline{\psfig{file=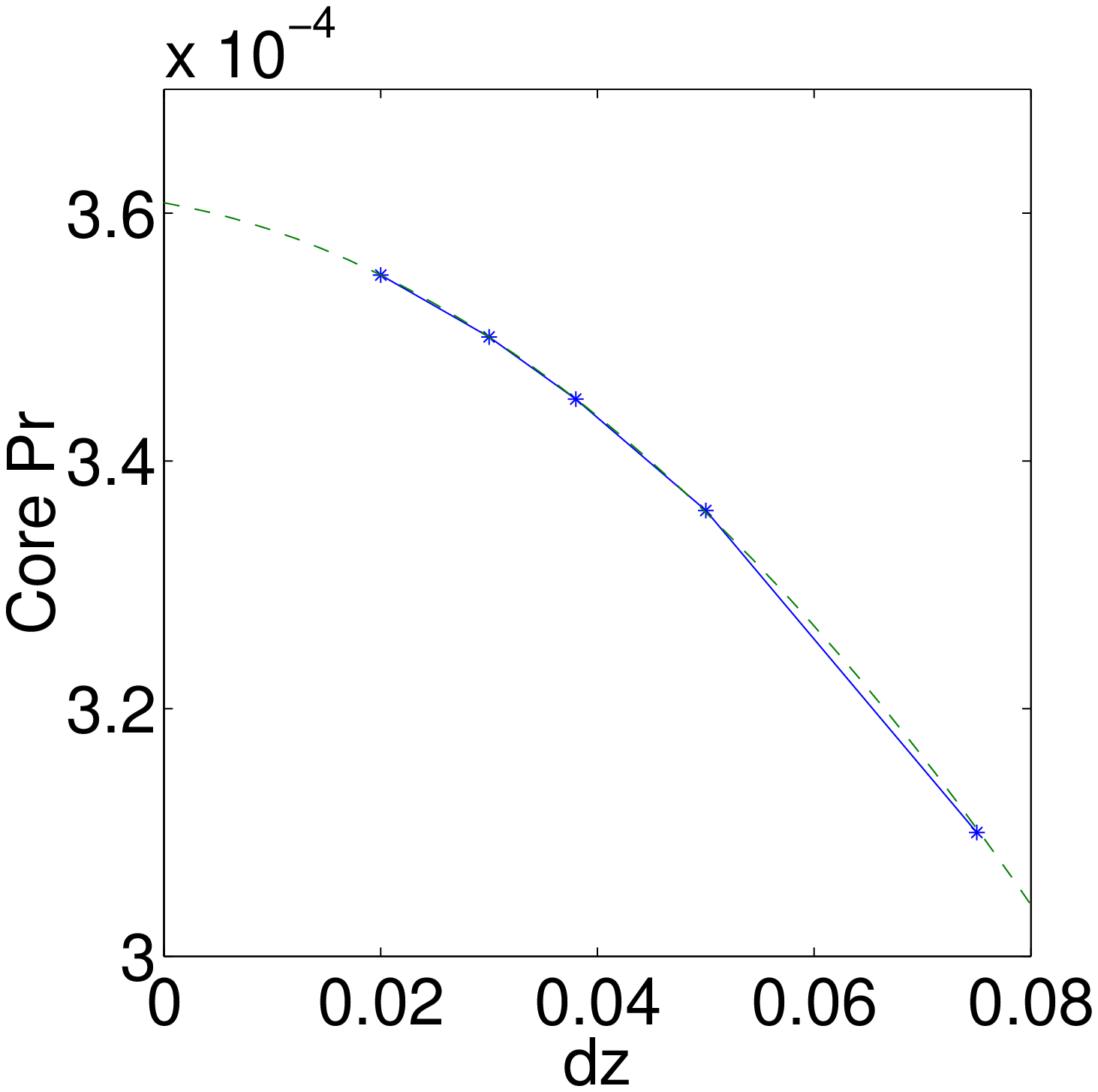,width=5.7cm}
  \hspace{0.1cm} \psfig{file=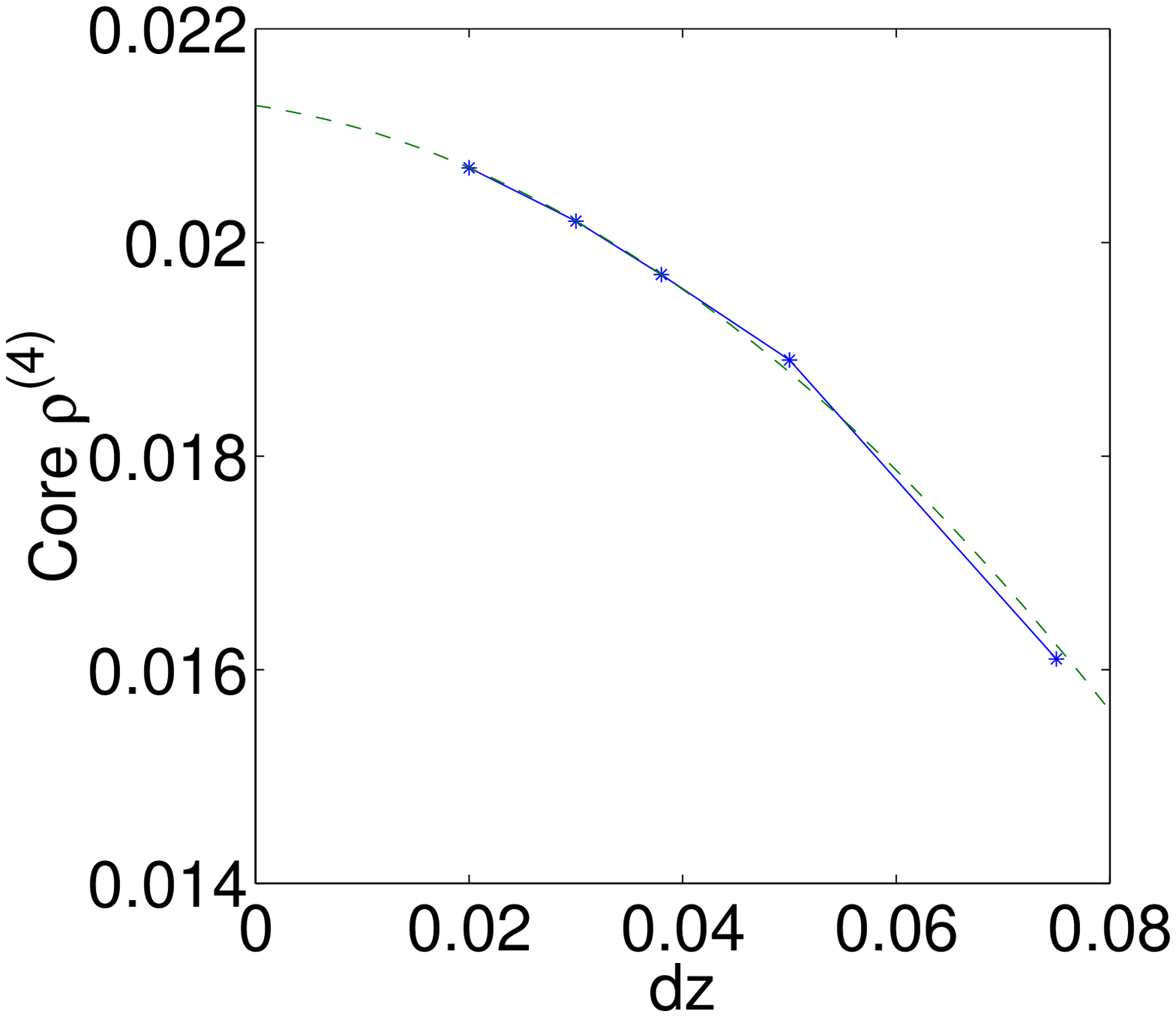,width=6cm}}
\vspace{0.2cm}
\caption[short]{ \figuremode  An illustration of convergence tests for the \emph{largest}
  star with $\xi = 3.0$ and a low density $\rho_0 = 0.02$. The lattice
  $z$ resolution is varied whilst keeping $z_{\rm max}$ fixed. Note
  that $r$ resolution variation behaves in a similar manner giving
  second order convergent quantities. The $z$ resolution is more
  interesting in its behavior as the brane has a coordinate distance
  of one unit from $z = 0$. Thus in order to capture behavior well one
  requires much better resolution, $dz << 1$. The core values ($r=0,
  z=1$) for the metric functions $T, A$ and the core pressure and
  effective density are plotted against varying $dz$.  Second order
  behavior is seen in $dz$ as expected and quadratic fits are
  performed (shown as the dotted line) using the smallest 3 values of
  $dz$, the extrema of the quadratic fits each being consistent with
  $dz = 0$, as required by second order scaling.  We see a deviation
  from the quadratic fit at large $dz$. The largest variation is seen
  in $T$ and for $dz = 0.075$ the value is over a factor of two
  different from its apparent continuum value. In both $T, A$ the
  largest point appears to deviate from the quadratic scaling. For $dz
  = 0.15$ no convergence was possible, the method behaving unstably.
  When $dz$ is much larger than $\sim 0.1$ the scaling deviates from
  its asymptotic form as the lattice no longer resolves the gap
  between the brane and singular boundary $z = 0$. In fact comparing
  the $dz = 0.03$ values with the continuum extrapolation we find the
  largest error for $T$ gives a $\sim 10 \%$ variation, the other
  quantities having smaller errors $\lesssim 4 \%$. $dz = 0.03$ is the
  resolution used elsewhere in this paper for \emph{large} star
  calculations.  (lattice: $dr = 0.2$, $r_{\rm max} = 20$, $dz = {\rm
    various}$, $z_{\rm max} = 46$)
\label{fig:converge} 
}
\end{figure}

The convergence of solutions with increasing $r$ and $z$ resolution
was tested, figure \ref{fig:converge} showing various intrinsic brane
quantities for differing $dz$. Second order convergence is clearly
seen, and also for varying $dr$ which is not plotted here. For $dr$
the resolution is relatively higher, the extrapolated continuum values
for the metric functions on the brane differing from those at the
resolutions used elsewhere in the paper, by a maximum of $\sim 0.3
\%$, for $\xi = 3.0$ stars. This is clearly much smaller than any
systematic expected from section \ref{sec:linear_check}. The case of
varying $dz$ is more interesting than $dr$ as there is a fundamental
bound on resolution from the coordinate distance of the brane to the
boundary of AdS, at $z = 0$. For \emph{small} stars the total lattice
size may be a few AdS lengths and the brane is positioned at $z = 1$.
Thus even at low resolutions the lattice does `see' that the brane is
not located at the very singular boundary. Similarly in $r$, where the
lattice is chosen to be a factor of $\sim 10$ wider than the star
itself, and only on extremely coarse lattices would the star fail to
be resolved.  Whilst the situation in $r$ remains the same for
\emph{large} stars, such as $\xi = 3$, the $z$ lattice extends out to
$z \sim 45$, and the distance between the boundary and $z = 1$ must
still be well resolved or the solution will behave in a singular
manner at the brane, and not relax. Thus large $z$ resolutions are
critical in order to get convergence for \emph{large} stars. Whilst
the $r$ resolution can scale with the star radius, the $z$ resolution
must be $dz \lesssim 0.1$ regardless of the star radius or the
solutions will not relax. For the largest stars computed, the typical
resolution used, $dz = 0.03$, gives a maximum difference of $\simeq
10\%$ from the extrapolated continuum value for the metric functions.
This is of order the differences with the independent linear method of
section \ref{sec:linear_check}, expected to give a measure of
systematic error.  Thus in section \ref{sec:non-linear} we extrapolate
to a continuum value using two different $dz$ lattices to calculate
the same solution. The details of the procedure are described in
appendix \ref{app:numerical}. It is not exact, as with less resolution
the solutions for denser stars do not converge, and thus the
extrapolation must itself be extrapolated. However the error in this
is expected to be much less than the systematic indicated in section
\ref{sec:linear_check}.

It is crucial to test the sensitivity of the solutions to the physical
size of the lattice. Table \ref{tab:origin} shows averages of the
absolute values of both the constraints, and in addition the equation
$\{rr+zz\}$ for various lattice sizes. As discussed, $\{rr+zz\}$ is an
elliptic equation which is satisfied to machine precision, but with
the singular term approximation. The $\{rr+zz\}$ given here is without
this singular term replacement and therefore indicates the error
involved in this approximation. The averages in the tables are
computed over the common inner region of the lattices, excluding the
innermost 5 grid points which, whilst being well behaved and regular
even with the $\frac{1}{r}$ terms in the constraints, would weight the
regularization errors near the axis very strongly.

We find that varying the size of the lattice in the $r$ direction
keeping $z_{\rm max}$ fixed has little effect on the constraints. Thus
the size changes displayed only involve the $z$ extent, which for our
configurations and resolutions does influence the quality of the
solution. One clearly sees that increasing $z_{\rm max}$, keeping the
$z$ resolution and $r_{\rm max}$ fixed, decreases both the constraints
and also $\{rr+zz\}$. One sees that for the \emph{larger} star, $\xi =
1$, the effect is very dramatic if $z_{max} < r_{max}$. Empirically we
find that the lattice must be as large or larger in $z$ than in $r$ to
get good quality solutions, the effect being more marked the
\emph{larger} the star. In terms of absolute constraint violation we
may compare these averaged values to the physical curvature computed
in the rescaled Weyl tensor components $C1, C2, C3, C4$ (appendix
\ref{app:einstein_eqns}) and again we see that both the constraint
violation, and the error due to the singular source term
regularization using $B2$, are again several orders of magnitude
smaller than the curvatures generated in these solutions. We see the
averaged error $<| \{rr+zz\} |>$ also becomes less, together with the
constraints. This is because $B2$ is integrated, using the $\{rz\}$
constraint, in from $r = 0$ and the $z_{max}$ boundary. The better the
boundary conditions at $z_{\rm max}$, the closer $B$ is to $B2$
globally and the better the approximation.

\begin{table}
\begin{center}
\begin{tabular}{llllllll}
$\rho_0$ & $\xi$ & $r_{\rm max}$ & $z_{\rm max}$ & $< | \{rz\} |>$ & $<|
  \{rr-zz\} |>$ & $<| \{rr+zz\} |>$ &
  $|C2|_{peak}$ \\ 
\hline \hline
0.1 & 0.3  & 2.0 & 1.5    & 3.2\e{-4}  & 1.5\e{-3}  & 2.0\e{-3}  & 0.32 \\
(\emph{linear}) & (\emph{small})
            &     & 2.0   & 1.0\e{-4}  & 0.79\e{-3} & 1.1\e{-3}  &      \\
     &      &     & 3.0   & 0.30\e{-4} & 0.31\e{-3} & 0.38\e{-3} &      \\ \hline
5.0  & 0.3  & 2.0 & 1.5   & 2.7\e{-2}  & 1.3\e{-1}  & 1.7\e{-1}  & 14.1 \\
(\emph{non-linear}) & (\emph{small})
            &     & 2.0   & 0.83\e{-2} & 0.62\e{-2} & 0.89\e{-1} &      \\
     &      &     & 3.0   & 0.24\e{-2} & 0.24\e{-2} & 0.29\e{-1} &     \\ \hline
0.01 & 1.   & 10. & 6.0   & 6.1\e{-5}  & 2.3\e{-4}  & 1.6\e{-4}  & 1.1\e{-2} \\
(\emph{linear}) & (\emph{medium})
            &     & 10.   & 0.93\e{-5} & 0.27\e{-4} & 0.30\e{-4} &     \\
     &      &     & 15.   & 0.29\e{-5} & 0.12\e{-4} & 0.12\e{-4} &     \\
\end{tabular}
\caption[short]{ \figuremode  
  Table showing constraint violation for 3 configurations; two sizes,
  $\xi = 0.3, 1$, and both linear and non-linear densities for one
  size, $\rho_0 = 0.1, 5.0$ respectively for $\xi = 0.3$. One finds
  that for $r$ resolutions high enough to capture the density profile
  accurately, increasing or decreasing the lattice size makes little
  difference to the constraints. It is the extent of the lattice in
  the $z$ direction that determines the constraint satisfaction. The
  absolute values of the constraints are averaged over lattices with
  different $z_{\rm max}$. The average is taken only over the common
  portion of the lattices for a given $\xi, \rho_0$. The first 5
  lattice points in $r$ are excluded as $\frac{1}{r}$ contributions
  receive an unphysically large weighting there. In all cases the
  constraints are better satisfied as $z_{\rm max}$ is increased for
  fixed $z$ resolution. The effect is more marked for the
  \emph{larger} star with $\xi = 1$. This is because the metric falls
  away less quickly in $z$ the \emph{larger} the star, and thus the
  lattice must be large enough in $z$ to ensure asymptotic AdS is well
  approximated. Typically for $z_{\rm max} \gtrsim r_{\rm max} \gtrsim
  6 \xi$ the lattice is large enough to accommodate the star.  We see
  in the $\xi = 1$ case that there is a huge improvement in increasing
  $z_{\rm max}$ from $0.5 \, r_{\rm max}$ to $r_{\rm max}$, but then
  the improvement is much less for the next increase to $1.5 \, r_{\rm
    max}$, showing that the perturbation was not captured for $z_{\rm
    max} = 0.5 \, r_{\rm max}$, but is for $z_{\rm max} \geq r_{\rm
    max}$.  Note that the peak value of the Weyl component $C2$ (see
  appendix \ref{app:einstein_eqns}) is shown.  This allows one to
  compare the average constraint violation against a quantity which
  characterizes the curvature. We see that the average constraint
  violations are considerably smaller, $\sim 1000$ times, for the two
  linear configurations and for the non-linear one even more. Thus the
  constraints appear to be even better satisfied relative to the
  perturbation for dense stars, the regime of interest for this paper.
  The elliptic equation $\{rr+zz\}$ is also plotted, but now without
  the singular source terms replaced using the integrated $B2$
  solution.  Of course with the $B2$ replacement this equation is
  satisfied to very high precision as specified by the relaxation
  procedure.  Without the replacement, the quality of the
  approximation can be assessed.  Note that the violation of this
  equation has similar behavior with grid size to that of $\{rz\}$.
  This is to be expected as $B2$ is integrated using the $\{rz\}$
  constraint equation.  Therefore when this equation is better
  satisfied, the $B2$ source term replacement is also a better
  approximation. (For $\xi = 0.3$; $dr = 0.01$, $dz = 0.01$. For $\xi
  = 1$; $dr = 0.1$, $dz = 0.01$)
\label{tab:origin}}
\end{center}
\end{table}

In table \ref{tab:origin2} we plot the core metric function values
against lattice size in $z$. Again we see convergence of the solution
as the physical $z$ size increases, although the variations in the
values are much smaller, a maximum variation of $\sim 4 \%$ over the
range, compared with the averaged constraints of table
\ref{tab:origin} which vary by an order of magnitude over the range.
Similarly calculating the observables on the brane, outlined in
section \ref{sec:small_stars}, the density and pressures, both actual
and induced, and in addition the induced Weyl component, all give the
same small variations over the range of lattice sizes, as they are
constructed from these metric functions. Thus we find that the brane
geometry, our primary interest, is not sensitive to the lattice size,
and thus the exact form of the horizon metric, although the
constraints are.

\begin{table}
\begin{center}
\begin{tabular}{llllll}
$\rho_0$ & $\xi$ & $r_{\rm max}$ & $z_{\rm max}$ & $T_0$ & $A_0$ \\ 
\hline \hline
0.1 & 0.3  & 2.0  & 1.5   & -0.007500  & 0.004190  \\
(\emph{linear}) & (\emph{small})
            &     & 2.0   & -0.007512 & 0.004154  \\
     &      &     & 3.0   & -0.007513 & 0.004137  \\ \hline
5.0  & 0.3  & 2.0 & 1.5   & -0.7030   & 0.3047  \\
(\emph{non-linear}) & (\emph{small})
            &     & 2.0   & -0.6969   & 0.2989  \\
     &      &     & 3.0   & -0.6942   & 0.2965  \\ \hline
0.01 & 1.   & 10. & 6.0   & -0.003698 & 0.002328 \\
(\emph{linear}) & (\emph{medium})
            &     & 10.   & -0.003692 & 0.002240 \\
     &      &     & 15.   & -0.003691 & 0.002237 \\
\end{tabular}
\caption[short]{ \figuremode  
  Table showing the variation of metric functions at the star core for
  the different size lattices used in table \ref{tab:origin}. We see
  that for the resolution used the core values of the metric
  functions, and indeed their profiles (not shown here), are very
  stable to changing the physical lattice size. This shows that the
  boundary conditions on the asymptotic $z$ boundary are accurately
  forcing the metric to AdS. The largest variation is found for the
  $\xi=1$ configuration, going from $z_{\rm max}$ being only $5$ to
  $10$. Although this has a large effect on the constraints which show
  a 6 fold decrease, the maximum change for the metric functions is in
  $A$ which varies by $4 \%$.  Note that the metric functions vary
  most at the core of the star as this is furthest from a boundary
  where they are fixed.
\label{tab:origin2}}
\end{center}
\end{table}

%
\subsection{Asymptotic Behavior}
%

\begin{figure}
\centerline{\psfig{file=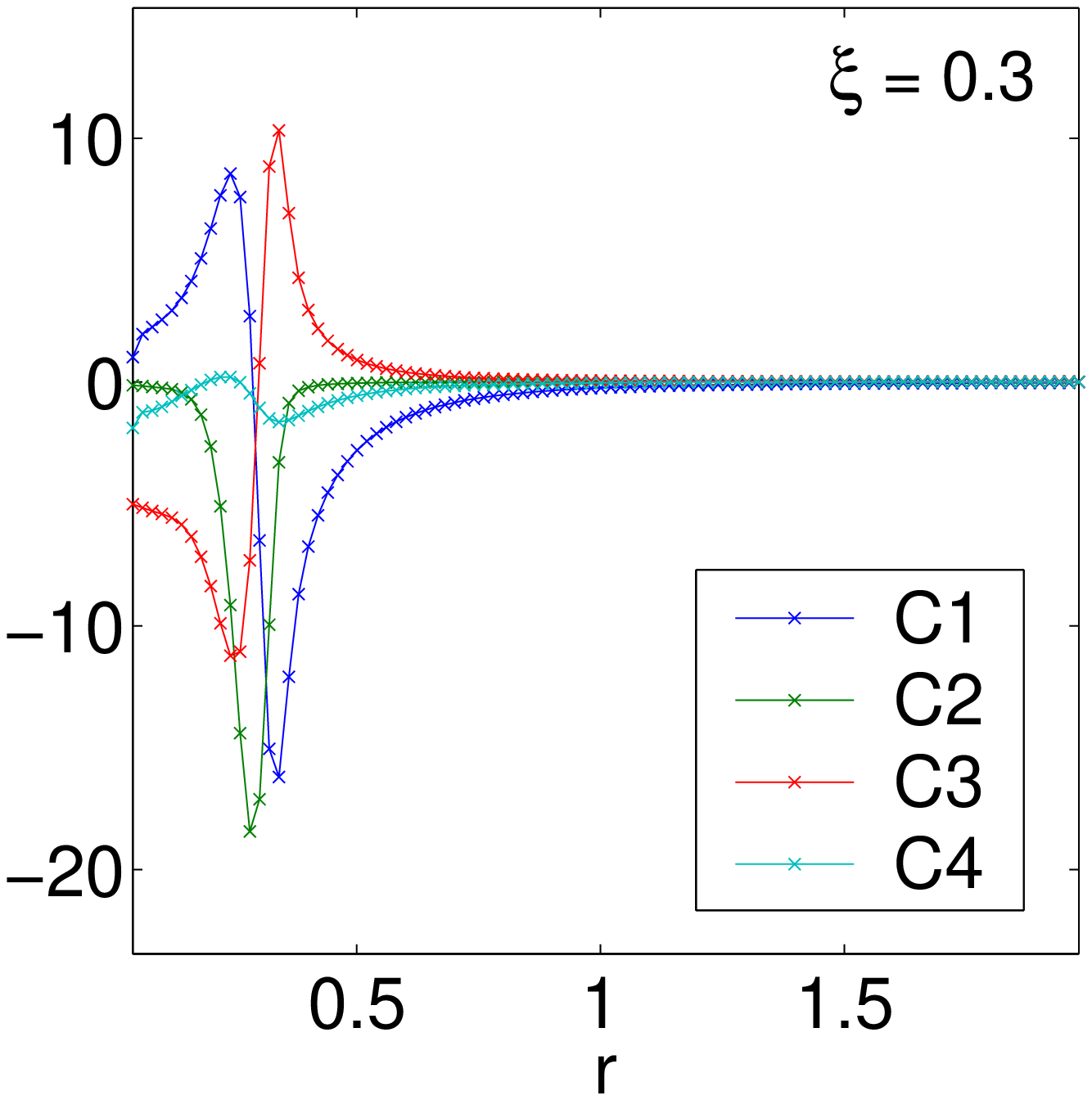,width=6.3cm} \hspace{0.1cm}
  \psfig{file=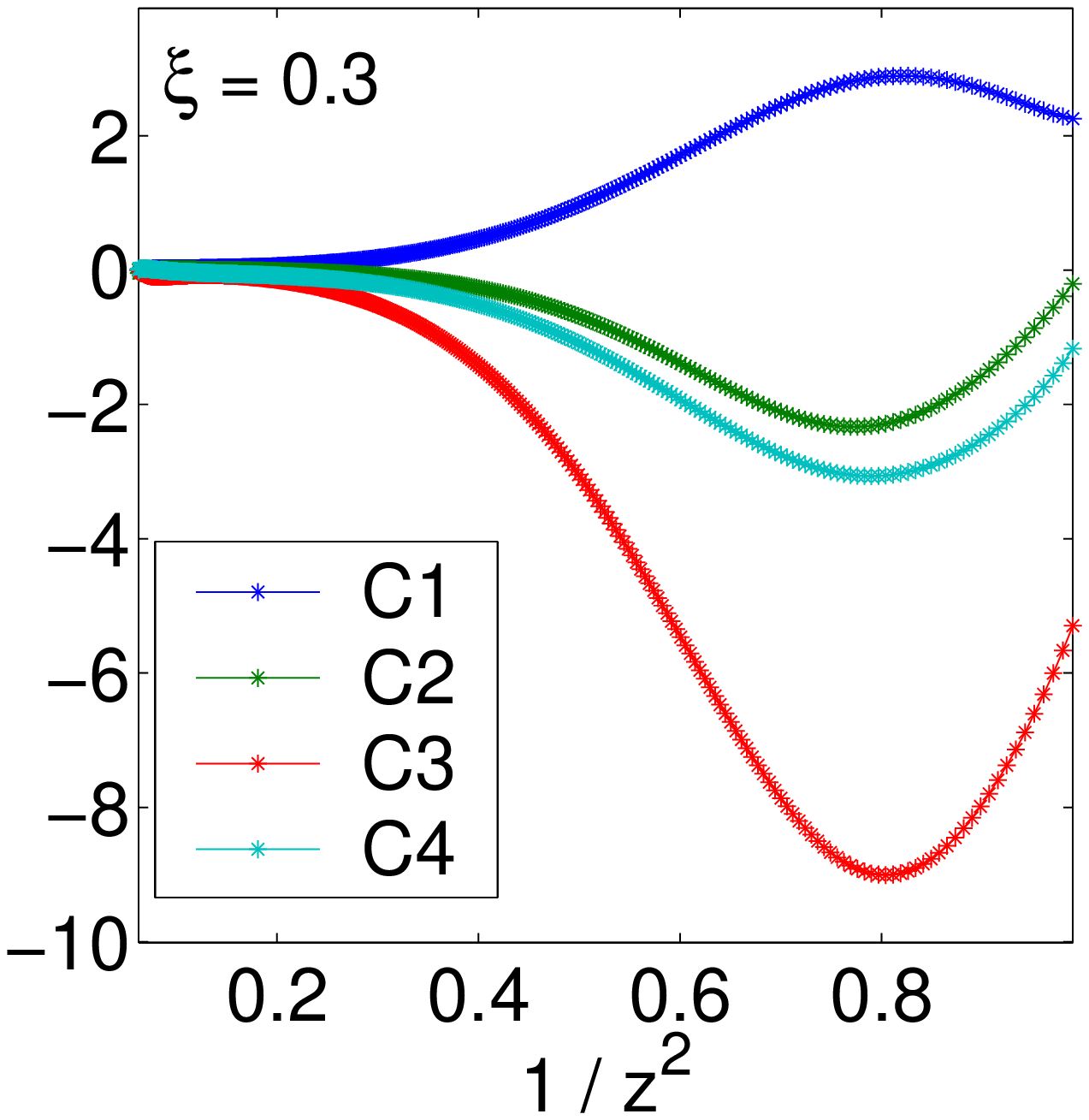,width=6.1cm}}
\centerline{\psfig{file=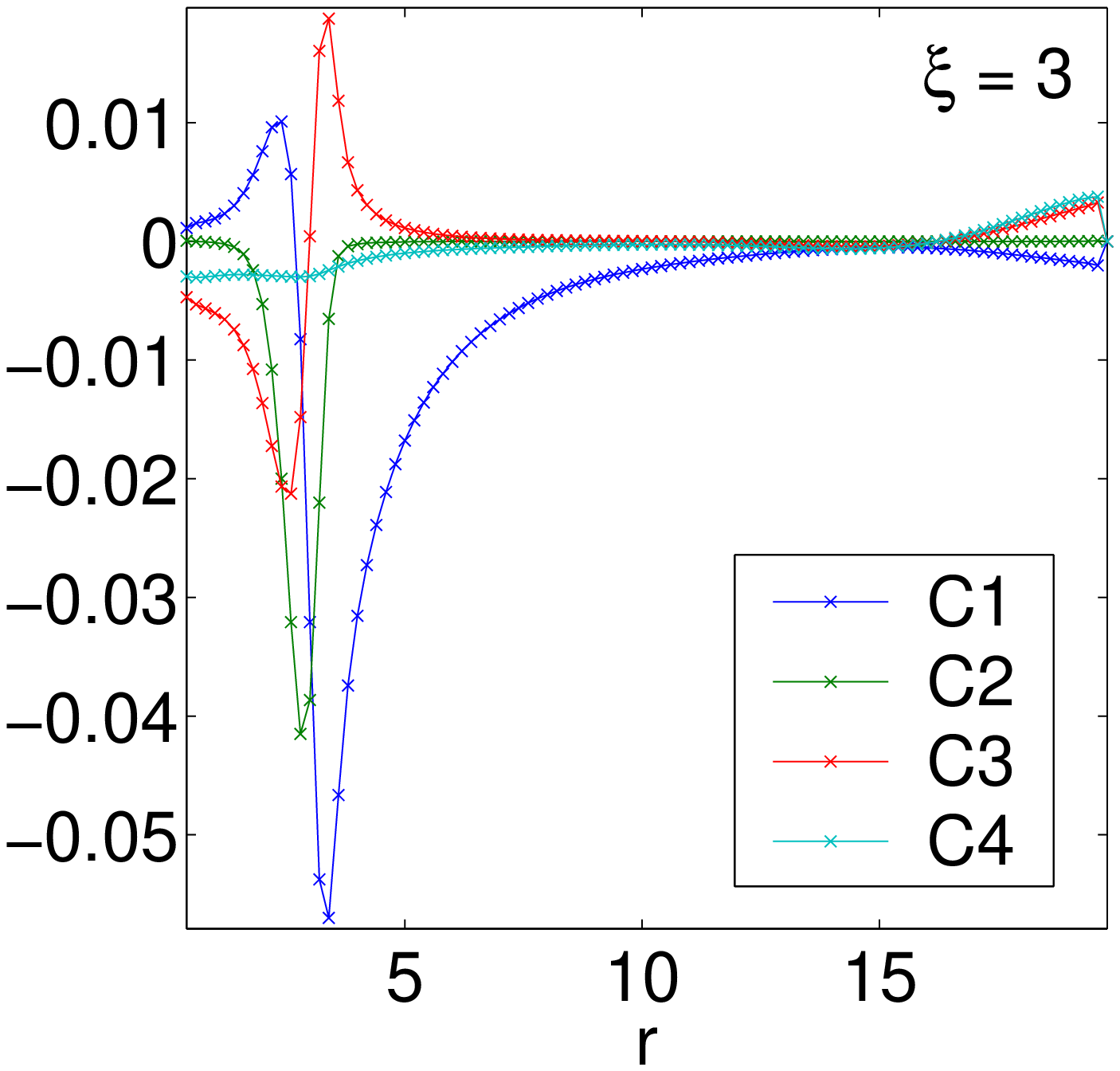,width=6.5cm} \hspace{0.1cm}
  \psfig{file=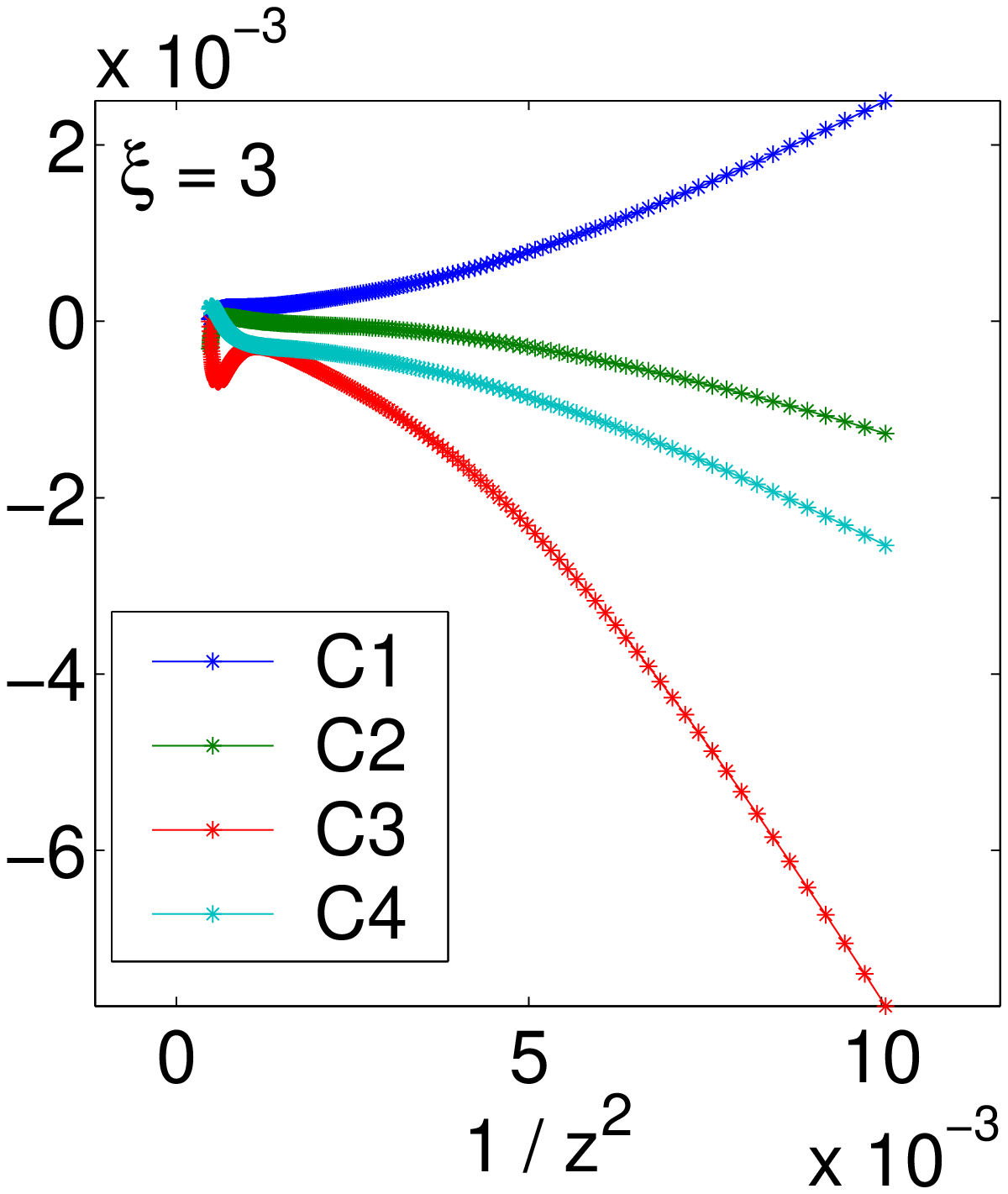,width=5.8cm}}
\caption[short]{ \figuremode  
  An illustration of $C1, C2, C3, C4$ for the densest \emph{small} and
  \emph{large} configurations used, top, $\xi = 0.3$ with source
  $\rho_0 = 7.0$ and bottom, $\xi = 3$ with $\rho_0 = 0.149$. On the
  left the plots show the components on the brane at $z=1$. Here $C1,
  C2, C3, C4$ give physical curvatures. Curvature invariants would be
  combinations of these quantities with no further $r,z$ dependence.
  These curvatures all decay to zero for large $r$. This occurs very
  cleanly for the \emph{small} star, and slightly less cleanly for the
  \emph{large} star with small curvature being induced at the edge of
  the lattice by the lattice boundaries. This is still only a maximum
  $5 \%$ error compared to the peak curvatures. We conclude that the
  boundary condition $T, A, B \rightarrow 0$ as $r \rightarrow r_{\rm
    max}$ is therefore correctly reproducing the AdS behavior. We
  expect the finite boundaries to give some error for the
  \emph{larger} stars as the correlation functions decay as
  $\frac{1}{r}$ rather than $\frac{1}{r^2}$ as for \emph{small} stars,
  and so the perturbation decays more slowly. On the right are plots
  showing how the components $C1, C2, C3, C4$ decay towards the
  horizon for a fixed value of $r$ near the axis (for $\xi = 0.3$, $r
  = 0.1$ and $\xi = 3.0$, $r = 1.0$). As discussed in the appendix
  \ref{app:einstein_eqns}, these functions have the blue-shifting
  factor $\frac{1}{z^2}$ removed so that errors appropriate to the
  numerical solution can be seen. For the true curvature invariants to
  tend to zero, $C1, C2, C3, C4$ must decay faster than
  $\frac{1}{z^2}$, and therefore the functions are plotted against
  $\frac{1}{z^2}$. For both the \emph{small} and \emph{large} star it
  is clear that the functions do decay faster, the functions being
  shallower than linear as $\frac{1}{z^2} \rightarrow 0$. This implies
  that $T,A,B \rightarrow 0$ on the large $z$ boundary is reproducing
  asymptotic AdS.  We also note that for the \emph{large} star there
  are small errors at the boundary. If we had plotted the Weyl
  components with blue-shifting $z^2$ factor included, these points at
  $z \sim 40$ would completely dominate the plot, but we see here that
  they are simply small numerical errors confirming that the rescaled
  $C1, C2, C3, C4$ are the correct measures to use.  ($\xi = 0.3$
  lattice: $dr = 0.02$, $r_{\rm max} = 2$, $dz = 0.005$, $z_{\rm max}
  = 4$; $\xi = 3$ lattice: $dr = 0.2$, $r_{\rm max} = 20$, $dz =
  0.03$, $z_{\rm max} = 46$)
\label{fig:weyl} 
}
\end{figure}

We see that the asymptotic behavior is correct by examining the Weyl
components. Whilst the metric functions may have a non-zero
perturbation, some component of this may well be gauge. The Weyl
components are non-zero only for physical curvatures. Figure
\ref{fig:weyl} firstly shows that the 4 Weyl components decay quickly
on the brane as $r$ approaches the asymptotic boundary, for both a
typical \emph{small} and \emph{large} star solution. It also shows a
magnified version of these components in the asymptotic $z$ regions of
the same solution for fixed $r$ near the axis. This confirms that the
curvature is dying away and asymptotic AdS is being approached at
large $z$, implying the boundary conditions are as required.  It is
important to note that $C1, C2, C3, C4$ are not the Weyl components
but rather have been scaled by $\frac{1}{z^2}$ to make them a better
numerical estimator of the solution accuracy. Note that the fall off
in $C1, C2, C3, C4$ for both the \emph{small} and \emph{large} stars
shown in figure \ref{fig:weyl} is faster than $\frac{1}{z^2}$ for
large $z$, which they are plotted against, implying the Weyl
components, and therefore curvature invariants, do indeed go to zero.
We see small boundary effects for the \emph{large} star which imply
that the actual Weyl components are very large at this boundary due to
the $z^2$ factor, which is $\sim 10^3$ there. The advantage of
plotting the rescaled $C1, C2, C3, C4$ against $\frac{1}{z^2}$ is that
this is clearly seen to be a numerical artifact of the boundary, and
the correct asymptotic AdS behavior is observed. Thus for both the
asymptotic $r$ and $z$ boundaries the solutions are seen to correctly
approach AdS.

%
\section{Appendix B: Technical Details}
%

%
\subsection{Einstein Equations and Matching Conditions}
\label{app:einstein_eqns}
%

In this appendix we give the Einstein tensor components, the matching
equations and the Weyl tensor components for the metric
\eqref{eq:metric_conf}. Expressing the metric in its natural one form
basis,
\begin{equation}
ds^2 = - w_{(t)}^{\mu} w_{(t) \mu} + w_{(r)}^{\mu} w_{(r) \mu} +
w_{(\Omega)_i}^{\mu} w_{(\Omega)_i \mu} + w_{(z)}^{\mu} w_{(z) \mu}
\end{equation}
where $i= \theta, \phi$, giving the one forms on an $S^2$. Firstly,
the non zero Einstein tensor components are, \vspace{5pt}
\begin{align}
- w^{\mu}_{(t)} w^{\nu}_{(t)} G_{\mu\nu} = G^{t}_{~t} = & z^2
  \,e^{-2\,( A + B )} \Big[ \, \left[ 3 \, \partial_{r}^2 A -
  \partial_{r}^2 B + 3\,{({\partial_{r} A})^2} + 3\,{({\partial_{r} B})^2} - 6\,{({\partial_{r} A})\,({\partial_{r} B})} 
   \right] + \left[ \, \partial_r
  \leftrightarrow \partial_z \, \right] 
\notag \\ & 
\notag \\ & 
 \qquad  
+ \frac{6}{r} \partial_{r} ( A - B ) - 
  \frac{1}{r^2} \left( e^{4 B} - 1 \right) 
-   \frac{6}{z} \,\partial_{z} ( A - B ) 
+ {\frac{6}{{z^2}}} \Big]
\notag \\ \notag
\end{align} 
\begin{align}
w^{\mu}_{(r)} w^{\nu}_{(z)} G_{\mu\nu} = G^{r}_{~z}  = & z^2
  \,e^{-2\,( A + B )} \Big[
- 2\,\partial_{r} \partial_{z} A + 2\,\partial_{r} \partial_{z} B -
  \partial_{r} \partial_{z} T + 
  2\,(\partial_{r} A)\,(\partial_{z} A) + 2\,(\partial_{r} A)\,(\partial_{z}
  B) 
\notag \\ & 
\notag \\ &  
 \qquad + 2\,(\partial_{r} B)\,(\partial_{z} A) - 6\,(\partial_{r} B)\,(\partial_{z} B) + (\partial_{z} A)\,(\partial_{r} T) + (\partial_{z} B)\,(\partial_{r} T) + (\partial_{r} A)\,(\partial_{z} T) 
\notag \\ &  
\notag \\ &  
 \qquad  + (\partial_{r} B)\,(\partial_{z} T) - 
  (\partial_{r} T)\,(\partial_{z} T) + \frac{4}{r} \,\partial_{z} B - 
  \frac{3}{z} \,r\,\partial_{r} ( A + B ) \Big]
\notag \\ \notag 
\end{align} 
\begin{align}
w^{\mu}_{(r)} w^{\nu}_{(r)} G_{\mu\nu} = G^{r}_{~r}  = & z^2
  \,e^{-2\,( A + B )} \Big[  + 
  \partial_{z}^2 T + 2\,\partial_{z}^2 A - 2\,\partial_{z}^2 B 
- 2\,(\partial_{r} A)\,(\partial_{r} B) - 6\,(\partial_{z} A)\,(\partial_{z} B) 
+ 3\,{({\partial_{r} A})^2} 
\notag \\ &
\notag \\ & 
 \qquad   - {({\partial_{r} B})^2} + {({\partial_{z} A})^2} + 5\,{({\partial_{z} B})^2}  + 
  3\,(\partial_{r} A)\,(\partial_{r} T) - 
  (\partial_{r} B)\,(\partial_{r} T) +
  (\partial_{z} A)\,(\partial_{z} T) 
\notag \\ & 
\notag \\ & 
 \qquad - 3\,(\partial_{z} B)\,(\partial_{z} T) + {({\partial_{z} T})^2} + \frac{1}{r} \left( 4\,\partial_{r} A + 2\,\partial_{r} T \right)
- \frac{1}{r^2} \left( e^{4 B} - 1 \right) 
\notag \\ & 
\notag \\ & 
\qquad
+ \frac{1}{z} \left( - 
  3\,\partial_{z} A + 
  9\,\partial_{z} B - 
  3\,\partial_{z} T \right)
+ {\frac{6}{{z^2}}}
\Big]
\notag \\ \notag 
\end{align} 
\begin{align} 
w^{\mu}_{(\Omega_{\theta})} w^{\nu}_{(\Omega_{\theta})} G_{\mu\nu} =
  G^{\theta}_{~\theta}  = &  z^2 \,e^{-2\,( A + B )} \Big[ 
\, \left[\, 2 \partial_{r}^2 A  + {({\partial_{r} A})^2}  +
  {({\partial_{r} B})^2} - 2 {({\partial_{r} A})\,({\partial_{r} B})} 
 \right. 
\notag \\ &
\notag \\ & 
 \qquad \left. + (\partial_{r} A)\,(\partial_{r} T) - (\partial_{r} B)\,(\partial_{r} T) + 
  {({\partial_{r} T})^2} + \partial_{r}^2 T \right] + \left[ \, \partial_r
  \leftrightarrow \partial_z \, \right] 
\notag \\ &
\notag \\ & 
 \qquad + \frac{1}{r} \left( + 2\,\partial_{r} A - 2\,\partial_{r} B + 
  \partial_{r} T \right)
+ \frac{1}{z} \left( - 
  3\,\partial_{z} A + 3\,\partial_{z} B - 3\,\partial_{z} T \right) 
+ {\frac{6}{{z^2}}} \Big]
\notag \\ \notag 
\end{align} 
\begin{align}
w^{\mu}_{(z)} w^{\nu}_{(z)} G_{\mu\nu} = G^{z}_{~z}  = & z^2
  \,e^{-2\,( A + B )} \Big[ 
2\,\partial_{r}^2 A - 2\,\partial_{r}^2 B + \partial_{r}^2 T
- 6\,(\partial_{r} A)\,(\partial_{r} B) - 2\,(\partial_{z} A)\,(\partial_{z} B)  
+ {({\partial_{r} A})^2} 
\notag \\ &
\notag \\ &
 \qquad + 5\,{({\partial_{r} B})^2} + 3\,{({\partial_{z} A})^2} - {({\partial_{z} B})^2} + (\partial_{r} A)\,(\partial_{r} T) - 
  3\,(\partial_{r} B)\,(\partial_{r} T)   
\notag \\ &
\notag \\ &
 \qquad + 3\,(\partial_{z}
  A)\,(\partial_{z} T) - (\partial_{z} B)\,(\partial_{z} T)  + {({\partial_{r} T})^2} + \frac{1}{r} \left( 4\,\partial_{r} A - 8\,\partial_{r}
  B + 2\,\partial_{r} T \right)
\notag \\ &
\notag \\ &
 \qquad
- \frac{1}{r^2} \left( e^{4 B} - 1 \right) + \frac{1}{z} \left( - 
  9\,\partial_{z} A + 
  3\,\partial_{z} B - 
  3\,\partial_{z} T \right) 
+ {\frac{6}{{z^2}}}
\Big]
\label{eq:einstein_eqns}
\end{align}
where in our units the Einstein equations are $G^{t}_{~t} = G^{r}_{~r}
+ G^{z}_{~z} = G^{\theta}_{~\theta} = 6$ together with the constraint
equations $G^{r}_{~z} = 0$ and $G^{r}_{~r} - G^{z}_{~z} = 0$. The
matching equation \eqref{eq:isreal} has non trivial components,
\vspace{5pt}
\begin{align}
-6 + 6\,z\,\partial_{z} A - 2\,z\,\partial_{z} B + 
  \left( 6 + \rho \right) \, e^{A + B} & = 0 
\notag \\
\notag \\
-6 + 4\,z\,\partial_{z} A - 4\,z\,\partial_{z} B + 2\,z\,\partial_{z} T + 
  \left( 6 - P_r \right) \, e^{A + B} & = 0
\notag \\
\notag \\
-6 + 4\,z\,\partial_{z} A + 2\,z\,\partial_{z} T + \left( 6 - P_{\theta}
  \right) \, e^{A + B} & = 0
\label{eq:matching_eqns}
\end{align}
with $P_r, P_{\theta}$ being the radial and angular pressure components. Isotropy on the brane requires that these be equal and the last two component equations therefore imply that $B_{,z} = 0$.

Together with the Einstein tensor, the Weyl tensor specifies the
geometry.  Given the symmetries of the metric, and of the Weyl tensor
itself, it has 4 independent components, which can be taken as the
following, \vspace{5pt}
\begin{align}
z^2 C1 & = w_{(t)}^{\mu} w_{(r)}^{\nu} w_{(t)}^{\alpha} w_{(r)}^{\beta} C_{\mu
  \nu \alpha \beta} = - C_{t r}^{~~~t r} 
\notag \\ &
\notag \\
& = z^2 e^{- 2 (A+B)} \Big[ - \frac{1}{2} \partial_{r} \partial_{r}
  A + \frac{1}{6} \partial_{r}^2 B  + \frac{1}{2} \partial_{r}^2 T + \frac{1}{2} {({\partial_{r} A})^2} + \frac{1}{3} (\partial_{r} A)\,(\partial_{r} B) - 
  \frac{5}{6} \,{({\partial_{r} B})^2}   
\notag \\ &
\notag \\ & \qquad \qquad
- (\partial_{r} A)\,(\partial_{r} T) - \frac{1}{3} (\partial_{r} B)\,(\partial_{r} T)  + 
  \frac{1}{2} {({\partial_{r} T})^2} + \frac{1}{6} \partial_{z}^2 A - \frac{1}{2} \partial_{z}^2 B
\notag \\ &
\notag \\ & \qquad \qquad
  - \frac{1}{6} \partial_{z}^2 T - \frac{1}{6} {({\partial_{z} A})^2}
  - (\partial_{z} A)\,(\partial_{z} B) 
+ \frac{7}{6} \,{({\partial_{z} B})^2} + \frac{1}{3} (\partial_{z} A)\,(\partial_{z} T) 
+ (\partial_{z} B)\,(\partial_{z} T) 
\notag \\ &
\notag \\ & \qquad \qquad
- \frac{1}{6} {({\partial_{z} T})^2} + \frac{1}{r} \left( + {\frac{\partial_{r} A}{3\,r}} + 
  {\frac{\partial_{r} B}{r}} - {\frac{\partial_{r} T}{3\,r}} \right)
- \frac{1}{6 r^2} \left( e^{4\,B} - 1 \right) \Big]
\notag \\ 
\notag 
\end{align}
\begin{align}
z^2 C2 & = w_{(t)}^{\mu} w_{(r)}^{\nu} w_{(t)}^{\alpha} w_{(z)}^{\beta} C_{\mu
  \nu \alpha \beta} = - C_{t r}^{~~~t z} 
\notag \\ &
\notag \\
& = z^2 e^{- 2 (A+B)} \Big[ -\frac{2}{3} \,\partial_{r} \partial_{z} A +
  \frac{2}{3} \,\partial_{r} A\,\partial_{z} A + 
  \frac{2}{3} \,\partial_{z} A\,\partial_{r} B + \frac{2}{3} \,\partial_{r} \partial_{z} B + 
  \frac{2}{3} \,\partial_{r} A\,\partial_{z} B - 2\,\partial_{r}
  B\,\partial_{z} B 
\notag \\ &
\notag \\ & \qquad \qquad
 - \frac{2}{3} \,(\partial_{z} A)\,(\partial_{r} T) - 
  \frac{2}{3} \,(\partial_{z} B)\,(\partial_{r} T) + \frac{2}{3}
  \,\partial_{r} \partial_{z} T - 
  \frac{2}{3}\,(\partial_{r} A)\,(\partial_{z} T) 
\notag \\ &
\notag \\ & \qquad \qquad
- \frac{2}{3}
  \,(\partial_{r} B)\,(\partial_{z} T) + 
  \frac{2}{3}\,(\partial_{r} T)\,(\partial_{z} T) +
  \frac{4}{3\,r}\,\partial_{z} B \Big]
\notag \\ 
\notag 
\end{align}
\begin{align}
z^2 C3 & = w_{(t)}^{\mu} w_{(z)}^{\nu} w_{(t)}^{\alpha} w_{(z)}^{\beta} C_{\mu
  \nu \alpha \beta} = - C_{t z}^{~~~t z} 
\notag \\ &
\notag \\
& = z^2 e^{- 2 (A+B)} \Big[ + \frac{1}{6} \partial_{r}^2 A -
  \frac{1}{2} \partial_{z}^2 A - \frac{1}{2} \partial_{r}^2 B + 
  \frac{1}{6} \partial_{z}^2 B - \frac{1}{6} \partial_{r}^2 T + 
  \frac{1}{2} \partial_{z}^2 T  
- \frac{1}{6} {({\partial_{r} A})^2} - (\partial_{r} A)\,(\partial_{r} B) 
\notag \\ &
\notag \\ & \qquad \qquad 
+ \frac{7}{6} \,{({\partial_{r} B})^2}  
+ \frac{1}{3} (\partial_{r} A)\,(\partial_{r} T) + (\partial_{r} B)\,(\partial_{r} T)  - 
  \frac{1}{6}{({\partial_{r} T})^2}  
+ \frac{1}{2} {({\partial_{z} A})^2} 
\notag \\ &
\notag \\ & \qquad \qquad
+ \frac{1}{3} (\partial_{z} A)\,(\partial_{z} B) - 
  \frac{5}{6} \,{({\partial_{z} B})^2} - (\partial_{z} A)\,(\partial_{z} T)  
- \frac{1}{3} (\partial_{z} B)\,(\partial_{z} T) +
  \frac{1}{2} {({\partial_{z} T})^2} 
\notag \\ &
\notag \\ & \qquad \qquad
+ \frac{1}{r} \left( + \frac{1}{3} \partial_{r} A - \frac{5}{3}
  \,\partial_{r} B - \frac{1}{3} \partial_{r} T \right) 
- \frac{1}{6\,r^2} \left( e^{4\,B} - 1 \right) \Big]
\notag \\ 
\notag 
\end{align}
\begin{align}
z^2 C4 & = w_{(r)}^{\mu} w_{(z)}^{\nu} w_{(r)}^{\alpha} w_{(z)}^{\beta} C_{\mu
  \nu \alpha \beta} = C_{r z}^{~~~r z} 
\notag \\ &
\notag \\
& = z^2 e^{- 2 (A+B)} \Big[ -\frac{1}{6} \partial_{r}^2 A - \frac{1}{6}
  \partial_{z}^2 A - \frac{5}{6} \,\partial_{r}^2 B - 
  \frac{5}{6}\,\partial_{z}^2 B + \frac{1}{6} \partial_{r}^2 T + 
  \frac{1}{6} \partial_{z}^2 T  
+ \frac{1}{6} {({\partial_{r} A})^2} - \frac{1}{3} (\partial_{r} A)\,(\partial_{r} B) 
\notag \\ &
\notag \\ & \qquad \qquad
+ \frac{1}{6}{({\partial_{r} B})^2} - \frac{1}{3} (\partial_{r}
  A)\,(\partial_{r} T) + \frac{1}{3} (\partial_{r} B)\,(\partial_{r} T)  + 
  \frac{1}{6} {({\partial_{r} T})^2} 
+ \frac{1}{6} {({\partial_{z} A})^2} 
\notag \\ &
\notag \\ & \qquad \qquad
- \frac{1}{3} (\partial_{z}
  A)\,(\partial_{z} B) + \frac{1}{6} {({\partial_{z} B})^2} - \frac{1}{3}
  (\partial_{z} A)\,(\partial_{z} T) 
+ \frac{1}{3} (\partial_{z} B)\,(\partial_{z} T) + \frac{1}{6} {({\partial_{z}
  T})^2} 
\notag \\ &
\notag \\ & \qquad \qquad
+ \frac{1}{r} \left( + \frac{1}{3} \partial_{r} A - \frac{1}{3}
  \partial_{r} B - \frac{1}{3} \partial_{r} T \right)
+ \frac{1}{6\,r^2} \left( e^{4\,B} - 1 \right) \Big]
\label{eq:weyl_cmpts}
\end{align}
Note that in both the Einstein and Weyl tensor components above, the
indices are arranged as they appear in curvature invariants, an
overall blue-shifting factor of $z^2$ appearing.  This blue-shifting
factor will amplify numerical errors on the lattice. We therefore
define the rescaled Weyl tensor quantities $C1, C2, C3, C4$ above, and
Einstein equations $\{AB\}$,
\begin{equation}
\{AB\} = \frac{1}{z^2} w_{(A)}^{\mu} w_{(B)}^{\nu} ( G_{\mu\nu} - 6 g_{\mu\nu} )
\end{equation}
which have the factor removed. The elliptic equations and constraints,
solved by relaxation and integration, have no such factor as it simply
multiplies the whole Einstein tensor components homogeneously. Thus we
shall use these rescaled quantities to assess the numerical errors,
the terms now having no $z$ prefactor, as for the equations that are
numerically solved. Using the original scaling would blow up tiny
errors at large $z$, and as we expect there to be such errors from
finite boundary effects this is less than satisfactory. For
\emph{small} stars this is no problem, but we find for the
\emph{large} stars that there do exist small boundary errors, which
are hugely amplified when multiplied by $z^2$ which may be $\sim
10^{3}$.  We check that the metric solutions are insensitive to these
boundary conditions (appendix \ref{app:testing}), and finding that
they are, must conclude that the geometric quantities are not a good
measure of solution quality as tiny changes in the metric at large $z$
give huge changes in the Weyl components due to the blue-shifting. The
quantities that reside on the brane at $z=1$ are totally unaffected.

We conclude the section by giving the 3 independent components of the
4-dimensional Einstein tensor, and the 1 independent component of the
4-dimensional Weyl tensor, for the induced metric
\eqref{eq:metric_induced}. These are,
\begin{align}
\rho^{(4)} = G^{(4)t}_{~~~~t} & = e^{-2 (A+B)} \Big[ 2 \partial^2_r B - 2 \partial^2_r A - 5 (\partial_r B)^2 - (\partial_r A)^2 + 6 \partial_r A
\partial_r B 
\notag \\ &
\notag \\ &
\qquad \qquad + \frac{1}{r} \left(8 \partial_r B - 4 \partial_r
    A \right) + \frac{1}{r^2} \left(e^{4 B}-1\right) \Big]
\notag \\
\notag \\
P_r^{(4)} = G^{(4)r}_{~~~~r} & = e^{-2 (A+B)} \Big[ - 2 (\partial_r T) (\partial_r B) - 2 (\partial_r A) (\partial_r B) + 2 (\partial_r A) (\partial_r T) + (\partial_r A)^2
+ (\partial_r B)^2 
\notag \\ &
\notag \\ &
\qquad \qquad + \frac{2}{r} \left( \partial_r T + \partial_r A - \partial_r B \right) - \frac{1}{r^2} \left( e^{4 B} - 1 \right) \Big]
\notag
\end{align}
\begin{align}
P_{\theta}^{(4)} = G^{(4)\theta}_{~~~~\theta} & = e^{-2 (A+B)} \Big[ \partial^2_r T + \partial^2_r A - \partial^2_r B + 2
(\partial_r B)^2 + (\partial_r T)^2 - 2 (\partial_r T) (\partial_r B)
- 2 (\partial_r A) (\partial_r B) 
\notag \\ &
\notag \\ &
\qquad \qquad + \frac{1}{r} \left( \partial_r A + \partial_r T - 3 \partial_r B \right) \Big]
\end{align}
and the Weyl component,
\begin{align}
C^{(4)} = C_{t r}^{~~~t r} & = - \frac{1}{3} e^{-2 (A+B)} \Big[ + \partial^2_r T - \partial^2_r A + \partial^2_r B - 2 (\partial_r A) (\partial_r T) + 
   (\partial_r T)^2 + (\partial_r A)^2 - (\partial_r B)^2 
\notag \\ &
\notag \\ &
\qquad \qquad + \frac{1}{r} \left( \partial_r B + \partial_r A - \partial_r T \right) + \frac{1}{r^2}
\left( 1 - e^{4 B} \right) \Big]
\end{align}

%
\subsection{Asymptotic Linear Behavior}
\label{app:asym_lin}
%

%
%

One can solve the linear theory for a point source. The $a$ equation
\eqref{eq:lin_laplace} solution is constructed from the modes
satisfying the radiation boundary conditions, discussed in section
\ref{sec:RS_gravity}, as,
\begin{equation}
a(r,z) = \sqrt{\frac{2}{\pi}} \int_0^\infty dp \, a(p) \, (p
r)^{-\frac{3}{2}} \, J_{\frac{3}{2}}(p r) \, (p z)^2 \, K_2(p z)
\end{equation}
for some coefficients $a(p)$. Let us consider $a(p) = a_0$. Then,
\begin{equation}
a(r,z) = \frac{a_0}{\sqrt{r^2 + z^2}}
\end{equation}
and we can solve \eqref{eq:lin_metric_transform} and
\eqref{eq:lin_metric_transform2}, and find,
\begin{align}
  g(r,z) & = a_0 \frac{z}{\sqrt{r^2 + z^2}} + \int_0^\infty dp F(p)
  \cos{p r} \, e^{-p z}
  \notag \\
  f(r,z) & = - a_0 \frac{r}{\sqrt{r^2 + z^2}} - \int_0^\infty dp F(p)
  \sin{p r} \, e^{-p z}
\end{align}
and imposing the matching condition, $\partial_z a = - \frac{2}{r}
\partial_r g$ at $z = 1$, from \eqref{eq:lin_matching}, we obtain
\begin{equation}
F(p) = - \frac{3 a_0}{\pi} K_0(p) e^{p}
\end{equation}
On the brane we calculate,
\begin{align}
g & = - \frac{a_0}{2 \sqrt{1 + r^2}} \sim - \frac{a_0}{2} \frac{1}{r}
    + O( \frac{a_0}{r^3} )
\notag \\
f & =  - \frac{a_0}{\sqrt{1 + r^2}} ( r - \frac{3}{\pi} \sinh^{-1}{r} )
    \sim - a_0 + \frac{3 a_0}{\pi} \frac{\log{2 r}}{r} + O(
    \frac{a_0}{r^2} )
\end{align}
Then, $g \sim - a_0 / 2 r + O(1/r^3)$ which to leading order in large
$r >> 1$ (ie. large compared to the AdS length) is the inverse $r$
response to a point density source on the brane at $r = 0$ from
equation \eqref{eq:lin_matching}. From \eqref{eq:lin_metric_cmpts}, we
see that at large $r$ on the brane, $T, A, B \rightarrow 0$.

At large $z >> 1$ we receive contributions from modes with $k << 1$
and therefore expand $K_0(k)$, in large $z$ and large $r$ such that $r
<< z$, to yield,
\begin{align}
a & = \frac{a_0}{z} ( 1 + O(\frac{r^2}{z^2}) )
\notag \\
f & = a_0 \frac{r}{z} ( - 1 + \frac{3}{\pi} \frac{1}{z} ( \log(2) - 1
+ \log(z) ) + O( |\frac{\log{z}}{z^2}| + \frac{r^2}{z^2}) )
\notag \\ 
g & = a_0 ( 1 - \frac{3}{\pi} \frac{1}{z} ( \log{2} + \log{z} ) + O(
| \frac{\log{z}}{z^2} | + \frac{r^2}{z^2} ) )
\label{eq:lin_afg}
\end{align}
implying, from \eqref{eq:lin_metric_cmpts}, that,
\begin{align}
T & = a_0 \frac{1}{z} ( - 4  + O( |\frac{\log{z}}{z}| +
\frac{r^2}{z^2} ) )
\notag \\
A & = a_0 \frac{1}{z} ( - 1  + O( |\frac{\log{z}}{z}| +
\frac{r^2}{z^2} ) )
\notag \\
B & = a_0 \frac{1}{z} \frac{r^2}{z^2} ( \frac{3}{4} +
O(\frac{r^2}{z^2}) )
\label{eq:lin_TAB}
\end{align}
and therefore indeed $T, A, B \sim 1 / z \rightarrow 0$ as $z
\rightarrow \infty$. Note that this is a good approximation for $r >>
1$ with $r < z$, and represents the asymptotic form for a point
density source.

Examining the Weyl components of the previous appendix section,
\ref{app:einstein_eqns}, we also see that these in fact decay as $\sim
1 / z$ for large $z$. The way the functions $T, A$ enter is always
with derivatives taken. Derivatives involving $r$ remove the leading
$1 / z, \log{z} / z$ terms, and the resulting $1 / z^3$ behavior
cancels the overall $z^2$ factor in these Weyl components, with
indices arranged as they occur in curvature invariants, to yield $1 /
z$ asymptotically. Derivatives in $z$ simply reduce the leading $1 /
z$ to higher inverse powers which again combine with the overall $z^2$
to give $1 / z$ asymptotically. There are more subtle terms in $B$,
such as those involving no derivatives, $z^2 (e^{4 B} - 1) / r^2$ or
only one $z$ derivative, $z^2 \partial_z B / r$. However, we note that
$B$ in fact scales as $1 / z^3$ unlike $T, A$ and thus these terms
have the decaying behavior too. Thus, in the linear theory,
asymptotically all the Weyl components do indeed decay and the horizon
geometry is regular, and that of AdS.

%
\subsection{Numerical Details: Finite Differencing, Relaxation and
  Integration Schemes}
\label{app:numerical}
%

The lattice used to cover the $rz$ plane is rectangular, with even
spacing $dr, dz$ in the $r, z$ directions.  We discretize the fields
over the lattice as $X_{i,j}$ where $i, j$ are the $r$ and $z$
positions respectively and run as $i=0,1,\ldots i_{\rm max}$,
$j=0,1,\ldots j_{\rm max}$. The lattice is truncated at a maximum size
in both $r$ and $z$ and the boundary conditions $T,A,B = 0$ are
applied there.

The elliptic equations are differenced using standard second order
templates, and the second derivative terms imply they are naturally
evaluated at $(i,j)$. The constraint equation $\{rz\}$ contains terms
taking the generic form,
\begin{equation}
\partial_r \partial_z X + a \, \frac{1}{r} \partial_r Y + b \, \partial_r X
\partial_z Y + \ldots = 0
\end{equation}
where $X, Y$ are some metric functions. This is then evaluated at the
center of a lattice cell $(i+\frac{1}{2},j+\frac{1}{2})$, compatible
with the second order mixed derivative operators.  This equation is
used to integrate $T$ along the brane boundary by evaluating it at
$(i+\frac{1}{2},\frac{1}{2})$ solving it for $T_{i,0}$ which allows an
integration in from $r=\infty$ iteratively if $T_{i+1,0}$ is known and
the bulk values $T_{i,1}, T_{i+1,1}$ are also known.  Similarly the
Einstein tensor components $G^{r}_{~r}$ and $G^{z}_{~z}$ reside
naturally at $z = z_{i+\frac{1}{2},j}$ and $z = z_{i,j+\frac{1}{2}}$
respectively, as indicated by their highest derivative operators, the
first having only first order $r$ derivatives and the second, only
first order $z$ derivatives.  The matching conditions are of generic
form,
\begin{equation}
z \, \partial_{z} X + \rho e^{Y} = 0 
\end{equation}
where again $X, Y$ schematically represent metric functions. The first
order derivative operators imply the brane naturally resides midway
between $j = 0, 1$ at $z = z_{i,\frac{1}{2}}$.

The elliptic equations are solved by splitting the Laplacian term from
the source term in \eqref{eq:TAB_poisson},
\begin{equation}
\laplace X = {\rm Src}_X (T,A,B)
\end{equation}
where ${\rm Src}_X$ are the source functionals and $X = T,A,B$. An
over-relaxation scheme is used to partially relax the 3 Poisson
equations holding the sources constant. The sources are then updated
with the new values of $T,A,B$ calculated from the Poisson equations.
At each step, only a few cycles of the over-relaxation are computed.
This is iterated, and at each update step the boundary conditions are
imposed.  Relaxation is continued until machine precision is reached.
More sophisticated Poisson solvers were implemented but the highly
local behavior of the over-relaxation scheme appeared to give quicker
and more reliable overall convergence.

The $\{rz\}$ constraint can be used to integrate $B$ out from the
origin $r=0$, determining $B_{i+1,j}$ in terms of
$B_{i,j},B_{i,j+1},B_{i+1,j+1}$. As discussed in section
\ref{sec:origin}, this is used to calculate some of the source terms
in ${\rm Src}_{A,B}$, the source functional for the $A$ and $B$ bulk
equations. The terms we replace are the ones that have singular
coefficients as $r \rightarrow 0$ in front of $B$, and its
derivatives.  There are no such terms in the $T$ equation. For the $A$
and $B$ equations we rewrite the sources as
\begin{align}
\laplace A & = {\rm Src}^{(2)}_A + \frac{\partial_{r} B2}{r}
\notag \\ 
\laplace B & = {\rm Src}^{(2)}_B + 
 - {\frac{3\,\partial_{r} B2}{r}} + \frac{1}{r^2} \left( 1 - e^{4\,B2} \right)
\label{eq:TAB_poisson_Bsrc} 
\end{align}
where $B2$ is the solution of $B$ calculated from integrating the
$\{rz\}$ equation. The terms ${\rm Src}^{(2)}_{A,B}$ refer to the
remainder of the source terms as in equations \eqref{eq:TAB_poisson}.

In section \ref{app:testing} the magnitude of the $B2$ source terms,
compared to the other non-singular source terms, are calculated for a
typical solution. These sources are suppressed by inverse powers of
$r$ so that they contribute locally near the axis, and not
asymptotically at large $r$. Thus although the integration for $B2$ is
non-local over the lattice, the sources it generates for the bulk
equations only have a local contribution. This appears to be local and
small enough that the `Poisson' equations can indeed be simultaneously
relaxed. As stated earlier, it was found that directly determining $B$
from the constraints did not allow the other two functions $T, A$,
which are sourced by $B$, to be relaxed.

A salient feature of the scheme is the resolution requirements. The
behavior of solutions, discussed in detail in section
\ref{sec:confinement}, is that the radial extent of the metric
perturbation, $\Delta r$ is of order the radial extent of the source.
The protrusion of the perturbation from the brane into the bulk in the
$z$ direction, $\Delta z$, is such that $\Delta z \sim \Delta r$ in
this conformal gauge. Thus for large stars, $\xi >> 1$ so $\Delta z >>
1$, one finds that $dz$, the $z$ lattice spacing must be such that $dz
<< 1$ in order to resolve the near brane behavior and get convergent
solutions. Thus for \emph{large} stars, whilst the number of points in
the $r$ direction can be kept the same as for \emph{small}, with a
correspondingly larger $r$ lattice spacing $dr$, the number in the $z$
direction must grow, and lattices become prohibitively large.  The
convergence time scales badly with the number of lattice points, and
therefore we find that we cannot generate solutions for $\xi > 3$ in a
reasonable time.  However, as we see in section \ref{sec:large_stars},
this is already \emph{large} enough to observe 4-dimensional intrinsic
behavior emerging. Variable lattice spacing in the $z$ direction was
implemented, with a larger density of points near the brane, but
little or no improvement was found so regularly spaced lattices were
used to produce the results displayed here.

In section \ref{sec:non-linear} we use the largest stars relaxed, with
$\xi = 3$, and compare results with smaller stars. Solutions are
relaxed for two values of $dz$, and quadratic extrapolation is used to
calculate the $dz = 0$ continuum value. This is necessary as, for the
\emph{largest} stars with $\xi = 3$, $dz$ is still quite coarse as
shown in the convergence test in figure \ref{fig:converge}, the
coarseness being required in order to have convergence of the most
non-linear configurations in a reasonable computer time.  Calculations
performed on the finer grids were repeated for the courser $dz$ value.
The denser stars, near their upper mass limit, were not convergent on
the coarser grid, and therefore only the lower density stars of the
series could be used to perform corrections. For higher densities the
values were linearly extrapolated. Typical corrections to the $\xi =
1.5$ configurations were $\sim 1-2 \%$, for $\xi = 2.0$ they were
$\sim 3-4 \%$ and for $\xi = 3.0$ corrections of $\sim 5-15 \%$ were
implied depending on the quantity measured.  From section
\ref{sec:linear_check}, we estimate the systematic errors in the
\emph{largest} stars, $\xi = 3.0$, to be of order $\sim 10 \%$ in the
metric function $T$ from comparison with the linear theory.  Thus the
systematic error in the extrapolation due to only having a few coarser
solutions to use, expected to be considerably less than the correction
itself, is still likely to be smaller than this estimated systematic
error.  Even for $\xi = 3.0$ stars, this total estimated systematic
error is certainly small enough for the purposes of section
\ref{sec:non-linear}.

In section \ref{sec:small_stars} one clearly sees diverging core
pressure for finite core density in \emph{small} stars. For densities
above the critical value, the scheme diverges rather than finding a
convergent solution. For \emph{large} stars the scheme fails to
produce converging solutions before the critical density is reached,
where the pressure is still finite. We believe this is an artifact of
the numerical method rather than that solutions do not exist. As
discussed in section \ref{sec:large_stars} we expect that the
diverging behavior of \emph{large} stars in Randall-Sundrum is very
similar to that in standard 4-dimensional gravity.

%
\newpage 
%

\newcommand{\href}[1]{}%
\newcommand{\dhref}[1]{}%
\newenvironment{hpabstract}{%
  \renewcommand{\baselinestretch}{0.2}
  \begin{footnotesize}%
}{\end{footnotesize}}%
\newcommand{\hpeprint}[1]{%
  \href{http://arXiv.org/abs/#1}{\texttt{#1}}}%
\newcommand{\hpspires}[1]{%
  \dhref{http://www.slac.stanford.edu/spires/find/hep/www?#1}{\ (spires)}}%

%

\begin{thebibliography}{999}
\bibitem{Arkani-Hamed:1998rs}
\textsc{N.~Arkani-Hamed, S.~Dimopoulos {\upshape and} G.~Dvali}: The hierarchy
  problem and new dimensions at a millimeter.
\newblock \textsl{Phys. Lett.} \textbf{B429} (1998) 263,
  \hpeprint{hep-ph/9803315}.

\bibitem{Antoniadis:1998ig}
\textsc{I.~Antoniadis, N.~Arkani-Hamed, S.~Dimopoulos {\upshape and} G.~Dvali}:
  New dimensions at a millimeter to a Fermi and superstrings at a TeV.
\newblock \textsl{Phys. Lett.} \textbf{B436} (1998) 257--263,
  \hpeprint{hep-ph/9804398}.

\bibitem{Randall:1999vf}
\textsc{L.~Randall {\upshape and} R.~Sundrum}: An alternative to
  compactification.
\newblock \textsl{Phys. Rev. Lett.} \textbf{83} (1999) 4690,
  \hpeprint{hep-th/9906064}.

\bibitem{Lykken:1999nb}
\textsc{J.~Lykken {\upshape and} L.~Randall}: The shape of gravity.
\newblock \textsl{JHEP} \textbf{06} (2000) 014, \hpeprint{hep-th/9908076}.

\bibitem{Bergshoeff:2000zn}
\textsc{E.~Bergshoeff, R.~Kallosh {\upshape and} A.~Van Proeyen}: Supersymmetry
  in singular spaces.
\newblock \textsl{JHEP} \textbf{10} (2000) 033, \hpeprint{hep-th/0007044}.

\bibitem{Duff:2000az}
\textsc{M.~Duff, J.~Liu {\upshape and} K.~Stelle}: A supersymmetric type IIB
  Randall-Sundrum realization (2000).
\newblock Preprint \hpeprint{hep-th/0007120}.

\bibitem{Giddings:2000mu}
\textsc{S.~Giddings, E.~Katz {\upshape and} L.~Randall}: Linearized gravity in
  brane backgrounds.
\newblock \textsl{JHEP} \textbf{03} (2000) 023, \hpeprint{hep-th/0002091}.

\bibitem{Garriga:1999yh}
\textsc{J.~Garriga {\upshape and} T.~Tanaka}: Gravity in the brane-world.
\newblock \textsl{Phys. Rev. Lett.} \textbf{84} (2000) 2778--2781,
  \hpeprint{hep-th/9911055}.

\bibitem{Kudoh:2001wb}
\textsc{H.~Kudoh {\upshape and} T.~Tanaka}: Second order perturbations in the
  Randall-Sundrum infinite brane-world model (2001).
\newblock Preprint \hpeprint{hep-th/0104049}.

\bibitem{Giannakis:2000zx}
\textsc{I.~Giannakis {\upshape and} H.~Ren}: Recovery of the Schwarzschild
  metric in theories with localized gravity beyond linear order.
\newblock \textsl{Phys. Rev.} \textbf{D63} (2001) 024001,
  \hpeprint{hep-th/0007053}.

\bibitem{Sigurdsson:2001wz}
\textsc{S.~Sigurdsson}: Experimental hints of gravity in large extra
  dimensions? (2001).
\newblock Preprint \hpeprint{astro-ph/0107169}.

\bibitem{Wald}
\textsc{R.~Wald}: General Relativity.
\newblock University of Chicago Press.

\bibitem{Hartle:1978}
\textsc{J.~Hartle}: Bounds on the Mass and Moment of Inertia of Non-Rotating
  Neutron Stars.
\newblock \textsl{Phys. Rept.} \textbf{46C} (1978) 201--247.

\bibitem{Israel:1966rt}
\textsc{W.~Israel}: Singular hypersurfaces and thin shells in general
  relativity.
\newblock \textsl{Nuovo Cim.} \textbf{B44S10} (1966) 1.

\bibitem{Brecher:1999xf}
\textsc{D.~Brecher {\upshape and} M.~Perry}: Ricci-flat branes.
\newblock \textsl{Nucl. Phys.} \textbf{B566} (2000) 151,
  \hpeprint{hep-th/9908018}.

\bibitem{Chamblin:1999by}
\textsc{A.~Chamblin, S.~Hawking {\upshape and} H.~Reall}: Brane-world black
  holes.
\newblock \textsl{Phys. Rev.} \textbf{D61} (2000) 065007,
  \hpeprint{hep-th/9909205}.

\bibitem{Gregory:1993vy}
\textsc{R.~Gregory {\upshape and} R.~Laflamme}: Black strings and p-branes are
  unstable.
\newblock \textsl{Phys. Rev. Lett.} \textbf{70} (1993) 2837,
  \hpeprint{hep-th/9301052}.

\bibitem{Gregory:2000gf}
\textsc{R.~Gregory}: Black string instabilities in anti-de Sitter space.
\newblock \textsl{Class. Quant. Grav.} \textbf{17} (2000) L125,
  \hpeprint{hep-th/0004101}.

\bibitem{Charmousis:1999rg}
\textsc{C.~Charmousis, R.~Gregory {\upshape and} V.~Rubakov A.}: Wave function
  of the radion in a brane world.
\newblock \textsl{Phys. Rev.} \textbf{D62} (2000) 067505,
  \hpeprint{hep-th/9912160}.

\bibitem{Chamblin:1999cj}
\textsc{A.~Chamblin {\upshape and} G.~Gibbons}: Nonlinear supergravity on a
  brane without compactification.
\newblock \textsl{Phys. Rev. Lett.} \textbf{84} (2000) 1090,
  \hpeprint{hep-th/9909130}.

\bibitem{Arkani-Hamed:1998nn}
\textsc{N.~Arkani-Hamed, S.~Dimopoulos {\upshape and} G.~Dvali}: Phenomenology,
  astrophysics and cosmology of theories with sub-millimeter dimensions and TeV
  scale quantum gravity.
\newblock \textsl{Phys. Rev.} \textbf{D59} (1999) 086004,
  \hpeprint{hep-ph/9807344}.

\bibitem{Hoyle:2000cv}
\textsc{C.~Hoyle et~al}: Sub-millimeter tests of the gravitational
  inverse-square law: A search for 'large' extra dimensions.
\newblock \textsl{Phys. Rev. Lett.} \textbf{86} (2001) 1418--1421,
  \hpeprint{arXiv:hep-ph/0011014}.

\bibitem{Argyres:1998qn}
\textsc{P.~Argyres, S.~Dimopoulos {\upshape and} J.~March-Russell}: Black holes
  and sub-millimeter dimensions.
\newblock \textsl{Phys. Lett.} \textbf{B441} (1998) 96,
  \hpeprint{hep-th/9808138}.

\bibitem{Shiromizu:1999wj}
\textsc{T.~Shiromizu, K.~Maeda {\upshape and} M.~Sasaki}: The Einstein
  equations on the 3-brane world.
\newblock \textsl{Phys. Rev.} \textbf{D62} (2000) 024012,
  \hpeprint{gr-qc/9910076}.

\bibitem{Sasaki:1999mi}
\textsc{M.~Sasaki, T.~Shiromizu {\upshape and} K.~Maeda}: Gravity, stability
  and energy conservation on the Randall- Sundrum brane-world.
\newblock \textsl{Phys. Rev.} \textbf{D62} (2000) 024008,
  \hpeprint{hep-th/9912233}.

\bibitem{Emparan:1999wa}
\textsc{R.~Emparan, G.~Horowitz {\upshape and} R.~Myers}: Exact description of
  black holes on branes.
\newblock \textsl{JHEP} \textbf{01} (2000) 007, \hpeprint{hep-th/9911043}.

\bibitem{Giannakis:2000ss}
\textsc{I.~Giannakis {\upshape and} H.~Ren}: Possible extensions of the 4-D
  Schwarzschild horizon in the brane world.
\newblock \textsl{Phys. Rev.} \textbf{D63} (2001) 125017,
  \hpeprint{hep-th/0010183}.

\bibitem{Shiromizu:2001jm}
\textsc{T.~Shiromizu {\upshape and} D.~Ida}: Anti-de Sitter no hair, AdS/CFT
  and the brane-world.
\newblock \textsl{Phys. Rev.} \textbf{D64} (2001) 044015,
  \hpeprint{hep-th/0102035}.

\bibitem{Kraus:1999it}
\textsc{P.~Kraus}: Dynamics of anti-de Sitter domain walls.
\newblock \textsl{JHEP} \textbf{12} (1999) 011, \hpeprint{hep-th/9910149}.

\bibitem{Binetruy:1999hy}
\textsc{P.~Binetruy, C.~Deffayet, U.~Ellwanger {\upshape and} D.~Langlois}:
  Brane cosmological evolution in a bulk with cosmological constant.
\newblock \textsl{Phys. Lett.} \textbf{B477} (2000) 285,
  \hpeprint{hep-th/9910219}.

\bibitem{Binetruy:1999ut}
\textsc{P.~Binetruy, C.~Deffayet {\upshape and} D.~Langlois}: Non-conventional
  cosmology from a brane-universe.
\newblock \textsl{Nucl. Phys.} \textbf{B565} (2000) 269,
  \hpeprint{hep-th/9905012}.

\bibitem{Chamblin:1999ya}
\textsc{A.~Chamblin {\upshape and} H.~Reall}: Dynamic dilatonic domain walls.
\newblock \textsl{Nucl. Phys.} \textbf{B562} (1999) 133,
  \hpeprint{hep-th/9903225}.

\bibitem{Bowcock:2000cq}
\textsc{P.~Bowcock, C.~Charmousis {\upshape and} R.~Gregory}: General brane
  cosmologies and their global spacetime structure.
\newblock \textsl{Class. Quant. Grav.} \textbf{17} (2000) 4745--4764,
  \hpeprint{hep-th/0007177}.

\bibitem{Gregory:2001xu}
\textsc{R.~Gregory {\upshape and} A.~Padilla}: Nested braneworlds and strong
  brane gravity (2001).
\newblock Preprint \hpeprint{hep-th/0104262}.

\bibitem{Gregory:2001dn}
\textsc{R.~Gregory {\upshape and} A.~Padilla}: Braneworld instantons (2001).
\newblock Preprint \hpeprint{hep-th/0107108}.

\bibitem{Verlinde:1999fy}
\textsc{H.~Verlinde}: Holography and compactification.
\newblock \textsl{Nucl. Phys.} \textbf{B580} (2000) 264--274,
  \hpeprint{hep-th/9906182}.

\bibitem{Gubser:1999vj}
\textsc{S.~Gubser}: AdS/CFT and gravity.
\newblock \textsl{Phys. Rev.} \textbf{D63} (2001) 084017,
  \hpeprint{hep-th/9912001}.

\bibitem{deBoer:1999xf}
\textsc{J.~de~Boer, E.~Verlinde {\upshape and} H.~Verlinde}: On the holographic
  renormalization group.
\newblock \textsl{JHEP} \textbf{08} (2000) 003, \hpeprint{hep-th/9912012}.

\bibitem{Verlinde:1999xm}
\textsc{E.~Verlinde {\upshape and} H.~Verlinde}: RG-flow, gravity and the
  cosmological constant.
\newblock \textsl{JHEP} \textbf{05} (2000) 034, \hpeprint{hep-th/9912018}.

\bibitem{Shiromizu:2001ve}
\textsc{T.~Shiromizu, T.~Torii {\upshape and} D.~Ida}: Brane-world and
  holography (2001).
\newblock Preprint \hpeprint{hep-th/0105256}.

\bibitem{Shiromizu:2000pg}
\textsc{T.~Shiromizu {\upshape and} M.~Shibata}: Black holes in the brane
  world: Time symmetric initial data.
\newblock \textsl{Phys. Rev.} \textbf{D62} (2000) 127502,
  \hpeprint{hep-th/0007203}.

\bibitem{Chamblin:2000ra}
\textsc{A.~Chamblin, H.~Reall, H.~Shinkai {\upshape and} T.~Shiromizu}: Charged
  brane-world black holes.
\newblock \textsl{Phys. Rev.} \textbf{D63} (2001) 064015,
  \hpeprint{hep-th/0008177}.

\bibitem{Germani:2001du}
\textsc{C.~Germani {\upshape and} R.~Maartens}: Stars in the braneworld (2001).
\newblock Preprint \hpeprint{hep-th/0107011}.

\bibitem{Weyl:1917}
\textsc{H.~Weyl}: \textsl{Ann. Phys. (Leipzig)} \textbf{54} (1917) 117.

\bibitem{Myers:1987rx}
\textsc{R.~Myers}: Higher dimensional black holes in compactified space-times.
\newblock \textsl{Phys. Rev.} \textbf{D35} (1987) 455.

\end{thebibliography}
\end{document}